\documentclass[a4paper,nobind]{ociamthesis} 
\correctionstrue
\usepackage[style=numeric-comp, sorting=none, backend=bibtex, doi=false, isbn=false, url=false]{biblatex}
\newcommand*{\bibtitle}{References}

\usepackage{amssymb}
\usepackage{graphicx}% Include figure files
\usepackage{bm}% bold math
\usepackage{physics}
\usepackage{amsthm}
\usepackage{amsmath}
\usepackage{tabularx}
\usepackage{hhline}
\usepackage{graphicx}
\usepackage{footmisc}
\usepackage{setspace}
\usepackage{mathtools}
\usepackage{enumitem}
\usepackage{afterpage}
\usepackage[FIGTOPCAP, normalsize]{subfigure}
\usepackage{tikz}
\usepackage{cleveref}
\usepackage[compat=1.1.0]{tikz-feynman}
\usepackage{feynmf}

\usepackage{slashed}
\usepackage{float} % for forced here [H] in figures
\usepackage{fancyvrb} % for mycode Verbatime environment
\usepackage{bbding,pifont} % for my slides link symbol \HandRight

\newcommand{\MeV}{\;\mathrm{MeV}}
\newcommand{\GeV}{\;\mathrm{GeV}}

\newcommand{\ep}{\epsilon}
\newcommand{\oT}{\overline{T}}
\newcommand{\oF}{\overline{\mathcal{F}}}
\newcommand{\lambdavec}{{\vec{\lambda}}}

\newcommand{\T}{\mathcal{T}}
\newcommand{\I}{\mathcal{I}}
\newcommand{\Li}{\text{Li}}
\newcommand{\Ic}{\textbf{I}}
\newcommand{\Tc}{\textbf{T}}

\newcommand{\dLIPS}{d\text{LIPS}}
\newenvironment{mycode}{\VerbatimEnvironment\begin{Verbatim}[formatcom=\scriptsize\baselinestretch=0.1]}{\end{Verbatim}}

\usepackage{color}
\usepackage{xcolor}
\usepackage{hyperref}
\usepackage{xspace}
\usepackage[utf8]{inputenc}
\usepackage{multirow}
\usepackage{nccmath} % \mfrac
\usepackage[clock]{ifsym}

\numberwithin{equation}{subsection}

\title{High-precision scattering amplitudes \\ for LHC phenomenology}
\author{\textsc{Piotr Bargie\l{}a}}
\college{Rudolf Peierls Centre for Theoretical Physics}

\degree{Doctor of Philosophy in Theoretical Physics}
\degreedate{Trinity 2023}

\begin{document}
\setlength{\textbaselineskip}{22pt plus2pt}
\setlength{\frontmatterbaselineskip}{17pt plus1pt minus1pt}
\setlength{\baselineskip}{\textbaselineskip}
\setcounter{secnumdepth}{2}
\setcounter{tocdepth}{2}

\begin{romanpages}

\maketitle

\begin{dedication}
\textit{To my family}
\end{dedication}

\begin{acknowledgements}

Most of all, I would like to thank my supervisor Fabrizio Caola for his guidance throughout my DPhil studies.
He taught me the scientific method, which provides a firm canvas for the work described here.
Due to his boundless knowledge, I greatly benefited from our broad discussions.

I also thank my collaborators Federico Buccioni, Amlan Chakraborty, Federica Devoto, Herschel Chawdhry, Giulio Gambuti, Xiao Liu, Andreas von Manteuffel, and Lorenzo Tancredi for their contribution to our projects.
They maintained the highest research standards in our work.
In addition, I thank my examiners Gavin Salam and Simon Badger for pointing out minor corrections to the draft after my DPhil viva.

I am grateful to the scientific community of the University of Oxford Department of Physics for providing a stimulating research environment.
Due to a rich variety of fields pursued here, my interactions in the Beecroft Building proved fruitful every day.
Especially those with my officemates, who always guaranteed an open-minded discussion.

Moreover, I would like to thank the vibrant community of Mansfield College.
Particularly, to my friends from the Middle Common Room for sharing memorable experiences.
Together with the students encountered in a variety of societies, they allowed me to exchange ideas across multiple fields.

My DPhil was fully supported by the European Research Council Starting Grant 804394 \textsc{HipQCD}.
Besides funding my research, it also allowed for presenting my work to the scientific community at multiple conferences, as well as for expanding my knowledge on the state-of-the-art methods in my field at graduate schools.
Networking at these events with experts in my field provided me with a invaluable feedback.

I would like to personally thank Iwo Bia\l{}ynicki-Birula for inspiring me to pursue Theoretical Physics.
Working for my first physicist role model helped me decide to leave Engineering and choose a career path that led me here.

Finally, I am eternally grateful to my family members, without whom none of this would be possible.
To my grandparents, for being my inspiration.
To my parents and sister for providing me with a home during the nomadic times of the pandemic.
Together with my friends, they always supported me in my endeavors.

\end{acknowledgements}

\begin{abstract}

In this work, we consider scattering amplitudes relevant for high-precision Large Hadron Collider (LHC) phenomenology.
They provide the most fundamental representation of the quantum probability amplitude for the scattering of elementary particles.
As such, they are a necessary ingredient of theoretical predictions for hadronic differential cross section distributions for any process.
Our predictions rely on the Quantum Field Theory of the Standard Model (SM) of Elementary Particles.
Any confirmed deviation found between theoretical predictions and experimental measurements would yield the discovery of New Physics.

We analyse the general structure of amplitudes, and we review state-of-the-art methods for computing them.
We discuss advantages and shortcomings of these methods, and we point out the bottlenecks in modern amplitude computations.
As a practical illustration, we present frontier applications relevant for multi-loop multi-scale processes.

We compute the helicity amplitudes for the processes $gg\to\gamma\gamma$ and $pp\to\gamma$+jet in three-loop massless Quantum Chromodynamics (QCD).
We have adopted a new projector-based prescription to compute helicity amplitudes in the 't Hooft-Veltman scheme.
We also rederived the minimal set of independent Feynman integrals for this problem using the differential equations method, and we confirmed their intricate analytic properties.
By employing modern methods for integral reduction, we provide the final results in a compact form, which is appropriate for efficient numerical evaluation.

Beyond QCD, we have computed the two-loop mixed QCD-electroweak amplitudes for $pp \to Z$+jet process in light-quark-initiated channels, without closed fermion loops.
This process provides important insight into the high-precision studies of the SM, as well as into Dark Matter searches at the LHC.
We have employed a numerical approach based on high-precision evaluation of Feynman integrals with the modern Auxiliary Mass Flow method.
The obtained numerical results in all relevant partonic channels are evaluated on a two-dimensional grid appropriate for further phenomenological applications.

\end{abstract}

\begin{originality}

The work described here is based on the following publications with shared authorship:
\begin{itemize}
	\item (1) P. Bargiela, F. Caola, A. von Manteuffel, L. Tancredi, “Three-loop helicity amplitudes for diphoton production in gluon fusion”, JHEP 02 (2022) 153, arXiv:2111.13595~\cite{Bargiela:2021wuy},
	\item (2) P. Bargiela, F. Buccioni, F. Caola, F. Devoto, A. von Manteuffel, L. Tancredi, “Signal-background interference effects in Higgs-mediated diphoton production beyond NLO”, Eur.Phys.J.C 83 (2023) 2, 174, arXiv:2212.06287~\cite{Bargiela:2022dla},
	\item (3) P. Bargiela, A. Chakraborty, G. Gambuti, “Three-loop helicity amplitudes for photon+jet production”, Phys.Rev.D 107 (2023) 5, L051502, arXiv:2212.14069~\cite{Bargiela:2022lxz},
	\item (4) P. Bargiela, F. Caola, H. Chawdhry, X. Liu, “Two-loop mixed QCD-electroweak amplitude for Z+jet production”, in preparation.
\end{itemize}

The part of the work summarised in paper (1) not performed by me was the Integration-By-Parts (IBP) reduction of required Feynman integrals.
This step was completed by A. von Manteuffel, and I describe it only briefly here.
Also, for a check of the three-loop integrand, L. Tancredi provided an independent calculation.

My only original contribution to paper (2) was to provide the scale dependent three-loop QCD scattering amplitude for diphoton production in gluon fusion.
I also rederived the phase space integrals required in the soft-virtual approximation at NNLO.
I do not elaborate on the computational details of this work.
I present it only as a phenomenological application of my three-loop amplitude calculation.

The work described in paper (3) required two separate calculations, in the quark annihilation channel, and in the gluon fusion channel.
I provided the bare amplitude in the gluon fusion channel, while A. Chakraborty provided the bare amplitude in the quark annihilation channel.
G. Gambuti performed the UV-IR subtraction to a finite remainder.
As in the paper (1), we used the IBP table of A. von Manteuffel.

The part of the work described in the manuscript (4) not performed by me was the numerical evaluation of relevant Feynman integrals.
This step was completed by X. Liu, and I describe it only briefly here.
Also, F. Buccioni provided us with renormalization factors up to the two-loop mixed QCD-electroweak order.

Throughout this work, the programs \texttt{qgraf-xml-drawer}~\cite{qraf:drawer}, \texttt{TikZ-Feynman}~\cite{Ellis:2016jkw}, \texttt{JaxoDraw}~\cite{Vermaseren:1994je,Binosi:2003yf}, and \texttt{Inkscape}~\cite{Inkscape} were extensively used to generate graphics.

\bigskip

I presented the work described here at the following conferences:
\begin{itemize}
	\item{28 Jun 2023, }{\textit{Two-loop mixed QCD-electroweak amplitudes for Z+jet production}, LoopFest 2023, SLAC, \href{https://indico.cern.ch/event/1227237/contributions/5366579/}{\HandRight} }
	\item{30 May 2023, }{\textit{Three-loop four-particle QCD amplitudes}, RadCor 2023, Crieff, poster and presentation, \href{https://indico.ph.ed.ac.uk/event/118/contributions/2362/}{\HandRight} }
	\item{28 Apr 2023, }{\textit{Three-loop four-particle QCD amplitudes}, University of Edinburgh, Higgs Centre Amplitudes Meeting seminar}
	\item{2 Dec 2022, }{\textit{Two-loop mixed QCD-electroweak amplitudes for Z+jet production}, QCD@LHC 2022, Orsay, \href{https://indico.cern.ch/event/1150707/contributions/5114809/}{\HandRight} }
	\item{20 Sep 2022, }{\textit{Two-loop mixed QCD-electroweak amplitudes for Z+jet production}, HP2 2022, Newcastle upon Tyne, \href{https://conference.ippp.dur.ac.uk/event/1100/contributions/5779/}{\HandRight} }
	\item{\,8 Aug 2022, }{\textit{Three-loop QCD helicity amplitudes for $gg\to\gamma\gamma$}, Amplitudes 2022, Prague, poster}
	\item{\,\,8 Jul 2022, }{\textit{Three-loop four-particle QCD amplitudes}, ICHEP 2022, Bologna, \href{https://agenda.infn.it/event/28874/contributions/169965/}{\HandRight} }
	\item{14 May 2022, }{\textit{Three-loop helicity amplitudes for diphoton production in gluon fusion}, LoopFest 2022, Pittsburgh, \href{https://indico.cern.ch/event/1107840/contributions/4842778/}{\HandRight} }
	\item{16 Dec 2021, }{\textit{Three-loop helicity amplitudes for diphoton production in gluon fusion}, YTF 2021, Durham, online}
	\item{31 Mar 2021, }{\textit{High precision QCD amplitudes (towards higher-loop revolution)}, NExT PhD Workshop 2021, Sussex, online}
	\item{21 Jan 2021, }{\textit{High precision QCD amplitudes (in pursuit of removing redundancies)}, BUSSTEPP 2021, London, online}
\end{itemize}

\end{originality}

\dominitoc
\flushbottom

\tableofcontents
\begin{mclistof}{List of Abbreviations}{3.2cm}
	\item[LHC] Large Hadron Collider
	\item[HEP] High Energy Physics
	\item[SM] Standard Model
	\item[BSM] Beyond Standard Model
	\item[QFT] Quantum Field Theory
	\item[QCD] Quantum Chromodynamics
	\item[EWK] electroweak
	\item[LIPS] Lorentz invariant phase space
	\item[LO] leading order
	\item[NLO] next-to-leading order
	\item[NNLO] next-to-next-to-leading order
	\item[dimReg] dimensional regularization
	\item[ISP] irreducible scalar product
	\item[IBP] integration-by-parts
	\item[LIs] Lorentz invariance identities
	\item[MI] Master Integral
	\item[DEQ] differential equation
	\item[BC] boundary condition	
	\item[UV] ultraviolet
	\item[IR] infrared
	\item[GPL] Generalised Polylogarithm
	\item[HPL] Harmonic Polylogarithm
	\item[AMFlow] Auxiliary Mass Flow
\end{mclistof} 

\end{romanpages}

\flushbottom

\chapter{\label{ch:intro}Theoretical background}

\minitoc

\section{LHC Physics}
\label{sec:intro.LHC}

A pursuit of understanding the most fundamental structure of Nature has driven the curiosity of humankind for centuries.
Due to the Heisenberg's uncertainty principle, probing Nature at smaller distances requires higher energies to resolve the internal dynamics.
A frontier field in these studies is High Energy Physics (HEP).
Currently, the most accurate description at this elementary level is formalised in the Standard Model (SM) of Particle Physics.
Indeed, it provided 13 orders of magnitude agreement between theoretical prediction and experimental measurement of the electron magnetic dipole moment~\cite{Fan:2022eto}.
To date, it is the most precise agreement between theory and experiment for any quantity.

\begin{figure}[h]
	\centering
	\includegraphics[width=0.35\textwidth]{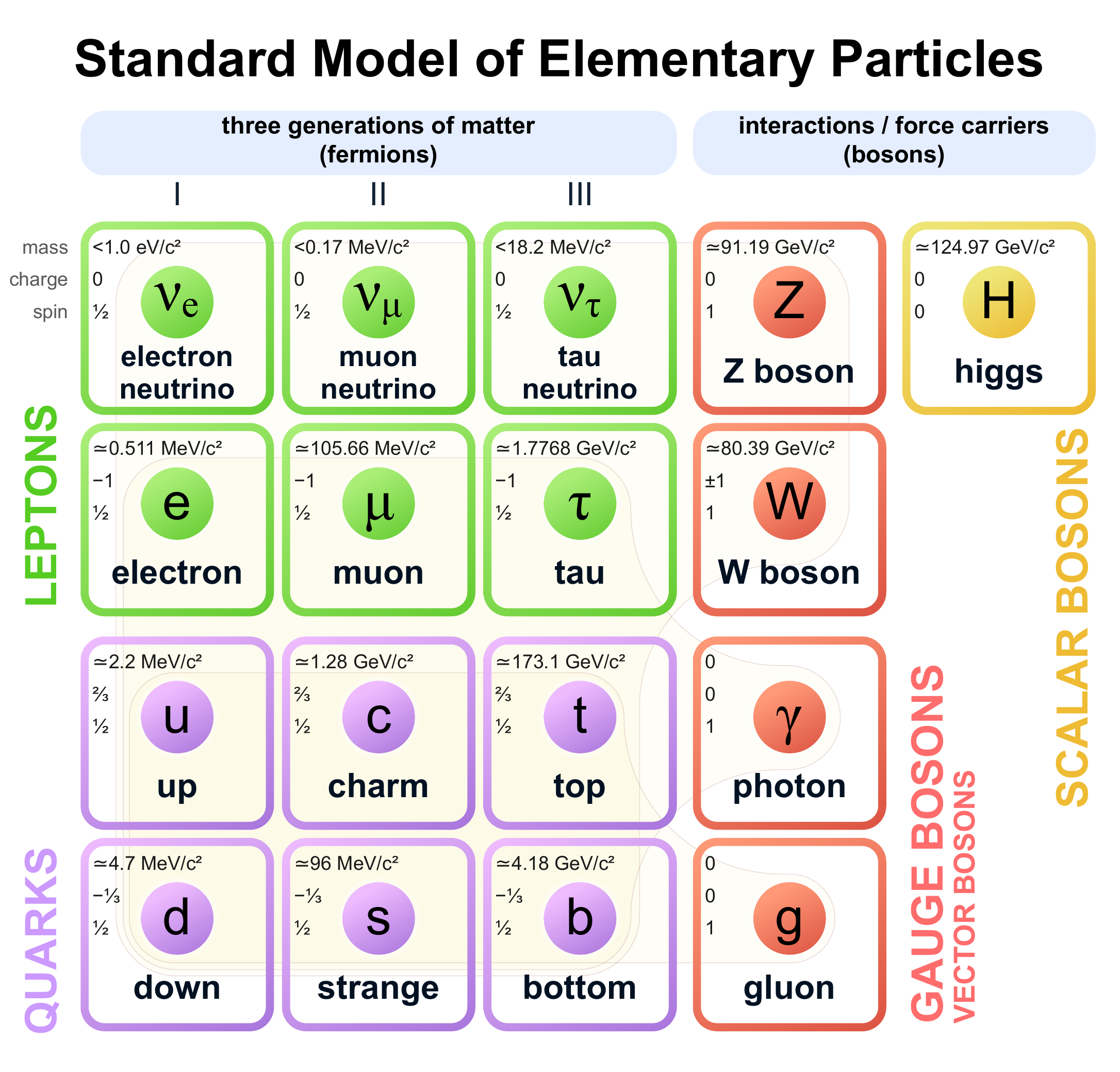}
	\includegraphics[width=0.55\textwidth]{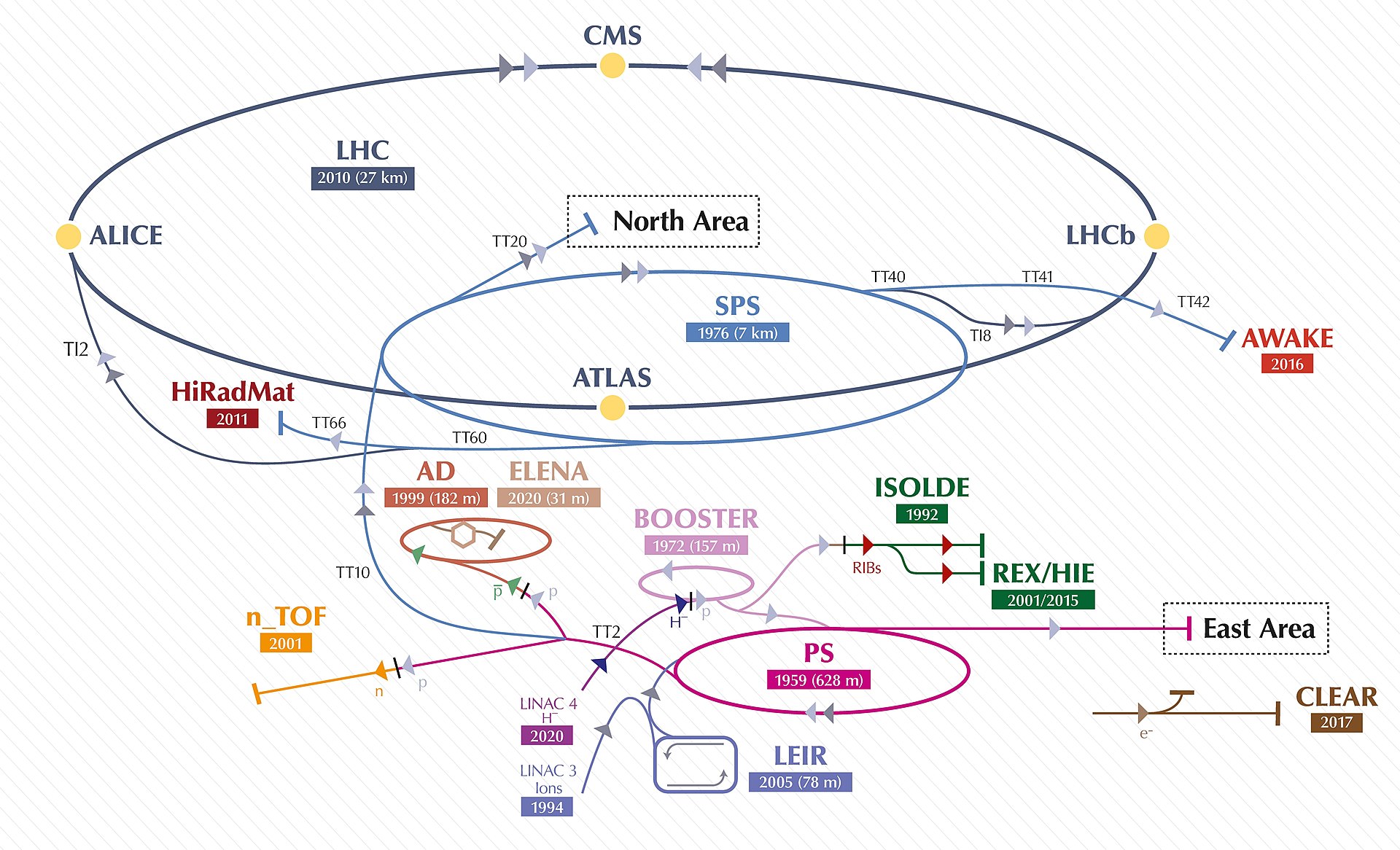}
	\caption{Particle content of the SM~\cite{wiki:SM}, as well as a schematic description of the LHC~\cite{wiki:LHC}.}
	\label{fig:smlhc}
\end{figure}

Multiple collider experiments contributed to confirming the particle content of the SM, which is summarised in Fig.~\ref{fig:smlhc}.
Most recently, the ATLAS and CMS experiments at the Large Hadron Collider (LHC) discovered the Higgs boson~\cite{ATLAS:2012yve,CMS:2012qbp}.
It was the last missing evidence for a particle predicted by the SM.
The LHC is the biggest collider experiment built to date, see its schematic description in Fig.~\ref{fig:smlhc}.
It consists of a tunnel ring of 27~km in circumference, where two beams of protons circulate.
Each beam consists of over 2000 bunches with $10^{11}$ protons which are scattered every 25~ns at energies reaching 13.6~TeV.
Each proton beam has a width of around 10~$\mu m$, and it is bent using powerful magnets of strength reaching 8~T.
This results in tens of collisions per proton bunch crossing, which corresponds to around a billion events per second.
This provides enough statistics to observe rare processes like the Higgs boson production.

\begin{figure}[h]
	\centering
	\includegraphics[width=0.9\textwidth]{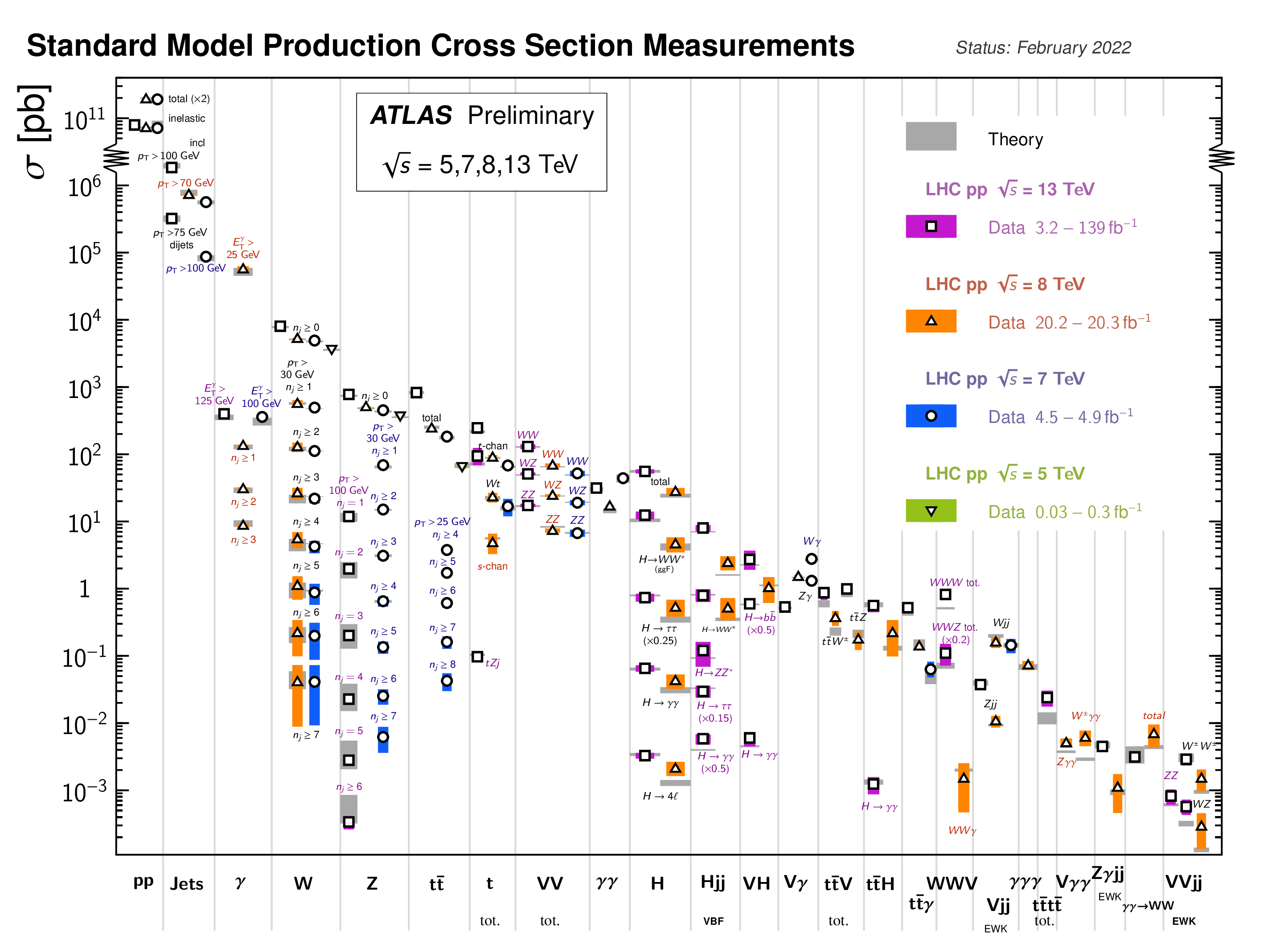}
	\caption{Cross section measurements for different processes at the LHC~\cite{ATLAS:2022djm}.}
	\label{fig:xsec}
\end{figure}

Since the confirmation of the particle content of the SM with the Higgs boson discovery, the LHC entered a precision HEP era.
Matching the high precision of the measurements with theoretical predictions requires accounting for higher order quantum corrections arising from interactions with the vacuum.
These corrections are the main topic of this work.
Any confirmed discrepancy between the theoretical predictions and experimental measurements would yield a discovery of a new Beyond Standard Model (BSM) Physics.
So far, all the measurements agree with SM predictions for variety of scattering processes of cross sections spanning 14 orders of magnitude, as summarised in Fig.~\ref{fig:xsec}.
Still, some of the properties of elementary particles are yet to be confirmed, e.g. couplings of Higgs boson to first and second generation fermions, as well as to itself.

\section{Quantum Chromodynamics}
\label{sec:intro.QCD}

One of the two sectors of the SM is Quantum Chromodynamics (QCD).
It describes the strong interactions which hold nucleons together.
At high energy, the degrees of freedom are point-like indivisible partons, contrarily to the low-energy hadronic states.
The model is formulated as a nonabelian Quantum Field Theory (QFT) with a SU($N_c$) gauge group with $N_c=3$ colours.
The corresponding Lagrangian~\footnote{In this work, we refer to the Lagrangian density as the Lagrangian.}
\begin{equation}
	\mathcal{L}_{\text{QCD}} =
	- \frac{1}{4} F^a_{\mu\nu} F^{a\mu\nu}
	+ \sum_{f=1}^{n_f} \bar{q}_f (i\slashed{D} - m_f) q_f
	- \frac{1}{2\xi} (\partial^\mu A^a_\mu)^2
	+ (\partial^\mu \bar{c}_g^a) \tilde{D}_\mu^{ac} c_g^c
	+ \theta \frac{g_s^2}{32\pi^2} \tilde{F}^a_{\mu\nu} F^{a\mu\nu}
\end{equation}
has a Yang-Mills gluon kinetic term, Dirac quark term, gauge fixing term, gluonic ghost term, and the CP violating term, respectively.
The gluon field strength reads
\begin{equation}
	F^a_{\mu\nu} = \partial_\mu A^a_\nu - \partial_\nu A^a_\mu - g_s f^{abc} A^b_\mu A^c_\nu \,,
\end{equation}
its dual is $\tilde{F}^a_{\mu\nu} = \frac{1}{2} \varepsilon_{\mu\nu\rho\sigma} F^{a\rho\sigma}$,
while the covariant derivatives are
\begin{equation}
\begin{split}
	D_\mu &= \quad\, \partial_\mu + i g_s T^a A_\mu^a \,, \\
	\tilde{D}_\mu^{ac} &= \delta^{ac} \partial_\mu + g_s f^{abc} A_\mu^b \,,
\end{split}
\end{equation}
where $g_s$ is the strong coupling constant.
The bosonic gluon field $A^a_\mu$ transforms in the adjoint representation of the gauge group, and it is a strong force carrier.
The fermionic quark field $q_f$ of flavour $f$ transforms in the fundamental representation, and it describes matter.
In QCD, there are $n_f=6$ quark flavours, up, down, charm, strange, top, and bottom, with corresponding masses $m_f$, as in Fig.~\ref{fig:smlhc}.
The fundamental generators $T^a$ of the gauge group are related to the adjoint $f^{abc}$ ones via the underlying Lie algebra
\begin{equation}
	[T^a,T^b] = i f^{abc} T^c \,, \qquad a=1,\dots,N_c^2-1 \,,
\end{equation}
with the normalization
\begin{equation}
	{\rm Tr}(T^a T^b) = T_F \delta^{ab} \,,~~~~ T_F = \frac{1}{2} \,.
\end{equation}
The gauge fixing parameter can take arbitrary values.
Throughout this work, we fix $\xi=1$ in the Feynman - 't Hooft gauge.
The anti-commuting complex scalar ghost field $c_g^a$ transforms in the adjoint representation, and it is introduced by the Faddeev-Popov gauge fixing procedure~\cite{Faddeev:1967fc}.
Since experimentally the value of the $\theta$ angle is bounded to be very small, we will treat our QCD theory as CP symmetric.

As for other QFTs, quantum corrections affect the bare parameters entering the Lagrangian.
Their physicality can be restored with a renormalization procedure, which relates the initial bare unmeasurable quantities to the observable ones via
\begin{equation}
	g_{s,b} = Z_{g_s} g_s \,,\qquad
	m_{f,b} = Z_{g_s} m_f \,,\qquad
	\psi_{i,b} = \sqrt{Z_{i}} \psi_i \,,
\end{equation}
where the renormalization factor can be written as $Z_j=1+\delta_j$, and $\psi_i$ refers the wave function of any field $A^a_\mu$, $q_f$, or $c_g^a$.
Formally, renormalization removes divergences arising from the bare description.
A renormalization scheme is defined by specifying the constraints on the two-point functions $\Sigma_i$, which fix the mass and wave function renormalization factors of particle $i$, as well as on the vertex functions $V$, which renormalize the coupling.
Due to an ambiguity in the subtracted convergent part, one can define multiple renormalization schemes.
For example, in Sec.~\ref{sec:amp.div.UV}, we elaborate on the modified minimal subtraction $\overline{ \rm MS}$ scheme, which removes only the divergent terms.

\begin{figure}[h]
	\centering
	\includegraphics[width=0.8\textwidth]{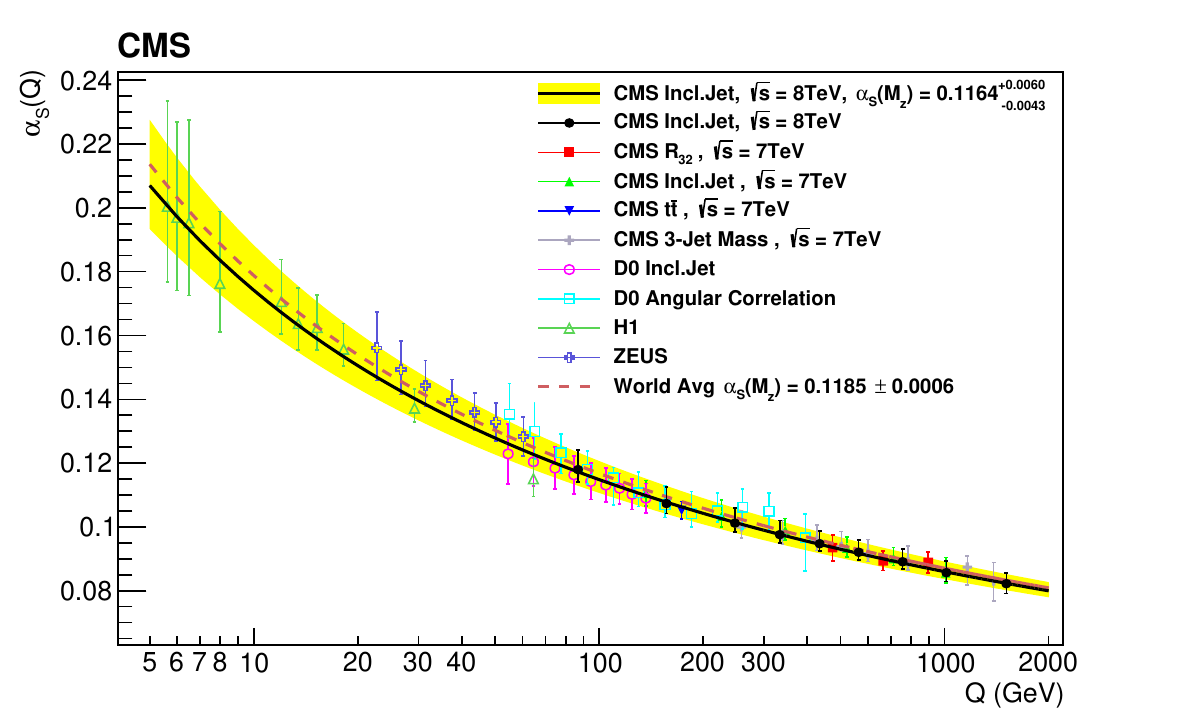}
	\caption{Measurements of the running of the strong coupling~\cite{CMS:2016lna}.}
	\label{fig:alphas}
\end{figure}

One of the consequences of renormalization is the running of the strong coupling $\alpha_s = \frac{g_s^2}{4\pi}$ with energy scale at which we are probing the high-energy system, called renormalization scale $\mu_R$,
\begin{equation}
	\frac{d \, \alpha_s(\mu_R^2)}{d \log \mu_R^2} = \beta(\alpha_s) 
	= - \beta_0 \frac{\alpha_s^2(\mu_R^2)}{2\pi} 
	+ \mathcal{O}(\alpha_s^3) \,.
\end{equation}
Above, in the so called leading logarithmic (LL) approximation, we neglect higher order dependence on $\alpha_s$.
The corresponding leading term in the beta function, $\beta_0$, is positive in QCD, as elaborated on in Sec.~\ref{sec:amp.div.UV}.
Therefore, the strong coupling exhibits asymptotic freedom, i.e. it decreases with energy.
In the LL approximation, it reads
\begin{equation}
	\alpha_s(\mu_R^2) \approx
	\frac{2\pi}{\beta_0\log\frac{\mu_R^2}{\Lambda_{\text{QCD}}^2}} \,,
\end{equation}
see Fig.\ref{fig:alphas}.
At low energy $\Lambda_{\text{QCD}}=\mathcal{O}(100)$~MeV, QCD reaches a Landau pole, where the strong coupling becomes infinite.
This corresponds to strong interactions which hold hadrons together.

Since at low energy the degrees of freedom of QCD change from partons to hadrons, their dynamics is described by an Effective Field Theory (EFT) called the Chiral Lagrangian, see review in Ref.~\cite{Scherer:2022foe}.
The theory is organized in a simultaneous expansion in derivatives and quark masses, which are treated as small perturbations around the chiral limit.
In the chiral limit of vanishing light-quark masses $m_q \to 0$, the chiral symmetry between the right and left-handed quarks is restored.
The masses of low-energy particles can be computed numerically in the lattice path integral formulation of QCD, see details in Ref.~\cite{FlavourLatticeAveragingGroupFLAG:2021npn}.
This approach also provides a numerical evidence of the colour confinement phenomenon, which forbids resolving the colour charge at low energies.
The corresponding analytic solution remains unknown, and it is a generalisation of one of the Millennium Prize Problems.

\begin{figure}[h]
	\centering
	\includegraphics[width=0.45\textwidth]{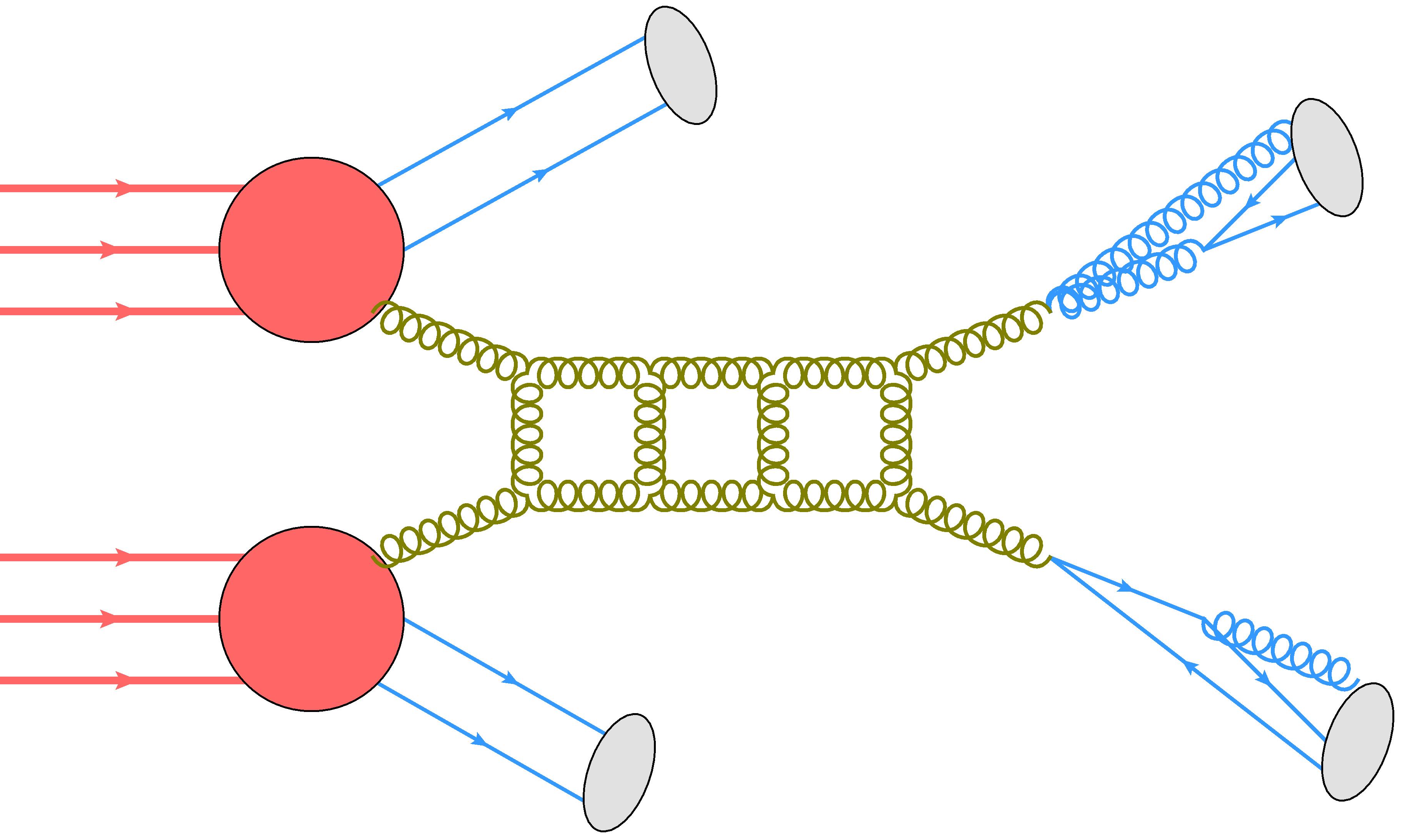}
	\hfill
	\includegraphics[width=0.45\textwidth]{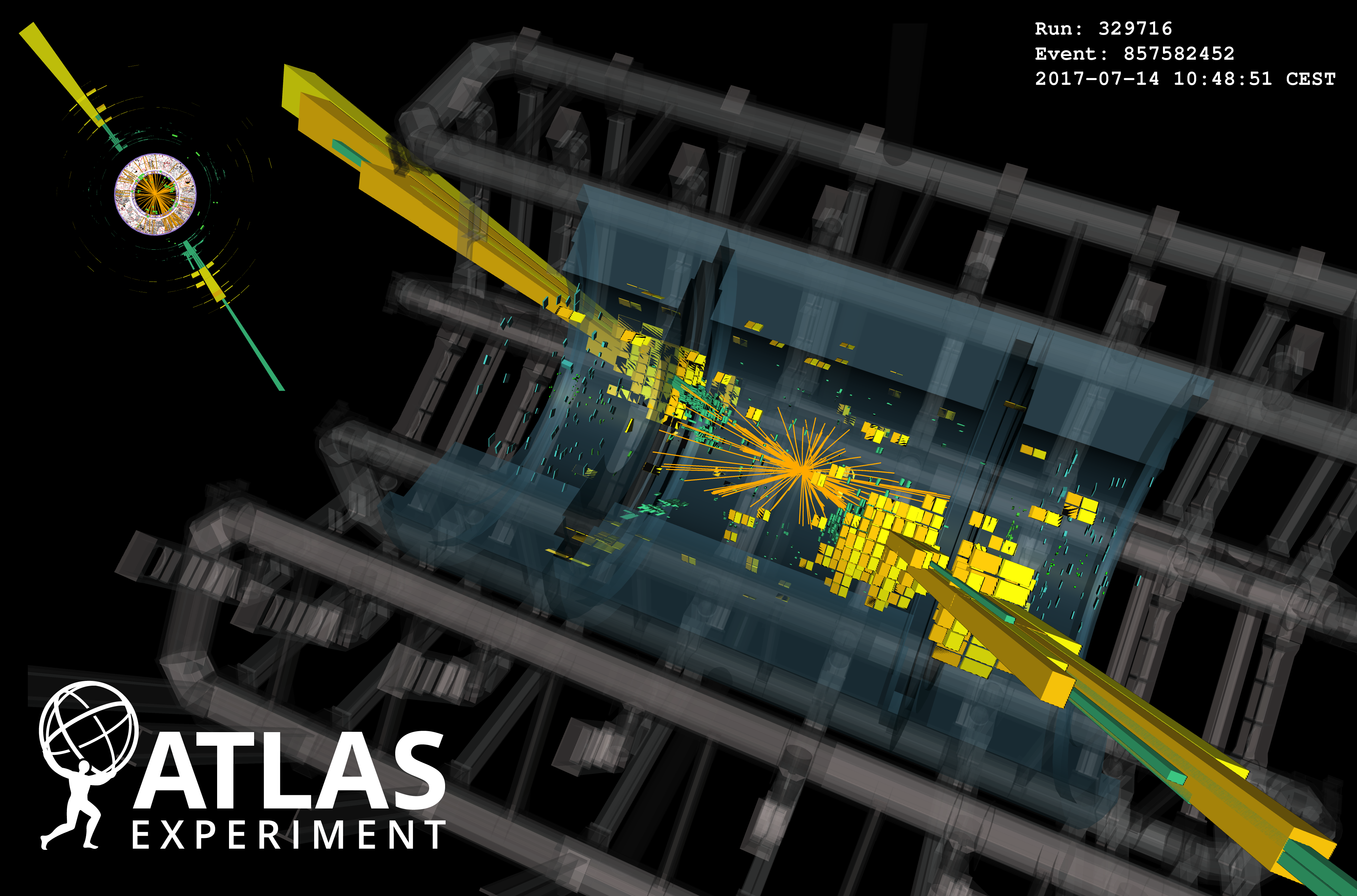}
	\caption{hadronic $\sigma$ = {\color{red} PDFs} $\otimes$ {\color{olive} hard scattering} $\otimes$ {\color{blue} real radiation} $\otimes$ {\color{gray} hadronization}
		Schematic description of the factorization theorem for theoretical prediction for the example dijet production, together with its experimental counterpart.}
	\label{fig:fact}
\end{figure}

In order to provide theoretical predictions for the hadronic cross section $\sigma$ for the scattering process $pp \to X$, we assume that it arises from a collision of two partons, $i$ and $j$, each stemming from a separate proton $p$.
This can be formulated in terms of the factorization theorem
%\\
%\scalebox{0.95}{\parbox{1.0\linewidth}{
\begin{equation}
\begin{split}
	\sigma_{pp \to X}(s_h) &=
	{\color{gray} \int_{\{K\}}^X}
	{\color{blue} \int_K^{\{K\}}} \,
	{\color{red} \sum_{i,j} \int_0^1 dx_1 \int_0^1 dx_2 \, f_i(x_1,\mu_F) f_j(x_2,\mu_F)} \,
	{\color{olive} \hat{\sigma}_{ij \to K}}\left(s,\mu_R,\mu_F\right) \\
	&+ \mathcal{O}\left(\frac{\Lambda^n_{\text{QCD}}}{Q^n}\right) \,,
\end{split}
\label{eq:intro.fact}
\end{equation}
%}}
%\\
schematically depicted in Fig.~\ref{fig:fact}.
The hadronic $s_h$ and partonic $s$ energies are related by $s=x_1x_2s_h$.
The probability for an incoming parton $i$ of a momentum fraction $x$ to be found at factorization scale $\mu_F$ in the proton $p$ is described by a Parton Distribution Function (PDF) ${\color{red} f_i(x,\mu_F)}$, see review in Ref.~\cite{Forte:2010dt}.
Their $x$-dependence has to be fitted from the experimental data, while the $\mu_F$-dependence is governed by the DGLAP evolution equation~\cite{Altarelli:1977zs,Gribov:1972ri,Dokshitzer:1977sg}.
The partonic cross section for the hard scattering $ij \to K$
\begin{equation}
	{\color{olive} \hat{\sigma}} = \frac{1}{2s} \int d\text{LIPS} \, |\mathcal{A}|^2
\label{eq:intro.xsec}
\end{equation}
consists of the flux factor, integral over the Lorentz Invariant Phase Space (LIPS), as well as the scattering amplitude $\mathcal{A}$, respectively.
The latter is at the center of this work.
By ${\color{blue} \int_K^{\{K\}}}$, we schematically denote the real radiation effects at intial and final partonic states.
They result in a much larger set of final partons $\{K\}$, with lower energies.
The evolution of the real radiation effects is described by Parton Showers, see details in Ref.~\cite{Buckley:2011ms}.
Accuracy of the shower can be understood in the framework of logarithmic resummation, detailed in Ref.~\cite{Banfi:2004yd}.
In order to arrive at infrared-safe final states, the partons $\{K\}$ have to be clustered together with jet algorithms, see review in Ref.~\cite{Salam:2010nqg}.
These algorithms are based on the sequential recombination, and the most widely used belong to the class of generalized-$k_t$ algorithms.
Besides the resilience of jet definition to soft and collinear radiation, a jet algorithm should also provide an easy implementation and detector calibration.
Finally, at $\Lambda_{\text{QCD}}$ energy, hadronization models, as detailed in Ref.~\cite{Buckley:2011ms}, need to be used to transition from the partonic $\{K\}$ to the hadronic final state $X$, which we schematically denote by ${\color{gray} \int_{\{K\}}^X}$.
The factorization theorem in Eq.~\ref{eq:intro.fact} is valid only up to power corrections at probe scale $Q$, see Ref.~\cite{Caola:2021kzt}.
The power $n$ depends on specific observable and process, and establishing it is currently researched~\cite{Makarov:2023ttq}.
Linear power corrections $n=1$ are the most important for phenomenology.

\section{Electroweak sector}
\label{sec:intro.EWK}

Besides QCD, the SM has also the electroweak (EWK) sector.
This model is formulated as a nonabelian QFT with a SU$(2)_L\otimes$U$(1)_Y$ gauge group with chiral fermionic matter and a complex scalar doublet.
The symmetry is spontaneously broken at a Weinberg weak mixing angle $\theta_W\approx$ 28$^\circ$ to electromagnetic U$(1)_E$ when one of the four real scalars in the complex scalar doublet, the Higgs boson $H$, acquires a nonzero vacuum expectation value (vev) $v\approx$~246~GeV.
This Higgs mechanism endows weak vector bosons $Z$ and $W^\pm$ with masses
\begin{equation}
	m_Z = \frac{ev}{2s_wc_w} \,, \qquad
	m_W = \frac{ev}{2s_w} \,,
\end{equation}
respectively, and leaves the photon $A$ massless.
For compactness, we defined $s_w = \sin\theta_W$ and $c_w = \cos\theta_W$.
Note that
\begin{equation}
	c_w = \frac{m_W}{m_Z} \,, \quad \text{and} \quad
	c_w^2+s_w^2=1 \,.
\end{equation}
Since the electric coupling is
\begin{equation}
	\alpha(0)=\frac{e^2(0)}{4\pi} \approx \frac{1}{137} \,,
\end{equation}
the electroweak interactions are indeed weak.

The spontaneously broken EWK Lagrangian reads
\begin{equation}
	\begin{split}
		\mathcal{L}_{\text{EWK}} &= \mathcal{L}_{c} + \mathcal{L}_{\text{YM}} + \mathcal{L}_{H} + \mathcal{L}_{\text{GB}} + \mathcal{L}_{\text{lep}} + \mathcal{L}_{\text{ghost}} + \mathcal{L}_{\text{GF}} \,,
	\end{split}
\end{equation}
\begin{equation}
	\begin{split}
		\mathcal{L}_{c} &=
		e \sum_{f=1}^6 Q_f \, \bar{q}_f \gamma_\mu q_f \, A^\mu \\
		&+ \frac{e}{\sqrt{2}s_w} \sum_{f,g=1}^3 \left( V_{fg} \, \bar{u}_{f,L} \gamma^\mu d_{f,L} \, W^+_\mu
		+ V^*_{fg} \, \bar{d}_{f,L} \gamma^\mu u_{f,L} \, W^-_\mu \right) \\
		&+ e \sum_{f=1}^6 \left( g_{f,L} \, \bar{q}_{f,L} \gamma^\mu q_{f,L}
		+ g_{f,R} \, \bar{q}_{f,R} \gamma^\mu q_{f,R} \right) Z^\mu \,,
	\end{split}
\end{equation}
\begin{equation}
	\begin{split}
		\mathcal{L}_{\text{YM}} &=
		- \frac{1}{4} A_{\mu\nu} A^{\mu\nu} 
		- \frac{1}{4} Z_{\mu\nu} Z^{\mu\nu}
		- \frac{1}{4} W^+_{\mu\nu} W^{-\mu\nu} \\
		&+ i e ( W^+_{\mu\nu} W^{-\mu} A^\nu 
		- W^-_{\mu\nu} W^{+\mu} A^\nu
		+ A_{\mu\nu} W^{-\mu} W^{-\nu} ) \\
		&+ \frac{iec_w}{s_w} ( W^+_{\mu\nu} W^{-\mu} Z^\nu 
		- W^-_{\mu\nu} W^{+\mu} Z^\nu
		+ Z_{\mu\nu} W^{-\mu} W^{-\nu} ) \\
		&- \frac{e^2}{2s_w^2} ( 2 g^{\mu\nu} g^{\rho\sigma} - g^{\mu\rho} g^{\nu\sigma} - g^{\mu\sigma} g^{\nu\rho} ) \\
		&\times \Bigg( \Bigg.  W^+_\mu W^-_\nu ( s_w^2 A_\rho A_\sigma 
		+ c_w^2 Z_\rho Z_\sigma 
		+ 2 s_w c_w A_\rho Z_\sigma )
		- \frac{1}{2} W^+_\mu W^+_\nu W^-_\rho W^-_\sigma \Bigg. \Bigg) \,,
	\end{split}
\end{equation}
\begin{equation}
	\begin{split}
		\mathcal{L}_{H} &= \frac{1}{2} (\partial^\mu H) \partial_\mu H
		- \frac{1}{2} m_H^2 H^2
		- \frac{2m_H^2}{v} H^3
		- \frac{m_H^2}{2v^2} H^4 \\
		&+ \left( m_W^2 W^{+\mu} W^-_\mu + \frac{1}{2} m_Z^2 Z^{\mu} Z_\mu \right) \left( 1 + \frac{1}{v}H \right)^2 
		- \frac{1}{v} \sum_{f=1}^6 m_f \, \bar{q}_f q_f \, H \,,
	\end{split}
\label{eq:LagrH}
\end{equation}
with field strengths
\begin{equation}
	A_{\mu\nu} = \partial_\mu A_\nu - \partial_\nu A_\mu \,, \qquad
	Z_{\mu\nu} = \partial_\mu Z_\nu - \partial_\nu Z_\mu \,, \qquad
	W^\pm_{\mu\nu} = \partial_\mu W^\pm_\nu - \partial_\nu W^\pm_\mu \,,
\end{equation}
left and right-handed quark states
\begin{equation}
	q_L = \frac{1-\gamma_5}{2} q \,, \qquad
	q_R = \frac{1+\gamma_5}{2} q \,,
\end{equation}
CKM flavour mixing matrix $V_{fg}$, the couplings
\begin{equation}
	\begin{split}
		g_{f,L} &= g_{f,-} = c_{f,V}+c_{f,A} = \frac{I^3_{f}-s_w^2 Q_{f}}{s_wc_w} \,,\\ 
		g_{f,R} &= g_{f,+} = c_{f,V}-c_{f,A} = -\frac{s_wQ_{f}}{c_w} \,,
	\end{split}
\end{equation}
electric charge $Q_{up} = \frac{2}{3}$ and $Q_{dn} = -\frac{1}{3}$, as well as the weak isospin generator $I^3_{up} = \frac{1}{2}$ and $I^3_{dn} = - \frac{1}{2}$.
We do not write here the explicit form of Lagrangian terms involving Goldstone bosons $\chi$ and $\varphi^\pm$, leptons $e_f$ and $\nu_f$, ghosts $c_{Z}$, $c_{W^\pm}$, and $c_{\gamma}$, as well as gauge fixing, since they do not appear in our further discussion.

\begin{figure}[h]
	\centering
	\includegraphics[width=0.9\textwidth]{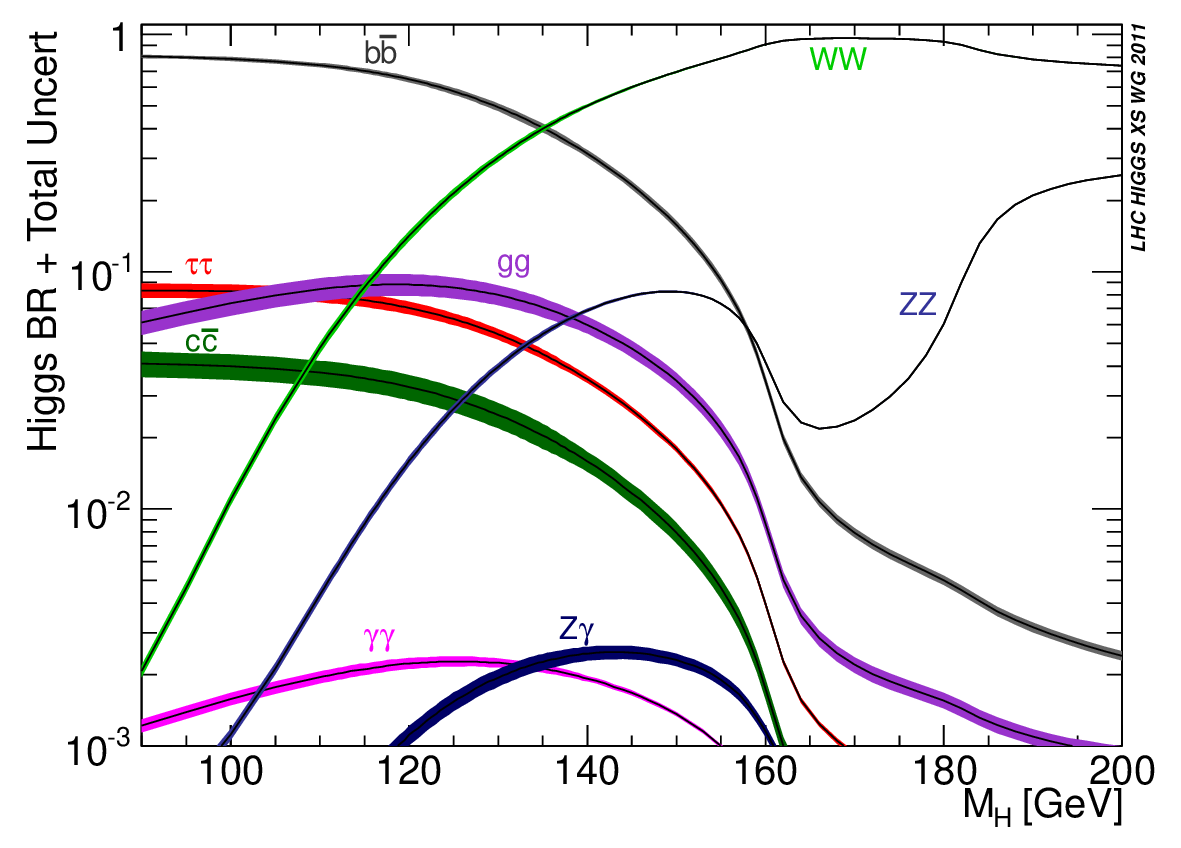}
	\caption{The SM Higgs branching ratios as a function of Higgs mass~\cite{Dittmaier:2012vm}.}
	\label{fig:HBR}
\end{figure}

Since the EWK sector of the SM has much more complicated structure then QCD, we only point out here some important facts.
Firstly, even though the EWK theory is not confining, it is not possible to directly detect the massive EWK bosons in a collider.
It is due to the fact that these particles decay before reaching calorimeters, see corresponding branching ratios in Fig.~\ref{fig:HBR}.
Thus, the discovery of any such particle was the most convenient in clean decay channels, i.e. those consisting of long-lived particles e.g. leptons and photons.
Secondly, since some of the EWK particles can be treated as asymptotic states, one can employ a more physical renormalization scheme than for QCD.
This on-shell renormalization scheme is defined for external states on the mass shell.
We elaborate on this scheme in Sec.~\ref{sec:ppjZ.uvir}.
Thirdly, contrarily to QCD, the EWK sector is manifestly CP violating.
Interestingly, this effect alone does not fully explain the resulting matter-antimatter asymmetry in the Universe~\cite{Farrar:1993hn}.
Together with other microscopically-unexplained cosmological phenomena e.g. Dark Matter, they show the shortcomings of the current formulation of the SM.
The modern precision phenomenology program provides a strong background for searches of any BSM signatures which would explain these effects.

\chapter{\label{ch:amp}Scattering amplitudes}

\minitoc

\section{Overview}
\label{sec:amp.overview}

Scattering amplitudes are the main subject of this thesis, as introduced in Sec.~\ref{sec:intro.QCD}.
They describe the quantum-mechanical probability amplitude for the scattering of particles in an initial state $| i \rangle$ resulting in a final state $\langle f |$.
In QFT, they are embedded in the scattering S-matrix
\begin{equation}
	\langle f | S | i \rangle = 1 + i \, \mathcal{A}(i \to f) \,.
	\label{eq:Smatrix}
\end{equation}

It is an open problem in Theoretical Physics to compute the scattering amplitude $\mathcal{A}$ for a generic process $i \to f$ in an interacting QFT, even numerically.
Since the amplitude is a function of the QFT coupling $g$, a systematic approximation can be designed around a small value of the coupling.
Phenomenologically, it is a good approximation in the EWK theory at all energy scales, while in QCD only in the high-energy region.
It is due to the fact that the electric coupling is small and it grows with energy very slowly in the range of energies reached by the collider.
Contrarily, the strong coupling decreases with energy and it reaches its Landau pole at an energy available to a collider, see Fig.~\ref{fig:alphas}.
Formally, we write a perturbative series
\begin{equation}
	\mathcal{A} = \sum_{n=0} \left( \frac{\alpha}{2\pi} \right)^n \mathcal{A}^{(n)} \,,
\label{eq:ampPert}
\end{equation}
where the bare coupling is $\alpha = \frac{g^2}{4\pi}$.

Due to the LSZ theorem, the fixed-order amplitude $\mathcal{A}^{(n)}$ can be computed as
\begin{equation}
	\mathcal{A}^{(n)} = \sum \text{Feynman diagrams} \,,
\end{equation}
where the Feynman diagrams are all the connected amputated graphs possible to construct from Feynman rules by connecting the vertices appearing in the Lagrangian.
Therefore, Eq.~\ref{eq:ampPert} corresponds to an expansion in higher loop corrections.
These diagrammatic loops carry intermediate unconstrained degrees of freedom.
According to the superposition principle in Quantum Mechanics, we need to sum over all of them.
Summing over the continuous degrees of freedom gives rise to the notion of \textit{Feynman loop integrals}, which are a major part of this work, as discussed in Sec.~\ref{sec:amp.int}.
Calculating the higher loop corrections still remains a challenge.
Nonetheless, perturbative series provides a method for a systematic improvement.
When using fixed-order amplitudes to construct the corresponding cross section corrections, as in Eq.~\ref{eq:intro.xsec}, we perturbatively arrive at precision levels referred to as \textit{leading order} (LO), \textit{next-to-leading order} (NLO), \textit{next-to-next-to-leading order} (NNLO), and higher.

It is worth pointing out the two major issues with the perturbative solution in Eq.~\ref{eq:ampPert}.

Firstly, the SM formulated naively in $d=4$ spacetime dimensions leads to divergent predictions.
These divergences can be avoided by various regularization schemes.
In this work we will use the dimensional regularization scheme (dimReg).
It means performing all the amplitude calculations in $d=4-2\ep$ dimensions.
The divergent poles in $\ep$ are universal and they can be subtracted, as explained in Sec.~\ref{sec:amp.div}.
This leads to a finite physical four-dimensional remainder.
Secondly, the perturbative series of the SM is convergent only asymptotically~\cite{Dyson:1952tj}.
Indeed, the factorial growth of higher-order perturbative coefficients originates in e.g. renormalon contributions, see review in Ref.~\cite{Beneke:1998ui}.
This effect can be accounted for by considering the power corrections, as introduced in Eq.~\ref{eq:intro.fact}.

The difficulty in computing higher order corrections is twofold.
Firstly, the growth in the number of required Feynman diagrams is  beyond-factorial.
Since usually some of the corresponding expressions are very complicated, evaluating them requires a lot of computational resources, thus increasing the complexity of the problem.
Secondly, the mathematical structure of the associated multi-loop Feynman integrals is in general unknown.
Recently, there have been major advances in both the complexity and integral frontier.
It is worth noting that a lot of these developments originate in the studies of more symmetric or toy model QFTs, see review in e.g. Refs~\cite{Elvang:2015rqa,Bern:2022jnl}.

In general, the scattering amplitude can be written in terms of three mathematical structures, gauge group colour $\mathcal{C}_c$, Lorentz tensor $T_t$, and Feynman integral $\mathcal{I}_{i}$
\begin{equation}
	\mathcal{A} = \sum_{c,t,i} \mathcal{C}_c \, T_t \, \mathcal{I}_{i} \, r_{c,t,i} \,.
	\label{eq:3struct}
\end{equation}
Importantly, they linearly decompose the whole amplitude.
Moreover, since they are independent to each other, they can be analysed and simplified separately in a gradual manner.
Firstly, the colour structure depends on colour indices $a_i$ of external particles and on the number of colours $N_c$, i.e. $\mathcal{C}_c(\vec{a},N_c)$.
Secondly, the tensor structure $T_t(\lambdavec,p_i^\mu)$ depends on helicities $\lambda_i$ and momenta $p_i^\mu$ of external particles in the process.
As we will see, it can be shown that these two structures are loop-independent for the processes considered here.
Thirdly, Feynman integrals $\mathcal{I}_{i}(d,p_i^\mu)$ depend on the spacetime dimension $d$ and on the kinematics $p_i^\mu$.
Finally, the coefficients $r_{c,t,i}(d,p_i^\mu)$ are algebraic functions.
In the following sections we will elaborate on the properties of these three structures.
As an illustrative example, we will focus on the two-loop QCD amplitude for the $q\bar{q} \to gg$ process.

\section{Kinematics}
\label{sec:amp.kin}

Let us fix the notation on kinematics of the $n$-particle or $n$-\textit{point} process $i \to f$ with $m$ incoming states $i$ and $n-m$ outgoing states $f$.
We denote the external momenta by $p_i$, while internal loop momenta by $k_l$.
We refer to external particles as \textit{legs}.
It is convenient to choose the all-incoming notation for the external momenta in the process
\begin{equation}
	\sum_{j=1}^m i_j(p_j) \to \sum_{l=m+1}^n f_l(-p_l)
\end{equation}
such that the momentum conservation is sign-symmetric
\begin{equation}
	\sum_{j=1}^n p_j = 0 \,,
\end{equation}
and we can treat all $n$ states as incoming~\footnote{Note that the incoming fermion is equivalent to an outgoing anti-fermion.}.
If all the external particles as physical, their momenta are on the mass-energy shell or \textit{on-shell}
\begin{equation}
	p_i^2=m_i^2 \,.
\end{equation}
The kinematics of the scattering amplitude $\mathcal{A}(p_i)$ is completely defined by kinematic Lorentz invariants of the process, i.e. masses $m_i$, Mandelstam variables
\begin{equation}
	s_{ij}=(p_i+p_j)^2 \,,
\end{equation}
and totally antisymmetric products
\begin{equation}
	\varepsilon(i,j,k,l) = \varepsilon_{\mu\nu\rho\sigma} p_i^\mu p_j^\nu p_k^\rho p_l^\sigma \,.
\end{equation}
We abbreviate
\begin{equation}
	p_{ij}=p_i+p_j \,,\qquad p_{ijk}=p_i+p_j+p_k \,.
\end{equation}
The number of independent Mandelstam invariants for a scattering of $n>3$ particles is constrained by an overall momentum conservation and Poincar\'e symmetry, and in four dimensions it yields
\begin{equation}
	\text{\# indept Mandelstams} = 3n-10+ \text{\# masses} \,.
\end{equation}
It is often convenient to factorize from the amplitude some overall mass-dimensionful factor $M$
\begin{equation}
	\mathcal{A}_n(s_{ij},m_k^2) = M^{[\mathcal{A}_n]} \mathcal{\tilde{A}}_n\left( \frac{s_{ij}}{M^2},\frac{m_k^2}{M^2} \right) \,,
\label{eq:ampMassDimFact}
\end{equation}
where the mass dimension of the amplitude
\begin{equation}
	[\mathcal{A}_n] = 4-n
\end{equation}
can be easily understood from by dimensional analysis of the cross section definition in Eq.~\ref{eq:intro.xsec}.

In this work, we focus on four-point processes.
Since there are not enough linearly independent momenta in the problem for the totally antisymmetric products $\varepsilon(i,j,k,l)$ not to vanish, the kinematics is fully parametrized by the three cyclic Mandelstam variables
\begin{equation}
	s_{12} = s = (p_1+p_2)^2 \,, \quad
	s_{23} = u = (p_2+p_3)^2 \,, \quad
	s_{13} = t = (p_1+p_3)^2 \,,
\end{equation}
which are related by a momentum conservation relation
\begin{equation}
	s+t+u = \sum_{i=1}^{4} m_i^2 \,.
\end{equation}
For a massless $m_i=0$ scattering, it is convenient to define one dimensionless ratio
\begin{equation}
	x = -\frac{t}{s}
	\label{eq:x}
\end{equation}
which carries all the nontrivial kinematic dependence.

\section{Colour structure}
\label{sec:amp.col}

\begin{figure}[h]
	\centering
	\includegraphics[width=0.65\textwidth]{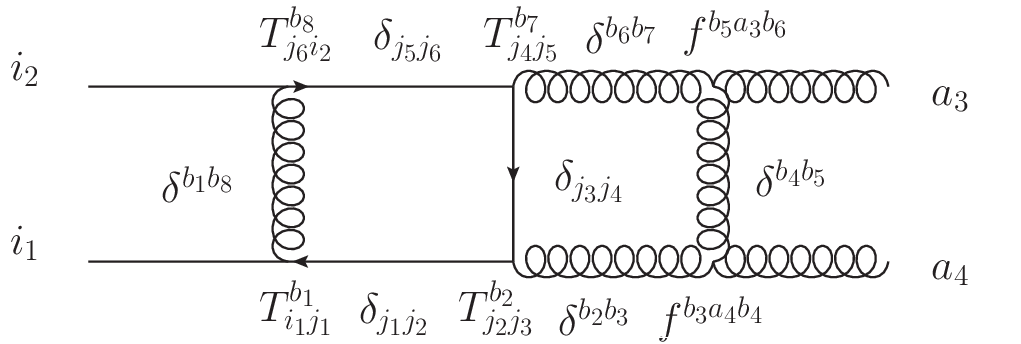}
	\caption{Colour structures in example Feynman diagram for the two-loop QCD amplitude for the $q \bar{q} \to gg$ process.}
	\label{fig:col}
\end{figure}

\noindent
We first describe the colour structures $\mathcal C_c$ of the scattering amplitude $\mathcal A$.
It stems from the SU(3) gauge group of QCD, as defined in Sec.~\ref{sec:intro.QCD}.
As an example of colour structures encountered in amplitude computation, consider the two-loop $q \bar{q} \to gg$ process, with sample Feynman diagram depicted in Fig.~\ref{fig:col}.
When combining multiple colour generators, it is convenient to express $f^{abc}$ in terms of fundamental generators
\begin{equation}
	f^{abc}	= -2i \Tr \left( T^a T^b T^c - T^c T^b T^a \right) \,,
\end{equation}
and then perform sums over duplicate indices
\begin{equation}
	T^a_{ij} T^a_{kl} = \frac{1}{2} \left( \delta_{il} \delta_{kl} - \frac{1}{N_c} \delta_{ij} \delta_{kl} \right) \,.
\label{eq:amp.col.tttodelta}
\end{equation}
We note that these colour identities can be interpreted diagrammatically in the so called \textit{'t Hooft double-line formalism}~\cite{tHooft:1973alw}.
We defined upper indices $\{a,b,c\}$ to be in the adjoint, while lower indices $\{i,j\}$ in the fundamental representation.
Applying these identities to the Feynman diagram in Fig.~\ref{fig:col} terminates at chains of fundamental generators
$(T^{a_3}T^{a_4})_{i_1i_2}$ and $\Tr(T^{a_3}T^{a_4})\delta_{i_1i_2}$,
as well as on Casimir invariants, which for our purposes will be only quadratic, i.e.
\begin{equation}
	T^a_{ij}T^a_{jk} = C_F \delta_{ik} \,,\qquad 
	f^{acd}f^{bcd} = C_A \delta^{a b} \,.
\end{equation}
In QCD, $C_A=N_c=3$ and $C_F=\frac{N_c^2-1}{2N_c}=4/3$.

For the Feynman diagram in Fig.~\ref{fig:col} modified by swapping the gluon legs, the colour chain $(T^{a_4}T^{a_3})_{i_1i_2}$ appears instead of $(T^{a_3}T^{a_4})_{i_1i_2}$.
Moreover, the diagrams with massless closed fermion loops are proportional to the number of corresponding quarks, i.e. $n_f=5$ in QCD.
Up to $L \leq 2$ loops, the QCD correction to the $q \bar{q} \to gg$ process involves monomials of degree $L$ in $\{C_A,C_F,T_F,n_f\}$.
Therefore, all the possible colour structures in the two-loop QCD amplitude for this process read
\begin{equation}
\begin{split}
	\mathcal{C}_c &\in \{ (T^{a_3}T^{a_4})_{i_1i_2} \,, (T^{a_4}T^{a_3})_{i_1i_2} \} \otimes \{ C_A^2, C_F^2, n_f^2, C_A C_F, C_A n_f, C_F n_f \} \\
	&\oplus \{ \Tr(T^{a_3}T^{a_4})\delta_{i_1i_2} \} \otimes \{ C_A, C_F, n_f \} \,.
\end{split}
\label{eq:color.chains}
\end{equation}
After extracting the colour structures $\mathcal{C}_c$, we can decompose our amplitude $\mathcal{A}$ in terms of colour-stripped amplitudes $A_c$
\begin{equation}
	\mathcal{A} = \sum_{c} \mathcal{C}_c A_c \,.
\end{equation}
Note that the fundamental colour chain basis in Eq.~\ref{eq:color.chains} is convenient to define colour-ordered scattering amplitudes.
It is explicit at tree-level, where the $\Tr(T^{a_3}T^{a_4})$ term vanishes, and one can collect Feynman diagrams into two colour-ordered subsets corresponding to
\begin{equation}
	\mathcal{A}_{\text{tree}}
	= (T^{a_3}T^{a_4})_{i_1i_2} \, A_{\text{tree}}(1_q,2_{\bar{q}},3_g,4_g)
	+ (T^{a_4}T^{a_3})_{i_1i_2} \, A_{\text{tree}}(1_q,2_{\bar{q}},4_g,3_g) \,.
\label{eq:tree.color}
\end{equation}
The colour-ordered amplitudes $A_{\text{tree}}$ are separately gauge invariant.
In fact, it is enough to directly compute only one of them, and the other can be related by a simple gluon exchange transformation.

\section{Lorentz tensor structure}
\label{sec:amp.Lorentz}

\begin{figure}[h]
	\centering
	\includegraphics[width=0.65\textwidth]{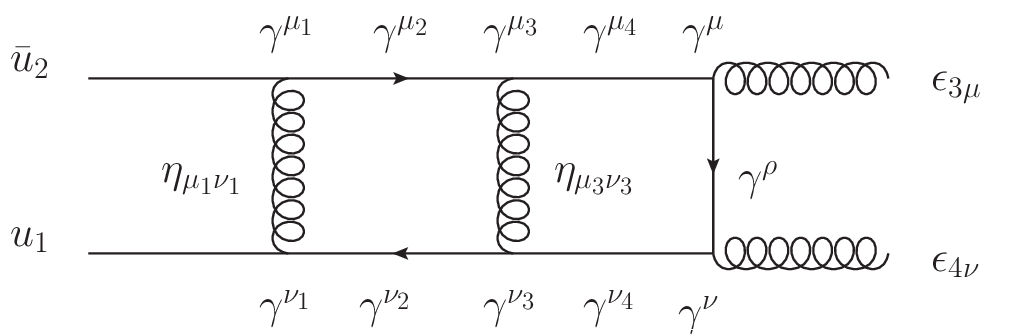}
	\caption{Lorentz tensor structures in example Feynman diagram for the  two-loop QCD $q \bar{q} \to gg$ amplitude. The open indices of $\gamma$ matrices are understood to be contracted with corresponding propagator momenta.}
	\label{fig:ten}
\end{figure}

\noindent
Secondly, we proceed to the description of the Lorentz tensor structure $T$ of the colour-stripped amplitude $A_c$.
It originates from gauge bosons, i.e. external polarization four-vectors, propagators, and vertices, as well as from fermions, i.e. external spinor states, propagators, and vertex Dirac matrices.
For instance, consider the two-loop $q \bar{q} \to gg$ process in QCD, with example Feynman diagram shown in Fig.~\ref{fig:ten}.
We can further decompose each of the corresponding colour-stripped amplitudes as
\begin{equation}
	A = \epsilon_{3,\mu} \epsilon_{4,\nu} \,\, \bar{u}_2 A^{\mu \nu} u_1 \,,
	\label{eq:ampten}
\end{equation}
where the amplitude tensor $A^{\mu \nu}$ consists of strings of Dirac $\gamma$ matrices.

\subsection{Independent Lorentz tensors}
\label{sec:tensors}

Contrarily to canonical textbook treatment relying on diagram-by-diagram analysis, beyond one loop, it is convenient to find all independent Lorentz tensors $\Gamma_i^{\mu\nu}$ depending only on external tensor indices of the process.
With this, we can write
\begin{equation}
	A^{\mu\nu} = \sum_{i=1}^{n_t} \mathcal F_i \, \Gamma_i^{\mu\nu} \,,
\end{equation}
where $\mathcal F_i$ are Lorentz scalar \textit{form factors}.
This decomposition is important for later steps in Feynman integral reduction, see Sec.~\ref{sec:amp.int.IBP}.

The number of independent Lorentz tensors $n_t$ in $d$ spacetime dimensions is process-dependent.
For example, at tree level $q \bar{q} \to gg$, it is possible to construct 12 tensors using the external momenta $\{p_i^\mu\}$, the metric tensor $g^{\mu\nu}$, and a single $\gamma$ matrix, provided that the external states are considered to be on-shell.
This means requiring transversality of massless gluons $p_3\cdot \epsilon_3 = 0$ and $p_4\cdot \epsilon_4 = 0$, quark states satisfying Dirac equations $\bar{u}_2 \, \slashed{p}_2 = 0$ and $\slashed{p}_1 u_1 = 0$, and with conserved momentum.
One may eliminate further redundancies by making a specific choice for the reference vectors of both external massless bosons e.g. by imposing
\begin{equation}
	\ep_3 \cdot p_2=0\,,\quad \ep_4 \cdot p_1=0\,,
	\label{eq:refmom}
\end{equation}
which is equivalent to choosing reference momenta of polarization vectors $\ep_3$ and $\ep_4$ to be $q_3=p_2$ and $q_4=p_1$, respectively.
This leaves us with 4 tensors.
Beyond tree level, there are also contributions from multiple strings of $\gamma$ matrices.
In $d$ dimensions, the Fierz basis grows, thus it is impossible to relate these strings to single $\gamma$ matrix tensors.
For the $q \bar{q} \to gg$ process, we have $n_t=5$ independent tensors in $d$ dimensions
\begin{align}
	\Gamma_1^{\mu \nu} &= p_2^{\nu} \gamma^{\mu}\,, \;\;\;\;\;
	\Gamma_2^{\mu \nu} = p_1^{\mu} \gamma^{\nu}\,, \nonumber \\
	\Gamma_3^{\mu \nu} &= p_1^{\mu} p_2^{\nu} \slashed{p}_3\,, \;\;
	\Gamma_4^{\mu \nu} = g^{\mu\nu}\slashed{p}_3\,, \nonumber \\
	\Gamma_5^{\mu \nu} &= \gamma^\mu \slashed{p}_3 \gamma^\nu \,,
	\label{eq:structs}
\end{align}
in agreement with Ref.~\cite{Peraro:2020sfm}.
It is worth stressing that Eq.~\ref{eq:structs} is valid at any perturbative order.
It is convenient to further refer to $T_i = \epsilon_{3,\mu} \epsilon_{4,\nu} \, \bar{u}_2 \Gamma_i^{\mu \nu} u_1$ as \textit{tensors}.

The $n_t=5$ does not coincide with the number of independent physical helicity states for the $q \bar{q} \to gg$ process.
Indeed, there are $(2^2 \cdot 2)/2 = 4$ independent helicity states in $d=4$ dimensions, since each massless gluon has 2 polarizations, while the 2 polarizations of massless fermion pair are related by parity.
This is consistent with the \textit{'t Hooft-Veltman scheme}, where all external particles are treated as four-dimensional.
Therefore, there is a redundancy in constructing the tensors $\Gamma^{\mu\nu}$ in $d$-dimensions.
According to a recently proposed argument in Refs~\cite{Peraro:2019cjj,Peraro:2020sfm}, it is possible to show which subset of $\Gamma^{\mu\nu}$ spans the whole space of purely four-dimensional tensors, as well as to directly project out the unphysical $-2\ep$-dimensional tensors.

Let us apply the method of Refs~\cite{Peraro:2019cjj,Peraro:2020sfm} to our $q \bar{q} \to gg$ example.
We introduce a new tensor basis $\oT_i$
\begin{equation}
	A = \sum_{i=1}^{n_t=5} \oF_i \, \oT_i\,,
	\label{eq:ampoT}
\end{equation}
where the first $\bar{n}_t=4$ tensors are identical to the ones introduced before
\begin{equation}
	\oT_i = T_i\,, \quad  i=1,\dots,\bar{n}_t \,.
	\label{eq:tens16}
\end{equation}
According to Ref.~\cite{Peraro:2020sfm}, these 4 tensors span the whole physical $d=4$ subspace, and they do not have any component in the $-2\ep$ directions.
The last tensor $\oT_5$ can then be chosen in such a way that it is constrained to live in the $-2\ep$ subspace.
This can be achieved by simply removing from the original $T_5$ its projection along $\oT_{1\dots4}$
\begin{equation}
	\oT_5 = T_5 - \sum_{j=1}^{\bar{n}_t=4}(\mathcal{P}_j T_5) \oT_j \,,
	\label{eq:tens7}
\end{equation}
where the projectors $\mathcal P_i$ are defined through
\begin{equation}
	\sum_{\rm pol} \mathcal P_i \oT_j = \delta_{ij} \,,
\end{equation}
and yield explicitly~\cite{Peraro:2020sfm}
\begin{equation}
	\mathcal P_i = \sum_{j=1}^{\bar{n}_t=4} \left( \sum_{\rm pol} \oT_i^\dagger \oT_j \right)^{-1}_{(4 \times 4)} \oT_j^\dagger \,,
\end{equation}
where
%\\
%\scalebox{0.9}{\parbox{1.0\linewidth}{
\begin{equation}
\renewcommand{\arraystretch}{1.2}
\left( \sum_{\rm pol} \oT_i^\dagger \oT_j \right)^{-1} =
\frac{1}{(3-d)t}
\left(
\begin{array}{ccccc}
	-\frac{u}{2 s^2} & 0 & \frac{u}{2 s^2 t} & 0 & 0 \\
	0 & -\frac{u}{2 s^2} & -\frac{u}{2 s^2 t} & 0 & 0 \\
	\frac{u}{2 s^2 t} & -\frac{u}{2 s^2 t} & -\frac{d u^2+4 s^2+4 s u}{2 s^2 u t^2} & -\frac{2 s+u}{2sut} & 0 \\
	0 & 0 & -\frac{2 s+u}{2sut} & -\frac{1}{2 u} & 0 \\
	0 & 0 & 0 & 0 & \frac{1}{4u\,\ep} \\
\end{array}
\right)
\end{equation}
%}}
%\\
%\noindent
Note the explicit decoupling of the unphysical $-2\ep$-dimensional block.
The new tensor $\oT_5$ reads
\begin{equation}
	\oT_5 = T_5
	- \frac{1}{s}\left( u(\oT_1-\oT_2) + 2\oT_3 + s\oT_4 \right) \,.
	\label{eq:barT}
\end{equation}
Constructed in this way tensor $\oT_5$ vanishes identically for all $\ep$ when evaluated at physical helicity states.
This suppresses the divergent $\ep^{-1}$ factor in the unphysical $-2\ep$-dimensional block.
Therefore, in the 't Hooft-Veltman scheme, the tensor $\oT_5$ can be dropped in the amplitude calculation.
In this manner, we restore the one-to-one correspondence between the physical amplitudes and the form factors $\oF_{1\dots4}$.

The polarization sums used to define the projectors $\mathcal P_i$ need have a structure consistent with our choice of the reference momenta in Eq.~\ref{eq:refmom}.
Thus, in the light-like axial gauge, they read
\begin{equation}
	\begin{split}
		\sum_{pol} u_1 \bar{u}_1 &= \slashed{p}_1 \,, \\
		\sum_{pol} u_2 \bar{u}_2 &= \slashed{p}_2 \,, \\
		\sum_{pol} \ep_3^{*\mu} \ep_3^\nu &= - g^{\mu\nu}
		+ \frac{p_2^\mu p_3^\nu + p_3^\mu p_2^\nu}{p_2 \cdot p_3} \,, \\
		\sum_{pol} \ep_4^{*\mu} \ep_4^\nu &= - g^{\mu\nu}
		+ \frac{p_1^\mu p_4^\nu + p_4^\mu p_1^\nu}{p_1 \cdot p_4} \,.
	\end{split}
	\label{eq:polsum}
\end{equation}
When performing polarization sums, the following identity for a trace of generic number of $\gamma$ matrices proves useful to be applied recursively
\begin{equation}
	\Tr(\gamma_{\mu_1} \cdots \gamma_{\mu_n}) = \sum_{i=2}^{n} g_{\mu_1\mu_i} \Tr(\gamma_{\mu_2} \cdots \gamma_{\mu_{i-1}} \gamma_{\mu_{i+1}} \cdots \gamma_{\mu_n}) \,.
	\label{eq:traces}
\end{equation}
The dimension of Lorentz indices $\mu_i$ is $d$ i.e. $g_\mu^\mu=d$, while the integer dimension of $\gamma$ matrices is $\Tr(\mathbb{I}) = 2^{\lfloor d/2 \rfloor}$.
In practice, setting $d=4$ in $\Tr(\mathbb{I})$ is sufficient because this trace factorizes from the whole amplitude.
The Dirac trace identities resulting from Eq.~\ref{eq:traces} have been implemented e.g. in the \texttt{FORM} program~\cite{Vermaseren:2000nd}, as well as in a \texttt{Mathematica} package \texttt{Tracer}~\cite{Jamin:1991dp}.

\subsection{Spinor-helicity formalism}
\label{sec:spinhel}

In order to efficiently evaluate Lorentz tensor structures $\oT_i$ on physical helicity states $\lambdavec$, we use the so called \textit{spinor-helicity formalism}.
We refer to the scattering amplitude evaluated at a fixed helicity state $\lambdavec$ as the \textit{helicity amplitude}
\begin{equation}
	A_\lambdavec = \sum_{i=1}^{\bar{n}_t} \oF_i \, \oT_{i,\lambdavec} \,.
	\label{eq:ampoTl}
\end{equation}
We provide here the definition of our notation, which agree with Ref.~\cite{Dixon:1996wi}.
Further details on the standard identities are provided in appendix~\ref{app:b}, following our notation.
These identites are algorithmic, and some of them have been implemented e.g. in a \texttt{Mathematica} package \texttt{Spinors}~\cite{Maitre:2007jq}.

We denote the polarized external fermionic states as
\begin{equation}
	\begin{split}
		u_+(p_i) &= \frac{1+\gamma_5}{2}u(p_i) \quad\,\, = v_-(p_i) = |p_i \rangle = |i^+ \rangle = |i \rangle \,, \\
		u_-(p_i) &= \frac{1-\gamma_5}{2}u(p_i) \quad\,\, = v_+(p_i) = |p_i ] = |i^- \rangle = |i ] \,, \\
		\bar{u}_+(p_i) &= u^\dagger(p_i)\gamma^0\frac{1-\gamma_5}{2} = \bar{v}_-(p_i) = [ p_i| = \langle i^+| = [ i| \,, \\
		\bar{u}_-(p_i) &= u^\dagger(p_i)\gamma^0\frac{1+\gamma_5}{2} = \bar{v}_+(p_i) = \langle p_i| = \langle i^-| = \langle i| \,,
	\end{split}
\end{equation}
and anti-fermionic states as $v_\pm(p_i)=u_\mp(p_i)$.
The spinor variables $\{\langle i|, |i \rangle, [ i|, |i ]\}$ are enough to completely parametrize the kinematics of physical helicity states.
We define the left-handed (or `-' helicity) current for a pair of incoming fermion $i$ and anti-fermion $j$ to be
\begin{equation}
	\mathcal{J}^\mu_L
	= \bar{u}_L(p_j) \gamma^\mu u_L(p_i)
	= \bar{u}_-(p_j) \gamma^\mu u_-(p_i)
	= \langle j \gamma^\mu i ] \,.
\end{equation}
The polarization vectors of a massless gauge bosons with momentum $p$ and a reference momentum $q$ read
\begin{equation}
	\epsilon^\mu_{-}(p,q) = - \frac{\langle p \gamma^\mu q ] }{ \sqrt{2} [ q p ]}\,, \qquad
	\epsilon^\mu_{+}(p,q) = \frac{[p \gamma^\mu q \rangle}{ \sqrt{2} \langle q p \rangle }
\end{equation}
with $\epsilon^*_\pm(p,q)=\epsilon_\mp(p,q)$.

As a result, we can express the helicity amplitude $A_\lambdavec$ in a compact form.
It is the most explicit at tree level, where for the colour-ordered amplitude defined in Eq.~\ref{eq:tree.color} is nonvanishing only in two helicity configurations
\begin{equation}
\begin{split}
	A(1_q^+,2_{\bar{q}}^-,3_g^+,4_g^-) &= i g_s^2
	\frac{\langle 24 \rangle^3 \langle 14 \rangle}{\langle 12 \rangle \langle 23 \rangle \langle 34 \rangle \langle 41 \rangle} \,, \\
	A(1_q^+,2_{\bar{q}}^-,3_g^-,4_g^+) &= i g_s^2
	\frac{\langle 23 \rangle^3 \langle 13 \rangle}{\langle 12 \rangle \langle 23 \rangle \langle 34 \rangle \langle 41 \rangle} \,.
\end{split}
\label{eq:tree.spin}
\end{equation}
They are referred to as \textit{maximally helicity violating} (MHV).
We point out that among major developments in simplifying higher-point tree level amplitudes, the momentum twistor parametrization is even less redundant then spinor-helicity~\footnote{
Indeed, as introduced in Ref.~\cite{Hodges:2009hk}, it implicitly implements Fierz identities between spinors.
In addition, it allows for a geometrical interpretation of the resulting amplitude expression, see review in Ref.~\cite{Elvang:2015rqa}.}.

An important consequence of expressing the helicity amplitude $A_\lambdavec$ in terms of the spinor-helicity variables is the explicit \textit{little group} covariance.
The little group is a subgroup of the Poincar\'e group, and it transforms only in the helicity space.
With the little group scaling
\begin{equation}
	|i \rangle \to \alpha^{-1/2} |i \rangle \,, \quad
	\langle i| \to \langle i| \alpha^{-1/2} \,, \quad
	|i ] \to \alpha^{1/2} |i ]  \,, \quad
	[ i| \to [ i| \alpha^{1/2} \,,
\end{equation}
one can extract from the amplitude $A_\lambdavec$ the helicity $\lambda_i$ of the particle $i$
\begin{equation}
	A_\lambdavec \to \alpha^{\lambda_i} A_\lambdavec \,.
\end{equation}

\section{Feynman integral structure}
\label{sec:amp.int}

\begin{figure}[h]
	\centering
	\includegraphics[width=0.6\textwidth]{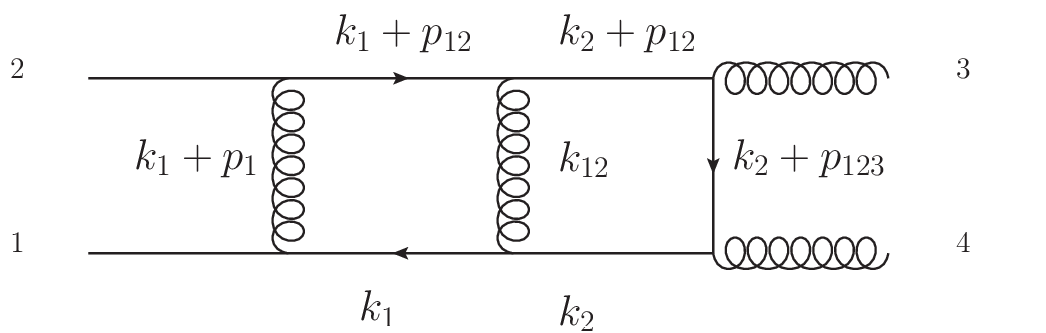}
	\caption{Feynman integral structure in example Feynman diagram for the two-loop $gg \to \gamma\gamma$ QCD amplitude.}
	\label{fig:prop}
\end{figure}

\noindent
We now describe the Feynman integral structure of the Lorentz scalar form factors $\oF_i$.
In our representation, the unconstrained intermediate continuous degrees of freedom are the \textit{loop momenta} $k_i$.
According to Sec.~\ref{sec:amp.overview}, integrating over them leads to the Feynman integral of the form
\begin{equation}
	\mathcal{I} = \int \frac{d^dk_1}{(2\pi)^d} \frac{d^dk_2}{(2\pi)^d} \,
	\frac{\mathcal N}{\mathcal D} \,.
\end{equation}
Consider for example the two-loop QCD $q\bar{q} \to gg$ process, with example Feynman diagram shown in Fig.~\ref{fig:prop}.
The denominator $\mathcal D$ of the corresponding Feynman integral $\mathcal{I}$ is a product of propagators of all intermediate virtual particles.
The numerator $\mathcal N$ depends on contractions of boson and fermion propagators with external states.
We will review now state-of-the-art methods of evaluating multiloop Feynman integrals.
For more details, see e.g. Refs~\cite{Smirnov:2006ry,Blumlein:2021ynm,Weinzierl:2022eaz}.

\subsection{Integral topologies}
\label{sec:amp.int.topo}

Consider again the example Feynman diagram in Fig.~\ref{fig:prop} for the two-loop massless $q\bar{q} \to gg$ process.
According to our discussion in Sec.~\ref{sec:tensors}, we can project out all independent Lorentz tensor structures, and obtain Lorentz scalar form factors for this Feynman diagram.
Consider one of these form factors
\begin{equation}
	\oF_{\text{FD}} =	\mathcal{P} \cdot \text{FD} = 
	\int \frac{d^dk_1}{(2\pi)^d} \frac{d^dk_2}{(2\pi)^d} \,
	\frac{\mathcal N_{\text{FD}}(d,s_{ij},\{p_i\cdot k_j\},\{k_i \cdot k_j\})}{\mathcal D_{1} \dots \mathcal D_{7}} \,,
\label{eq:INTbeforeLaporta}
\end{equation}
with denominator factors $\mathcal D_i$ belonging to the set of 7 propagators
\begin{equation}
\mathcal D_i \in \{(k_1)^2, (k_2)^2, (k_{12})^2, (k_1+p_1)^2, (k_1+p_{12})^2, (k_2+p_{12})^2, (k_2+p_{123})^2\} \,,
\end{equation}
as in Fig.~\ref{fig:prop}.
It is convenient to treat all Lorentz scalar products $\{p_i\cdot k_j\},\{k_i \cdot k_j\}$ in numerator $\mathcal N_{\text{FD},i}$ on the same footing as propagators $\mathcal D_i$.
Given the kinematics of the problem, there are 9 independent scalar products involving loop momenta i.e. 3 of the type $k_i \cdot k_j$, and 6 of the type $p_i \cdot k_j$.
These 9 scalar products can be linearly related to our 7 propagators $\mathcal D_i$ complemented with 2 so called \textit{irreducible scalar products} (ISPs) $k_2 \cdot p_1$ and $k_1 \cdot p_3$ which do not explicitly appear in the denominator $\mathcal D_{\text{FD}}$ of the Feynman diagram.
For convenience, we take linear combination of these ISPs such that they are of the propagator type, e.g. $\{\mathcal D_8=(k_2+p_1)^2, \mathcal D_9=(k_1+p_{123})^2\}$.
In this way, we define an \textit{integral topology}, i.e. a closed set of 9 independent generalised propagators.

With this definition, we can express all scalar products $\{p_i\cdot k_j\},\{k_i \cdot k_j\}$ in the numerator $\mathcal N_{\text{FD}}$ as linear combinations of the 9 generalised propagators $\mathcal D_i$ in our integral topology.
Therefore, we can linearly expand the considered form factor $\oF_{\text{FD}}$ in terms of purely propagator-type Lorentz-scalar Feynman integrals $\mathcal{I}_{\vec{n}}$ with coefficients $N_{\text{FD},\vec{n}}(d,s_{ij})$ independent of loop momenta, i.e.
\begin{equation}
	\oF_{\text{FD}} = \sum_{\vec{n} \, \in \, \text{int set}} \mathcal N_{\text{FD},\vec{n}}(d,s_{ij}) \, \mathcal{I}_{\vec{n}} \,.
\end{equation}
We introduced here the so called \textit{Laporta notation}
\begin{equation}
	\mathcal{I}_{\vec{n}} \equiv
	\int \frac{d^dk_1}{(2\pi)^d} \frac{d^dk_2}{(2\pi)^d} \,
	\frac{1}{\mathcal{D}_{1}^{n_1} \cdots \mathcal{D}_{9}^{n_9}} \,,
\end{equation}
where each Feynman integral $\mathcal{I}_{\vec{n}}$ in an integral topology is specified by its set of integer propagator exponents $\vec{n}$.
Contrarily to Eq.~\ref{eq:INTbeforeLaporta}, the exponents $n_i$ no longer have to be positive.
Indeed, scalar products arising from numerator $\mathcal N_{\text{FD}}$ are treated as denominators raised to negative powers.

\begin{figure}[h]
	\centering
	\includegraphics[width=0.9\textwidth]{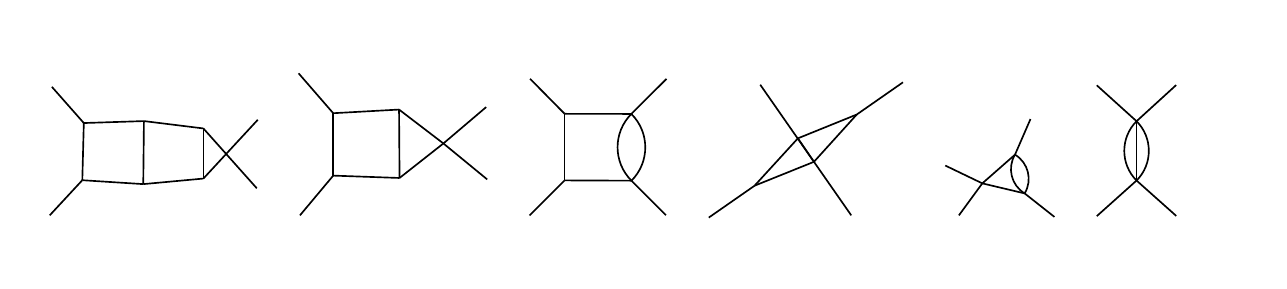}
	\caption{Example 1PI subsectors of the massless double-box integral.}
	\label{fig:subsec}
\end{figure}

At this point, it is convenient to introduce some standard quantities describing the properties of a Feynman integral $\mathcal{I}_{\vec{n}}$ for a given $\vec{n}$.
To setup the notation, we denote the total number of exponents in the integral topology as $N$, and we define the Heaviside step function $\Theta(x)$ such that $\Theta(x \leq 0)=0$ and $\Theta(x > 0)=1$.
In our two-loop four-point massless example, $N=9$.
A \textit{sector} $S$ is a binary representation of positive exponents in $\vec{n}$
\begin{equation}
	S = \sum_{i=1}^{N} \Theta(n_i) \cdot 2^{i-1} \,.
\end{equation}
It provides a systematic way to partially order all integrals in the topology.
Subsectors of sector $S$ correspond to Feynman integrals with a pinched subset of propagators with positive exponents, as depicted in Fig.~\ref{fig:subsec}.
Moreover, we denote by $r$ the total number of positive exponents, i.e.
\begin{equation}
	r = \sum_{i=1}^{N} \Theta(n_i) \,,
	\label{eq:rIBP}
\end{equation}
the sum of all negative exponents by $s$, i.e. 
\begin{equation}
	s = - \sum_{i=1}^{N} \Theta(-n_i) \cdot n_i \,,
	\label{eq:sIBP}
\end{equation}
and the sum of denominators raised to power greater then 1 by $d$, i.e. 
\begin{equation}
	d = \sum_{i=1}^{N} \Theta(n_i-1) \cdot (n_i-1) \,.
	\label{eq:dIBP}
\end{equation}
As mentioned above, $s$ arises from ISPs cancellations between numerator and denominator.
On the other hand, $d$ originates in Feynman diagrams with loop-corrected virtual propagators.
The larger the parameters $\{r,s,d\}$ are, the more complex the Feynman integral $\mathcal{I}_{\vec{n}}$ is.

We can extend our analysis to all Feynman diagrams contributing to a given scattering process, e.g. $q\bar{q} \to gg$ in two-loop QCD.
Indeed, one can find a minimal set of integral topologies onto which all relevant Feynman diagrams can be mapped
\begin{equation}
	A = \sum_{t \in \text{topo}} A_t \,.
\end{equation}
We note that for a given Feynman diagram, this mapping is not unique.
It is due to the fact that lower-sector Feynman diagrams can be subsectors of multiple integral topologies.
For example, the last integral in Fig.~\ref{fig:subsec}, the two-loop sunrise, is a subsector of the first, third, and fourth integral in Fig.~\ref{fig:subsec}.
For a generic two-loop four-point massless amplitude, there are 2 independent integral topologies, planar PL and nonplanar NPL
\begin{equation}
	\scriptsize
	\begin{split}
		\text{PL} &= \{(k_1)^2, (k_2)^2, (k_{12})^2, (k_1+p_1)^2, (k_2+p_1)^2, (k_1+p_{12})^2, (k_2+p_{12})^2, (k_1+p_{123})^2, (k_2+p_{123})^2\} \,, \\
		\text{NPL} &= \{(k_1)^2, (k_2)^2, (k_{12})^2, (k_1+p_1)^2, (k_2+p_1)^2, (k_1+p_{12})^2, (k_{12}-p_{3})^2, (k_2-p_{123})^2, (k_{12}+p_{12})^2\} \,.
	\end{split}
\end{equation}
Their names correspond to the genus $g$ of the underlying manifold they can be embedded in, here $g\in\{0,1\}$.
For fully-coloured external particles, there is a hierarchy in contributions from higher genus Feynman diagrams~\cite{tHooft:1973alw}.
They are suppressed as $\frac{1}{N_c^g}$, which can be used to classify the so called \textit{subleading colour} contributions.

In addition to the two independent integral topologies, PL and NPL, Feynman diagrams with crossed external kinematics are mapped onto crossed integral topologies e.g. PLx123=PL$|_{p_1 \to p_2 , p_2 \to p_3 , p_3 \to p_1}$ or PLx12x34=PL$|_{p_1 \to p_2 , p_2 \to p_1 , p_3 \to p_4 , p_4 \to p_3}$, abbreviated in the cyclic notation.
Since Feynman integrals are functions of kinematic invariants, the number of crossings is equal to the number of permutations of kinematic invariants.
In our two-loop four-point massless example, there are 6 independent crossings of each PL and NPL topologies, e.g. PL, PLx12, PLx123, PLx124, PLx1234, and PLx1243, making the total of 12 integral topologies for the process.
We also point out that with our definition of integral topology, in principle, multiple 7-propagator diagrams can belong to a common 9-propagator integral topology.
We will refer to these sets of 7 propagators as \textit{top sectors} of an integral topology.
\begin{figure}[h]
	\centering
	\includegraphics[width=0.9\textwidth]{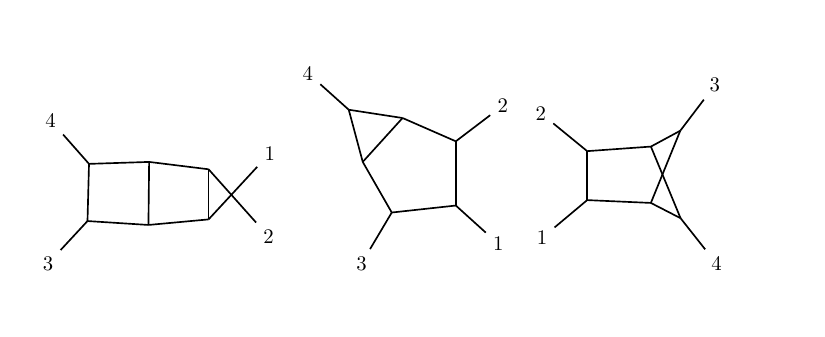}
	\caption{Example top sectors of integral topologies for a two-loop four-point massless amplitude. The first two belong to PL, while the last to NPL.}
	\label{fig:topsec}
\end{figure}
For example, in Fig.~\ref{fig:topsec}, we can see that our PL integral topology has 2 top sectors, while NPL has only 1.
We will now proceed to the methods of evaluating all the integrals $\mathcal{I}_{t,\vec{n}}$ required for the procees.

\subsection{Integration-by-parts identities}
\label{sec:amp.int.IBP}

Feynman integrals $\mathcal{I}_{\vec{n}}$ satisfy various properties.
Exploiting them significantly simplifies their calculation.
One of the simplest statements is that scaleless integrals vanish in the dimReg scheme.
Indeed, a Feynman integral $\mathcal{I}_{\vec{n}}$ has a mass dimension 
\begin{equation}
	[\mathcal{I}_{\vec{n}}]=Ld-2\sum_{i}n_i \,,
	\label{eq:massDim}
\end{equation}
following from a power counting argument.
Therefore, if there is no kinematic invariant in the considered sector, it is impossible to construct a mass dimensionful factor in a generic spacetime dimension $d$.
As a result, we can identify a vast number of \textit{zero sectors} in a given integral topology.
For example, the following two-loop double bubble integral is scaleless if $p_1^2=0$
\begin{align}
	\includegraphics[width=0.1\textwidth]{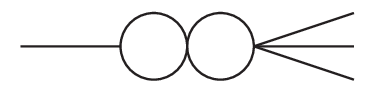} = 
	\int \frac{d^dk_1}{(2\pi)^d} \frac{d^dk_2}{(2\pi)^d}
	\frac{1}{(k_1)^2 (k_1+p_1)^2 (k_2)^2 (k_2-p_1)^2} \propto (p_1^2)^{d-4} = 0 \,.
\end{align}

Moreover, since Feynman integral $\mathcal{I}_{\vec{n}}$ has a uniform mass dimension $[\mathcal{I}_{\vec{n}}]$, one can factor it out with some arbitrary mass $M$
\begin{equation}
	\mathcal{I}_{\vec{n}}(s_{ij},m_k^2) = M^{[\mathcal{I}_{\vec{n}}]} \, \mathcal{\tilde{I}}_{\vec{n}}\left( \frac{s_{ij}}{M^2},\frac{m_k^2}{M^2} \right) \,,
\end{equation}
similarly as for the whole scattering amplitude in Eq.~\ref{eq:ampMassDimFact}.
This fact is reflected in the so called \textit{scaling relation}
\begin{equation}
	\left( \sum_{i<j} s_{ij} \, \partial_{s_{ij}} + \sum_k m_k^2 \, \partial_{m_k^2} \right) \mathcal{I}
	= \frac{1}{2}[\mathcal{I}] \, \mathcal{I} \,.
\label{eq:intScaleRel}
\end{equation}
As a consequence, a derivative of a Feynman integral in one of kinematics variables is redundant to all the other derivatives.

Furthermore, Feynman integrals are invariant under a shift of the loop momenta
\begin{equation}
	k_i \to k_i + q(p_j,k_l) \,.
	\label{eq:shiftLoopMom}
\end{equation}
The vanishing of corresponding infinitesimal transformation leads to the so called \textit{Integration-By-Parts identities} (IBPs)~\cite{Tkachov:1981wb,Chetyrkin:1981qh}
\begin{align}
	\int \frac{d^dk_1}{(2\pi)^d} \frac{d^dk_2}{(2\pi)^d}
	\frac{\partial}{\partial k^\mu_i}\frac{q^\mu}{\mathcal{D}_{t,1}^{n_1} \cdots \mathcal{D}_{t,9}^{n_9}} = 0\,.
	\label{eq:IBP}
\end{align}
They provide linear relations between different Feynman integrals.
For example, applying Eq.~\ref{eq:IBP} on the kite integral \includegraphics[width=0.1\textwidth]{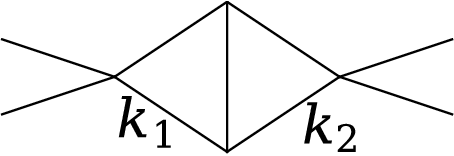} with $q=k_1-k_2$ leads to
\begin{figure}[H]
	\centering
	\includegraphics[width=0.7\textwidth]{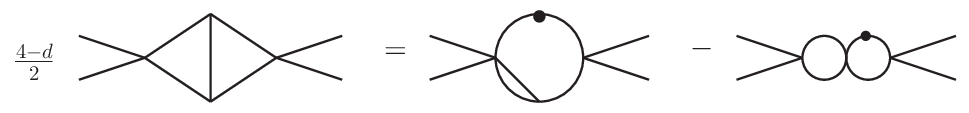}
\end{figure}
\noindent
In this instance, we have expressed a double triangle integral in terms of simpler double bubble interals for the cost of introducing squared propagators, indicated by a dot symbol.
The procedure of linearly relating all Feynman integrals in the problem to a minimal linearly independent set of so called \textit{Master Integrals} (MIs) $M_i$ is referred to as the \textit{integral reduction}, or simply the \textit{IBP reduction}.
It is one of the most intricate bottlenecks in modern scattering amplitude calculations.

In order to perform the IBP reduction in practice, one needs to guarantee closure of the system of Feynman integrals generated with IBPs in Eq.~\ref{eq:IBP}.
It can be achieved by exploiting the fact, that Lorentz-scalar Feynman integrals are also invariant under Lorentz transformation of the external momenta~\cite{Gehrmann:1999as}.
\begin{equation}
	p_i^\mu \to \Lambda^\mu_{\,\,\,\,\nu} \, p_i^\nu \,.
	\label{eq:shiftExtMom}
\end{equation}
Since the infinitesimal Lorentz parameters are antisymmetric $\omega^\mu_{\,\,\,\,\nu} = - \omega_\mu^{\,\,\,\,\nu}$, one arrives at the so called \textit{Lorentz Invariance identities} (LIs)
\begin{align}
	p_{j\,\mu} \, p_{l\,\nu} \sum_{i=1}^3 \left( p_i^\nu \frac{\partial}{\partial p_{i,\mu}} - p_i^\mu \frac{\partial}{\partial p_{i,\nu}} \right)	\mathcal{I}_{t,\vec{n}} = 0 \,.
	\label{eq:LI}
\end{align}
In Ref.~\cite{Laporta:2000dsw}, it has been proven, that when following the so called \textit{Laporta algorithm}, IBPs and LIs generate enough identities to close the linear system of integrals.
There are multiple public implementations of the Laporta algorithm, most recently available in e.g. \texttt{reduze}~\cite{Studerus:2009ye,vonManteuffel:2012np} and \texttt{kira}~\cite{Maierhofer:2017gsa,Maierhofer:2018gpa} programs, as well as in \texttt{Mathematica} packages, \texttt{LiteRed}~\cite{Lee:2013mka} and \texttt{FIRE}~\cite{Smirnov:2019qkx}.
As an example, we can IBP reduce the kite integral 
\includegraphics[width=0.1\textwidth]{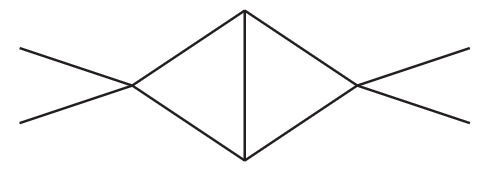}
onto a linearly independent basis of 2 MI bubbles without any squared propagators.
\begin{figure}[H]
	\centering
	\includegraphics[width=0.7\textwidth]{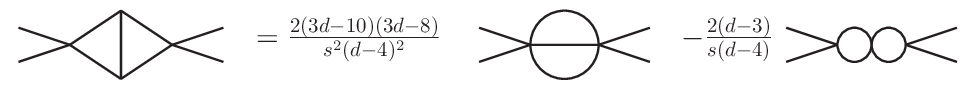}
\end{figure}
\noindent
In general, the coefficients $c_i$ of MIs $M_i$ are algebraic functions of kinematic invariants and the spacetime dimension $d$.
We point out that in a modern mathematical language, these coefficients may be interpreted as intersection numbers in a cohomology generated by the total derivative in the IBP Eq.~\ref{eq:IBP}~\footnote{As introduced in detail in Refs~\cite{Mizera:2017rqa,Mastrolia:2018uzb}, this intersection number is a specific type of a scalar product in a space spanned by the MIs.
The corresponding cohomology is an equivalence class of integrands to which a total derivative can be added.
Importantly, this intersection number can be computed with a direct recursive residue-based formula, rather than by solving a system of linear equations}.

It is important to note that the IBP identities in Eq.~\ref{eq:IBP} hold only for Lorentz scalar Feynman integrals.
For this reason, the tensor reduction, as described in Sec.~\ref{sec:amp.Lorentz}, is necessary for further simplification in the integral structure.
Importantly, the textbook approach of expressing tensors of the Feynman integrand in terms of scalars using symmetry arguments does not hold beyond one-loop.
It is due to the existence of ISPs.
We note that recently, there have been developments in extending the standard one-loop Passarino-Veltman reduction to higher orders~\footnote{As detailed in Ref.~\cite{Feng:2022uqp}, this method introduces an auxiliary vector to be contracted with any remaining open Lorentz index.
It provides closed formulae for higher-tensor integral reduction using a recursive approach.}.

\subsection{Divergences in Feynman integrals}
\label{sec:amp.int.div}

Before turning to the evaluation of MIs, we need to understand their divergent structure.
A Feynman integral $\I_{\vec{n}}(\ep,s_{ij},m_k)$ can have divergences stemming from some region dominated by a particular limiting behaviour of the virtual loop momentum $k$, or the real kinematic invariants $s_{ij}$ and $m_k$.
We call these divergences \textit{virtual} and \textit{real}, respectively.
For our purposes, we need to classify the virtual divergences into those stemming from the \textit{ultraviolet} (UV) region $k \to \infty$, and the \textit{infrared} (IR) region $k \to 0$.
They appear at the $k$-integrated level as poles in the dimensional regulator $\ep = \frac{4-d}{2}$.

Consider an example one-loop subsector of some $L$-loop integral with massless external legs $p_i$
\begin{align}
	\int \frac{d^dk}{(2\pi)^d}
	\frac{1}{ ((k-p_1)^2)^{n_2} \, (k^2)^{n_1} \, ((k+p_2)^2)^{n_3} \, ((k+p_{23})^2)^{n_4} \cdots } \,.
	\label{eq:ampDivEx}
\end{align}
In the limit $k \to \infty$, the dominant term yields
\begin{align}
	\int \frac{d^dk}{(2\pi)^d}
	\frac{1}{ (k^2)^{n_1} \, (k^2)^{n_2} \, (k^2)^{n_3} \, (k^2)^{n_4} \cdots } \,.
\end{align}
We can see that the integral is UV divergent if its \textit{superficial degree of divergence}
\begin{align}
	\text{SDOD} = [\mathcal{I}_{\vec{n}}]_{\ep=0}=4L-2\sum_{i}n_i
\end{align}
is non-negative.
In our one-loop example in Eq.~\ref{eq:ampDivEx}, this would be the case if $\sum_{i}n_i \leq 2$, e.g. for a bubble or a tadpole.
Since we can have one such UV divergence per loop order, we can have at most an $\ep^{-L}$ UV pole at $L$-loops.
We classify the UV divergence as logarithmic, linear, quadratic, etc. if SDOD=0,1,2,..., respectively.
In order to assure that an integral is UV finite, the SDOD of all its subsectors must be negative.

On the other hand, if the limit where all the denominators are small, in our example one-loop integral in Eq.~\ref{eq:ampDivEx}, the leading approximation is
\begin{align}
	\int \frac{d^dk}{(2\pi)^d}
	\frac{1}{ (-2k \cdot p_1)^{n_1} \, (k^2)^{n_2} \, (2k \cdot p_2)^{n_3} \, (p_{23}^2)^{n_4} \cdots } \,.
\end{align}
Contrarily to the UV divergences, the power-counting criterion SDOD~$\leq$~0 is a necessary but no longer sufficient condition for an integrals to be IR divergent.
In fact, one has to analyse all the subsectors of the integral.
In general, there are two different types of the arising IR divergences.
Indeed, each of the denominators of the type
$k \cdot q = |\vec{k}||\vec{q}|\left(1-\cos(\angle(\vec{k},\vec{q}))\right)$
can approach zero either at small energy of the loop particle $|\vec{k}| \to 0$, or at small angle $\angle(\vec{k},\vec{q}) \to 0$.
These behaviours are referred to as the \textit{soft}, and \textit{collinear} IR divergence, respectively.
Since both of these two effects are independent, they enhance the IR pole from single $\ep^{-1}$ to double $\ep^{-2}$.
Therefore, a $L$-loop Feynman integral can have up to $\ep^{-2L}$ IR pole.

In general, finding all the divergences of a Feynman integral is algorithmic.
For example, it was implemented in the \texttt{pySecDec} program~\cite{Borowka:2017idc,Borowka:2018goh} which is based on the \textit{sector decomposition} method.
Recently, an equivalent formulation based on tropical geometry was implemented in the program \texttt{feyntrop}~\cite{Borinsky:2023jdv}~\footnote{As introduced in Ref.~\cite{Arkani-Hamed:2022cqe}, this approach provides a geometrical interpretation to the rate of shrinking and expanding of the Schwinger parameters associated with integral propagators.}.

\subsection{Differential equations for Master Integrals}
\label{sec:amp.int.DEQ}

\begin{figure}[h]
	\centering
	\includegraphics[width=0.99\textwidth]{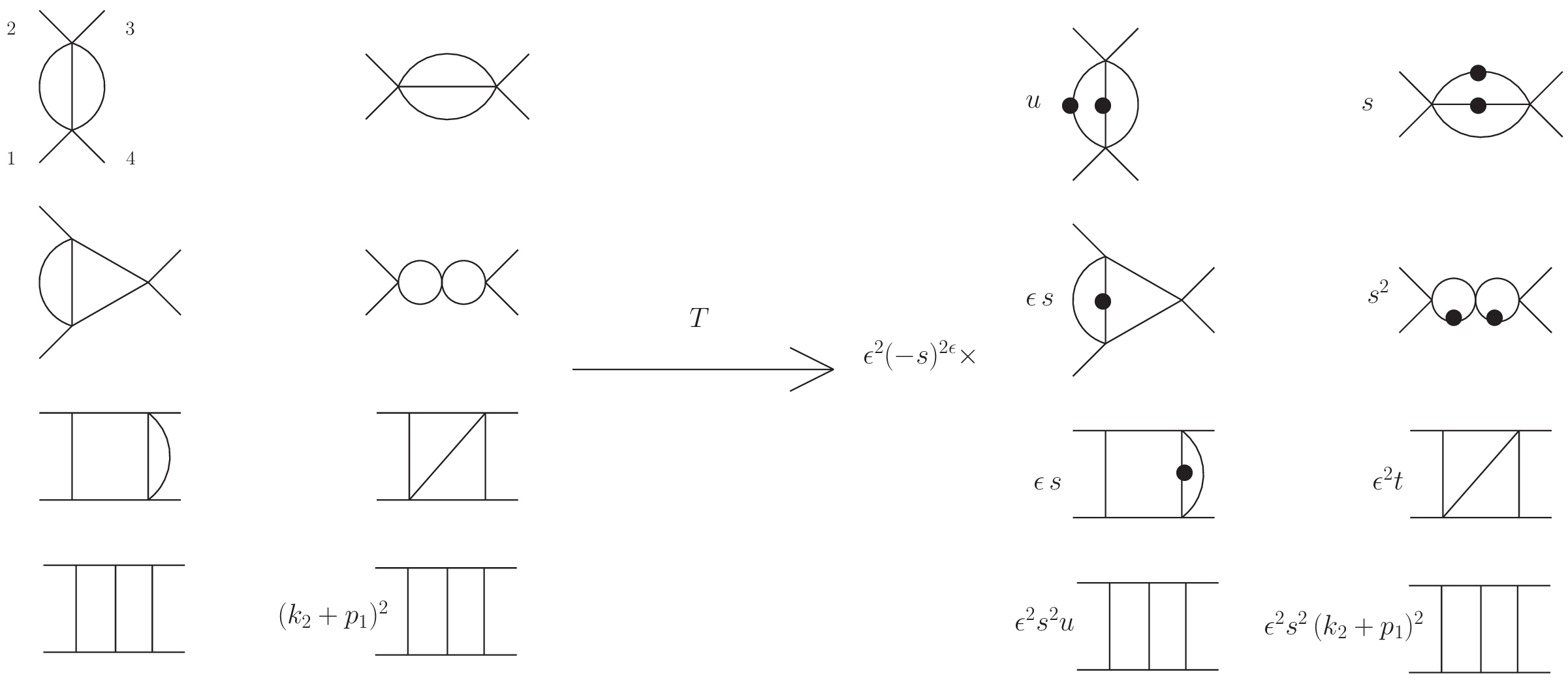}
	\caption{Integrands of MIs $M_i$ and $g_i$ for the PL topology before and after a canonical transformation $T$, respectively.}
	\label{fig:2L4p0mMIs}
\end{figure}

\noindent
After performing the IBP reduction of all Feynman integrals $\mathcal{I}_{t,\vec{n}}$ in the scattering problem, we are left with a smaller set of associated MIs $M_{t,i}$.
In general, finding the complete dependence of MIs on kinematic invariants $s_{ij}$ and $m_k$, as well as the spacetime dimension $d$, is an open problem in Mathematical Physics.
The difficulty originates in integrating complicated multivariate algebraic functions.
In order to postpone this cumbersome task, we will yet again use some general properties of Feynman integrals in order to simplify the problem.

Let us fix a top sector in a given integral topology $t$.
For instance, we will analyse the double box top sector in the planar topology PL, following Ref.~\cite{Henn:2014qga}.
An example set of MIs $M_i$ resulting from the IBP reduction is shown on the left-hand side of Fig.~\ref{fig:2L4p0mMIs}.
Since a sector of Feynman integrals is closed under the kinematic derivative operation, we can IBP reduce the derivatives of MIs, and express them as linear combinations of themselves.
In this way, we construct a first-order homogeneous linear differential equation (DEQ) satisfied by these MIs 
\begin{equation}
	\partial_x M_i(x,\ep) = b_{ij}(x,\ep) \, M_j(x,\ep) \,,
	\label{eq:deqMI}
\end{equation}
where $x$ was defined in Eq.~\ref{eq:x}.
This DEQ is impossible to solve analytically due to an intertwined dependence on both kinematic variable $x$ and spacetime dimension $d$ in the matrix $b$.
However, since in physical applications we are interested only in the solution for $M_i(x,\ep)$ expanded around $\ep=0$, it is convenient to transform the DEQ~\ref{eq:deqMI} into an $\ep$-factorized form.
Before elaborating on how to find it, let us further motivate seeking this transformation.

We transform the DEQ~\ref{eq:deqMI} with a change of basis
\begin{equation}
	M_i(x,\ep) = T_{ij}(x,\ep) \, g_j(x,\ep) \qquad \Rightarrow \qquad
	a(x) = T^{-1} \, b \, T - T^{-1} \, \partial_x T \,.
	\label{eq:deqCoB}
\end{equation}
According to Ref.~\cite{Henn:2013pwa}, one can find a specific transformation matrix $T$ which leads to the so called \textit{canonical system} of DEQ
\begin{equation}
	\partial_x g_i(x,\ep) = \ep \, a_{ij}(x) \, g_j(x,\ep) 
	= \ep \left( \frac{a_{0,ij}}{x} + \frac{a_{1,ij}}{1-x} \right) \, g_j(x,\ep) \,.
	\label{eq:deqCan}
\end{equation}
The integrals $g_i$ satisfying this DEQ form a \textit{canonical basis} of MIs, see the right-had side of Fig.~\ref{fig:2L4p0mMIs}.
Contrarily to DEQ~\ref{eq:deqMI}, the canonical DEQ~\ref{eq:deqCan} can be solved analytically order by order in $\ep$, and formally written as
\begin{equation}
\begin{split}
	g_i(x,\ep) &= \mathbb{P} e^{\ep \int_{x_0}^x dx' \, a_{ij}(x')} g_j(x_0,\ep) \\
	&= \left( \delta_{ij} + \ep \int_{x_0}^x dx_1 \, a_{ik}(x_1) \left( \delta_{kj} + \ep \int_{x_0}^{x_1} dx_2 \, a_{kj}(x_2) \right) + \mathcal{O}(\ep^3) \right) g_j(x_0,\ep) \,,
	\label{eq:deqGenSol}
\end{split}
\end{equation}
where the symbol $\mathbb{P}$ effectively discards $\frac{1}{n!}$ prefactor in the exponential series.
It is important to stress, that if one finds the transformation matrix $T$, then the problem of computing Feynman integrals reduces to evaluating only the boundary Feynman integrals $g_j(x_0,\ep)$, called \textit{boundary conditions} (BCs).
We note that for many known applications to various scattering processes, finding such a transformation matrix $T$ is indeed possible.
For processes with more scales, the canonical DEQ~\ref{eq:deqCan} generalises to be multivariate.
Moreover, for some of the complicated processes, an additional matrix is required, i.e.
\begin{equation}
	d g(\vec{x},\ep) = \left( \ep \, a(\vec{x}) + \tilde{a}(\vec{x}) \right) \, g(\vec{x}_0,\ep) \,.
	\label{eq:deqCanGen}
\end{equation}
This equation also can be solved perturbatively in $\ep$.

In order to understand how to find the canonical transformation matrix $T$, let us analyse the properties of the canonical basis $g_i$ in Fig.~\ref{fig:2L4p0mMIs}.
Firstly, we dress each one-loop bubble subsector with a single dot, i.e. a squared propagator, thus removing all the UV divergences.
Secondly, consider the $\ep^n$ prefactors.
They are introduced to cancel the highest $\ep^{-n}$ pole of each canonical MI.
In this way, different orders in $\ep$ of the canonical MIs do not interfere in the canonical DEQ.
Thirdly, the overall $\ep$-dependent part of the mass dimension $(-s)^{-2\ep}$ is factored out.
This allows not to propagate the redundant $\log^n(-s)$ terms in the expansion.
Fourthly, the four-dimensional part of the mass dimension of each integral is factored out based on the channel structure.
For example, the first diagram is a $u$-channel bubble, so we cancel its $u$ pole with a prefactor $u$.
The next-to-last diagram has a double ladder in the $s$ channel, and a single ladder in the $u$ channel, so the prefactor which cancels its poles is $s^2u$.
For more complicated diagrams, such a prefactor can be found by analysing their leading singularity.
Finally, the numerator of the last integrand $(k_2+p_1)^2$ plays the role of the $u$ prefactor in the next-to-last double-box diagram.

We can frame all these properties in a more mathematical language.
While the MI coefficients $c_i$ are algebraic functions, the MIs $M_i$ are \textit{transcendental functions}, i.e. they do not satisfy any polynomial equation.
In our two-loop four-point massless example, the simplest MI is the two-loop bubble in the $s$ channel
\begin{equation}
	M_{\text{2Lbub}}(s_{12})
	= \int \frac{d^dk_1}{i \pi^{d/2}} \frac{d^dk_2}{i \pi^{d/2}} \frac{1}{(k_1+p_{12})^2 \, k_2^2 \, k_{12}^2}
	= - (-s_{12})^{1-2\ep} \frac{\Gamma^3(1-\ep) \Gamma(2\ep-1) }{\Gamma(3-3\ep)} \,.
\end{equation}
Beyond the above power function of the form $x^\ep$, an example transcendental function arises after expanding it in $\ep$, i.e. $\log^n(x)$.
Following Ref.~\cite{Henn:2013pwa}, we can introduce a \textit{degree of transcendentality} $\mathcal{T}$, such that e.g. $\mathcal{T}(\log^n x)=n$ and $\mathcal{T}(\pi^n)=n$.
For convenience, we can also assign $\mathcal{T}(\ep)=-1$, such that $\mathcal{T}(x^\ep)=0$ to all orders in $\ep$.
A transcendental function is \textit{uniform} if all its summands have equal degree of transcendentality.
We also call a transcendental function $f$ \textit{pure} if $\mathcal{T}(df)=\mathcal{T}(f)-1$, i.e. is it uniform and its summands do not have any algebraic factors.
Importantly, the canonical MIs can be found by requiring that each $g_i$ is pure.

According to Ref.~\cite{Henn:2013pwa}, finding the transformation matrix $T$ is also equivalent to transforming our original DEQ~\ref{eq:deqMI} into the so called \textit{dlog} form
\begin{equation}
	d g(x,\ep) = \ep \left( a_\infty \, d\log(s_{12})
	+ a_1 \, d\log(s_{23})
	+ a_0 \, d\log(s_{13}) \right) \, g(x,\ep) \,,
	\label{eq:deq2Ldlog}
\end{equation}
with constant matrices $a_0$, $a_1$, and $a_\infty$.
Note, that the matrix $a_\infty$ is redundant to matrices $a_0$ and $a_1$.
This is a consequence of the scaling relation defined in Eq.~\ref{eq:intScaleRel}.
In more involved integral topologies, constructing the canonical basis may require multiple prefactors in the linear combination of scalar Feynman integrals, in comparison to our single-prefactor example in Fig.~\ref{fig:2L4p0mMIs}.
For simple topologies, the procedure of finding a dlog basis is algorithmic, and it is implemented in e.g. public packages to \texttt{Mathematica}, \texttt{DlogBasis}~\cite{Henn:2020lye} and \texttt{INITIAL}~\cite{Dlapa:2020cwj,Dlapa:2022wdu}.
For a more detailed introduction to the method of DEQ see Ref.~\cite{Henn:2014qga}.

\subsection{Iterated integrals}
\label{sec:amp.int.GPL}

We now introduce the types of special functions which solve a general canonical DEQ~\ref{eq:deqCan}.
A \textit{Chen's iterated integral}~\cite{Chen:1977oja}
\begin{equation}
	G(f_n,\dots,f_1;x) = \int_{x_n}^x dz \, f_n(z) \, G(f_{n-1},\dots,f_1;z) \,, \qquad
	G(x) = 1
	\label{eq:Chen}
\end{equation}
is defined recursively with general kernel functions $f_i(x)$, and a boundary $x_i$ of the integration interval, appropriate for a convergent definition.
It generates $\ep$-perturbative solutions to a canonical DEQ
\begin{equation}
	 \partial_x g(x,\ep) = \ep \left( \sum_{j \, : \, \beta_j \in \text{alphabet}} a_j \, \partial_x \log(\beta_j(x)) \right) \, g(x,\ep) \,,
	\label{eq:deqAlphabet}
\end{equation}
where $a_j$ are constant matrices, $\beta_j(x)$ are the so called \textit{letters} forming an \textit{alphabet}, specific for a corresponding integral topology, and the kernels $f_i(x) \in \{ \partial_x \log(\beta_j(x)) \,\, | \,\, \beta_j \in \text{alphabet}\}$.
In general, the letters $\beta_j(x)$ are algebraic functions.
There are different types of iterated integrals corresponding to specific types of polynomial equation satisfied by $\beta_j(x)$.
For example, if there are non-linearisable square roots of degree three or four in the alphabet, we refer to the iterated integral as \textit{elliptic}.
For our discussion, it is enough to focus on a linear alphabet, which generates the so called \textit{Generalised} or \textit{Goncharov Polylogarithms} (GPLs)~\cite{Goncharov:1998kja,Goncharov:2001iea}
\begin{equation}
	G(\alpha_n,\dots,\alpha_1;x) = \int_0^x \frac{dz}{z-\alpha_n} G(\alpha_{n-1},\dots,\alpha_1;z)\,,\,\,
	G(\underbrace{0,\dots,0}_{n};x) \equiv \frac{\ln^n x }{n!} \,,\,\,
	G(x)=1	
	\label{eq:GPL}
\end{equation}
where $\alpha_i$ are constant terms in kernels $f_i(x) = \partial_x \log(x-\alpha_i)$.
The number of letters $\alpha_i$ in a GPL is called its $weight$, and it coincides with its transcendental weight $\T$.

Iterated functions obey a so called \textit{shuffle algebra}.
For GPLs, it coincides with the \textit{Hopf algebra}~\footnote{This correspondence has been proven in Ref.~\cite{Goncharov:2005sla} and it allowed for a rigorous derivation of functional identities between GPLs.}
\begin{equation}
	G(\vec{\alpha};x) G(\vec{\beta};x) = \sum_{\vec{\gamma} \in \vec{\alpha} \sqcup \vec{\beta}} G(\vec{\gamma};x) \,,
	\label{eq:Hopf}
\end{equation}
where the sum runs over all shuffles i.e. permutations of $\alpha_i$ and $\beta_j$ preserving the order within $\vec{\alpha}$ and $\vec{\beta}$.
They lead to a lot of powerful weight-dropping identities e.g.
\begin{equation}
	G(\beta,\alpha,\alpha;x) - G(\alpha,\alpha,\beta;x) = 
	\frac{1}{2} G(\alpha;x)^2 G(\beta;x) - G(\alpha;x) G(\alpha,\beta;x) \,.
\end{equation}
It is clear that the Hopf algebra is closed within a fixed alphabet.
One of the most studied subalgebra of GPLs are the \textit{Harmonic Polylogarithms} (HPLs)~\cite{Remiddi:1999ew}, generated by an alphabet $\{-1,0,1\}$.~\footnote{In the literature, they are sometimes defined with a relative overall sign $(-1)^{\#(1s \,\text{in a GPL})}$.}
HPLs are also a generalisation of ordinary \textit{polylogarithms}, which are defined iteratively via
\begin{equation}
	\Li_n(x) = \int_{0}^{x} \frac{dz}{z} \Li_{n-1}(z) \,, \qquad
	\Li_1(x) = -\log(1-z) \,.
	\label{eq:Li}
\end{equation}
Indeed, we have
\begin{equation}
	\Li_{n+1}(x) = -G(\underbrace{0,\dots,0}_{n},1;x) \,.
\end{equation}
Note the well-known special values of polylogarithms
\begin{equation}
	\Li_{n}(1) = \zeta(n) \,.
\end{equation}

In addition, it is sometimes possible to express a GPL depending on an argument $y(x)$ in terms of GPLs depending explicitly on $x$.
This procedure is called \textit{fibration}, and it is algorithmic only for simple functions $y(x)$.
An example identity yields
\begin{equation}
	G(0,1;x) + G(0,1;1-x) = G(0;x) G(1;x) - \zeta(2) \,.
\end{equation}
When crossing branch cuts while changing variables, it is important to perform the analytic continuation correctly.
The analytic structure of a GPL is determined by its first inner letter $\alpha_1$, since further integrations do not change the location of the branch cut.
Hence, for example, $G(0;x)=\log(x)$ has a branch cut along $x \in (-\infty,0)$, while $G(1;x)=\log(1-x)$ has a branch cut along $x \in (1,\infty)$.
Depending on the sign of the imaginary part of $x$, one can arrive at different Riemann sheets, e.g.
\begin{equation}
	\log(-x \pm i\varepsilon) = \log(x e^{\pm i \pi}) = \log(x) \pm i\pi \,.
\end{equation}
Due to the ambiguity in the imaginary part, it is important not to cross more then one branch cut per a change of variables.
Note, that in more complicated identities, powers of $\pi^2$ may come from either analytic continuation $(i\pi)^{2n}$ or special values of HPLs e.g. $\zeta(2)=\frac{\pi^2}{6}$.
Some of the above identities have been algorithmically implemented e.g. in Mathematica packages \texttt{HPL}~\cite{Maitre:2005uu} and \texttt{PolyLogTools}~\cite{Duhr:2019tlz}.
For a detailed review of the properties of GPLs, see Ref.~\cite{Duhr:2019tlz}.

\subsection{Boundary Master Integrals}
\label{sec:amp.int.BC}

After discussing the types of functions appearing in a general perturbative solution to the canonical DEQ, introduced in Eq.~\ref{eq:deqGenSol}, we turn to solving a boundary value problem, specified by BCs $g_i(x_0,\ep)$.
For simple integral topologies, the BCs can be evaluated with direct integration methods, which we discuss in the next section.
However, for more complicated topologies, evaluating a large number of corresponding BCs may not be the most efficient.
Thankfully, we can again use the properties of Feynman integrals in order to overcome the complexity of the problem.
Indeed, we can require some \textit{regularity constraints} on our canonical MIs $g_i(x,\ep)$.

One type of the regularity constraints is to allow only kinematic branch cuts corresponding to physical multiparticle production channels, by the virtue of the optical theorem.
For example, a planar massless $s$-$u$ box cannot develop a branch cut in the $t$ channel, which imposes a constraint on the corresponding BC.
However, this argument does not suffice to constrain nonplanar integrals, since they can have branch cuts in all kinematic channels.

A more universal regularity constraint is based on a limiting behaviour of canonical MIs.
Consider for example the canonical DEQ in its dlog form for the planar double box topology in Eq.~\ref{eq:deq2Ldlog}.
In the limit of a small Mandelstam variable, we can solve it order by order in $\ep$, e.g. in the $t$ channel we have
\begin{equation}
	g(x,\ep) \overset{x \to 0}{\longrightarrow} x^{a_0 \, \ep} g(x \to 0,\ep) \,.
\end{equation}
Requiring this limiting solution to be regular is usually referred to as the \textit{UV constraint}, which refers to the suppression of the corresponding Mandelstam variable $t$ in a presence of some UV scale.
In general, the exponents $\kappa_m$ of eigenvalues $x^{\kappa_m \ep}$ of the matrix $x^{a_0 \, \ep}$ can have different signs.
Basing on the mass dimension of any Feynman integral in dimReg with $\ep>0$, only the negative exponents $\kappa_m<0$ are expected in the result~\cite{Henn:2020lye}.
This implies that the eigenvectors corresponding to positive exponents $\kappa_m>0$ must lead to a vanishing combination of $g_i$.
It imposes linear relations between the constants $c_{i,n}$ in $g_i(x_0,\ep) = \sum_{n=0} \ep^n \, c_{i,n}$.
A similar argument proceeds for all letters of the DEQ alphabet becoming small.
Importantly, this constraint holds to all orders in $\ep$ in each crossed channel.

For our two-loop four-point massless example, one can show that all the relevant BCs can be related to only one independent BC, which can be chosen to be a simple two-loop bubble integral.
Remarkably, this holds also at three loops, as we elaborate on in Sec.~\ref{sec:ggaa.calc.MI}.
Let us stress here how profound this statement is.
Indeed, among all the Lorentz-tensor Feynman integrals appearing in the scattering amplitude of any one, two, or three-loop four-point massless process, there is only one per loop order simple overall normalization boundary integral to be computed with a direct integration method.

\subsection{Direct integration with Symanzik polynomials}
\label{sec:amp.int.Symanzik}

Finally, we can no longer postpone evaluating some leftover independent Feynman integrals directly.
This stage is difficult not only because of the complexity of multi-loop multi-scale integrands, but also due to the lack of a unique algorithmic approach, as in any general integration problem.
Nonetheless, throughout the years, multiple methods have been successfully applied for specific classes of integrals.

As a starting point, we choose some parametrization of a scalar Feynman integral in order to avoid vector integration variables $k_l^\mu$.
We can write each denominator in terms of an integral over the so called \textit{alpha} or \textit{Schwinger parameter}
\begin{equation}
	\frac{1}{\mathcal{D}^{n}}
	= \frac{1}{(n-1)!} \int_{0}^{\infty} d\alpha \, \alpha^{n-1}
	e^{-\alpha\mathcal{D}} \,,
	\label{eq:SchwingerParams}
\end{equation}
which is also motivated by the worldline formulation of QFT, see e.g. a review in Ref.~\cite{Schubert:2001he}.
As a consequence, we can combine denominators together with the \textit{Feynman parametrization}
\begin{equation}
	\frac{1}{\mathcal{D}_{1}^{n_1} \mathcal{D}_{2}^{n_2}}
	= \int_{0}^{1} dx_1 dx_2 \, x_1^{n_1-1} x_2^{n_2-1} 
	\frac{\delta(1-x_1-x_2)}{(x_1\mathcal{D}_1+x_2\mathcal{D}_2)^{n_1+n_2}} \,.
	\label{eq:FeymParams}
\end{equation}
For completeness, we also mention a few other useful parametrizations, which we will not discuss in details here.
The \textit{Mellin-Barnes parametrization}
\begin{equation}
	\frac{1}{(k^2-m^2)^n}
	= \frac{1}{2 \pi i \Gamma(n)} \int_{-i\infty}^{i\infty} dz \, \Gamma(n+z) \Gamma(-z) \frac{(-m^2)^z}{(k^2)^{n+z}}
	\label{eq:MBparams}
\end{equation}
is useful for direct integration at fixed order in $\ep$, as automated in a \texttt{Mathematica} package \texttt{MB}~\cite{Czakon:2005rk}, see a review in Ref.~\cite{Dubovyk:2022obc}.
The \textit{Lee-Pomeransky parametrization}~\cite{Lee:2013hzt} is helpful for finding divergent regions of an integral.
Finally, the \textit{Baikov parametrization}~\cite{Baikov:1996iu} is appropriate for the intersection theory formulation in terms of the Baikov polynomial $\mathcal{B}$, as mentioned in Sec.~\ref{sec:amp.int.IBP}.

After combining all denominators with Feynman parametrization into a Lorentz-scalar Feynman integral, one can use loop momentum shift symmetry, as in Eq.~\ref{eq:shiftLoopMom}, to obtain an isotropic integral over the loop momenta.
Therefore, the generalised $d$-dimensional angular parameters defined via
\begin{equation}
\begin{split}
	d^dk &= d\Omega_d \, d|k| \, |k|^{d-1} \\
	d\Omega_d &= d\theta_{d-1} \sin^{d-2}(\theta_{d-1}) \cdots d\theta_{2} \sin(\theta_{2}) \, d\theta_{1} 
\end{split}
\label{eq:DdimAngles}
\end{equation}
can be easily integrated into
\begin{equation}
\begin{split}
	\Omega_d &= \int_{0}^{2\pi} d\theta_{1} \prod_{i=2}^{d-1} \left( \int_{0}^{\pi} d\theta_{i} \sin^{i-1}(\theta_{i}) \right)
	= \frac{2\pi^{d/2}}{\Gamma(d/2)} \,, \\
	\int \frac{d^dk}{(2\pi)^d} \frac{1}{(k^2-m^2)^n}
	&= (-1)^n \frac{i\Omega_d}{2(2\pi)^{d}} \frac{\Gamma(n-d/2)\Gamma(d/2)}{\Gamma(n)} (m^2)^{d/2-n} \,.
\end{split}
\label{eq:DdimAngleInt}
\end{equation}
In result, the Feynman parametrization of a general Feynman integral $\mathcal{I}_{\vec{n}}$ without nontrivial numerators reads
\begin{equation}
	\begin{split}
		\mathcal{I}_{\vec{n}}
		&= \int \left( \prod_{l=1}^L \frac{d^dk_l}{(2\pi)^d} \right)
		\frac{1}{\mathcal{D}_{1}^{n_1} \cdots \mathcal{D}_{r}^{n_r}} \\
		&= \int \left( \prod_{l=1}^L \frac{d^dk_l}{(2\pi)^d} \right)
		\frac{\Gamma(N)}{\prod_{i=1}^{r}\Gamma(n_i)}
		\int_{0}^{1} \left( \prod_{i=1}^r dx_i x_i^{n_i-1} \right)
		\frac{\delta(1-\sum_{i=1}^{r}x_i)}{(\sum_{i=1}^{r}x_i\mathcal{D}_i)^N} \\
		&= (-1)^N \left(\frac{i}{(4\pi)^{d/2}}\right)^L
		\frac{\Gamma(-[\mathcal{I}_{\vec{n}}]/2)}{\prod_{i=1}^{r}\Gamma(n_i)} \\
		& \qquad \times \int_{0}^{1} \left( \prod_{i=1}^r dx_i x_i^{n_i-1} \right)
		\delta(1-\sum_{i=1}^{r}x_i) \,
		\mathcal{U}^{-d/2-[\mathcal{I}_{\vec{n}}]/2} \, \mathcal{F}^{[\mathcal{I}_{\vec{n}}]/2} \,,
	\end{split}
	\label{eq:IntFeynParam}
\end{equation}
where $\Re(n_i)>0$, $N=\sum_{i=1}^{r}n_i$, $r$ is the rank the integral as introduced in Eq.~\ref{eq:dIBP}, the mass dimension $[\mathcal{I}_{\vec{n}}]=Ld-2\sum_{i}n_i$ is consistent with the definition in Eq.~\ref{eq:massDim}, while $\mathcal{U}$ and $\mathcal{F}$ are the \textit{first and the second Symanzik polynomials} in integration variables $x_i$, respectively.

\begin{figure}[h]
	\centering
	\includegraphics[width=0.5\textwidth]{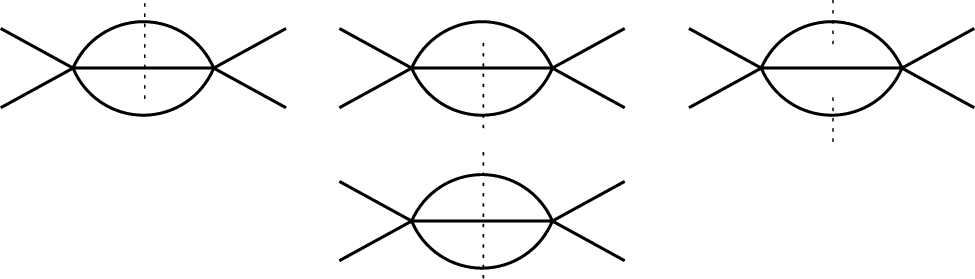}
	\caption{Symanzik graphs in the 1 and 2-forest for the $\mathcal{U}$ and $\mathcal{F}$ Symanzik polynomials, in the first and second row respectively, for the two-loop bubble subsector of the PL topology. The dashed lines denote the opening of loop corresponding lines.}
	\label{fig:Symanzik}
\end{figure}

It turns out that the Symanzik polynomials have a graph-theoretic representation, see a review in Ref.~\cite{Bogner:2010kv}.
Indeed, we can associate a connected graph $\mathcal{G}$ with the Feynman integral $\mathcal{I}_{\vec{n}}$.
We can define one of its \textit{spanning $k$-trees} as a result of deleting $L+k-1$ edges from the Feynman graph $\mathcal{G}$ such that we are left with exactly $k$ tree-level subgraphs.
A set of all spanning $k$-trees is called a \textit{$k$-forest}.
In this language, the Symanzik polynomials yield
\begin{equation}
\begin{split}
	\mathcal{U} &= \sum_{T \in \tau_1} \prod_{e \notin T} x_e \,, \\
	\mathcal{F} &= \sum_{\{T_1,T_2\} \in \tau_2} (p_{T_1} \cdot p_{T_2}) \prod_{e \notin \{T_1,T_2\}} x_e + \mathcal{U} \sum_{e\in\mathcal{G}} x_e m_e^2 \,,
\end{split}
\label{eq:Symanzik}
\end{equation}
where $\tau_1$ is a 1-forest of spanning 1-trees $T,T_1,T_2$, $\tau_2$ is a 2-forest of spanning 2-trees $\{T_1,T_2\}$, $e$ labels edges of the Feynman graph $\mathcal{G}$, with corresponding masses $m_e$, and $p_T$ is a momentum flowing into the tree $T$.
For the example two-loop s-channel bubble subsector of the PL topology in Fig.~\ref{fig:Symanzik}, the Symanzik polynomials read
\begin{equation}
	\begin{split}
		\mathcal{U} &= x_1 x_2 + x_2 x_3 + x_3 x_1 \,, \\
		\mathcal{F} &= -s \, x_1 x_2 x_3 \,.
	\end{split}
\label{eq:2LbubUF}
\end{equation}
It is clear, that $\mathcal{U}$ and $\mathcal{F}$ are homogeneous polynomials in $x_e$ of degree $L$ and $L+1$, respectively.
There is no graph-theoretical interpretation of nontrivial numerators in a Feynman integral.
However, in some cases, one can use the \textit{dimensional shift relations}~\cite{PhysRevD.54.6479,TARASOV1997455} e.g.
to decrease the power of a numerator by treating it as a denominator in higher $d+2$ spacetime dimensions.
These relations are expressed in terms of first Symanzik $\mathcal{U}$
\begin{equation}
	\begin{split}
		\mathcal{I}_{\vec{n}}^{(d-2)} &= \mathcal{U}(\textbf{1}^+,\dots,\textbf{N}^+) \mathcal{I}_{\vec{n}}^{(d)} \,, 
	\end{split}
	\label{eq:shiftDim}
\end{equation}
where we define the denominator power raising and lowering operators via
\begin{equation}
\textbf{j}^+ \mathcal{I}_{n_1,\dots,n_j,\dots,n_N}^{(d)} = n_j \mathcal{I}_{n_1,\dots,n_j+1,\dots,n_N}^{(d)} \,, \qquad
\textbf{j}^- \mathcal{I}_{n_1,\dots,n_j,\dots,n_N}^{(d)} = \mathcal{I}_{n_1,\dots,n_j-1,\dots,n_N}^{(d)} \,.
\label{eq:IBPpm}
\end{equation}

The Feynman parametrization in Eq.~\ref{eq:IntFeynParam} is convenient for noticing that the Feynman integral $\mathcal{I}_{\vec{n}}$ is \textit{projective}, i.e. invariant under
\begin{equation}
	x_i \to x_i \, \lambda_i \,.
	\label{eq:shiftFeynParam}
\end{equation}
Indeed, it can be easily observed by analysing the exponents of the homogeneous Symanzik polynomials.
One of the consequences is the \textit{Cheng-Wu theorem}~\cite{Cheng:1987ga}, which allows for constraining only a subset of the full sum $\sum_{i=1}^{r}x_i$ to be equal to 1.
When applied to only one variable, we can change the upper limit of Feynman parameter integrals to be equal and infinite i.e.
\begin{equation}
\int_{0}^{1} \left( \prod_{i=1}^r dx_i \right)
\delta(1-\sum_{i=1}^{r}x_i)
= \int_{0}^{\infty} \left( \prod_{i=1}^{r-1} dx_i \right) \,,
\label{eq:ChengWu}
\end{equation}
such that these multiple integrals become unordered.
The projective property is even more explicit in the language of differential forms~\footnote{Indeed, the integral over Feynman parameters is equivalent to an integral over the non-negative real projective space with an integration measure given by the differential form constructed from these Feynman parameters, see review in Ref.\cite{Weinzierl:2022eaz}}.

Having parametrized the integral $\mathcal{I}_{\vec{n}}$ in Eq.~\ref{eq:IntFeynParam}, we can finally attempt to perform the integration.
The integrals for which there exists an ordering of the underlying one-fold integrals which allows for explicit integration to GPLs at fixed order in $\ep$ are called \textit{linearly reducible}.
If it exists, finding such an ordering is algorithmic, and it has been implemented in a \texttt{Maple} package \texttt{HyperInt}~\cite{Panzer:2014caa}.
In general, however, square roots of polynomials of degree higher then 2 may appear in the integrand, which no longer integrate to GPLs.
These square roots usually arise in the two ways
\begin{equation}
	\int \frac{dx_1 \dots dx_n}{\sqrt{\text{poly}_{2(n+1)}(x_1 \dots x_2)}} \quad \text{or} \quad \int \frac{dx_1}{\sqrt{\text{poly}_{2(n+1)}(x_1)}} \,.
\label{eq:CY}
\end{equation}
It turns out that classifying the functions to which these expressions integrate is related to studying geometries within which the iterated integrals of multivariate rational functions of $x_i$ and $y=\sqrt{\text{poly}}$ are naturally defined.
Recent studies suggest~\cite{Bourjaily:2022bwx} that the integrals in Eq.~\ref{eq:CY} are related to Calabi-Yau $n$-fold geometries, and genus $n$ curves, respectively~\footnote{For an introduction to Calabi-Yau manifolds see Ref.\cite{Weinzierl:2022eaz}. Note that the Calabi-Yau 3-folds are well understood due to their application to compactification in String Theory, see review in Ref.~\cite{Blumenhagen:2013fgp}.}.
Proving this correspondence for a fixed Feynman integral topology can be very challenging.
For example, the underlying geometry of a family of higher-loop equal-mass banana-type Feynman integrals has been understood recently~\cite{Bonisch:2021yfw}.
For these integrals, it is enough to study the vanishing locus of the associated second Symanzik polynomial $\mathcal{F}=0$.
It leads to a spherical topology at $L=1$ loop, an elliptic curve at $L=2$ loops, a K3 surface at $L=3$ loops, and, in general, a Calabi-Yau $(L-1)$-fold at $L$ loops.
In general, classifying all possible geometries associated with Feynman integrals, as well as finding the corresponding special functions appearing after integration, are open problems in Mathematical Physics.

Mathematically, it is interesting to know what class of functions Feynman integrals evaluate to when keeping the full dependence on the dimension $d$ without expanding around small $\ep$.
It has been proven in Ref.~\cite{FeynmanGKZ}, that any generic Feynman integral $\mathcal{I}_{\vec{n}}$ in its full-$\ep$ form can be expressed in term of a \textit{Gel'fand-Kapranov-Zelevinsky (GKZ) hypergeometric function}.
Finding an explicit closed form solution for simple Feynman integrals is algorithmic, see e.g. a \texttt{Mathematica} package \texttt{FeynGKZ}~\cite{Ananthanarayan:2022ntm}, while in general, it still remains a challenge.
It is worth remembering, that in physical applications, we are only after an $\ep$-perturbative solution.
Therefore, beyond providing a closed form solution, one also needs to supply a systematic expansion method.
In practice, it usually means analysing the integrand of the closed form, which can be expanded in an easier manner.
For this reason, the particle phenomenology literature focuses mostly on the $\ep$-expanded form.
For simple generalisations of the Gauss hypergeometric function ${}_2 F_1$, expansion in $\ep$ can be automated, see e.g. a \texttt{Mathematica} package \texttt{HypExp}~\cite{Huber:2005yg} for ${}_n F_{n-1}$.

\section{Numerical methods for Feynman integrals}
\label{sec:amp.num.AMFlow}

In any phenomenological applications of scattering amplitude, the associated Feynman integrals have to be evaluated numerically at some kinematic phase space points.
Moreover, numerical samples also provide values for checks of any new analytic results.
There are multiple methods of evaluating Feynman integrals numerically.
All of them need to introduce a way of regulating the $\ep$ divergences in the integrals.
The most extensively used ones also have an efficient algorithmic public implementation.
An example of such a program is the \texttt{MB} package~\cite{Czakon:2005rk} for \texttt{Mathematica}, which is based on the Mellin-Barnes representation, as introduced in Sec.~\ref{sec:amp.int.Symanzik}.
Another instance is the \texttt{pySecDec}~\cite{Borowka:2017idc,Borowka:2018goh} program, which relies on the sector decomposition method, as mentioned in Sec.~\ref{sec:amp.int.div}.
Here, we describe a recent approach called the \textit{Auxiliary Mass Flow} (AMFlow) method~\cite{Liu:2017jxz,Liu:2020kpc,Liu:2021wks,Liu:2022chg,Liu:2022mfb,Liu:2022tji}, which is implemented in the \texttt{AMFlow} package~\cite{Liu:2022chg} for \texttt{Mathematica}.
Due to its high efficiency and precision, we use it for scattering amplitude computations later in this work.
For further details on other numerical methods for Feynman integrals see review in Ref.~\cite{Heinrich:2020ybq}.

The AMFlow method relies on promoting the small parameter $i\varepsilon$ in the Feynman propagator causality prescription to a free variable, denoted as \textit{auxiliary mass} $-\eta$, i.e.
\begin{equation}
	\frac{i}{\mathcal{D}_k+i\varepsilon} \to \frac{i}{\mathcal{D}_k-\eta} \,.
\end{equation}
Applying this substitution at the Feynman integral $\mathcal{I}$ level, one can treat the auxiliary mass $\eta$ as another kinematic variable in $\mathcal{I}(s_{ij},m_k,d,\eta)$.
With the use of IBP relations, one can find a set of integrals $\mathcal{I}_i$ with derivative in $\eta$ linearly related to themselves, i.e.
\begin{equation}
	\frac{\partial}{\partial\eta} \mathcal{I}_i(\eta) = A(\eta)_{ij} \, \mathcal{I}_j(\eta) \,.
\end{equation}
Such a differential equation requires an appropriate boundary condition.
It is convenient to fix this boundary condition at a rather exotic value $\eta=-i\infty$, since it allows for a simple evaluation of boundary integrals $\mathcal{I}_i(-i\infty)$.
Indeed, we can formally approximate each propagator at large $\eta$ with a Taylor series
\begin{equation}
	\frac{1}{((l+p)^2-m^2-\eta)^\nu} \approx \frac{1}{(l^2-\eta)^\nu} \sum_{i=0}^{N} \frac{\Gamma(\nu+i)}{i!\, \Gamma(\nu)} \left(-\frac{2l \cdot p+p^2-m^2}{l^2-\eta}\right)^i \,.
\end{equation}
By exploiting an iterative strategy, we can recursively reduce each boundary integral to a combination of vacuum bubbles with nontrivial numerators.
It substantially decreases the complexity of the boundary integral evaluation.

\begin{figure}[h]
	\centering
	\includegraphics[width=0.5\textwidth]{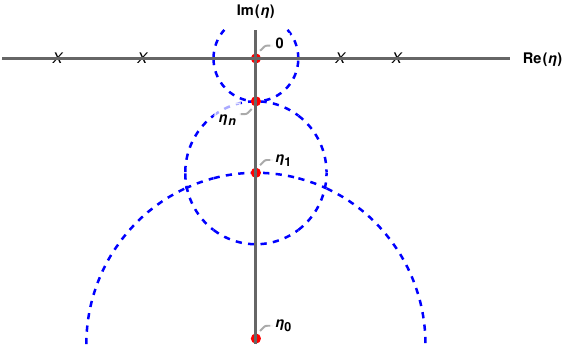}
	\caption{Schematic analytic continuation path $\eta \in \{-i\infty,\vec{\eta},-i0^-\}$ in the AMFlow method.}
	\label{fig:AMFlow}
\end{figure}

The last step of the AMFlow method is to relate this convenient boundary value to a physical one around $\eta=-i0^-$.
It can be realised via an analytic continuation along a chosen path $\eta \in \{-i\infty,\eta_1,\dots,\eta_{n-1},-i0^-\}$ which avoids branch cut and pole singularities, as schematically depicted in Fig.~\ref{fig:AMFlow}.
Subsequent values at $\eta_{i+1}$ are matched to their neighbour evaluated at $\eta_i$ in their Laurent series representation, expanded to a very high order.
Given the choice of the order $N$ in the expansion by regions, as well as the numerical path $\eta_i$, one can control the precision of the AMFlow method.
In terms of computational efficiency, the time consumption $T$ is dominantly linearly proportional to the precision $p$~\cite{Liu:2022chg},
\begin{equation}
	T = c_2 p^2 + c_1 p + c_0 \,, \qquad c_2 \ll c_1 \,,
\end{equation}
allowing for obtaining high precision with reasonable resources.

\section{Universality of divergences}
\label{sec:amp.div}

We have completed the discussion on the three mathematical structures appearing in scattering amplitudes.
After performing the decomposition in colour and tensor space, and evaluating all required integrals, we arrive at the final expression for the bare amplitude $\mathcal{A}^{(n)}_b$ at fixed order $n$.
We denote all bare quantities in our example two-loop QCD $q\bar{q} \to gg$ amplitude explicitly with a lower index $b$, as in Eq.~\ref{eq:ampPert}
\begin{equation}
	\mathcal{A} = (4\pi\alpha_{s,b}) \sum_{L=0}^{2} \left( \frac{\alpha_{s,b}}{2\pi} \right)^L \mathcal{A}^{(L)}_b \,.
	\label{eq:ampPertBare}
\end{equation}
As discussed in Sec.~\ref{sec:amp.int.div}, the divergences in $\ep$ arising from Feynman integrals lead to poles at the level of bare amplitude
\begin{equation}
	\mathcal{A}_b^{(2)}
	= \frac{\mathcal{A}^{(2)}_{b,4,\text{IR}}}{\epsilon^4}
	+ \frac{\mathcal{A}^{(2)}_{b,3,\text{IR}}}{\epsilon^3}
	+ \frac{\mathcal{A}^{(2)}_{b,2,\text{IR}}+\mathcal{A}^{(2)}_{b,2,\text{UV}}}{\epsilon^2}
	+ \frac{\mathcal{A}^{(2)}_{b,1,\text{IR}}+\mathcal{A}^{(2)}_{b,1,\text{UV}}}{\epsilon^1}
	+ \mathcal{O}(\ep^0) \,.
	\label{eq:ampPoles}
\end{equation}
In this section, we will see that both UV and IR poles of a scattering amplitude are universal.
As such, they can be predicted before computing the bare amplitude.

\subsection{UV renormalization}
\label{sec:amp.div.UV}

\begin{figure}[h]
	\centering
	\includegraphics[width=0.9\textwidth]{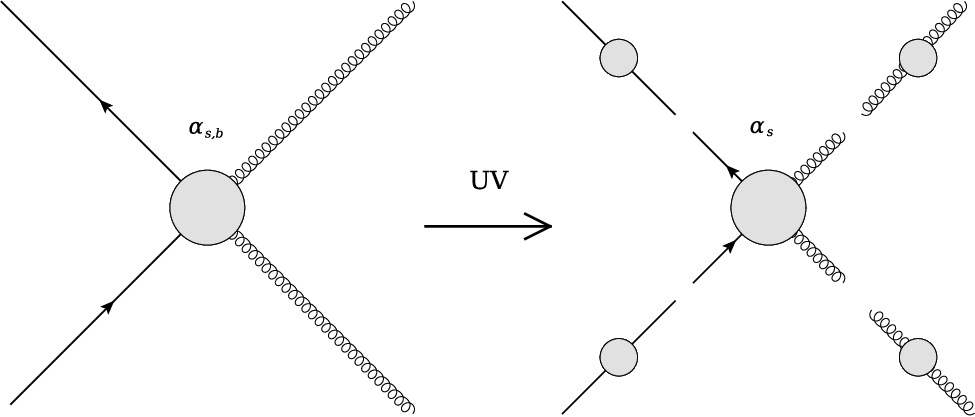}
	\caption{Schematic representation of the UV factorization for the $q\bar{q} \to gg$ process.}
	\label{fig:UV}
\end{figure}

\noindent
As argued in Sec.\ref{sec:intro.QCD}, bare quantities need to be renormalized in QFT, in order to correspond to physical predictions.
For a generic scattering amplitude, all three types of renormalization factors have to be considered, i.e. the wave function, coupling, and mass term.
For our example massless QCD correction to the $q\bar{q} \to gg$ process,
\begin{equation}
	\mathcal{A}(\alpha_{s,b}) \sqrt{Z_q Z_{\bar{q}} Z_g Z_g} = \mathcal{A}(\alpha_{s}) \, Z_{\mathcal{A}} \,,
	\label{eq:ampRenDef}
\end{equation}
depicted schematically in Fig.~\ref{fig:UV}, the mass renormalization is absent, and the renormalization factor of the whole amplitude reads
\begin{equation}
	Z_{\mathcal{A}}
	= 1 + \delta_{\mathcal{A}}
	= 1 + \frac{1}{2}(2\delta_{Z_u} + 2\delta_{Z_g}) + \delta_{\alpha_s} \,.
\end{equation}
By expanding the amplitude $\mathcal{A}$ and the counterterm $\delta_{\mathcal{A}}$ perturbatively in renormalized coupling $\alpha_{s}$ in Eq.~\ref{eq:ampRenDef}, we arrive at the renormalized perturbative series
\begin{equation}
	\mathcal{A} = (4\pi\alpha_{s}) \sum_{n=0}^{2} \left( \frac{\alpha_{s}}{2\pi} \right)^n \mathcal{A}^{(n)} \,.
	\label{eq:ampPertRen}
\end{equation}
The renormalized amplitudes $\mathcal{A}^{(n)}$ consist of bare amplitudes and counterterms mixed between orders, such that the UV poles in $\ep$ cancel exactly
\begin{equation}
	\mathcal{A}^{(2)}
	= \frac{\mathcal{A}^{(2)}_{4,\text{IR}}}{\epsilon^4}
	+ \frac{\mathcal{A}^{(2)}_{3,\text{IR}}}{\epsilon^3}
	+ \frac{\mathcal{A}^{(2)}_{2,\text{IR}}}{\epsilon^2}
	+ \frac{\mathcal{A}^{(2)}_{1,\text{IR}}}{\epsilon^1}
	+ \mathcal{O}(\ep^0) \,.
	\label{eq:ampIRpoles}
\end{equation}

Since the renormalization conditions of the two-point functions introduced in Sec.~\ref{sec:intro.QCD} are defined on the mass-shell, the wave function renormalization factors of massless particles vanish.
For the remaining strong coupling renormalization in massless QCD, it is convenient to choose the $\overline{\text{MS}}$ renormalization scheme, which cancels purely the $\ep$ poles without any finite remnant.
We renormalize the strong coupling constant via
\begin{equation}
	\alpha_{s,b} = \alpha_s(\mu)
	\,\,\times\,\,
	S_{\epsilon}^{-1}
	\left(\frac{\mu}{\mu_0}\right)^{2\ep}
	Z_{\alpha_s} \,,
	\label{eq:MSbar}
\end{equation}
with renormalization scale $\mu$, reference scale $\mu_0$, the $\overline{\text{MS}}$ factor
\begin{equation}
	S_{\epsilon} = (4\pi)^{\epsilon}e^{-\gamma_E \epsilon} \,,
	\label{eq:Sep}
\end{equation}
strong coupling renormalization factor
\begin{equation}
	Z_{\alpha_s} = 1-\frac{\beta_0}{\ep}\left(\frac{\alpha_s}{2\pi}\right) + 
	\left(\frac{\beta_0^2}{\epsilon^2}-\frac{\beta_1}{2\epsilon}\right)
	\left(\frac{\alpha_s}{2\pi}\right)^2+ \mathcal O(\alpha_s^3) \,,
	\label{eq:alphaRen}
\end{equation}
and the first two coefficients of the QCD beta function
\begin{equation}
	\beta_0 = \frac{11}{6}C_A - \frac{2}{3}T_F n_f\,,\qquad 
	\beta_{1} = \frac{17}{6}C_A^2 -  T_F n_f \left(\frac{5}{3}C_A+C_F\right) \,.
	\label{eq:beta}
\end{equation}

\subsection{IR regularization}
\label{sec:amp.div.IR}

After cancelling all UV poles in $\ep$ of the scattering amplitude $\mathcal{A}$ with the renormalization procedure, we are still left with IR poles in $\ep$, as in Eq.~\ref{eq:ampIRpoles}.
On the other hand, physical predictions cannot have any terms divergent in $\ep$.
In order to see that these IR poles are also spurious, consider an observable which a scattering amplitude leads to, i.e. the cross section in Eq.~\ref{eq:intro.xsec}.
The key observation is that the IR real radiation cannot be detected in a high-energy collision experiment, due to a finite resolution of detectors.
It means, that only real radiation above some cut-off energy and angular resolution, called \textit{hard radiation}, can be observed in a collider.
Therefore, when measuring a cross section for our example $q\bar{q} \to gg$ process, we also receive contributions from
$q\bar{q} \to gg+g_{\text{IR}}$,
$q\bar{q} \to gg+g_{\text{IR}}g_{\text{IR}}$,
$q\bar{q} \to gg+q_{\text{IR}}\bar{q}_{\text{IR}}$, etc.
Since these states belong to different Fock spaces, we should sum up their amplitudes after squaring their absolute value, as it represents the probability density
\begin{equation}
	\begin{split}
		2s\hat{\sigma} = & \int \dLIPS_2 \qquad |\mathcal{A}_{q\bar{q} \to gg}|^2 \\
		+ & \int \dLIPS_{2+1_{\text{IR}}} |\mathcal{A}_{q\bar{q} \to gg+g_{\text{IR}}}|^2 \\
		+ & \int \dLIPS_{2+2_{\text{IR}}} \left( |\mathcal{A}_{q\bar{q} \to gg+g_{\text{IR}}g_{\text{IR}}}|^2
		+ |\mathcal{A}_{q\bar{q} \to gg+q_{\text{IR}}\bar{q}_{\text{IR}}}|^2 \right) \\
		+ & (\text{more emissions}) + (\text{PDF counterterms}) \,.
		\label{eq:IRrealNonpert}
	\end{split}
\end{equation}
Together with the counterterms arising from the PDF evolution, these contributions mix at fixed order in perturbation theory.
In addition to $\ep$ poles originating in higher order virtual loop corrections, the phase space integrals over the IR degrees of freedom also introduce $\ep$ poles.
It has been proven by Kinoshita, Lee, and Nauenberg (KLN)~\cite{Kinoshita:1962ur,Lee:1964is} that these poles cancel order by order in perturbative SM.
\begin{equation}
	\begin{split}
		2s\hat{\sigma}
		& = \alpha_s^2 \left( 
		\int \dLIPS_2 \left| \mathcal{A}^{(0)}_{q\bar{q} \to gg} \right|^2 \right) \\
		& + \alpha_s^3 \left( 
		\int \dLIPS_2 \, 2\Re \left( \mathcal{A}^{(0)*}_{q\bar{q} \to gg} \mathcal{A}^{(1)}_{q\bar{q} \to gg} \right) 
		+ \int \dLIPS_{2+1_{\text{IR}}} \left| \mathcal{A}^{(0)}_{q\bar{q} \to gg+g_{\text{IR}}} \right|^2 \right) \\
		& + \alpha_s^4 \bigg(
		\int \dLIPS_2 \left( 2\Re \left( \mathcal{A}^{(0)*}_{q\bar{q} \to gg} \mathcal{A}^{(2)}_{q\bar{q} \to gg} \right) + \left| \mathcal{A}^{(1)}_{q\bar{q} \to gg} \right|^2 \right) \\
		& \qquad + \int \dLIPS_{2+1_{\text{IR}}} \left( 2\Re \left( \mathcal{A}^{(0)*}_{q\bar{q} \to gg+g_{\text{IR}}} \mathcal{A}^{(1)}_{q\bar{q} \to gg+g_{\text{IR}}} \right) + \left| \mathcal{A}^{(0)}_{q\bar{q} \to gg+g_{\text{IR}}} \right|^2 \right) \\
		& \qquad + \int \dLIPS_{2+2_{\text{IR}}} \left( \left| \mathcal{A}^{(0)}_{q\bar{q} \to gg+g_{\text{IR}}g_{\text{IR}}} \right|^2
		+ \left| \mathcal{A}^{(0)}_{q\bar{q} \to gg+q_{\text{IR}}\bar{q}_{\text{IR}}} \right|^2 \right) \bigg) \\
		& + \mathcal{O}(\alpha_s^5) + (\text{PDF counterterms}) = \mathcal{O}(\ep^0) \,.
		\label{eq:IRrealPert}
	\end{split}
\end{equation}
These real IR corrections factorize into universal expressions, and they can be predicted independently of the virtual contributions.
Therefore, in practice, we construct an IR-finite virtual quantity by subtracting all the predicted poles, since they have to cancel at the end.
Beyond NLO, constructing an explicit \textit{IR subtraction scheme} for a generic scattering process still remains a challenge.
For details on available NNLO slicing and subtraction schemes see e.g. Ref.~\cite{Heinrich:2020ybq}. 

\begin{figure}[h]
	\centering
	\includegraphics[width=0.9\textwidth]{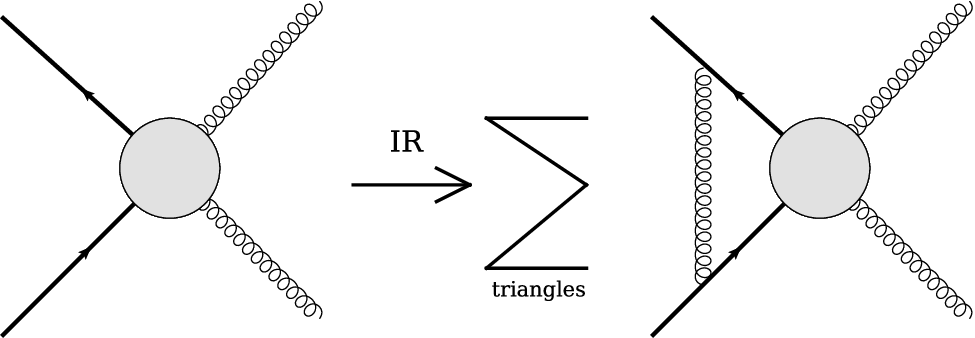}
	\caption{Schematic representation of the IR factorization.}
	\label{fig:IR}
\end{figure}

As discussed in Sec.~\ref{sec:amp.int.div}, the IR divergences of virtual loop corrections are also factorizable, and they mostly stem from the triangle subsectors.
The integral level analysis can be extended beyond Feynman diagrams to the whole amplitude, as depicted schematically in Fig.~\ref{fig:IR}.
As a result, we find a universal expression for the IR poles of the renormalized amplitude~\cite{Catani:1998bh,Catani:2000ef}
\begin{equation}
	\begin{split}
		\mathcal{A}^{(1)} &=
		\Ic_1(\ep) \, \mathcal{A}^{(0)} + \mathcal{O}(\ep^0) \\
		\mathcal{A}^{(2)} &=
		\Ic_2(\ep) \, \mathcal{A}^{(0)} + \Ic_1(\ep) \, \mathcal{A}^{(1)} + \mathcal{O}(\ep^0) \,,
	\end{split}
\end{equation}
which in principle can be generalised to higher orders.
For massless partons, the \textit{Catani operators} $\Ic_i$ read~\cite{Catani:1998bh}
\begin{align}
\begin{split}
	\Ic_1(\epsilon) &= \frac{e^{\gamma_E\epsilon}}{2\Gamma(1-\epsilon)}
	\sum_i \left( \frac{1}{\ep^2} + \frac{\gamma_i}{\Tc_i^2} \frac{1}{\ep} \right)
	\sum_{i \neq j \in \text{ext}} \Tc_i \circ \Tc_j \left(- \frac{\mu^2}{s_{ij}} \right)^\ep \,,	\\
	\Ic_2(\epsilon) &= -\frac{1}{2}\Ic_1(\epsilon)
	\left(\Ic_1(\epsilon)+\frac{2\beta_0}{\epsilon}\right)+
	\frac{e^{-\gamma_E\epsilon}\Gamma(1-2\epsilon)}{\Gamma(1-\epsilon)}
	\left(\frac{\beta_0}{\epsilon}+K\right)\Ic_1(2\epsilon) + \frac{H}{\ep} \,,
\end{split}
\label{eq:amp.Catani}
\end{align}
where the $\circ$ operator assumes an implicit sum over associated colour degrees of freedom.
The colour operator $\Tc_i$ of particle $i$ is equal to $if^{abc}$ for gluons, $T^a_{ij}$ for quarks, and $-T^a_{ij}$ for anit-quarks, while their squares evaluate to
\begin{equation}
	\Tc_q^2 = \Tc_{\bar{q}}^2 = C_F \,, \qquad
	\Tc_g^2 = C_A \,,
\end{equation}
the LO quark and gluon anomalous dimension yields
\begin{equation}
	\gamma_q = \gamma_{\bar{q}} = \frac{3}{2} C_F \,, \qquad
	\gamma_g = \beta_0 \,,
\end{equation}
respectively, the NLO cusp anomalous dimension is
\begin{equation}
	K = \left(\frac{67}{18}-\frac{\pi ^2}{6}\right)C_A - \frac{10}{9}n_f
	T_F,
\label{eq:amp.K}
\end{equation}
while the constant $H$ is process-specific and it contains the NLO anomalous dimensions, see e.g. Ref.~\cite{Becher:2009qa}.
This construction can be also extended to massive partons, as in Ref.~\cite{Catani:2000ef}.

Since the IR poles in $\ep$ arising from real and virtual corrections cancel exactly, it is reasonable to define a \textit{finite part} of the amplitude $\mathcal{A}^{(n,{\rm fin})}$ via
\begin{equation}
\begin{split}
	\mathcal{A}^{(1,{\rm fin})} &= \mathcal{A}^{(1)} - \Ic_1 \,  \mathcal{A}^{(0)} \\
	\mathcal{A}^{(2,{\rm fin})} &=
	\mathcal{A}^{(2)} - \Ic_2 \, \mathcal{A}^{(0)} - \Ic_1 \, \mathcal{A}^{(1)}
\end{split}
\end{equation}
Indeed, if we perform a consistent subtraction of the real corrections to the process, the finite part $\mathcal{A}^{(n,{\rm fin})}$ carries all the physical information about the virtual amplitude $\mathcal{A}^{(n)}$.
This finite part will be our final objective when providing results for scattering amplitudes.

\chapter{\label{ch:3L}Three-loop four-point massless QCD amplitudes}

\minitoc

\section{Massless QCD amplitudes frontier}
\label{sec:ggaa.frontier}

The reason for exploring the massless QCD sector of the SM is twofold.
Firstly, it deepens our understanding of not only the underlying background for scattering processes in hadron colliders, but also of the jet production, which is ubiquitous in most LHC analyses.
Secondly, it extends our knowledge about the mathematical structure of the amplitudes, due to the availability of analytic solutions.
As mentioned in Sec.~\ref{sec:amp.overview}, recently, there have been major advances in overcoming both computational and mathematical complexity of the problem.
\begin{figure}[h]
	\centering
	\includegraphics[width=0.45\textwidth]{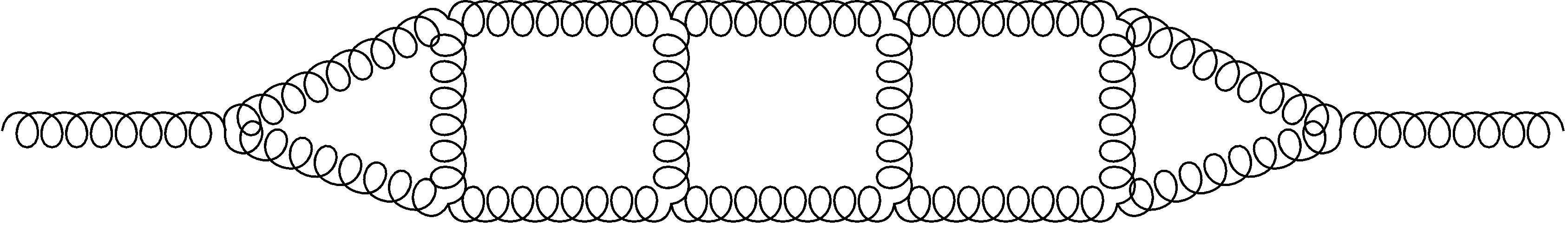}
	\includegraphics[width=0.45\textwidth]{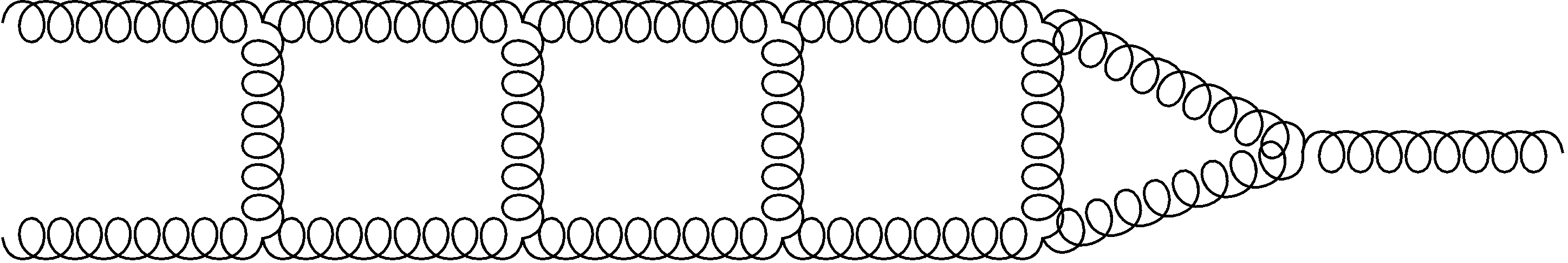}
	\includegraphics[width=0.35\textwidth]{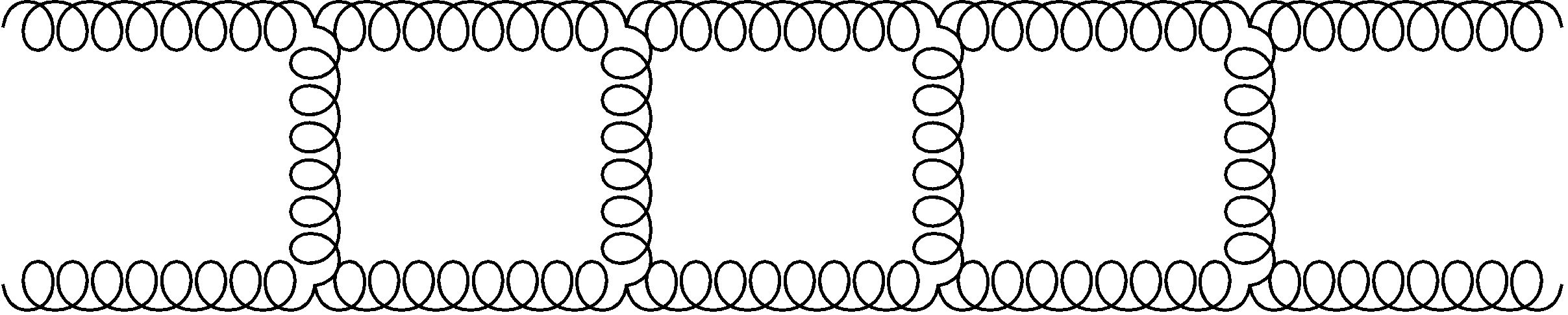}
	\includegraphics[width=0.3\textwidth]{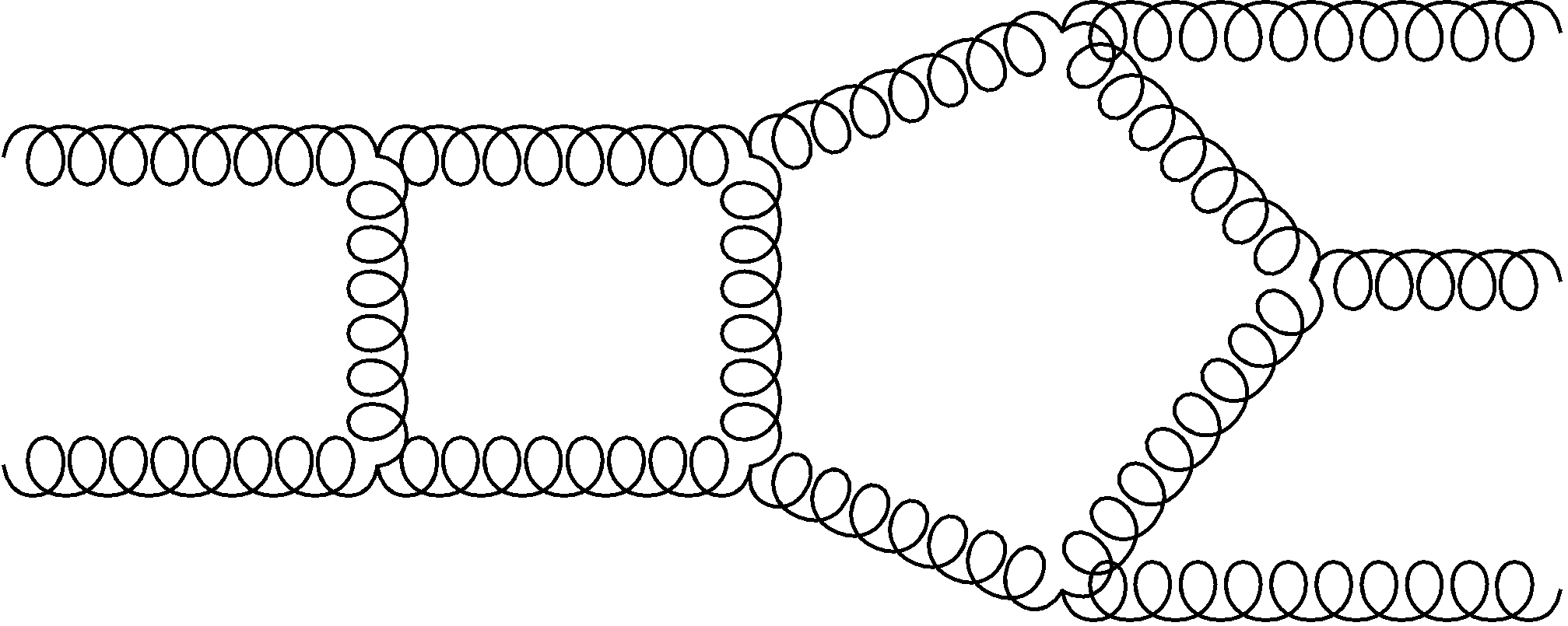}
	\includegraphics[width=0.25\textwidth]{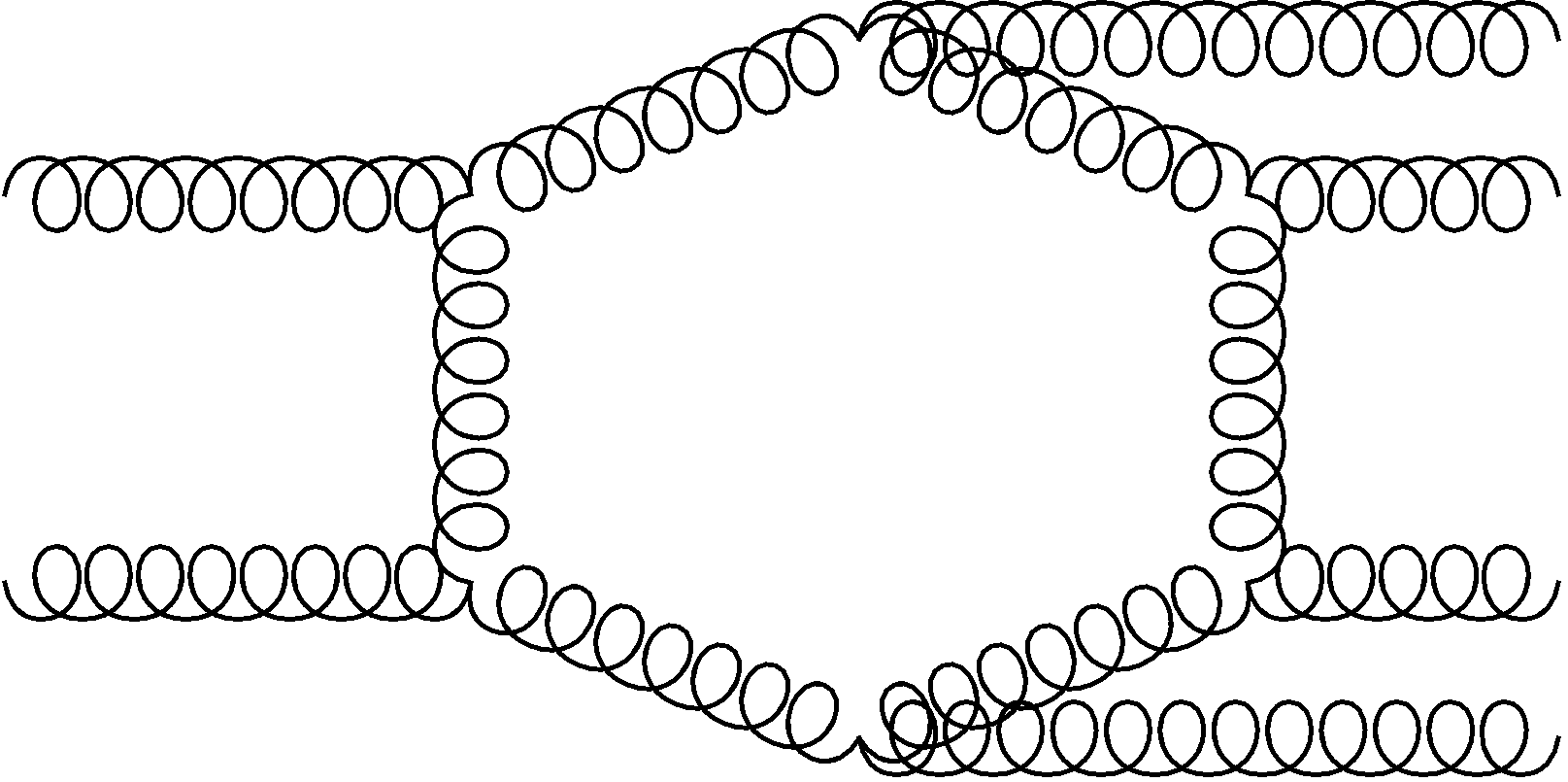}
	\caption{Example Feynman diagrams for the state-of-the-art massless QCD amplitudes, i.e. two-point five-loop, three-point four-loop, four-point three-loop, five-point two-loop, and six-point one-loop, respectively.}
	\label{fig:recentQCDres}
\end{figure}
Example Feynman diagrams for the different state-of-the-art analytic corrections are shown in Fig.~\ref{fig:recentQCDres}, and they correspond to two-point five-loop~\cite{Herzog:2017ohr}, three-point four-loop~\cite{Lee:2022nhh}, four-point three-loop~\cite{Caola:2020dfu,Caola:2021rqz,Bargiela:2021wuy,Caola:2021izf,Bargiela:2022lxz}, five-point two-loop~\cite{Abreu:2018zmy,Badger:2019djh,Chawdhry:2021mkw,Agarwal:2021vdh,Badger:2022ncb}, and six-point one-loop (\cite{Dunbar:2009uk} and references therein) amplitudes.
It is worth pointing out that for the number of legs beyond six, available results are mainly numerical, due to the complicated kinematics of these processes, see e.g. twenty gluon scattering in Ref.~\cite{Giele:2008bc}.
Nonetheless, at one-loop order, these corrections has been implemented in multiple public programs e.g. \texttt{OpenLoops}~\cite{Cascioli:2011va, Buccioni:2019sur}.
We also note that the tree-level amplitudes are well-understood for any $n$-point process.
For a review of recent results and automated tools, see Ref.~\cite{Heinrich:2020ybq}.
In this work, we are interested in three-loop four-point amplitudes.

Historically, the first three-loop full-colour massless QCD result was computed over 40 years ago for the two-point beta function~\cite{TARASOV1980429}.
It took 25 years to extend this calculation to the three-point form factor~\cite{Moch:2005tm}.
The first tree-loop four-point amplitude was computed a few years ago in a maximally supersymmetric gauge theory i.e. the $\mathcal{N}$=4 super-Yang-Mills~\cite{Henn:2016jdu}.
Recently, the first tree-loop four-point full-colour QCD amplitude frontier has been computed for the process $q\bar{q} \rightarrow \gamma\gamma$~\cite{Caola:2020dfu}, and later for $q\bar{q} \rightarrow q\bar{q}$~\cite{Caola:2021rqz},
$gg \rightarrow \gamma\gamma$~\cite{Bargiela:2021wuy},
$gg \rightarrow gg$~\cite{Caola:2021izf},
and $pp \rightarrow \text{j}\gamma$~\cite{Bargiela:2022lxz}.
\begin{table}[h]
	\renewcommand{\arraystretch}{1.2}
	\centering
	%\begin{tabular}{ p{4.0cm}||p{1.2cm}|p{1.2cm}|p{1.2cm}|p{1.2cm} }
	\begin{tabular}{ l||r|r|r|r }
		Number of diagrams & 0L & 1L & 2L & 3L \\
		\hline
		\hline
		$q\bar{q} \rightarrow \gamma\gamma$ & 2 & 10 & 143 & 2922 \\
		\hline
		$q\bar{q} \rightarrow q\bar{q}$ & 1 & 9 & 158 & 3584 \\
		\hline
		$gg \rightarrow \gamma\gamma$ & 0 & 6 & 138 & 3299 \\
		\hline
		$gg \rightarrow gg$ & 4 & 81 & 1771 & 48723 \\
		\hline
		$q\bar{q} \rightarrow g\gamma$ & 2 & 13 & 229 & 5334 \\
		\hline
		$gg \rightarrow g\gamma$ & 0 & 12 & 264 & 7356 \\
	\end{tabular}
\caption{Number of Feynman diagrams for available QCD amplitudes at different loop orders.}
\label{tab:3LdiagNum}
\end{table}
See the number of Feynman diagrams for these processes at different loop orders in Tab.~\ref{tab:3LdiagNum}, which reaches up to 48723 for the three-loop four-gluon scattering.
We summarise here our contribution to this frontier i.e. $gg \rightarrow \gamma\gamma$~\cite{Bargiela:2021wuy}, and $pp \rightarrow \text{j}\gamma$~\cite{Bargiela:2022lxz} process.
They both provide key ingredients for new high-precision phenomenological predictions.

\section{Overview of $gg \rightarrow \gamma\gamma$}
\label{sec:ggaa.complexity}

In this chapter, we describe our two calculations contributing to the tree-loop four-point frontier.
Let us focus on the $gg \rightarrow \gamma\gamma$ process~\cite{Bargiela:2021wuy}.
Since the calculation flow for $pp \rightarrow \text{j}\gamma$~\cite{Bargiela:2022lxz} is similar, we will return to it only when discussing the results in Sec.~\ref{sec:ggaa.ppja}.
The kinematics of the massless four-point process
\begin{equation}
	g (p_1) + g(p_2)  \to \gamma(-p_3) + \gamma(-p_4)
	\label{eq:ggaa}
\end{equation}
is the same as introduced in Sec.~\ref{sec:amp.kin}.
In the physical region, $s>0$, $t<0$, and $u<0$, thus the dimensionless ratio $x$ defined in e.q.~\ref{eq:x} is positive
\begin{equation}
	\text{(Regge, forward)} \qquad 0 < x = -\frac{t}{s} < 1 \qquad \text{(backward)} \,.
	\label{eq:ggaa.x}
\end{equation}
We write the perturbative expansion in the bare strong coupling similarly to Eq.~\ref{eq:ampPert}
\begin{equation}
	\mathcal{A} =
	(4 \pi \alpha) \left(\frac{\alpha_s}{2\pi}\right) \left[ \mathcal{A}^{(1)} + \left(\frac{\alpha_s}{2\pi}\right)
	\mathcal{A}^{(2)} + 
	\left(\frac{\alpha_s}{2\pi}\right)^2
	\mathcal{A}^{(3)} + \mathcal O(\alpha_s^3)\right] \,,
	\label{eq:ggaa.pert}
\end{equation}
where the superscript indicates the number of loops.

\begin{figure}
\scalebox{0.9}{\parbox{1.0\linewidth}{
		\centering
		\includegraphics[width=0.4\textwidth]{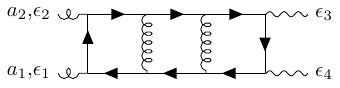}
		\begin{equation*}
			\begin{split}
				& \qquad\qquad\qquad\qquad = \, \frac{1}{2} \, g_s^6 \, e^2 \, {\color{red} \, n_f^{(V_2)} \, C_f^2 \, \delta^{a_1,a_2}} \,
				{\color{blue} \int \frac{d^dk_1}{(2\pi)^d} \, \frac{d^dk_2}{(2\pi)^d} \, \frac{d^dk_3}{(2\pi)^d}} \\
				\times& \frac{\color{gray} \text{tr}\left[
					\slashed{\ep}_1 (\slashed{k}_1) \slashed{\ep}_2 (\slashed{k}_1+\slashed{p}_2) \gamma^\mu (\slashed{k}_{13}+\slashed{p}_2) \gamma^\nu (\slashed{k}_2+\slashed{p}_4) \slashed{\ep}_4 (\slashed{k}_2) \slashed{\ep}_3 (\slashed{k}_2-\slashed{p}_3) \gamma_\nu (\slashed{k}_{13}-\slashed{p}_1) \gamma_\mu (\slashed{k}_1-\slashed{p}_1)
					\right]}
				{\color{blue} (k_1)^2 (k_1+p_2)^2 (k_{13}+p_2)^2 (k_2+p_4)^2 (k_2)^2 (k_2-p_3)^2 (k_{13}-p_1)^2 (k_1-p_1)^2 (k_3)^2 (k_{123}-p_{13})^2}
			\end{split}
		\end{equation*}
		\\
		\raggedright
		\qquad\quad {\color{red} colour} \qquad\qquad\qquad\qquad\quad {\color{gray} tensors} \qquad\qquad\qquad\qquad\qquad {\color{blue} integrals} \\
		\centering
		\includegraphics[width=0.2\textwidth]{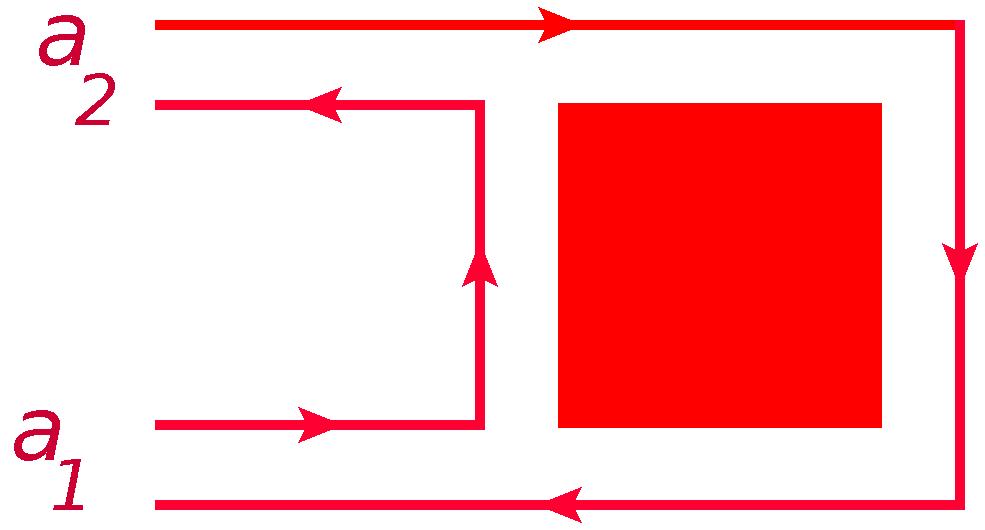}
		\hfill
		\includegraphics[width=0.3\textwidth]{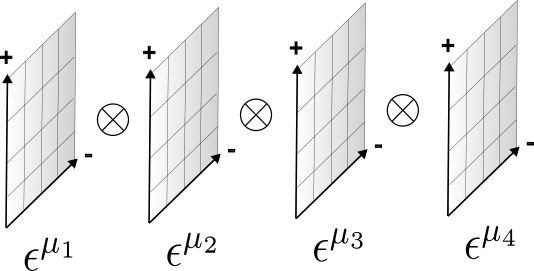}
		\hfill
		\includegraphics[width=0.4\textwidth]{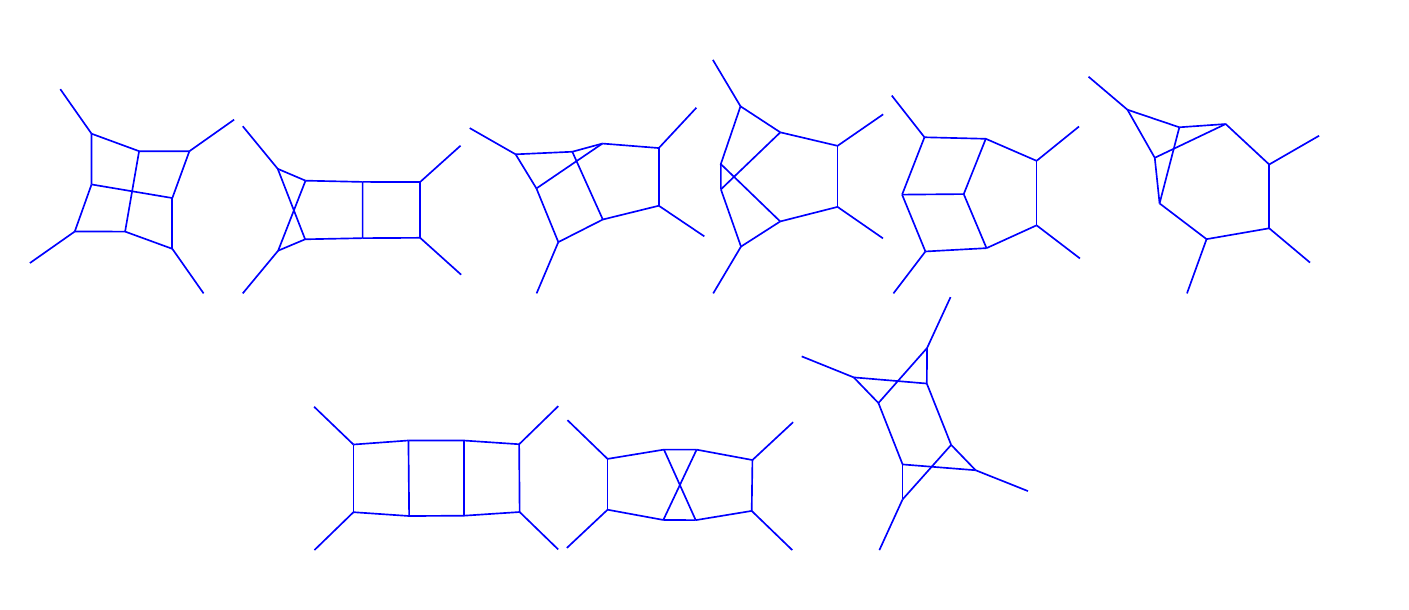}
		\begin{equation*}
			\mathcal{A}^{\vec{a},\lambdavec} = \sum_{c,t,i} {\color{red} \mathcal{C}_c^{\vec{a}}} \, {\color{gray} T_t^{\lambdavec}} \, {\color{blue} \mathcal{I}_{i}} \, r_{c,t,i}
		\end{equation*}
}}
\caption{Example Feynman diagram for $gg \rightarrow \gamma\gamma$ together with the three basis underlying structures.}
\label{fig:ggaa.diag}
\end{figure}
Consider an example Feynman diagram for the $gg \rightarrow \gamma\gamma$ process in Fig.~\ref{fig:ggaa.diag}.
We recognise the three underlying mathematical structures introduced in Ch.~\ref{ch:amp}, colour, tensor, and integral.
We also schematically depict in Fig.~\ref{fig:ggaa.diag} the basis elements onto which we will decompose these structures.
\begin{figure}[h]
	\centering
	\includegraphics[width=0.9\textwidth]{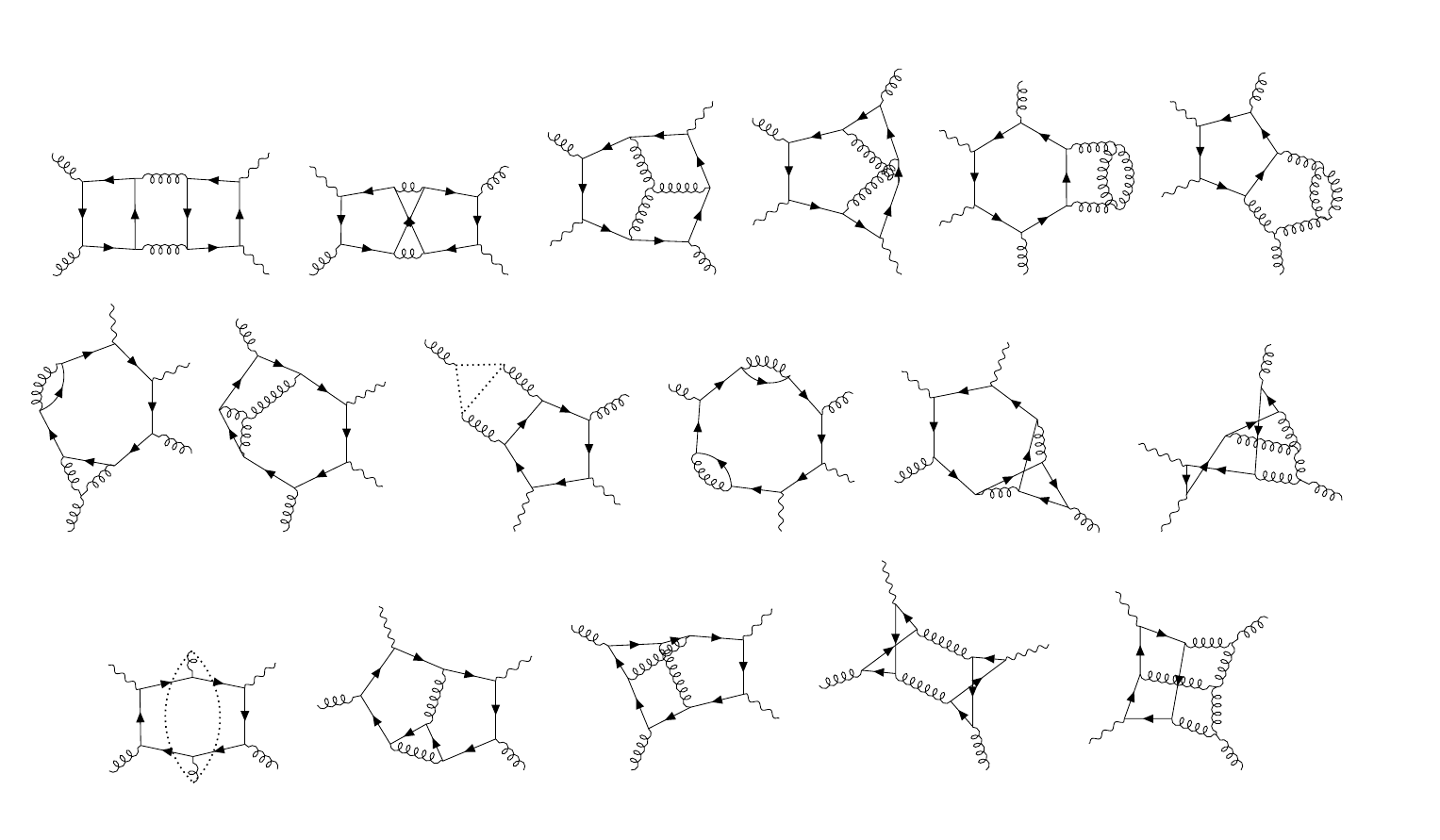}
	\caption{Example Feynman diagrams for the three-loop $gg \rightarrow \gamma\gamma$ process.}
	\label{fig:ggaa.diags}
\end{figure}
Similarly to this example diagram, one has to analyse all 3299 Feynman diagrams for this process, see their illustrative subset in Fig.~\ref{fig:ggaa.diags}.
\begin{figure}[h]
	\includegraphics[width=0.9\textwidth]{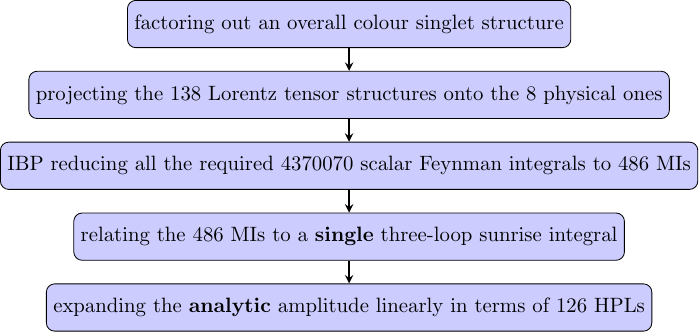}
	\caption{Computational flow chart for the amplitude.}
	\label{fig:ggaa.flowAmp}
\end{figure}
As we explain in the following sections, the methods introduced in Ch.~\ref{ch:amp} lead to a computational flow summarised in Fig.~\ref{fig:ggaa.flowAmp}.
\begin{table}[h]
	\renewcommand{\arraystretch}{1.2}
	\centering
	%\begin{tabular}{ p{9.5cm}||p{1.cm}|p{1.cm}|p{1.5cm} }
	\begin{tabular}{ l||r|r|r }
		& 1L & 2L & 3L \\
		\hline
		\hline
		Number of diagrams & 6 & 138 & 3299 \\
		\hline
		Number of integral topologies & 1 & 2 & 3 \\
		\hline
		Number of integrals before IBPs and symmetries & 209 & 20935 & 4370070 \\
		\hline
		Number of master integrals & 6 & 39 & 486 \\
		\hline
		\hline
		Size of the Feynman diagrams list [kB] & 4 & 90 & 2820 \\
		\hline
		Size of the result before integral reduction [kB] & 276 & 54364 & 19734644 \\
		\hline
		Size of the result in terms of MIs [kB] & 12 & 562 & 304409 \\
		\hline
		Size of the result in terms of HPLs [kB] & 136 & 380 & 1195 \\
	\end{tabular}
	\caption{Some indicators of the complexity of the result at various stages of the calculation at different loop orders.
	The size of the files with formulae in a consistent text format should be analysed relative to the corresponding leading order.}
	\label{tab:stats}
\end{table}
We employ various modern techniques to perform reductions in the three amplitude structures.
In Tab.~\ref{tab:stats}, we summarise the corresponding complexity of the result at various stages of the amplitude computation.
In addition, transforming the amplitude along these stages required hundreds of CPU hours and gigabytes of RAM memory.

\section{Amplitude structure}
\label{sec:ggaa.amp}

Before discussing the details of the calculation, we describe the three mathematical structures appearing in the $gg \rightarrow \gamma\gamma$ amplitude, colour, tensor, and integral, as introduced in Ch.~\ref{ch:amp}.

\subsection{Colour}
\label{sec:ggaa.amp.col}

\begin{figure}[h]
	\centering
	\includegraphics[width=0.3\textwidth]{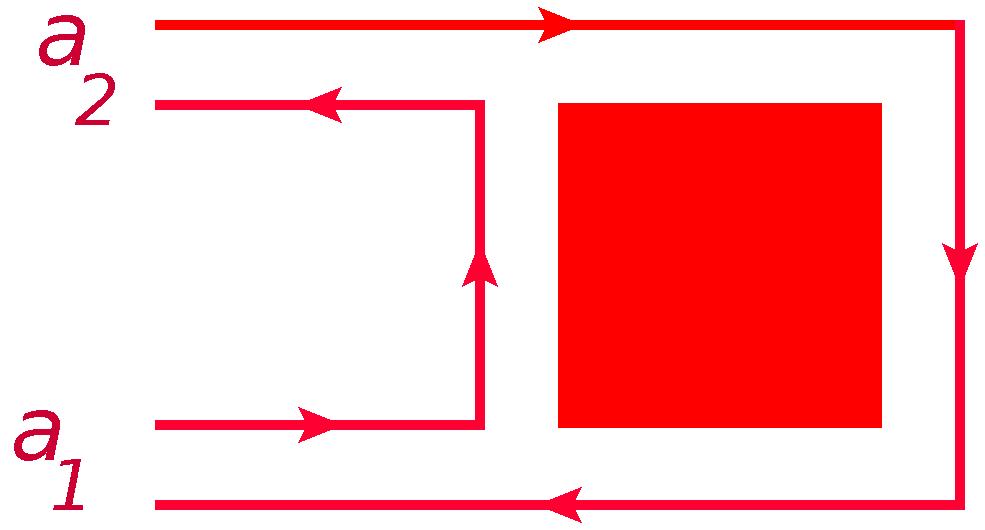}
	\caption{Schematic colour structure of the $gg \to \gamma\gamma$ amplitude.}
	\label{fig:ggaa.col}
\end{figure}

\noindent
As follows from Sec.~\ref{sec:amp.col}, the $gg \rightarrow \gamma\gamma$ process has a simple dependence on the colour indices $a_{1,2}$ of gluons $g_{1,2}$, which factorizes from the amplitude
\begin{equation}
\begin{split}
	\mathcal{A}^{a_1 a_2} &= \delta^{a_1 a_2} A \\
	&= \delta^{a_1 a_2} \sum_c C^{(3)}_c \, A_c \,.
\end{split}
\label{eq:ggaa.colDecomp}
\end{equation}
See the corresponding 't Hooft double line colour diagram for this amplitude in Fig.~\ref{fig:ggaa.col}.
The remaining colour-scalar structures $C^{(3)}_c$ are monomials in $\{C_A,C_F,n_f,n_f^V,n_f^{V_2}\}$ of degree 3.
\begin{figure}[h]
	\centering
	\includegraphics[width=0.22\textwidth]{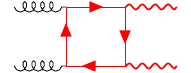}
	\includegraphics[width=0.35\textwidth]{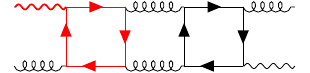}
	\includegraphics[width=0.35\textwidth]{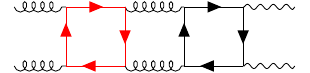}
	\caption{The three types of closed fermion loop factors denoted in red, $n_f^{V_2}=\sum_f Q_f^2$,  $n_f^V=\sum_f Q_f$, and $n_f$.}
	\label{fig:ggaa.nfV}
\end{figure}
The two factors, $n_f^V$ and $n_f^{V_2}$, are defined as
\begin{equation}
	n_f^V = \sum_f Q_f,
	\quad
	n_f^{V_2}=\sum_f Q_f^2 \,,
	\label{eq:nfdef}
\end{equation}
and in QCD with $n_f=5$ light quarks, they evaluate to $n_f^V = 1/3$ and $n_f^{V_2} = 11/9$, where the charge of quarks is $Q_{u,c} = 2/3$, $Q_{d,s,b} = -1/3$ in units of $e$.
These two factors correspond to attaching either one or two photons to the same closed fermion loop, respectively, as depicted in Fig.~\ref{fig:ggaa.nfV}.
These two types contribute differently than $n_f$ because the photon resolves the electric charge, instead of the strong coupling.
It is worth noting that the Feynman diagrams proportional to $n_f^V$ do not appear below the three-loop order for this process.

\subsection{Lorentz tensors}
\label{sec:ggaa.amp.ten}

\begin{figure}[h]
	\centering
	\includegraphics[width=0.7\textwidth]{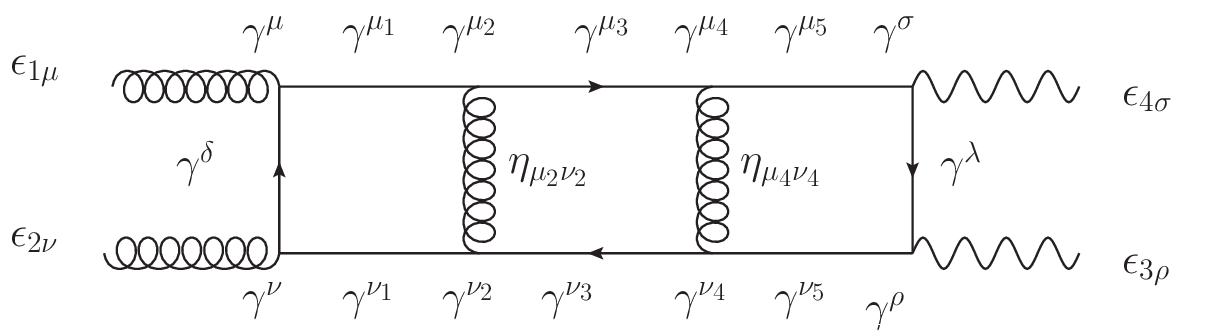}
	\caption{Lorentz tensor structures in example Feynman diagram for three-loop $gg \to \gamma\gamma$. The open indices of $\gamma$ matrices are understood to be contracted with corresponding propagator momenta.}
	\label{fig:ggaa:ten}
\end{figure}

\noindent
In order to describe the Lorentz tensor structure of the $gg \to \gamma\gamma$ amplitude, we will closely follow the argument in Sec.~\ref{sec:amp.Lorentz}.
We write the colour-stripped amplitude $A_c$ for any fixed colour factor $C^{(3)}_c$ as
\begin{align}
	A = A^{\mu \nu \rho \sigma} \, \epsilon_{1,\mu}(p_1) \epsilon_{2,\nu}(p_2) \epsilon_{3,\rho}(p_3) \epsilon_{4,\sigma}(p_4) \,.
	\label{eq:ggaa.tenDecomp}
\end{align}
The amplitude tensor $A^{\mu \nu \rho \sigma}$ can be expanded in a basis of Lorentz tensors $\Gamma_i^{\mu\nu\rho\sigma}$ as
\begin{equation}
	A^{\mu\nu\rho\sigma} = \sum_{i=1}^{n_t} \mathcal F_i \, \Gamma_i^{\mu\nu\rho\sigma} \,.
\end{equation}
The $138$ tensors $\Gamma_i^{\mu\nu\rho\sigma}$ are constructed from the independent external momenta $p_{1,2,3}^\mu$, as well as from the metric tensor $g^{\mu\nu}$~\cite{Binoth:2002xg}.
Since $A^{\mu\nu\rho\sigma}$ is contracted with the external polarisation vectors $\epsilon_i^\mu$, we can impose physical transversality conditions $p_i\cdot \epsilon_i = 0$, which eliminate 81 tensors proportional to $p_1^\mu$, $p_2^\nu$, $p_3^\rho$, and $p_4^\sigma$.
We also choose reference vectors $q_i = p_{i+1}$ for all the massless external gauge bosons such that
\begin{align}
	\epsilon_i \cdot p_{i+1} = 0 \,,\;\; \mbox{where} \;\; i=1,\dots,4 \;\; \mbox{and} \;\; p_5\equiv p_1 \,.
	\label{eq:ggaa.refMom}
\end{align}
As a result, we are left with $n_t=10$ independent Lorentz tensors in $d$ dimensions
\begin{align}
	\Gamma_1^{\mu \nu \rho \sigma} &= p_3^{\mu}p_1^{\nu}p_1^{\rho}p_2^{\sigma}\,, \;\;
	\Gamma_2^{\mu \nu \rho \sigma} = p_3^{\mu}p_1^{\nu}g^{\rho\sigma}\,, \nonumber \\
	\Gamma_3^{\mu \nu \rho \sigma} &= p_3^{\mu}p_1^{\rho}g^{\nu\sigma} \,\,\, \,, \;\;
	\Gamma_4^{\mu \nu \rho \sigma} = p_3^{\mu}p_2^{\sigma}g^{\nu\rho}\,, \nonumber \\
	\Gamma_5^{\mu \nu \rho \sigma} &= p_1^{\nu}p_1^{\rho}g^{\mu\sigma} \,\,\, \,, \;\;
	\Gamma_6^{\mu \nu \rho \sigma} = p_1^{\nu}p_2^{\sigma}g^{\mu\rho}\,, \nonumber \\
	\Gamma_7^{\mu \nu \rho \sigma} &= p_1^{\rho}p_2^{\sigma}g^{\mu\nu} \,\,\, \,, \;\;
	\Gamma_8^{\mu \nu \rho \sigma} = g^{\mu\nu}g^{\rho\sigma}\,, \nonumber \\
	\Gamma_9^{\mu \nu \rho \sigma} &= g^{\mu\sigma}g^{\nu\rho} \,\,\,\,\,\, \,, \;\;
	\Gamma_{10}^{\mu \nu \rho \sigma} = g^{\mu\rho}g^{\nu\sigma}.
	\label{eq:ggaa.Gammas}
\end{align}
After contracting with external polarization vectors, we arrive at
\begin{equation}
	T_i =  \Gamma_i^{\mu \nu \rho \sigma}\, \epsilon_{1,\mu} \epsilon_{2,\nu} \epsilon_{3,\rho} \epsilon_{4,\sigma} \,,
	\label{eq:ggaa.tensors}
\end{equation}
such that the colour-stripped amplitude can be decomposed as
\begin{equation}
	A = \sum_{i=1}^{10} \mathcal F_i(s,t) \, T_i \,.
	\label{eq:ggaa.ampT}
\end{equation}

\begin{figure}[h]
	\centering
	\includegraphics[width=0.4\textwidth]{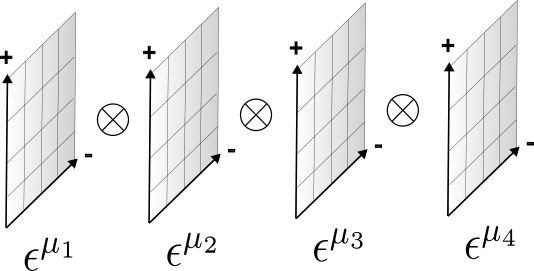}
	\caption{Spanning Lorentz tensor basis in $d=4$ for the $gg \to \gamma\gamma$ amplitude.}
	\label{fig:ggaa.ten}
\end{figure}

In $d=4$, there are $2^4/2 = 8$ independent helicity states, since each massless boson has 2 polarizations, and there is an overall QCD parity symmetry.
According to Ref.~\cite{Peraro:2020sfm}, the corresponding $\bar{n}_t=8$ tensors which span the physical four-dimensional subspace can be chosen as
\begin{equation}
	\oT_i = T_i\,,~  i=1,\dots,7,
	\qquad
	\oT_8 = T_8 + T_9 +T_{10} \,.
	\label{eq:ggaa.tens4}
\end{equation}
Following the discussion in Sec.~\ref{sec:amp.Lorentz}, the remaining two tensors can be projected out into the purely $-2\ep$ subspace with
\begin{equation}
	\oT_i = T_i - \sum_{j=1}^{8}(\mathcal{P}_j T_i) \oT_j \,,~ i=9,10 \,, \qquad
	\sum_{\text{pol}} \mathcal P_i \oT_j = \delta_{ij} \,,
	\label{eq:ggaa.tensEps}
\end{equation}
which expands to
\begin{equation}
	\begin{split}
		& \oT_9 \,\, = {T_{9}} \,\,\, - \frac{1}{3}\left( -\frac{2 {\oT_1}}{s u} - \,\,\, \frac{{\oT_6}}{s} - \frac{{\oT_2}+{\oT_3}+2 {\oT_4}-2 {\oT_5}-{\oT_6}-{\oT_7}}{t} + \,\,\frac{{\oT_3}}{u} + {\oT_8} \right) \,, \\
		&\oT_{10} = {T_{10}} - \frac{1}{3} \left( \,\,\,\,\, \frac{4 {\oT_1}}{s u} + \frac{2 {\oT_6}}{s} - \frac{{\oT_2}-2 {\oT_3}-{\oT_4}+{\oT_5}+2 {\oT_6}-{\oT_7}}{t} - \frac{2 {\oT_3}}{u} + {\oT_8} \right) \,.
		\label{eq:ggaa.tensEpsExp}
	\end{split}
\end{equation}
The explicit form of the projectors $\mathcal{P}_j$ is given in Ref.~\cite{Peraro:2020sfm}.

By evaluating the physical tensors $\oT_i$ at a fixed helicity state $\lambdavec = (\lambda_1,\lambda_2,\lambda_3,\lambda_4)$ we arrive at the helicity amplitude
\begin{equation}
	A_\lambdavec = \sum_{i=1}^{8} \oF_i \, \oT_{i,\lambdavec} \,,
\end{equation}
where the tensors $\oT_{9,\lambdavec}$ and $\oT_{10,\lambdavec}$ vanish.
Helicity amplitudes factorize
\begin{equation}
	A_\lambdavec = \mathcal S_\lambdavec \, f_\lambdavec
\end{equation}
into the overall spinor phases
\begin{align}
	\mathcal{S}_{++++} &= \frac{[1 2][3 4]}{\langle1 2\rangle\langle3
		4\rangle} \,, & %~ \,\,\,\,\,\,\,\,\,\,\,\,
	\mathcal{S}_{-+++} &=
	\frac{\langle1 2\rangle\langle1 4\rangle[2 4]}{\langle3
		4\rangle\langle2 3\rangle\langle2 4\rangle} \,, &%~
	\mathcal{S}_{+-++} &= \frac{\langle2 1\rangle\langle2 4\rangle[1
		4]}{\langle3 4\rangle\langle1 3\rangle\langle1 4\rangle} \,,
	\notag\\
	\mathcal{S}_{++-+} &= \frac{\langle3 2\rangle\langle3 4\rangle[2
		4]}{\langle1 4\rangle\langle2 1\rangle\langle2 4\rangle} \,,&%~
	\mathcal{S}_{+++-} &= \frac{\langle4 2\rangle\langle4 3\rangle[2
		3]}{\langle1 3\rangle\langle2 1\rangle\langle2 3\rangle} \,,&%&
	\mathcal{S}_{--++} &= \frac{\langle1 2\rangle[3 4]}{[1
		2]\langle3 4\rangle} \,,%&%~ \,\,\,\,\,\,\,\,\,\,\,\,\,\,
	\notag\\ 
	\mathcal{S}_{-+-+} &= \frac{\langle1 3\rangle[2 4]}{[1 3]\langle2
		4\rangle} \,,&%~ \,\,\,\,\,\,\,\,\,\,\,\,\,
	\mathcal{S}_{+--+} &= \frac{\langle2 3\rangle[1 4]}{[2 3]\langle1 4\rangle} \,.&&
	\label{eq:ggaa.S}
\end{align}
and the little-group-scalar helicity amplitudes
\begin{align}
	f_{++++} &=  \frac{t^2}{4}\left(\frac{2\oF_{6}}{u}-\frac{2\oF_{3}}{s}-\oF_{1}\right)+\oF_{8}\left(\frac{s}{u}+\frac{u}{s}+4\right)+\frac{t}{2}(\oF_{2}-\oF_{4}+\oF_{5}-\oF_{7})\,, \notag\\ 
	f_{-+++} &=  \,\,\,\, \frac{t^2}{4}\left(\frac{2\oF_{3}}{s}+\oF_{1}\right)+t\left(\frac{\oF_{8}}{s}+\frac{1}{2}(\oF_{4}+\oF_{6}-\oF_{2})\right)\,, \notag\\ 
	f_{+-++} &=  -\frac{t^2}{4}\left(\frac{2\oF_{6}}{u}-\oF_{1}\right)+t\left(\frac{\oF_{8}}{u}-\frac{1}{2}(\oF_{2}+\oF_{3}+\oF_{5})\right)\,, \notag\\ 
	f_{++-+} &= \,\,\,\, \frac{t^2}{4}\left(\frac{2\oF_{3}}{s}+\oF_{1}\right)+t\left(\frac{\oF_{8}}{s}+\frac{1}{2}(\oF_{6}+\oF_{7}-\oF_{5})\right)\,, \notag\\ 
	f_{+++-} &=  -\frac{t^2}{4}\left(\frac{2\oF_{6}}{u}-\oF_{1}\right)+t\left(\frac{\oF_{8}}{u}+\frac{1}{2}(\oF_{4}+\oF_{7}-\oF_{3})\right)\,, \notag\\ 
	f_{--++} &=  -\frac{t^2}{4}\oF_{1}+\frac{1}{2}t(\oF_{2}+\oF_{3}-\oF_{6}-\oF_{7})+2\oF_{8}\,, \notag\\ 
	f_{-+-+} &=  t^2\left(\frac{\oF_{8}}{su}-\frac{\oF_{3}}{2s}+\frac{\oF_{6}}{2u}-\frac{\oF_{1}}{4}\right)\,, \notag\\ 
	f_{+--+} &=  -\frac{t^2}{4}\oF_{1}+\frac{1}{2}t(\oF_{3}-\oF_{4}+\oF_{5}-\oF_{6})+2\oF_{8}
	\,.
	\label{eq:ggaa.fofF}
\end{align}
For the opposite helicity configurations, we define the helicity states using parity symmetry, i.e.
\begin{equation}
	A_{-\lambdavec} = A_{\lambdavec}
	\left( \langle ij \rangle \leftrightarrow [ji] \right) \,,
\end{equation}
where $-\lambda_i$ indicates the opposite helicity of $\lambda_i$.
We also note that due to the Bose symmetry, helicity amplitudes for $gg \to \gamma\gamma$ are symmetric under the exchange of $1\leftrightarrow 2$ or $3\leftrightarrow 4$, i.e.
\begin{equation}
	\begin{split}
		f_{\lambda_2 \lambda_1 \lambda_3 \lambda_4}(s,t) = f_{\lambda_1
			\lambda_2 \lambda_3 \lambda_4}(s,u)\,, \\ f_{\lambda_1 \lambda_2
			\lambda_4 \lambda_3}(s,t) = f_{\lambda_1 \lambda_2 \lambda_3
			\lambda_4}(s,u) \,.
	\end{split}
	\label{eq:ggaa.Bose}
\end{equation}
This leaves us with 6 fully independent helicity configurations.

\subsection{Feynman integrals}
\label{sec:ggaa.amp.int}

We proceed to the introduction of the last remaining amplitude structure, the Feynman integrals.
For the three-loop four-point massless process, there are three independent integral topologies, planar PL, single nonplanar NPL1, and double nonplanar NPL2
\begin{equation}
	\scriptsize
	\begin{split}
		\text{PL} = \{
		&(k_1)^2, (k_2)^2, (k_3)^2, (k_1-p_1)^2, (k_2-p_1)^2, (k_3-p_1)^2, (k_1-p_{12})^2, (k_2-p_{12})^2, (k_3-p_{12})^2, \\
		&(k_1-p_{123})^2, (k_2-p_{123})^2, (k_3-p_{123})^2, (k_1-k_2)^2, (k_1-k_3)^2, (k_2-k_3)^2
		\} \,, \\
		\text{NPL1} = \{
		&(k_1)^2, (k_2)^2, (k_3)^2, (k_1-p_1)^2, (k_2-p_1)^2, (k_3-p_1)^2, (k_1-p_{12})^2, (k_2-p_{12})^2, (k_3-p_{12})^2, \\
		&(k_1-p_{123})^2, (k_2-p_{123})^2, (k_3-p_{123})^2, (k_1-k_2)^2, (k_2-k_3)^2, (k_1-k_2+k_3)^2
		\} \,, \\		
		\text{NPL2} = \{
		&(k_1)^2, (k_2)^2, (k_3)^2, (k_1-p_1)^2, (k_2-p_1)^2, (k_3-p_1)^2, (k_1-p_{12})^2, (k_3-p_{12})^2, (k_1-k_2)^2, \\
		&(k_2-k_3)^2, (k_1-k_2+p_3)^2, (k_2-k_3+p_{123})^2, (k_2+p_3)^2, (k_1-k_3)^2, (k_2-p_{12})^2
		\} \,,
	\end{split}
\label{eq:ggaa.topos}
\end{equation}
each with 15 generalised propagators.
\begin{figure}[h]
	\centering
	\includegraphics[width=0.99\textwidth]{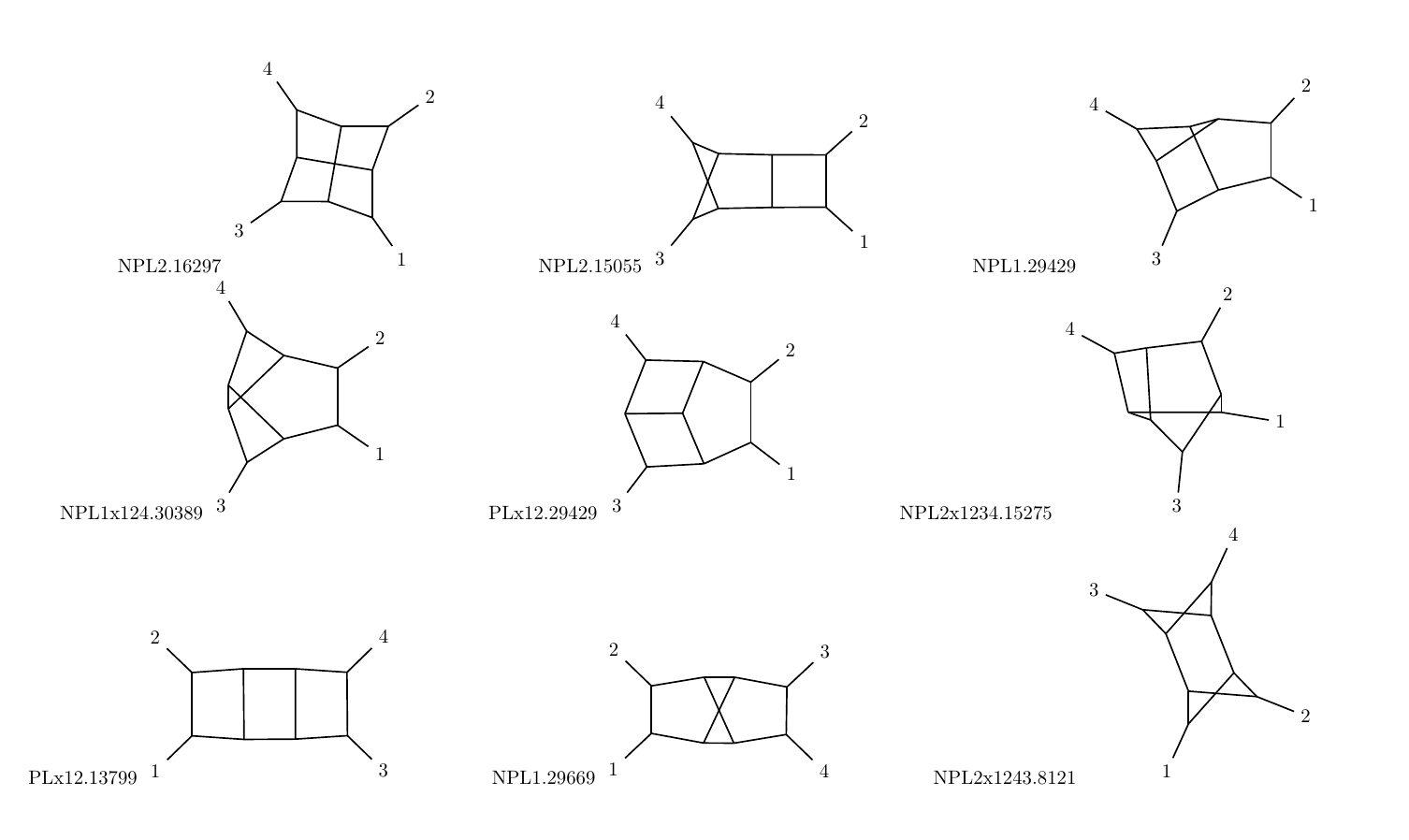}
	\caption{Top sectors of the 3 integral topologies PL, NPL1, and NPL2  defined in Eq.~\ref{eq:ggaa.topos}, in the format TOPOLOGYxCROSSING.SECTOR.
	They correspond to the top sector labels H, B, G, F, E, I, A, C, D of Ref.~\cite{Henn:2020lye}, respectively, up to crossings.}
	\label{fig:ggaa.topsec}
\end{figure}
See the corresponding top sectors in Fig.~\ref{fig:ggaa.topsec}.
In the Laporta notation introduced in Sec.~\ref{sec:amp.int.topo}, we write a Feynman integral $\mathcal{I}_{\vec{n}}$ as
\begin{equation}
	\mathcal{I}_{\vec{n}} 
	= \int \left(\prod_{i=1}^3 \mathcal{D}^d k_i \right) \,
	\frac{\mathcal D_{11}^{-n_{11}} \cdots \mathcal D_{15}^{-n_{15}}}{\mathcal D_{1}^{n_1} \dots \mathcal D_{10}^{n_{10}}} \,,
\end{equation}
where the integration measure is
\begin{equation}
	\mathcal{D}^d k_i = e^{\epsilon \gamma_E} \frac{d^d k_i}{i \pi^{d/2}} \,.
\end{equation}
This choice is convenient for compact expressions for the MIs.
Since it is an overall prefactor, one can reconstruct the physical Feynman measure $d^d k_i / (2\pi)^d$ at the end of the calculation.
\begin{figure}[h]
	\centering
	\includegraphics[width=0.9\textwidth]{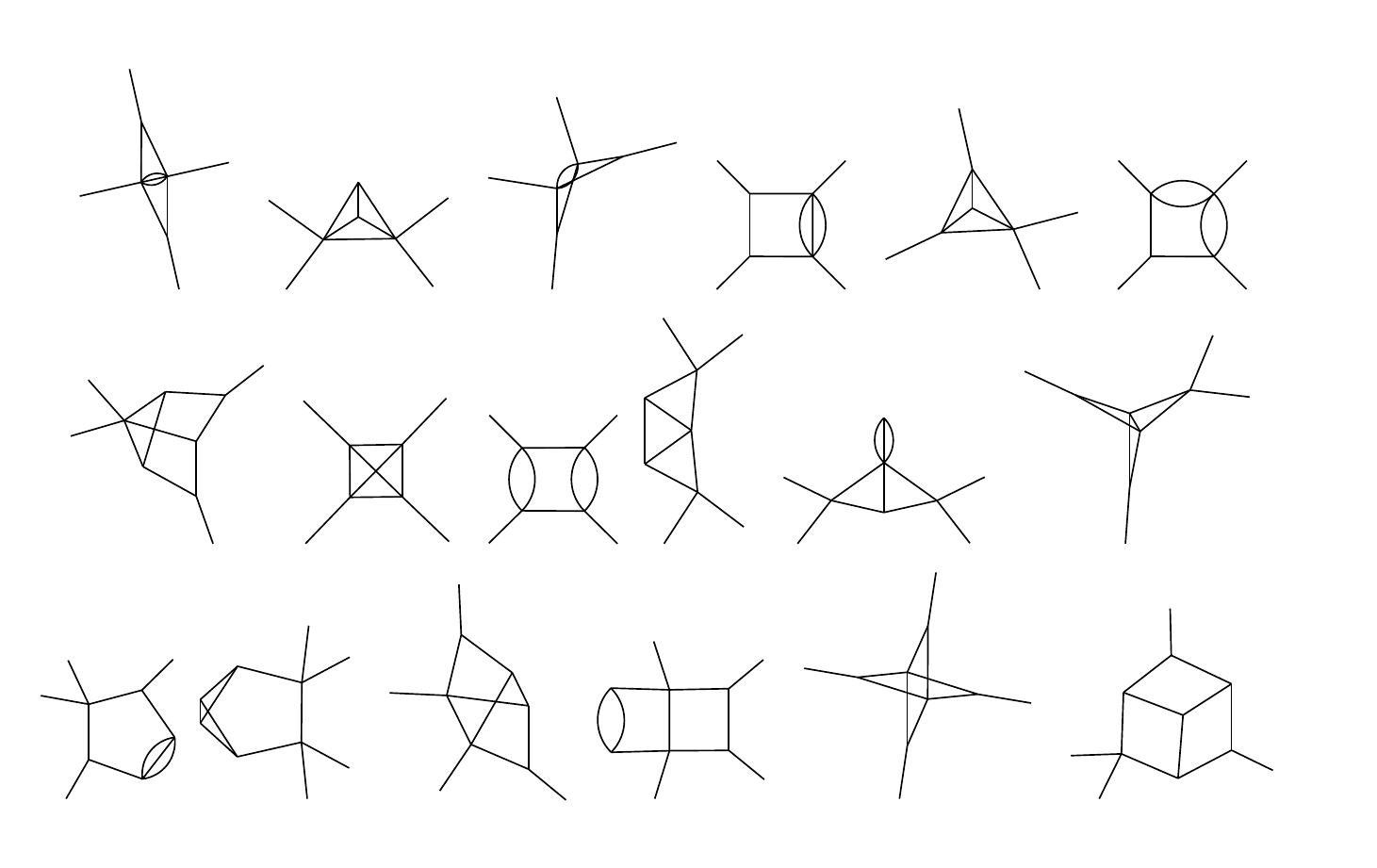}
	\caption{Example subsector Feynman integrals for the three-loop four-point process.}
	\label{fig:ggaa.ints}	
\end{figure}

As mentioned in Sec.~\ref{sec:ggaa.complexity}, after the Lorentz tensor decomposition, there are 4370070 Feynman integrals to be computed, see example graphs in Fig.~\ref{fig:ggaa.ints}.
The required IBP reduction, has to overcome the integrals with the complexity reaching $r=10$, $s=6$, and $d=2$, in the notation defined in Sec.~\ref{sec:amp.int.topo}.
It is especially cumbersome in the double nonplanar topology, which appears for the first time at the three-loop order.
In general, the more nonplanar the integrals are, the more complicated the reduction is.
In Sec.~\ref{sec:ggaa.calc.int}, we describe the modern methods that allowed us to solve the IBP system.
As a result, we related all the required Feynman integrals for this process to a set of 486 MIs.

\section{The calculation}
\label{sec:ggaa.calc}

Here, we give some technical details about the amplitude calculation.
Following an overall procedure outlined in Fig.~\ref{fig:ggaa.flowAmp}, we discuss the computational flow, and we briefly introduce the exploited computer programs.
We also elaborate on some particularities of linear systems of Feynman integrals.

\subsection{Integrand-level manipulations}
\label{sec:ggaa.calc.intd}

\begin{figure}[h]
	\includegraphics[width=0.9\textwidth]{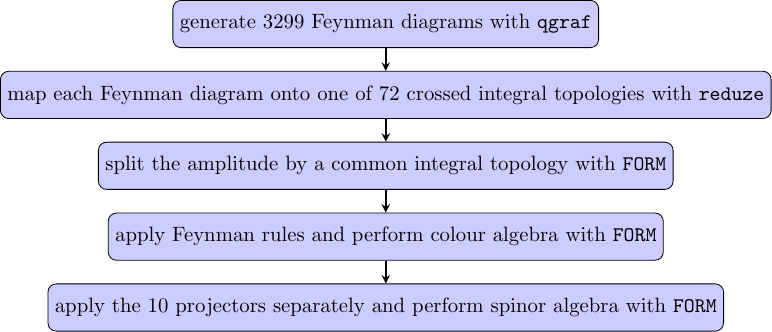}
	\caption{Computational flow for the integrand.}
	\label{fig:ggaa.flowIntd}
\end{figure}

\noindent
We start by describing manipulations of the amplitude performed at the integrand level, as outlined in Fig.~\ref{fig:ggaa.flowIntd}.
We begin from generating all the required 3299 Feynman diagrams for this three-loop process with the \texttt{qgraf}~\cite{Nogueira:1991ex} program.
The run card
\begin{mycode}
output = 'ggaa_3L.frm' ;
style = 'FORM.sty' ;
model = 'qcd';
in = g[p1], g[p2], a[p3], a[p4] ;
out = ;
loops = 3;
loop_momentum = k ;
options = onshell,notadpole,nosnail ;
\end{mycode}
results in a output file \texttt{ggaa\_3L.frm} with expressions in a format specified in \texttt{FORM.sty}, which is readable by the \texttt{FORM}~\cite{Vermaseren:2000nd} program.
The diagrams are generated according to a list \texttt{qcd} of available external states, propagators, and vertices in the theory.
The options \texttt{onshell,notadpole,nosnail} refer to requiring no corrections to external legs, no tadpole subdiagrams, and no snail~\footnote{Tadpole without a tail.} diagrams, respectively.
Applying the first option is justified by the LSZ theorem, while the remaining two follow from the lack of a scale in corresponding diagrams, thus making them vanish in dimReg.
An example output corresponding to the last Feynman diagram in Fig.~\ref{fig:ggaa.diags} yields
\begin{mycode}
G m764=(-1)*FeynDiag(764)*
g(iE1,p1,CiE1)*
g(iE3,p2,CiE3)*
a(iE5,p3,CiE5)*
a(iE7,p4,CiE7)*
gg(i1,i2,Ci1,Ci2,-k1)*
gg(i3,i4,Ci3,Ci4,k1-p1)*
gg(i5,i6,Ci5,Ci6,-k2)*
gg(i7,i8,Ci7,Ci8,k2-p2)*
uU(i9,i10,Ci9,Ci10,-k3)*
uU(i11,i12,Ci11,Ci12,-k3+p3)*
uU(i13,i14,Ci13,Ci14,k1+k2-k3-p1-p2)*
uU(i15,i16,Ci15,Ci16,k1+k2-k3+p3)*
gg(i17,i18,Ci17,Ci18,-k1-k2)*
uU(i19,i20,Ci19,Ci20,k1-k3-p1)*
ggg(i-1,i1,i3,Ci-1,Ci1,Ci3,-p1,k1,-k1+p1)*
ggg(i-3,i5,i7,Ci-3,Ci5,Ci7,-p2,k2,-k2+p2)*
Uau(i12,i-5,i9,Ci12,Ci-5,Ci9,-k3+p3,-p3,k3)*
Uau(i14,i-7,i15,Ci14,Ci-7,Ci15,k1+k2-k3-p1-p2,-p4,-k1-k2+k3-p3)*
ggg(i2,i6,i17,Ci2,Ci6,Ci17,-k1,-k2,k1+k2)*
Ugu(i10,i4,i19,Ci10,Ci4,Ci19,-k3,k1-p1,-k1+k3+p1)*
Ugu(i20,i8,i13,Ci20,Ci8,Ci13,k1-k3-p1,k2-p2,-k1-k2+k3+p1+p2)*
Ugu(i16,i18,i11,Ci16,Ci18,Ci11,k1+k2-k3+p3,-k1-k2,k3-p3);
\end{mycode}
Single letter functions refer to external states, double letter to propagators, while triple letter to vertices.
Lorentz indices are denoted by \texttt{i}, while colour indices by \texttt{Ci}.

Similarly, we generate an output \texttt{ggaa\_3L.yaml} in the style \texttt{reduze.sty}, appropriate for the \texttt{reduze}~\cite{Studerus:2009ye,vonManteuffel:2012np} program.
Then, we use \texttt{reduze} with a run card
\begin{mycode}
jobs:
  - find_diagram_shifts:
      qgraf_file: "ggaa_3L.yaml"
      output_file: "ggaa_3L.matched"
      info_file_form: "ggaa_3L.shifts"
\end{mycode}
to generate a file \texttt{ggaa\_3L.shifts} containing a set of substitution rules in a format readable by \texttt{FORM}.
These identities are shifts of loop momenta $k_{1,2,3}$ for each Feynman diagram, such that its propagators explicitly match with one of the integral topologies.
For our example last Feynman diagram in Fig.~\ref{fig:ggaa.diags}, \texttt{reduze} yields
\begin{mycode}
id FeynDiag(764) = Sector(NPL2, 10, 16297) * Shift(k3, p3-k1+k2, [], k1, p1-k1, [], k2, -p1+k3, []);
\end{mycode}
which corresponds to the shift $k_3 \to p_3-k_1+k_2$, $k_1 \to p_1-k_1$, and $k_2 \to -p_1+k_3$ onto a 10-propagator sector 16297 of the uncrossed topology NPL2.
Note, that in general, \texttt{reduze} maps the diagrams onto all possible 24 crossings \{, x12, x123, x1234, x124, x1243, x12x34, x13, x132, x1324, x134, x1342, x13x24, x14, x142, x1423, x143, x1432, x14x23, x23, x234, x24, x243, x34\} of the 3 integral topologies PL, NPL1, and NPL2, which results in 72 crossed integral topologies.
The first empty entry in this crossings list corresponds to the primary uncrossed kinematics.
The uncrossed integral topologies are specified in \texttt{config/integralfamilies.yaml}, e.g.
\begin{mycode}
integralfamilies:
  - name: "PL"
    loop_momenta: [k1, k2, k3]
    propagators:
      - [ "k1", 0 ]
      - [ "k2", 0 ]
      - [ "k3", 0 ]
      - [ "k1-p1", 0 ]
      - [ "k2-p1", 0 ]
      - [ "k3-p1", 0 ]
      - [ "k1-p1-p2", 0 ]
      - [ "k2-p1-p2", 0 ]
      - [ "k3-p1-p2", 0 ]
      - [ "k1-p1-p2-p3", 0 ]
      - [ "k2-p1-p2-p3", 0 ]
      - [ "k3-p1-p2-p3", 0 ]
      - [ "k1-k2", 0 ]
      - [ "k1-k3", 0 ]
      - [ "k2-k3", 0 ]
\end{mycode}
while the kinematics is defined in \texttt{config/kinematics.yaml} as
\begin{mycode}
kinematics :
  incoming_momenta: [ p1, p2, p3, p4 ]
  outgoing_momenta: [  ]
  momentum_conservation: [p4, "-p1 - p2 - p3"]
  kinematic_invariants:
    - [s23, 2]
    - [s13, 2]
  scalarproduct_rules:
    - [[p1,p1],  0]
    - [[p2,p2],  0]
    - [[p3,p3],  0]
    - [[p1,p2],  "-(s23+s13)/2"]
    - [[p1,p3],  "s13/2"]
    - [[p2,p3],  "s23/2"]
  symbol_to_replace_by_one: s23
\end{mycode}
The second entry of \texttt{kinematics\_invariants} refers to the mass dimension, e.g. $[s_{23}]=2$.
The second entry of \texttt{scalarproduct\_rules} denotes the result of performing a Lorentz scalar product of the first two subentries, e.g. $p_1 \cdot p_1 = 0$.
The \texttt{symbol\_to\_replace\_by\_one} option allows to choose a kinematic invariant to be scaled out from all the expressions in order to make them dimensionless.
It is important to use consistent incoming/outgoing notation for external momenta $p_i$ in both \texttt{qgraf} and \texttt{reduze} to map the diagrams correctly~\footnote{The difference may become explicit when comparing the results of two alternative \texttt{reduze} jobs, \texttt{find\_diagram\_shifts} and \texttt{find\_diagram\_shifts\_alt}.}.

With the two outputs of \texttt{qgraf} and \texttt{reduze} in the \texttt{FORM} format, we are ready to perform the mapping of Feynman diagrams onto integral topologies.
Our main tool, \texttt{FORM}, works similarly to \texttt{Mathematica}, however it implements only a minimal set of computer algebra rules designed for HEP symbolic manipulations.
There are various types of variables to be used, e.g.
\begin{mycode}
Symbol d, Nc, gs, s12, s23, s13 ;
Autodeclare Vector p, k, eps ; 
Autodeclare Index i = d, Ci = Nc ;
CFunction g, a, gg, uU, ggg, Uau, Ugu ;
\end{mycode}
They are easy to understand based on the earlier \texttt{qgraf} output example.
The \texttt{Symbol} type denotes scalar parameters.
The Lorentz vectors are declared automatically with arbitrary postfix, similarly as the Lorentz and colour indices, which take values up to \texttt{d} and \texttt{Nc}, respectively.
The commutative functions can take arguments in multiple types of variables.
As a result, we express our Feynman diagrams in terms of all these variables using \texttt{qgraf}.
Then we apply momentum shift identities generated by \texttt{reduze}.

At this point, our list of Feynman diagrams has already a size of about 3~MB of symbolic expressions, see Tab.~\ref{tab:stats}.
Further steps of colour and spinor algebra will only expand these expressions.
Therefore, a crucial step, which we use multiple times throughout the calculation, is to split the amplitude $\mathcal{A}$ into smaller independent pieces.
It not only allows for parallel evaluation of each amplitude piece $\mathcal{A}_{t}$ on separate CPU cores, but it also prevents performance issues of \texttt{FORM} for too large files.
Here, we split all Feynman diagrams among all the 72 crossed integral topologies.

Next, we apply Feynman rules and perform colour algebra with \texttt{FORM} at each of the 72 amplitude pieces $\mathcal{A}_{t}$ in parallel.
Example identities for the Feynman rules using pattern matching symbol \texttt{?}, delta function \texttt{d\_} and the \texttt{do} loop read
\begin{mycode}
#do N=1,`Nmax'
   id g(i?,p'N',Ci?) = eps'N'(imu'N')*d_(imu'N',i) * d_(Ci,CiExt'N');
   id a(i?,p'N',Ci?) = eps'N'(imu'N')*d_(imu'N',i);
#enddo
id uU(i1?,i2?,Ci1?,Ci2?,p?) = d_(Ci1,Ci2) * i_ * gamma(i1,i2,p) * propagator(p);
id Ugu(i1?,i2?,i3?,Ci1?,Ci2?,Ci3?,p1?,p2?,p3?) = i_ * gs * T(Ci2,Ci1,Ci3)/sqrt2 * gamma(i1,i3,i2);
.sort
\end{mycode}
The colour algebra follows from the identities in Sec.~\ref{sec:amp.col}, e.g.
\begin{mycode}
id T(Ci1?,Ci2?,Ci2?) = 0;
id T(Cia1?,Cii1?,Cii2?) * T(Cia1?,Cij1?,Cij2?)
   = d_(Cii1,Cij2)*d_(Cij1,Cii2) - 1/Nc * d_(Cii1,Cii2)*d_(Cij1,Cij2);
\end{mycode}
The first identity corresponds to the total traceless property of the fundamental generators of SU($N_c$), while the second implements Eq.~\ref{eq:amp.col.tttodelta}.
We then combine all gamma matrices \texttt{gamma(i,j,p)} into spinor strings e.g. \texttt{spst(k1 + p2, imu, eps1, imu, eps4, k2 + p3)}, and we linearly expand them with respect to the momenta.
We point out that since our process has external photons, it is important to identify at this stage the 3 types of closed fermion loops $n_f$, $n_f^V$ and $n_f^{V_2}$, as defined in Sec.~\ref{sec:ggaa.amp.col}.

Finally, we apply one of the 10 projectors $\mathcal{P}_i$ on each amplitude piece $\mathcal{A}_{t}$ in parallel, which leads to 720 form factors $\mathcal{F}_{t,i}$.
As argued in Sec.~\ref{sec:ggaa.amp.ten}, only 8 projectors are physical in the 't Hooft-Veltman scheme.
We consider all 10 projectors only for checking purposes.
We perform the $d$-dimensional trace identities of Dirac $\gamma$ matrices in Eq.~\ref{eq:traces} with an implemented command of \texttt{FORM}, \texttt{Tracen}.
We also sum over polarizations in $d$ dimensions, as in Eq.~\ref{eq:polsum}.
Throughout these steps, it is worth frequently applying kinematic identities e.g.
\begin{mycode}
id p1.p1 = 0;	id p1.eps1 = 0;
\end{mycode}
Since at this point all loop momenta are contracted either with each other $k_i \cdot k_j$ or external momenta $k_i \cdot p_j$, we can linearly relate all these scalar products to propagators in the corresponding integral topology.
In this way, we arrive at a list of around $4 \cdot 10^6$ scalar Feynman integrals in a symbolic expression for the amplitude of a total size of about 20~GB, see Tab.~\ref{tab:stats}.

\subsection{Integral reduction}
\label{sec:ggaa.calc.int}

\begin{figure}[h]
	\includegraphics[width=0.9\textwidth]{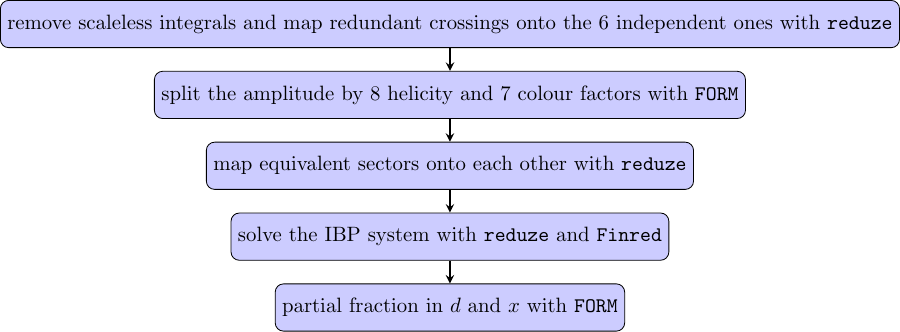}
	\caption{Computational flow for the integral reduction.}
	\label{fig:ggaa.flowInt}
\end{figure}

\noindent
In order to reduce the number of Feynman integrals in the process to a minimal MI basis, we proceed as outlined in Fig.~\ref{fig:ggaa.flowInt}.
We begin by finding scaleless integrals, e.g.
\begin{mycode}
id INT(PLx12x34,3,28,4,0,[],0,0,1,2,1,0,0,0,0,0,0,0,0,0,0) = 0;
\end{mycode}
which corresponds to a cubed one-loop vacuum bubble, which for a massless internal particle is indeed scaleless.
The corresponding \texttt{reduze} job has a structure
\begin{mycode}
jobs:
  - select_reductions:
      input_file: "list.0.NPL1"
      output_file: "list.0.NPL1.tmp"
  - reduce_files:
      equation_files: ["list.0.NPL1.tmp"]
      output_file: "list.0.NPL1.sol"
  - export:
      input_file: "list.0.NPL1.sol"
      output_file: "list.0.NPL1.sol.inc"
      output_format: "form"
\end{mycode}
It finds all the scaleless sectors in the input list of integrals \texttt{list.0.NPL1} and returns an output with resulting identities in the \texttt{FORM} format.
It also maps all the 24 crossings onto the 6 independent ones \{, x12, x123, x124, x1234, x1243\}, e.g.
\begin{mycode}
id INT(PLx12x34,5,8390,5,0,[],0,1,1,0,0,0,1,1,0,0,0,0,0,1,0) = 
  + INT(PL,5,4422,5,0,[],0,1,1,0,0,0,1,0,1,0,0,0,1,0,0)
	* (1);
\end{mycode}
Note the \texttt{reduze} integral notation
\begin{mycode}
INT(TOPOLOGYxCROSSING,r,SECTOR,N,s,[],n1,\dots,n15)
\end{mycode}
with the parameters defined as in Sec.~\ref{sec:amp.int.topo}.
We run this \texttt{reduze} job in parallel on lists of integrals belonging to a common crossed topology.
For performance reasons, it is important to divide longer lists of integrals into files containing at most $4 \cdot 10^4$ integrals, hence the index \texttt{.0.} in the above example.
After applying these identities with \texttt{FORM} on each form factor $\mathcal{F}_{t,i}$ separately, we reduce the list of integrals by $\mathcal{O}(10)$.

Since the above step mixes crossings of integrals, there may be some cancellations between the form factors.
In order to reduce this redundancy, we collect them back into the whole amplitude $\mathcal{A}$.
As the next steps mix the topologies even more, it is convenient to split with \texttt{FORM} the amplitude $\mathcal{A}$ by 7 colour factors $\{n_f^{V_2}$, $N_c (n_f^V)^2$, $N_c^{-1} (n_f^V)^2$, $N_c n_f n_f^{V_2}$,  $N_c^{-1} n_f n_f^{V_2}$, $N_c^2 n_f^{V_2}$, $N_c^{-2} n_f^{V_2}\}$ and 8 helicities $\{++++$, $-+++$, $+-++$, $++-+$, $+++-$, $--++$, $-+-+$, $+--+\}$, as defined in Eq.~\ref{eq:ggaa.fofF}.
Since these structures are independent, there are no cancellations between them.
In this way, we arrive at 56 amplitude pieces $f_{c,h}$.

The next step is to map equivalent sectors between topologies onto each other in order to arrive at as small a number of sectors as possible before the IBP reduction.
The \texttt{reduze} job is the same as in the previous example, however the \texttt{sectormappings} file should be generated with a job \texttt{setup\_sector\_mappings} with an option \texttt{allow\_general\_crossings: true}.
This allows \texttt{reduze} to find shifts of loop momenta which map sectors onto each other beyond a common topology.
For example, the identity
\begin{mycode}
id INT(NPL1,6,21020,6,2,[],-1,0,1,1,1,0,0,0,-1,1,0,0,1,0,1) =
  + INT(PLx123,6,12423,6,2,[],1,1,1,0,-1,0,0,1,0,0,0,-1,1,1,0)
    * (1)
  + INT(PLx123,6,12423,6,2,[],1,1,1,0,-1,0,0,1,-1,0,0,0,1,1,0)
    * (-1)
  + INT(PLx123,6,12423,6,2,[],1,1,1,0,-1,-1,0,1,0,0,0,0,1,1,0)
    * (1)
  + INT(PLx123,6,12423,6,1,[],1,1,1,0,0,0,0,1,-1,0,0,0,1,1,0)
    * (-s23)
  + INT(PLx123,6,12423,6,1,[],1,1,1,0,0,-1,0,1,0,0,0,0,1,1,0)
    * (2*s23)
  + INT(PLx123,6,12423,6,1,[],1,1,1,0,-1,0,0,1,0,0,0,0,1,1,0)
    * (2*s23+s13)
  + INT(PLx123,6,12423,6,0,[],1,1,1,0,0,0,0,1,0,0,0,0,1,1,0)
    * (s23*(2*s23+s13))
  + INT(PLx123,5,12421,5,1,[],1,0,1,0,0,0,0,1,-1,0,0,0,1,1,0)
    * (1)
  + INT(PLx123,5,12421,5,1,[],1,0,1,0,0,-1,0,1,0,0,0,0,1,1,0)
    * (-2)
  + INT(PLx123,5,12421,5,0,[],1,0,1,0,0,0,0,1,0,0,0,0,1,1,0)
    * (-2*s23-s13);
\end{mycode}
expresses a higher 21020 sector integral in a more difficult NPL1 topology in terms of lower sectors 12423 and 12421 of a simpler PLx123 topology.
Contrarily to the previous \texttt{reduze} output, these identities contain linear combinations of integrals with prefactors being polynomials in kinematic invariants.
Since \texttt{FORM} expands all the integral coefficients when substituting these rules into the amplitude, it can be prevented by changing the brackets of the type \texttt{* (s23*(2*s23+s13))} into \texttt{* (s23*num(2*s23+s13))}.
Due to a strict formatting rules of the \texttt{reduze} output, it can be easily achieved with e.g. \texttt{bash} command \texttt{sed}.
For similar complexity-motivated reason, it is useful to apply these identities with \texttt{FORM} only once per integral, e.g. with the following commands
\begin{mycode}
Bracket INT;
.sort
keep brackets;
#include reduze.sector.map.output
\end{mycode}
At this point, some of the numerators \texttt{num()} may cancel with propagators $s_{ij}^{-1}$.
Since \texttt{FORM} does not directly support momentum-conservation-induced fractions e.g. $(s_{23}+s_{13})^{-1}$, one may exploit the built-in command \texttt{polyratfun} to simplify the resulting rational functions.
Overall, this stage reduces the number of remaining integrals by $\mathcal{O}(2)$.

We now describe the IBP method for reducing the $\sim 2 \cdot 10^5$ remaining integrals onto a basis of corresponding 486 MIs.
As introduced in Sec.~\ref{sec:amp.int.IBP}, we rely on the Laporta algorithm~\cite{Laporta:2000dsw}.
We apply it separately to each of the top sectors in Fig.~\ref{fig:topsec}.
After generating the IBP relations, we arrive at up to $\mathcal{O}(10^8)$ linear identities for at most $\mathcal{O}(10^6)$ integrals per top sector.
Due to the sheer size of the resulting linear system, it is computationally challenging to solve it.
We exploit two main methods to overcome this complexity.

Firstly, we use $\mathbb{F}_p$ \textit{finite field} arithmetic to reconstruct all the coefficients of MIs in the decomposition of each given Feynman integral~\cite{vonManteuffel:2014ixa,vonManteuffel:2016xki,Peraro:2016wsq,Peraro:2019svx}.
These coefficients are rational functions of $d$ and $x$, for which we can write a complete ansatz of the form
\begin{equation}
	\frac{\text{poly}(x,d)}{(d-\#)^\# (x-\#)^\#} \,,
	\label{eq:ffansatz}
\end{equation}
following an algorithm introduced in Refs~\cite{vonManteuffel:2014ixa,Heller:2021qkz}.
We then evaluate these rational functions at fixed integer values of $d$ and $x$, modulo a 64 bit $\sim 10^9$ prime number called the \textit{modulus}.
With a large enough number of such numerical probes, it is possible to reconstruct the rational function in its analytic form.
A special case of this approach is a reconstruction of the coefficients of a univariate polynomial of degree $n$ with $n+1$ numerical probes.
In our case, at most $\mathcal{O}(10^4)$ probes per MI coefficient were sufficient.
This approach was implemented in a code \texttt{Finred}~\cite{vonManteuffel:2016xki}.

Secondly, we impose syzygy-based constraints~\cite{Gluza:2010ws,Ita:2015tya,Larsen:2015ped,Bohm:2017qme,Schabinger:2011dz,Agarwal:2020dye} to decrease the number of new integrals with squared denominators introduced by the IBP relations.
It substantially decreases the complexity of the reduction since squared denominators are naturally introduced by the derivative operator.
The term \textit{syzygy} refers to a linear relation between propagators $\mathcal{D}$ over a space of polynomials~\cite{Gluza:2010ws,Ita:2015tya,Larsen:2015ped,Bohm:2017qme,Schabinger:2011dz,Agarwal:2020dye}
\begin{equation}
	v_j^{\mu}(k_l,p_i) \frac{\partial}{\partial k_j^{\mu}} \mathcal D + \text{poly}(k_l,p_i) \mathcal D = 0 \,.
	\label{eq:syzygy}
\end{equation}
It is a well-studied object in the field of algebraic geometry and we can exploit it algorithmically.
In our case, we impose up to $\mathcal{O}(10^2)$ syzygy relations per top sector.
In result, the analytic formulae for all the required IBPs have total size of about 30~GB.
One of the simplest example identities reads
\begin{figure}[H]
	\centering
	\includegraphics[width=0.9\textwidth]{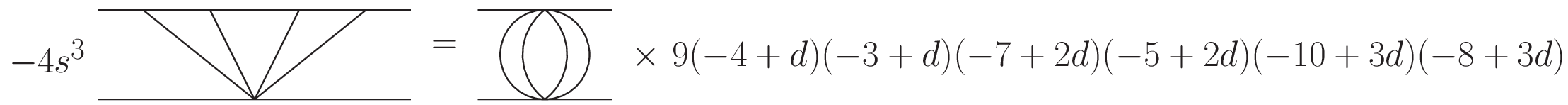}
\end{figure}
\noindent
The implementation of the IBP reduction was based on a mixture of \texttt{C++} and \texttt{Mathematica} code \texttt{Finred}~\cite{vonManteuffel:2016xki}.

In addition, we point out that it is convenient to expose the analytic structure of both IBP relations and the amplitude.
Since for this process the coefficients of MIs are rational functions, they have no branch cuts.
Therefore, the only analytic structure to expose are poles.
In practice, we rely on the method of partial fractioning in both $d$ and $x$, which was performed with \texttt{FORM}.
In fact, the partial-fractioned form provides a systematic way of simplifying the amplitude after substituting the IBP identities.

\subsection{Master Integrals}
\label{sec:ggaa.calc.MI}

\begin{figure}[h]
	\includegraphics[width=0.9\textwidth]{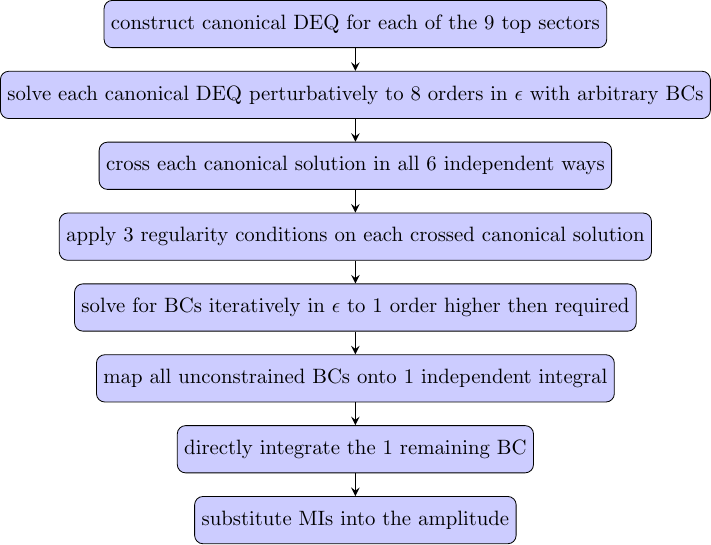}
	\caption{Computational flow for the Master Integrals.}
	\label{fig:ggaa.flowMI}
\end{figure}

\noindent
The three-loop four-point massless MIs have been computed for the first time in Ref.~\cite{Henn:2020lye}.
We independently re-derived these MIs in Ref.~\cite{Bargiela:2021wuy} and found perfect agreement, provided that the result in Ref.~\cite{Henn:2020lye} is analytically continued to the physical Riemann sheet.
We follow the same method as Ref.~\cite{Henn:2020lye}, outlined in Fig.~\ref{fig:ggaa.flowMI} and introduced in Secs~\ref{sec:amp.int.DEQ} and~\ref{sec:amp.int.BC}.

It is convenient to focus just on 1 crossing, since then the number of all 486 MIs drops to 221.
This number does not drop 6 times because some MIs may be symmetric under some crossings.
The remaining 5 crossings can be easily extracted afterwards.
There are 9 top sectors in the 3 uncrossed integral topologies PL, NPL1, and NPL2, as depicted in Fig.~\ref{fig:ggaa.topsec}.
We consider them separately.
It was shown in Ref.~\cite{Henn:2020lye} how to define a canonical MI dlog basis in momentum representation for each integral topology.
The resulting number of canonical MIs in each top sector ranges from 26 to 113.
Their sum does not coincide simply with 221 because some subsectors may be shared by different top sectors.
We provide the definition of canonical MIs in each top sector in terms of Feynman integrals in the Laporta notation in Ref.~\cite{Bargiela:2021wuy}.
The canonical DEQ is constructed by relating the derivative of canonical MIs to themselves using the already available IBP reduction table.
It is worth stressing, that for all 9 top sectors, the resulting DEQ has exactly the same structure as at lower loop orders, as defined in Eq.~\ref{eq:deqCan}, without any new letters.
The constant canonical matrices $a_i$ contain up to 11-digit fractions.

As mentioned in sec~\ref{sec:amp.int.DEQ}, the canonical differential equation can be solved perturbatively in $\ep$, as in Eq.~\ref{eq:deqGenSol}, in terms of HPLs, defined in Eq.~\ref{eq:GPL}, with letters 0 and 1.
Since we define our canonical MIs to be $\mathcal{O}(\ep^0)$, and we expect $\ep^{-2}$ poles per each loop order, we seek a perturbative solution at orders $\ep^0$,\dots,$\ep^6$.
However, as explained below, in order to generate all regularity constraints on the BCs, it is necessary to find a general solution to the canonical DEQ to one order higher, i.e. till $\ep^7$.
This leads to $2^1+\dots+2^7=254$ HPLs up to weight 7.
In practice, we do not have to perform any explicit integration in Eq.~\ref{eq:deqGenSol}, but we rather treat its result as a definition of an HPL with either $\frac{1}{x}$ or $\frac{1}{1-x}$ kernel in Eq.~\ref{eq:GPL}.
An example general solution looks like
\begin{equation}
\begin{split}
	\ep^6 s_{13} \times \,\, &\includegraphics[width=0.3\textwidth]{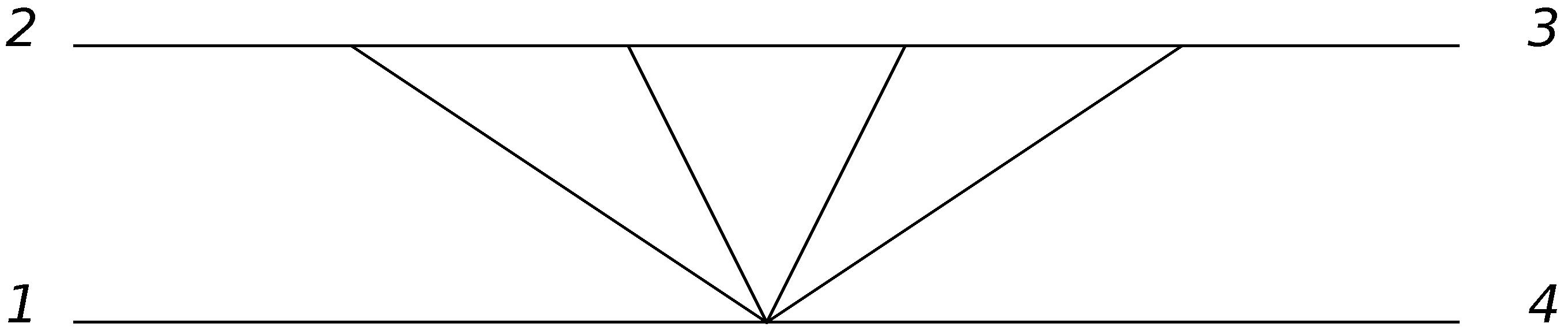}
	= \mathbb{P} e^{\ep \int_{x_0}^x dx' \, \frac{-3}{x'}} \left( \sum_{n=0} \ep^n \, c_n \right) \\
	&= c_0 + \ep \left( c_1 - 3 c_0 G(0;x) \right) \\
	&+ \ep^2 \left( c_2 - 3 c_1 G(0;x) + 9 c_0 G(0,0;x) \right) + \mathcal{O}(\ep^3) \,.
\end{split}
\end{equation}
More complicated solutions depend on both letters 0 and 1, as well as they mix the arbitrary boundary constants $c_{n,i}$ of all other canonical MIs $i$ in a given canonical system.

In order to find all regularity constraints on the BCs for the uncrossed MIs, as introduced in Sec.~\ref{sec:amp.int.BC}, it is necessary to consider all 6 crossed general solutions.
The 6 crossings \{, x12, x123, x124, x1234, x1243\} correspond to the change of variables $x \rightarrow \left\{x,1-x,\frac{1}{1-x},\frac{x-1}{x},\frac{x}{x-1},\frac{1}{x}\right\}$, respectively.
HPLs with crossed arguments can be related to HPLs of the explicit argument $x$ via the fibration procedure, as mentioned in Sec.~\ref{sec:amp.int.GPL}.
This fibration was implemented only up to weight 5 in the \texttt{ToFibrationBasis[GPL[fLinRed(x)], {x}, FitValue $\to$ \{x $\to$ \#Re $\pm$ \#Im\}]} function of the \texttt{PolyLogTools}~\cite{Duhr:2019tlz} package to \texttt{Mathematica}.
For this reason, we implemented our own fibration method up to weight 7.
Let us illustrate it with an example.

Consider a weight 6 HPL $G(1, 0, 0, 0, 0, 0; 1 - x)$.
Assuming that we know all relevant fibration identities at previous weight, i.e. 5, we apply them on a derivative of this HPL, since it lowers the weight by 1, i.e.
\begin{equation}
	\frac{d}{dx} G(1, 0, 0, 0, 0, 0; 1 - x) = \frac{1}{x} G(0, 0, 0, 0, 0; 1 - x) = \frac{1}{x} G(1, 1, 1, 1, 1; x) \,.
\end{equation}
If we now integrate back to weight 6, the only remaining ambiguity is in the integration constant
\begin{equation}
	G(1, 0, 0, 0, 0, 0; 1 - x) = \int \frac{dx}{x} G(1, 1, 1, 1, 1; x) = G(0, 1, 1, 1, 1, 1; x) + C \,.
\end{equation}
The constant $C$ can be reconstructed by probing the resulting relation at $x=0$ or $x=1$, since the special values of HPLs at these two points are well studied.
All the required HPLs with letters 0 and 1 in the relation are always regular at one of these fixed points.
For example, at $x=0$, we have
\begin{equation}
	\zeta_6 = 0 + C \,,
\end{equation}
which leads to the final identity
\begin{equation}
G(1, 0, 0, 0, 0, 0; 1 - x) = G(0, 1, 1, 1, 1, 1; x) + \zeta_6 \,.
\end{equation}
Similarly, we iteratively obtain all the identities in $1-x$ at weight 7 based on the ones at weight 6.

As mentioned in Sec.~\ref{sec:amp.int.GPL}, when changing variables in $x$ for the other 4 crossings, it is crucial not to cross more then one branch cut, in order to correctly perform the analytic continuation to the physical Riemann sheet defined by $s_{ij}+i\varepsilon$, 
The fibration in $1-x$ corresponds to the exchange of $t$ and $u$ Mandelstam variables, and it does not cross any branch cut, contrarily to the remaining 4 crossings.
Consider for example the $\frac{1}{x}$ variable, which corresponds to the exchange of $s$ and $t$.
\begin{figure}[h]
	\centering
	\includegraphics[width=0.9\textwidth]{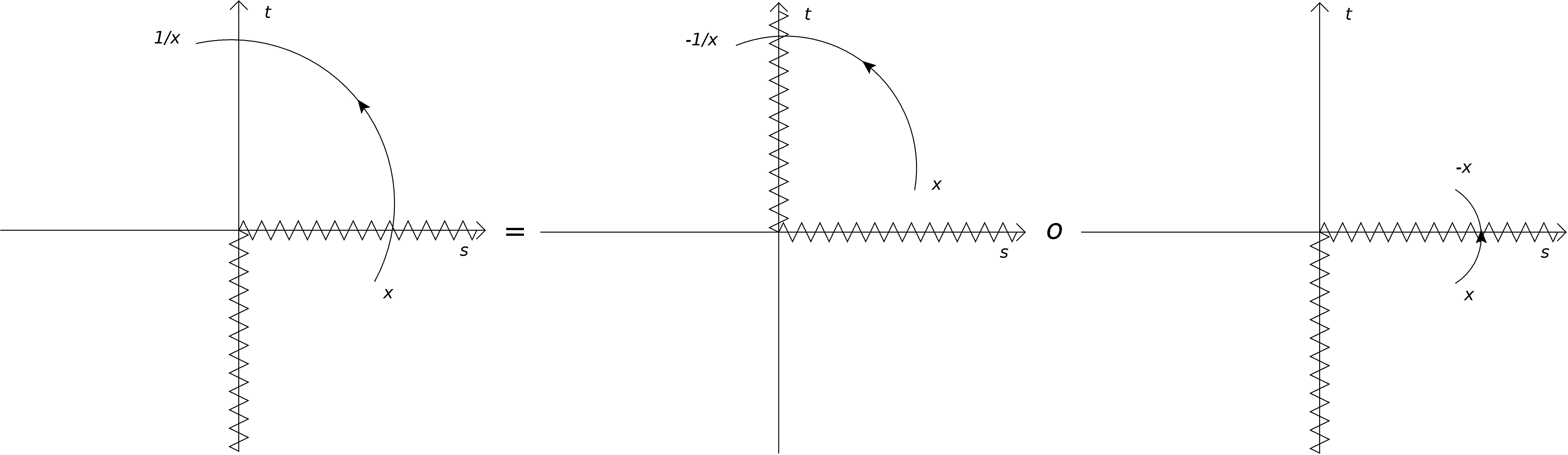}
	\caption{Example analytic continuation analysis.}
	\label{fig:ggaa.fib}
\end{figure}
Naively, we cross only one original branch cut along $s>0$, see Fig.~\ref{fig:ggaa.fib}, however, as soon as we cross it, the physical region $t<0$ becomes $t>0$, which corresponds to the branch cut in $t>0$.
In order to take this into account, we compose the transformation $x \to \frac{1}{x}$ from two steps, $\quad x \to -x$ and $x \to -\frac{1}{x}$, as in Fig.~\ref{fig:ggaa.fib}, following Ref.~\cite{Caola:2021rqz}.
Each of these two steps crosses only one branch cut at a time.
By analysing the causal Feynman prescription $s_{ij}+i\varepsilon$ in each of the 3 regions, one sees that in both transformations, the analytic continuation should have a positive sign $+i\varepsilon$.
Moreover, one can realise that all of the remaining 3 crossings can be composed from the 3 basic transformations $1-x$, $-x$ and $-\frac{1}{x}$.
Following the above analysis, we arrive at all 6 crossed general solutions to the canonical DEQ in each top sector.
It is important to notice that since we generate all crossed canonical solutions from the primary uncrossed one, the boundary constants $c_{n,i}$ stem only from the primary solution.

We now apply the 3 regularity conditions, as introduced in Sec.~\ref{sec:amp.int.BC}
\begin{equation}
	\lim\limits_{x \to 0,1,\infty} g(x,\ep) = x^{a_{0,1,\infty} \, \ep} g\left( \lim\limits_{x \to 0,1,\infty} x,\ep \right) \qquad \text{regular} \,,
\end{equation}
on each crossed canonical solution in each top sector.
For example, one of the simplest resulting identities yields
\begin{figure}[H]
	\centering
	\includegraphics[width=0.8\textwidth]{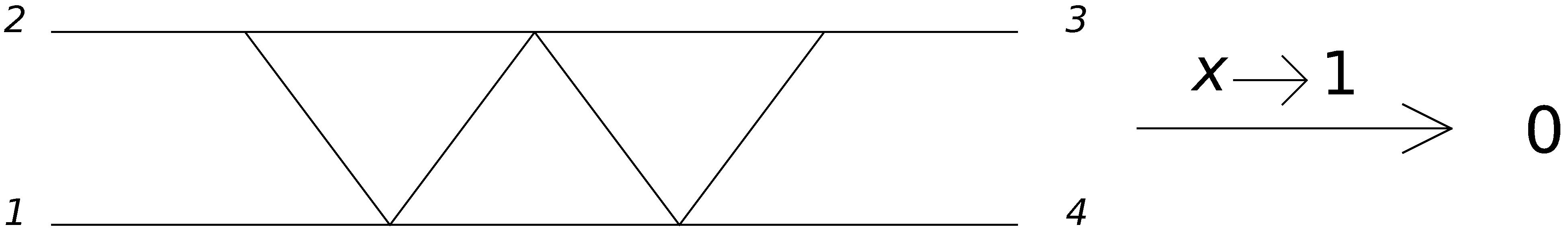}
\end{figure}
\noindent
When we combine the constraints resulting from the 3 limiting behaviours of the 6 crossed solutions in a fixed top sector, we obtain a system of linear relations between the boundary constants holding at any order in $\ep$.

In order to apply a maximal number of constraints at fixed order in $\ep$, we solve the corresponding linear system iteratively to one order higher.
It can be understood by noting that higher orders bring higher-weight HPL functions, which in general may have different analytic structure, thus leading to new constraints.
For example, at the leading $\ep^0$ order, we need to apply all the conditions resulting from both orders $\ep^0$ and $\ep^1$, which correspond to constant terms, as well as weight 1 HPLs $G(0,x)$ and $G(1,x)$.
In this way, we generate enough constraint to fix all but at most two constants at the $\ep^0$ order, depending on the top sector.
We proceed similarly up to order $\ep^6$, which is fixed by constraints at orders $\ep^6$ and $\ep^7$, and it requires evaluating the HPLs in the 3 limits $x \to 0,1,\infty$ up to weight 7.
It is important to stress, that this approach relates even the constant integrals to the few unconstrained boundary integrals.
Even though their derivative vanishes, they can be constrained because they are intertwined with higher sector integrals in the canonical system.
We also note that for more complicated top sectors, the relevant linear system of up to $\mathcal{O}(200)$ variables with rational number coefficients may be computationally nontrivial to invert.
For this reason, in order to improve the performance of the built-in \texttt{Solve[]} function of \texttt{Mathematica}, we implemented our own linear reductor based on the Gauss elimination method and the \texttt{RowReduce[matrix, Method $\to$ "OneStepRowReduction"]} function.

As mentioned in Sec.~\ref{sec:amp.int.BC}, it is possible to relate all the unconstrained integrals in each top sector to 1 overall integral.
It can be done by shifting the loop momenta such that the lower rank subsectors are mapped between the top sectors, as described in Sec.~\ref{sec:ggaa.calc.int}.
In this way, we relate all the boundary integrals to 1 overall normalization, which we can choose to be the simplest subsector in this kinematics, i.e. the three-loop sunrise.
An example sector mapping onto the sunrise subsector is
\begin{equation}
	I^{(\text{PL})}_{2,0,0,0,0,0,0,0,2,0,0,0,2,0,1} = I^{(\text{NPL1})}_{2,0,0,0,0,0,0,0,2,0,0,0,2,1,0} = \includegraphics[width=0.08\textwidth]{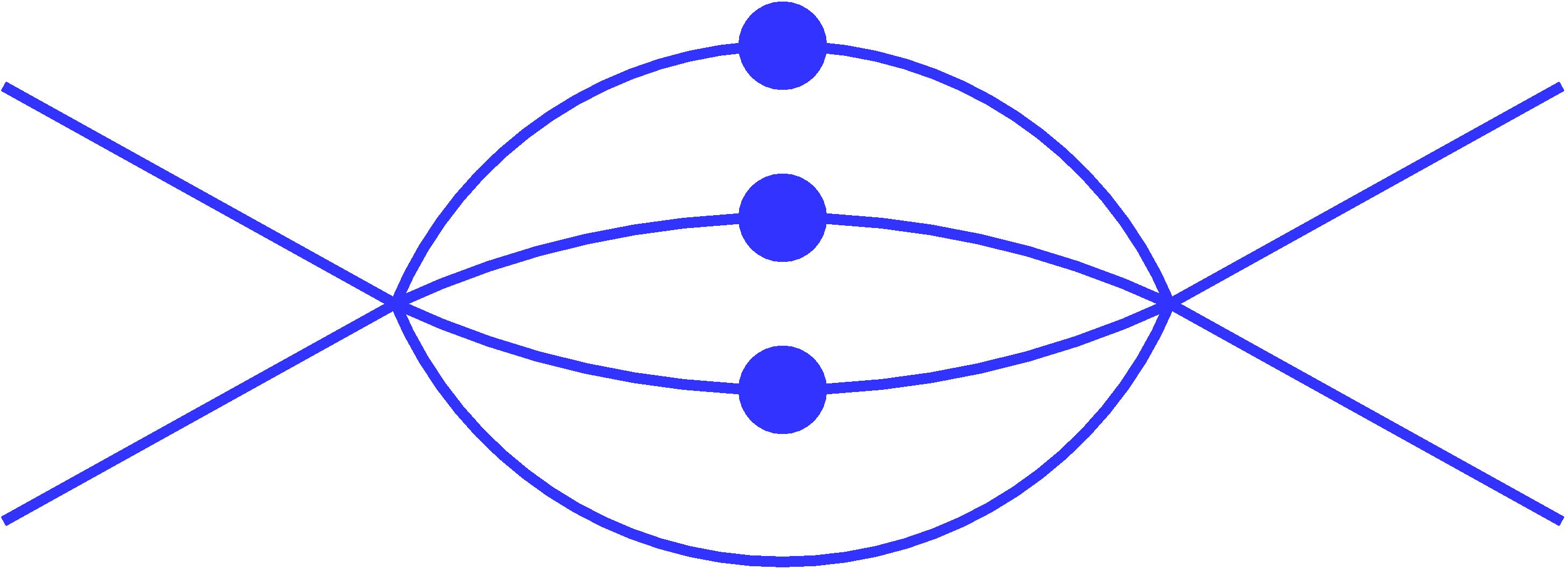} \,.
\end{equation}
Let us stress again how profound the reduction in complexity is.
Indeed, after projecting our tensor structures, IBP reducing scalar integrals to MI basis, finding kinematic dependence of MIs with DEQs, and finally constraining BCs for DEQs with regularity properties, there is only 1 independent constant integral to be computed using direct integration methods.
For completeness, we also recomputed and verified this property for all two-loop four-point massless MIs up to weight 6 in Ref.~\cite{Bargiela:2021wuy}.

\begin{figure}[H]
	\centering
	\includegraphics[width=0.9\textwidth]{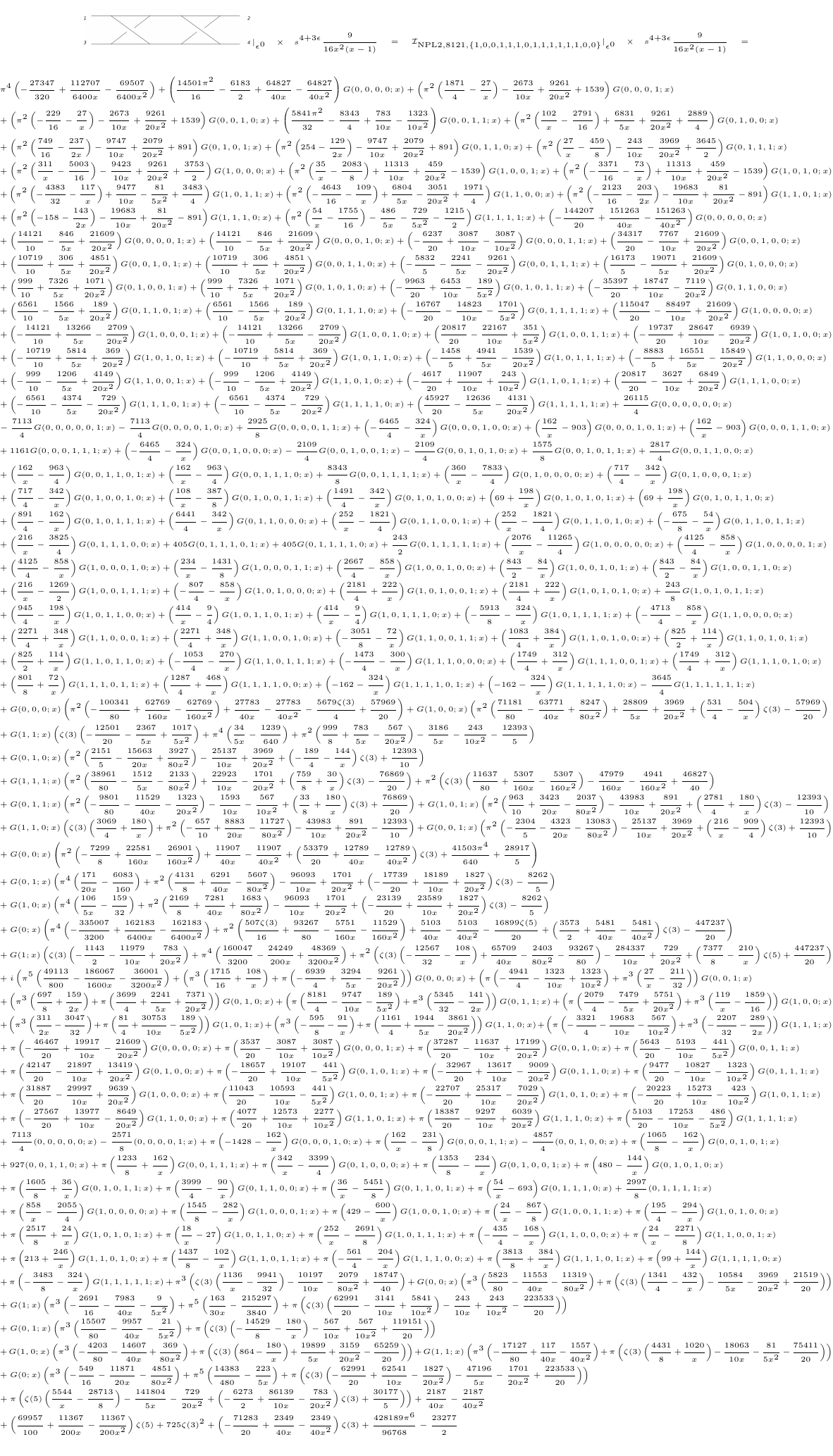}
	\caption{$\ep^0$ coefficient of the double nonplanar triple box integral.}
	\label{fig:ggaa.NPL3box}
\end{figure}

Let us now explicitly compute the three-loop sunrise integral
\begin{equation}
	\includegraphics[width=0.08\textwidth]{figures/jaxoDiags/3Lsunrise.jpg} \, (s)
	= \int \left(\prod_{i=1}^3 \mathcal{D}^d k_i \right) \,
	\frac{1}{(k_3)^2(k_{12})^4(k_{23})^4(k_1+p_{12})^4} \,.
\end{equation}
Amazingly, even at this stage, there are simplifications.
Indeed, the sunrise integral can be factorized in terms of one-loop bubbles, with arbitrary powers
\begin{equation}
	\int \frac{d^dk}{i \pi^{d/2}} \frac{1}{(-k^2)^{n_1}(-(k+q)^2)^{n_2}}
	= \frac{1}{(-q^2)^{n_{12}-d/2}} \,\,\times\,\, \frac{\Gamma(n_{12}-d/2) \Gamma(d/2-n_1) \Gamma(d/2-n_2)}{\Gamma(n_1) \Gamma(n_2) \Gamma(d-n_{12})} \,.
	\label{eq:1Lbubpowers}
\end{equation}
We see that this two-propagator expression integrates to a single propagator with a $\Gamma$-function-dependent prefactor
\begin{figure}[H]
	\centering
	\includegraphics[width=0.4\textwidth]{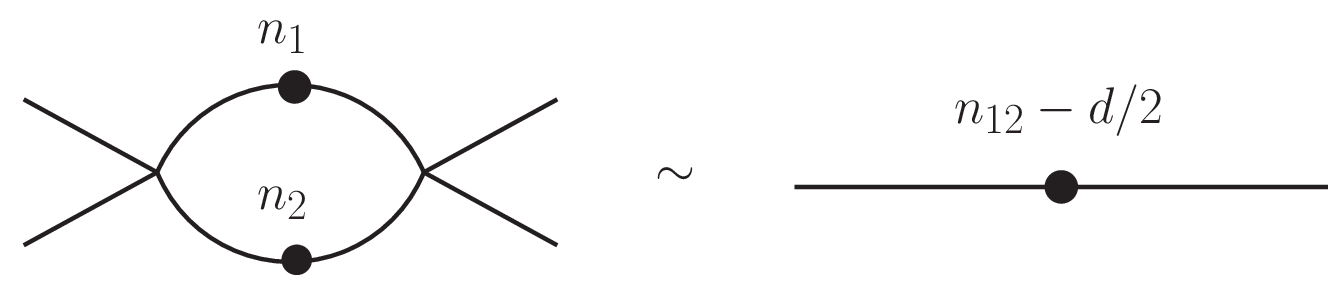}
\end{figure}
\noindent
Iteratively, we can apply this factorization three times
\begin{figure}[H]
	\centering
	\includegraphics[width=0.45\textwidth]{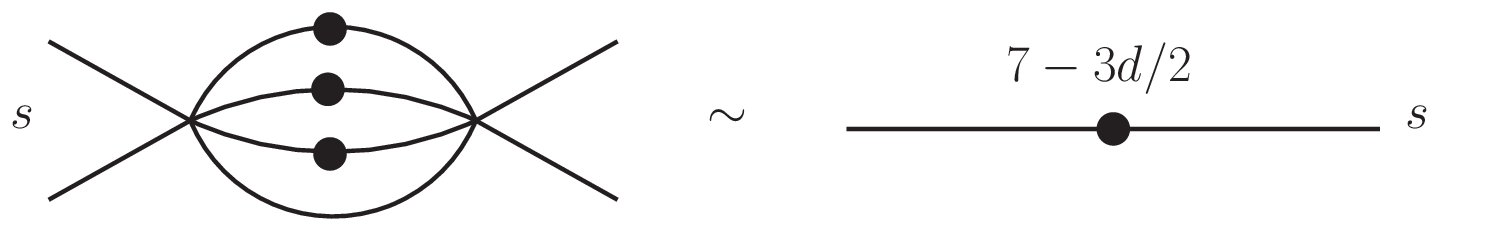}
\end{figure}
\noindent
to finally arrive at
\\
\scalebox{0.8}{\parbox{1.0\linewidth}{
\begin{equation}
	\begin{split}
		&-\ep^3 s^{1+3\ep} \int \left( \prod_{l=1}^{3} \frac{e^{\ep \gamma_E} d^dk_l}{i \pi^{d/2}} \right) \frac{1}{(k_3)^2(k_{12})^4(k_{23})^4(k_1+p_{12})^4} \\
		&= (-s-i\varepsilon)^{-3\ep} e^{3\ep\gamma_E} \frac{\Gamma(1-\ep)^4 \Gamma(1+3\ep)}{\Gamma(1-4\ep)} \\
		&=1
		+3i\pi\epsilon
		-\frac{19\pi^2}{4}\epsilon^2
		+\left(-29\zeta_3-\frac{21i\pi^3}{4}\right)\epsilon^3
		+\left(\frac{649\pi^4}{160}-87i\pi\zeta_3\right)\epsilon^4 \\
		&+\left(\frac{551\pi^2\zeta_3}{4}-\frac{1263\zeta_5}{5}+\frac{291i\pi^5}{160}\right)\epsilon^5
		+\left(\frac{609}{4}i\pi^3\zeta_3+\frac{841\zeta_3^2}{2}-\frac{3789i\pi\zeta_5}{5}-\frac{8137\pi^6}{24192}\right)\epsilon^6
	\end{split}
\label{eq:ggaa.sunrise}
\end{equation}
}}
\\
Since it is the only independent BC, we use it to obtain all the MIs in the problem.
See for example Fig.~\ref{fig:ggaa.NPL3box} for resulting formula for the NPL2 double nonplanar triple box top sector integral at $\ep^0$ order.
We stress, that computing this integral with direct integration methods is currently out of reach.

The only remaining step is to include all the obtained expressions for MIs into the amplitude using \texttt{FORM}.
Note, that since the canonical MIs are finite, all the poles in $\ep$ are carried by the IBP coefficients.
In order to correctly truncate the amplitude at $\ep^0$ order, the IBP coefficients need to be expanded from $\ep^{-6}$ to $\ep^0$.

\section{Checks and results}
\label{sec:ggaa.res}

\begin{figure}[H]
	\centering
	\includegraphics[width=0.45\textwidth]{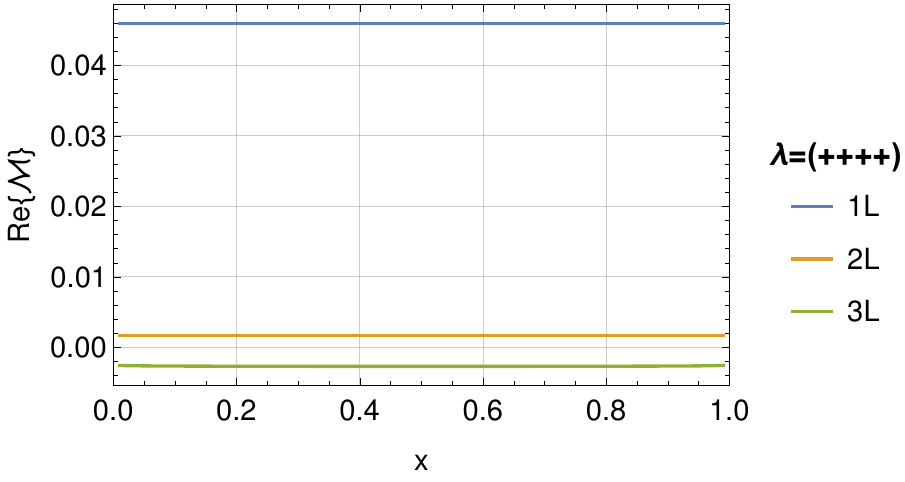}
	%\hfill
	\includegraphics[width=0.40\textwidth]{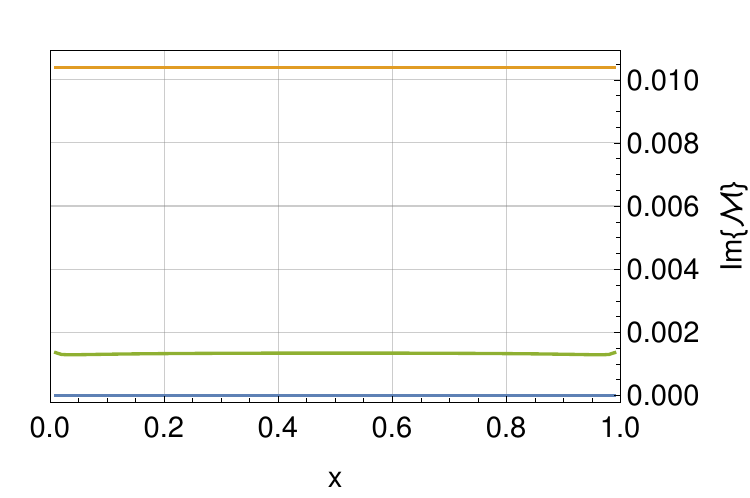}
	\includegraphics[width=0.45\textwidth]{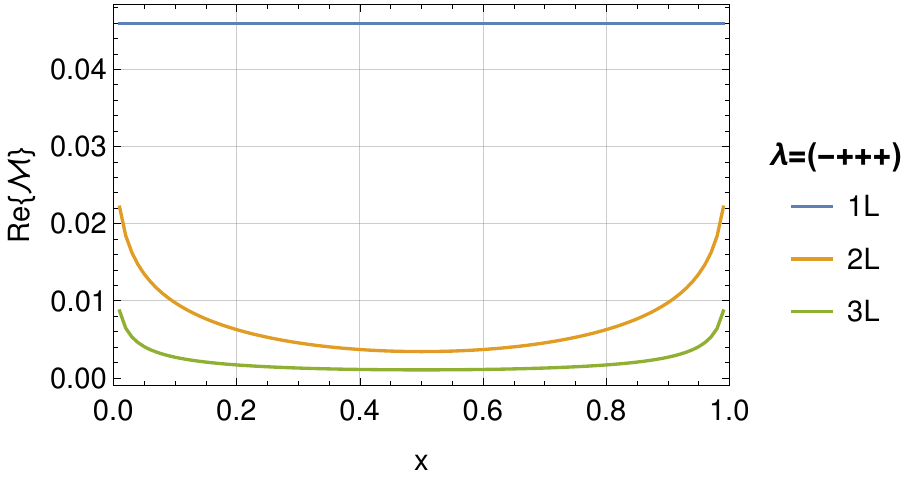}
	%\hfill
	\includegraphics[width=0.40\textwidth]{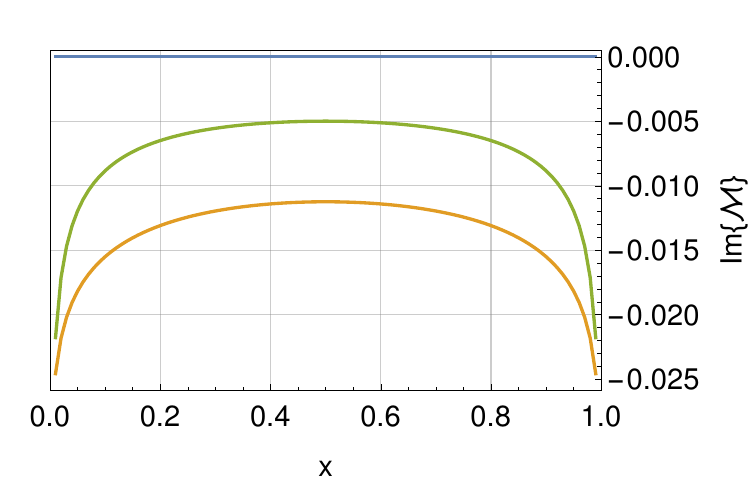}
	\includegraphics[width=0.45\textwidth]{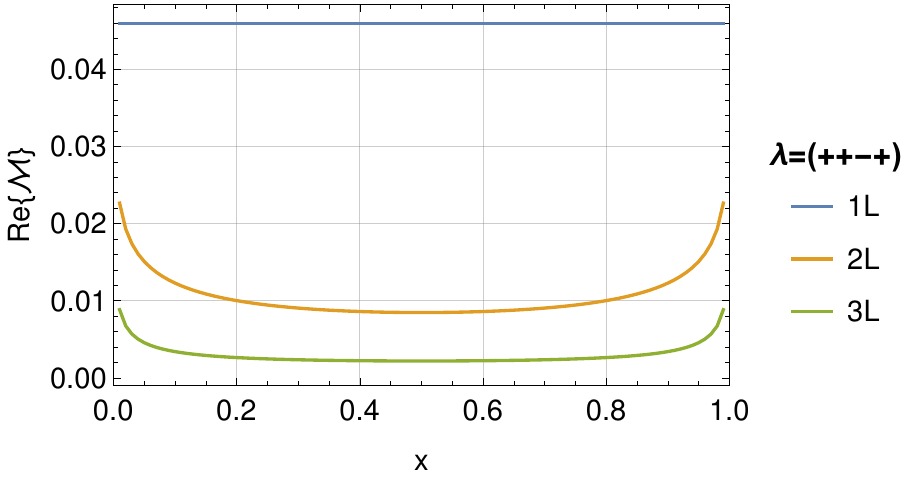}
	%\hfill
	\includegraphics[width=0.40\textwidth]{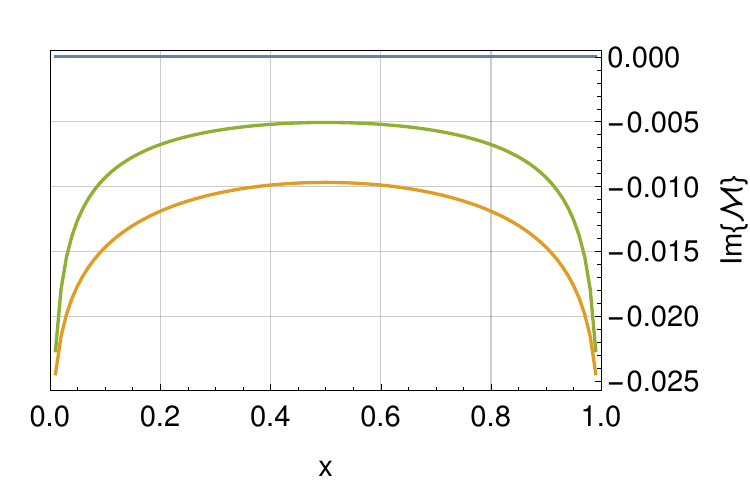}
	\includegraphics[width=0.45\textwidth]{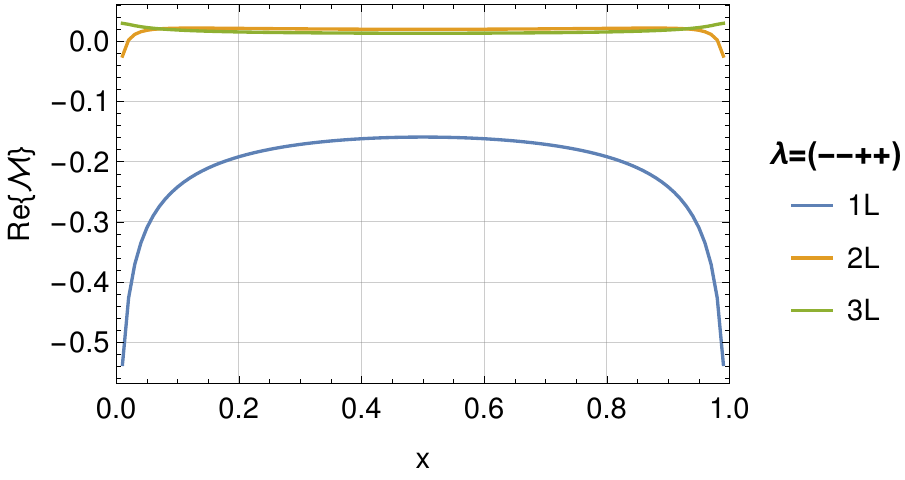}
	%\hfill
	\includegraphics[width=0.40\textwidth]{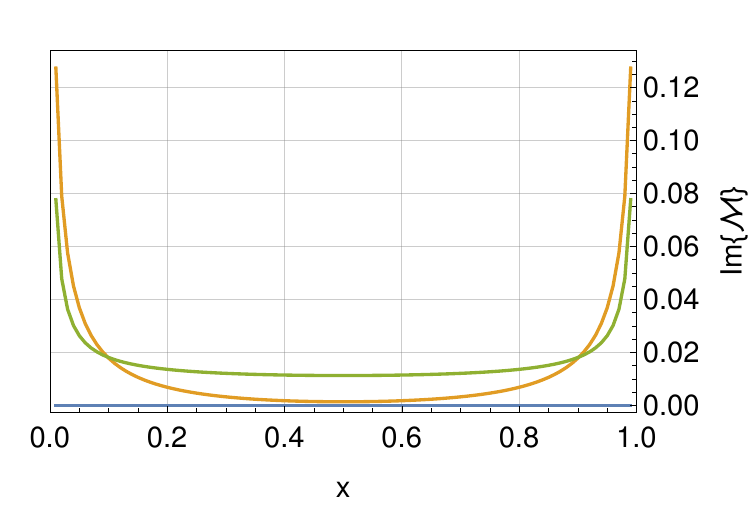}
	\includegraphics[width=0.45\textwidth]{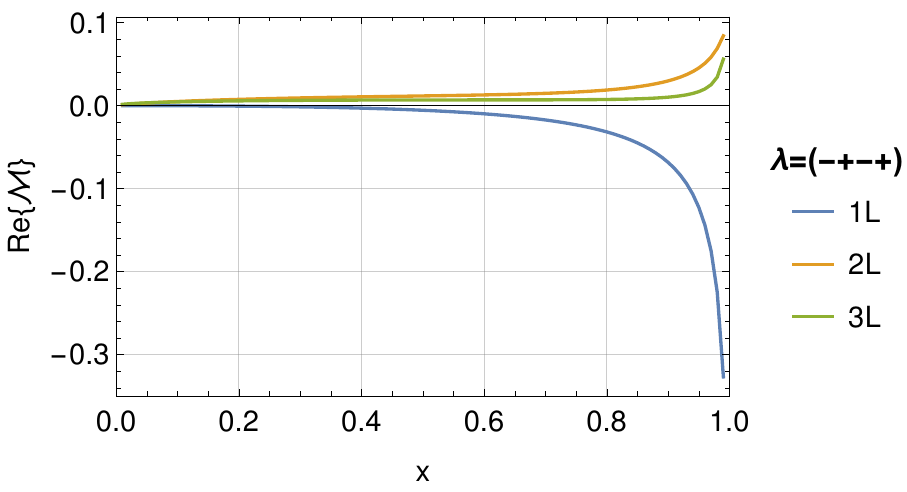}
	%\hfill
	\includegraphics[width=0.40\textwidth]{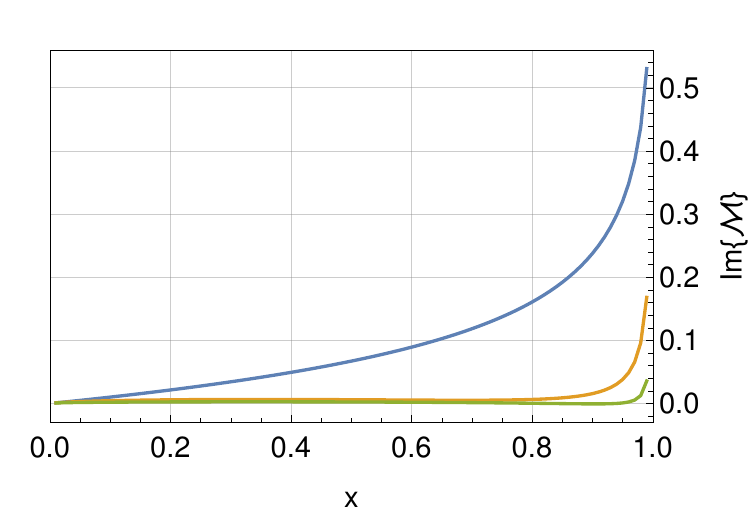}
	\caption{Finite part of amplitudes $\mathcal{M}^{(L)}_\lambdavec \equiv \left(\frac{\alpha_s}{2\pi}\right)^L \, f^{(L,{\rm fin})}_\lambdavec$ as functions of $x=-t/s$.}
	\label{fig:ggaa.res}
\end{figure}

\noindent
Having described the computational flow, we now move to the discussion of the results.
As argued in Sec.~\ref{sec:amp.div}, all the new physical information coming from the higher order correction is embedded in an appropriately defined finite part.
Following the procedure in Sec.~\ref{sec:amp.div}, we subtract both UV and IR poles in $\ep$.
Since the $gg \to \gamma\gamma$ amplitude does not have a tree level contribution, the bare three-loop amplitude has an $\ep$ pole structure of a genuine two-loop bare amplitude.

We renormalize the bare amplitudes in the $\overline{ \rm MS}$ scheme, as defined in Sec.~\ref{sec:amp.div.UV}.
As a consequence, the bare perturbative series in Eq.~\ref{eq:ggaa.pert} for the spinor-stripped helicity amplitudes $f_\lambdavec$ can be written in terms of the renormalized strong coupling as
\begin{equation}
	f_\lambdavec =
	(4 \pi \alpha) \left(\frac{\alpha_s}{2\pi}\right) \left[ f_\lambdavec^{(1)} + \left(\frac{\alpha_s}{2\pi}\right)
	f_\lambdavec^{(2)} + 
	\left(\frac{\alpha_s}{2\pi}\right)^2
	f_\lambdavec^{(3)} + \mathcal O(\alpha_s^3)\right] \,,
	\label{eq:ggaa.pert.ren}
\end{equation}
We provide the results at fixed renormalization scale $\mu^2=s$.
The full dependence on $\mu$ can be obtained using renormalisation group equation arguments.

Following the discussion in Sec.~\ref{sec:amp.div.IR}, we regularize the remaining IR divergences in $\ep$ with Catani operators defined in Eq.~\ref{eq:amp.Catani}.
For the $gg \to \gamma\gamma$ process, they have a colour-singlet form
\begin{align}
	\mathcal I_1(\epsilon) &= -\frac{e^{i\pi\epsilon}e^{\gamma_E\epsilon}}{\Gamma(1-\epsilon)}
	\left(\frac{C_A}{\epsilon^2}+\frac{\beta_0}{\epsilon}\right),
	\notag\\
	\mathcal I_2(\epsilon) &= -\frac{1}{2}\mathcal I_1(\epsilon)
	\left(\mathcal I_1(\epsilon)+\frac{2\beta_0}{\epsilon}\right)+
	\frac{e^{-\gamma_E\epsilon}\Gamma(1-2\epsilon)}{\Gamma(1-\epsilon)}
	\left(\frac{\beta_0}{\epsilon}+K\right)\mathcal I_1(2\epsilon) + 2\frac{e^{\epsilon\gamma_E}}{\Gamma(1-\epsilon)}H_g \,,
	\label{eq:ggaa.Catani}
\end{align}
with the NNLO cusp anomalous dimension $K$ introduced in Eq.~\ref{eq:amp.K}, and~\cite{Harlander:2000mg}
\begin{equation}
	H_g = \frac{1}{2\epsilon} \left[ \left(\frac{\zeta
		(3)}{4}+\frac{5}{24}+\frac{11 \pi ^2}{288}\right) C_A^2 +T_F n_f
	\left(\frac{C_F}{2}-\left(\frac{29}{27}+\frac{\pi^2}{72}\right)C_A\right)+\frac{10
	}{27}T_F^2 n_f^2 \right] \,.
	\label{eq:ggaa.H}
\end{equation}
As a consequence, we define the finite part $f_\lambdavec^{(L,{\rm fin})}$ via
\begin{align}
	f_\lambdavec^{(1)} &= f_\lambdavec^{(1,{\rm fin})}, \notag\\  f_\lambdavec^{(2)} &=
	\mathcal I_1\, f_\lambdavec^{(1)} + f_\lambdavec^{(2,{\rm fin})}, \notag\\ 
	f_\lambdavec^{(3)} &= \mathcal I_2\, f_\lambdavec^{(1)} + \mathcal
	I_1\,f_\lambdavec^{(2)} + f_\lambdavec^{(3,{\rm fin})} \,.
	\label{eq:ggaa.fin}
\end{align}

We finally proceed to the presentation of the results for $f_\lambdavec^{(L,{\rm fin})}$.
Despite huge complexity at intermediate stages, the final result is remarkably compact.
The simplest helicity configuration is $\lambdavec =(++++)$.
It is a consequence of the optical theorem.
Indeed, all possible combinations of tree-level helicity amplitudes resulting from the unitarity cuts vanish.
Therefore, the one-loop result is just a rational number.
The finite part of the all-plus $gg \to \gamma\gamma$ amplitude at one, two, and three loops reads
\begin{align}
	f^{(1,{\rm fin})}_{++++} &= 2 n_f^{V_2} \,,
	\label{ampppppLO}\\
	f^{(2,{\rm fin})}_{++++} &= 2 n_f^{V_2} \left(2 C_A - 3 C_F + i\pi\beta_0 \right) \,,
	\label{ampppppNLO}
	\\
	f^{(3,{\rm fin})}_{++++} &= 
	\Delta_1(x)\, n_f^{V_2} C_A^2
	+ \Delta_2(x)\, n_f^{V_2} C_A C_F
	+ \Delta_3(x)\, n_f n_f^{V_2}  C_A
	+ \Delta_4(x)\, (n_f^V)^2 C_A \nonumber \\
	&\quad + \Delta_5(x)\, n_f^{V_2} C_F^2 
	+ \Delta_6(x)\, (n_f^V)^2 C_F
	+ \Delta_7(x)\, n_f n_f^{V_2} C_F
	+ \Delta_8(x)\, n_f^2 n_f^{V_2} \nonumber \\
	&\quad + \{(x)\leftrightarrow(1-x)\} \,,
	\label{amppppp}
\end{align}
with
\\
\scalebox{0.8}{\parbox{0.9\linewidth}{
\allowdisplaybreaks
\begin{align}
	\Delta_1(x) &=
	-
	\mfrac{23 L_1 (L_1+2 i \pi )}{9 x^2}+
	\mfrac{32 L_1 (L_1+2 i \pi )-
		46 (L_1+i \pi )}{9x}-
	\mfrac{17}{36} L_0^2-
	\mfrac{19}{36} L_0 L_1+
	\mfrac{1}{9}L_0-2 i \pi L_0
	\nonumber \\ & \quad
	+\mfrac{1}{288}\pi ^4
	-\mfrac{373}{72} \zeta_3
	-\mfrac{185}{72} \pi ^2
	+\mfrac{4519}{324}
	+\mfrac{1}{2}i \pi  \zeta_3
	+\mfrac{11}{144} i \pi ^3
	+\mfrac{157}{12} i \pi 
	+ \mfrac{43}{9} L_0 x
	\nonumber \\ & \quad
	-\mfrac{7}{9} x^2 \left((L_0-L_1)^2
	+\pi ^2\right)
	\,, \nonumber \\
	%%%
	\Delta_2(x) &=
	\mfrac{8 L_1 (L_1+2 i \pi )}{3 x^2}
	+\mfrac{16 (L_1+i \pi )
		-{8} L_1 (L_1+2 i \pi )}{3x}
	-\mfrac{1}{3}L_0^2
	+\mfrac{5 }{6}L_0 L_1
	+\mfrac{17}{3}L_0+i \pi  L_0
	-\mfrac{5 }{12}\pi ^2
	\nonumber \\ & \quad
	-\mfrac{199}{6}
	-{8} i \pi 
	-\mfrac{16}{3} L_0 x
	+\mfrac{4}{3} x^2 \left((L_0-L_1)^2+\pi ^2\right)
	\,, \nonumber \\
	%%%
	\Delta_3(x) &= 
	\mfrac{L_1 (L_1+2 i \pi )}{18 x^2}
	+\mfrac{2(L_1+i \pi )
		- L_1 (L_1+2 i \pi )}{18x}
	-\mfrac{1}{36}L_0^2
	+\mfrac{1}{36}L_0 L_1
	-\mfrac{1}{9}L_0
	-\mfrac{61 }{36}\zeta_3
	+\mfrac{475}{432} \pi ^2
	\nonumber \\ & \quad
	-\mfrac{925}{324}
	-\mfrac{1}{72}i \pi ^3
	-\mfrac{175 }{54}i \pi 
	+\mfrac{2}{9} L_0 x
	+\mfrac{1}{36} x^2 \left((L_0-L_1)^2+\pi ^2\right)
	\,, \nonumber \\
	%%%
	\Delta_4(x) &=
	-\mfrac{5 L_1 (L_1+2 i \pi )}{4 x^2}
	+\mfrac{ L_1 (L_1+2 i \pi )-8 (L_1+i \pi )}{2x}
	+\mfrac{1}{4}L_0^2
	-\mfrac{1}{4}L_0 L_1
	-{2} L_0
	-{6} \zeta_3
	+\mfrac{1}{8}\pi ^2
	-\mfrac{1}{2}
	\nonumber \\ & \quad 
	+{4} L_0 x
	-x^2 \left((L_0-L_1)^2+\pi ^2\right)
	\,, \nonumber \\
	%%%
	\Delta_5(x) &= 
	-\mfrac{L_1 (L_1+2 i \pi )}{x^2}
	+\mfrac{L_1 (L_1+2 i \pi )-2 (L_1+i \pi )}{x}
	-\mfrac{1}{2}L_0^2
	-i \pi  L_0
	+\mfrac{39}{4}
	+i \pi 
	+{2} L_0 x
	\nonumber \\ & \quad
	-\mfrac{1}{2} x^2 \left((L_0-L_1)^2+\pi ^2\right)
	\,, \nonumber \\
	%%%
	\Delta_6(x) &= 
	\mfrac{10 L_1 (L_1+2 i \pi )}{3 x^2}
	+\mfrac{32 (L_1+i \pi )
		-{4} L_1 (L_1+2 i \pi )}{3x}
	-\mfrac{2}{3} L_0^2
	+\mfrac{2}{3} L_0 L_1
	+\mfrac{16}{3} L_0
	+{16} \zeta_3
	-\mfrac{1}{3}\pi ^2
	\nonumber \\ & \quad
	+\mfrac{4}{3}
	-\mfrac{32}{3} L_0 x
	+\mfrac{8}{3} x^2 \left((L_0-L_1)^2+\pi ^2\right)
	\,, \nonumber \\
	%%%
	\Delta_7(x) &= 
	\mfrac{5 L_1 (L_1+2 i \pi )}{3 x^2}
	+\mfrac{10 (L_1+i \pi )
		-8 L_1 (L_1+2 i \pi )}{3x}
	+\mfrac{2}{3} L_0^2
	+\mfrac{1}{3}L_0 L_1
	-\mfrac{10}{3}L_0+2 i \pi L_0
	+{4} \zeta_3
	\nonumber \\ & \quad
	-\mfrac{\pi ^2}{6}
	+{5}
	-{3} i \pi 
	-\mfrac{10 }{3}L_0 x
	+\mfrac{1}{3} x^2 \left((L_0-L_1)^2+\pi ^2\right)
	\,, \nonumber \\
	%%%
	\Delta_8(x) &= 
	-\mfrac{23 }{216}\pi ^2
	+\mfrac{5 }{27}i \pi 
	\,,
	\label{ampppppNNLO}
\end{align}
}}
\\
and $L_0=\ln(x)$ and $L_1=\ln(1-x)$.
Remarkably, these are the only transcendental functions required to describe our all-plus amplitude.
Note that the transcendental weight drops from the expected 2 to resulting 0 at one loop, from 4 to 1 at two loops, and from 6 to 4 at three loops.
In comparison, the analogous expressions for both the $gg \to gg$~\cite{Caola:2021izf} and $gg \to g\gamma$~\cite{Bargiela:2022lxz} amplitudes satisfy a more symmetric weight-drop pattern $2L \to 2(L-1)$.
The stronger weight drop from 4 to 1 in two-loop $gg \to \gamma\gamma$ amplitude is independent of the IR subtraction scheme because of the simple logarithmic structure of this colour singlet process at $\mu^2=s$.

The less helicity-violating the configuration is, the more complex results we obtain.
Nonetheless, the increase in complexity in the finite part $f_\lambdavec^{(3,\rm fin)}$ for single and double-minus helicity configurations $\lambdavec$ is at most $\mathcal{O}(10)$.
We provide complete results for all helicity configurations in a computer-readable format in Ref.~\cite{Bargiela:2021wuy}.
In Fig.~\ref{fig:ggaa.res} we plot the finite part at one-, two-, and three-loop order for $\lambdavec =(++++)$, $(-+++)$, $(++-+)$, $(--++)$, and $(-+-+)$ helicity configurations.
All the other helicity amplitudes can be obtained from them using Bose symmetry ($x\leftrightarrow 1-x$) and parity.

Interestingly, there is a specific new feature appearing at the three-loop order comparing to lower orders.
In turns out, that the three-loop $gg \to \gamma\gamma$ amplitude is no longer integrable.
Indeed, when constructing the total cross section as in Eq.~\ref{eq:intro.xsec}, the $x$ variable has to integrated over in the whole physical region $0 < x < 1$ as in Eq.~\ref{eq:ggaa.x}.
However, in the $x \to 0$ forward limit, we obtain the power-like $1/x$ kinematic divergence
\begin{equation}
	\begin{split}
		f^{(3,{\rm fin})}_{\text{all}+} &\overset{x \to 0}{\longrightarrow} \frac{1}{x}i\pi(8C_F-3C_A)(n_f^V)^2 \cdot \frac{1}{2} \,, \\
		f^{(3,{\rm fin})}_{\text{single}-} &\overset{x \to 0}{\longrightarrow} \frac{1}{x}i\pi(8C_F-3C_A)(n_f^V)^2 \cdot \frac{1}{90}(75-14\pi^2) \,, \\
		f^{(3,{\rm fin})}_{--++} &\overset{x \to 0}{\longrightarrow} \frac{1}{x}i\pi(8C_F-3C_A)(n_f^V)^2 \cdot \frac{1}{810}(-375+556\pi^2-1440\zeta_3) \,, \\
		f^{(3,{\rm fin})}_{-+-+} &\overset{x \to 0}{\longrightarrow} \frac{1}{x}i\pi(8C_F-3C_A)(n_f^V)^2 \cdot \frac{1}{30}(45-4\pi^2) \,,
	\end{split}
\label{eq:ggaa.Regge}
\end{equation}
for all the independent helicity configurations.
This issue has a natural explanation for the processes with partonic final states, since the partons are not well-defined asymptotic states.
The problem can be resolved by imposing kinematic cuts on $x$, corresponding to the detector limitations.
Even though it recovers phenomenological consistency, it is still important to understand the theoretical origin of this issue.
Notice, that all the limits in Eq.~\ref{eq:ggaa.Regge} are proportional to the colour factor $(n_f^V)^2$.
If fact, they all stem from Feynman diagrams like in the middle of Fig.~\ref{fig:ggaa.nfV}, i.e. with each external photon attached to a separate closed fermion loop.
Specifically, the $1/x$ behaviour comes from the unitarity cut across the two internal gluons.
This corresponds to the exchange of the two soft gluons in the high-energy $-t \ll s$ limit, which can be studied in the framework Regge theory, see Ref.~\cite{Gribov:2009cfk} for a review. 

Let us discuss the structure of final results.
As mentioned in Sec.~\ref{sec:ggaa.amp.col}, the $L$-loop $gg \to \gamma\gamma$ amplitude is homogeneous polynomial of order $L$ in $C_A$, $C_F$, $n_f$, $n_f^V$, and $n_f^{V_2}$.
At one-loop, the whole amplitude is proportional to $n_f^{V_2}$, since the two photons must both couple to the same closed fermion loop.
At two loops, the structures $n_f^{V_2}\times \{C_F,C_A\}$ appear in the bare amplitude.
The finite remainder contains in addition a term proportional to $n_f^{V_2} n_f$ stemming from $\beta_0$ in the UV/IR regularisation.
Note that there is no $(n_f^V)^2$ contribution.
Indeed, the $(n_f^V)^2$ colour factor only appears if the two photons are attached to two different closed fermion lines.
Such diagrams do appear at two loops, but they are of the form of two $\gamma g g^*$ one-loop triangles connected through an off-shell gluon $g^*$ propagator.
Due to an argument analogous to Furry's theorem, these diagrams give no net contribution to the amplitude.
A similar argument allows one to conclude that there is no net contribution from Feynman diagrams with colour factors $n_f (n_f^{V})^2$ at three loops.
Furthermore, it is easy to see that the structure $n_f^2 n_f^{V_2}$ is absent in the three-loop bare amplitude.
We note however that the $n_f^2 n_f^{V_2}$ structure contributes to our finite remainders, since it is induced by the $n_f$ dependence of the UV/IR counterterms.
Since there is no $(n_f^V)^2$ contribution at lower loops, the $(n_f^V)^2$ term in the bare three loop amplitude must be finite.
We observe, however, that it is non-zero.
Indeed, at three loops this colour factor appears in
triple-box diagrams for which the Furry argument outlined above is no longer applicable.

In terms of the kinematic $x$-dependence of the final results, they contain terms of the form $G(a_1,\dots,a_n;x)/x^k$ ($-2\leq k \leq 2$) and $G(a_1,\dots,a_n;x)/(1-x)^k$ ($1\leq k \leq 2$), where $a_i\in\{0,1\}$ and $0\le n \le 6$.
Instead of the HPLs, we found it useful to also consider the alternative functional basis described in Ref.~\cite{Caola:2020dfu} to speed up the numerical evaluation of the final result.
Using the algorithm of Ref.~\cite{Duhr:2011zq}, we have constructed a basis of logarithms, classical polylogarithms and multiple polylogarithms to rewrite the HPLs without introducing any new spurious singularities.
We used products of lower weight functions whenever possible, and we preferred functions whose series representation requires a small number of nested sums.
In this way, we found that 23 independent transcendental functions and their products suffice to represent all our HPLs up to weight 6.
The new basis consists of 2 logarithms, $\ln(x)$, $\ln(1-x)$, 12 classical polylogarithms, $\Li_2$ of $x$, $\Li_3$ of $x$ and $1-x$, and $\Li_4$, $\Li_5$, $\Li_6$ of $x, 1-x$ and $-x/(1-x)$, as well as 9 multiple polylogarithms $\Li_{3,2}(1,x),\Li_{3,2}(1-x,1),\Li_{3,2}(x,1),\Li_{3,3}(1-x,1),\Li_{3,3}(x,1),\Li_{3,3}\left(\frac{-x}{1-x},1\right),\Li_{4,2}(1-x,1),\Li_{4,2}(x,1),\Li_{2,2,2}(x,1,1)$.
Here, we follow the notation of Ref.~\cite{Vollinga:2004sn} and define
\begin{equation}
	\label{eq:Linm}
	\Li_{m_1,\dots,m_k}(x_1,\dots,x_k) = \sum_{i_1>\dots>i_k>0} \frac{x_1^{i_1}}{i_1^{m_1}} \, \dots \, \frac{x_k^{i_k}}{i_k^{m_k}} \,.
\end{equation}
In Ref.~\cite{Bargiela:2021wuy}, we provide our results expressed also in terms of multiple polylogarithms.
For convenience, we also provide the results for the finite part of the one- and two-loop helicity amplitudes up to weight 6.

\section{Application to Higgs width determination}
\label{sec:ggaa.Higgs}

In this section, following our publication~\cite{Bargiela:2022dla}, we describe a particular application of the three-loop $gg \to \gamma\gamma$ amplitude to the determination of the decay width of the Higgs boson $\Gamma_H$.

As mentioned in Sec.~\ref{sec:intro.LHC}, the Higgs boson discovery confirmed the particle content of the Standard Model of Elementary Particles.
Higgs is the heaviest boson in the SM, with the averaged experimental mass value is $m_H=125.25\pm0.17\GeV$~\cite{ParticleDataGroup:2022pth}.
The SM prediction for the Higgs decay width is $\Gamma_H \approx 4.1\MeV$~\cite{LHCHiggsCrossSectionWorkingGroup:2016ypw}, which corresponds to a short lifetime of $\approx 1.6 \cdot 10^{-22}$~s.
We can read off from Eq.~\ref{eq:LagrH} that the Higgs couples to all the massive particles in the SM.
Specifically, its self-interacting qubic and quartic vertices are characterised by a coupling constant $\lambda_{\text{SM}}=\frac{m_H^2}{2v^2} \approx 0.13$, where $v \approx 246 \GeV$ is the vacuum expectation value of the Higgs field.
In order to measure this self-coupling, multiple Higgs bosons have to be produced.
Although it is out of the current reach of the LHC energies, the measured absolute value of the Higgs self-coupling has been bounded to at most $6.7$ times the SM value $\lambda_{\text{SM}}$~\cite{ATLAS:2021ifb,CMS:2022dwd}.
Testing theoretical predictions against experimental measurements of various Higgs boson properties is crucial for a better understanding of the structure of fundamental interactions.

In this section, we are interested in the Higgs width $\Gamma_H$ parameter.
The direct measurement of this parameter is limited by the finite
resolution of the detectors, specifically the fact that for a given
energy of photons and charged leptons, the reconstructed energy
fluctuates from one event to another, e.g. due to the stochastic
nature of calorimetric energy measurements. The channels with the
best resolution are $H \to \gamma\gamma$ and $H \to $~4leptons, where the resolution is
of the order of 0.5-1.5~GeV. This precludes a direct kinematic
measurement of the much smaller Higgs decay width.
Therefore, we need to rely on indirect methods.
Assuming that the New Physics particles are much heavier then the Higgs mass, we can work in the framework of the \textit{Standard Model Effective Field Theory} (SMEFT), see a review in Ref.~\cite{Brivio:2017vri}.
Basing on this formulation, the Higgs width can be extracted from a global fit of all the SM parameters~\cite{Ethier:2021bye}.
Since this approach is very broad, it does not reach high precision for specific parameters.
The best current bound of the Higgs width bases on the so called \textit{off-shell method}~\cite{Caola:2013yja}, and it yields $\Gamma_H = 3.2^{+2.4}_{-1.7} \MeV$~\cite{CMS:2022ley}.
Since it is an indirect method, it relies on some additional assumptions.
For this reason, it is important to confront the results of the off-shell method against some other method based on a different set of assumptions.
This can be achieved by considering the method of the \textit{signal-background interference} in the diphoton channel, which we henceforth focus on.

\begin{figure}[h]
	\centering
	\includegraphics[width=0.56\textwidth]{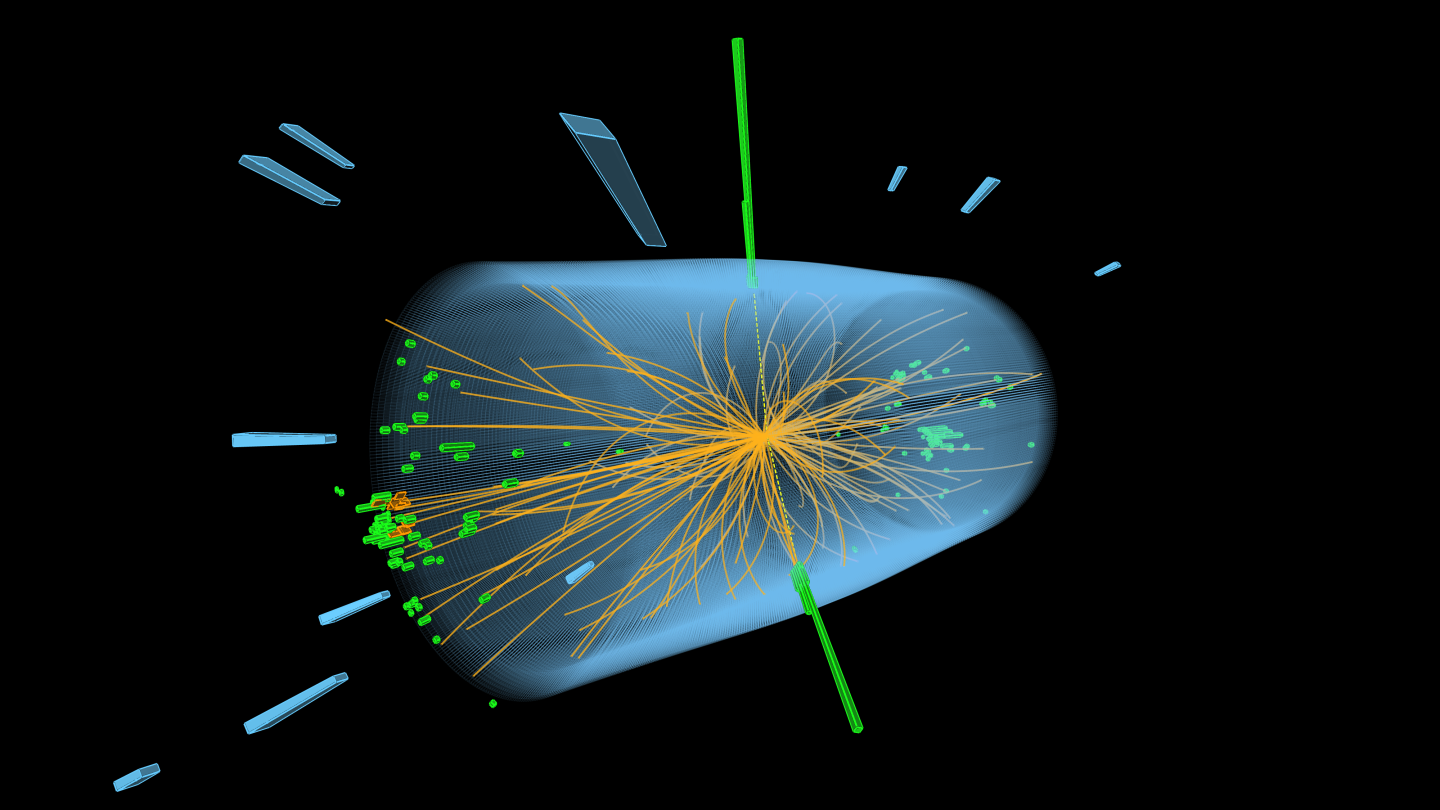}
	\includegraphics[width=0.34\textwidth]{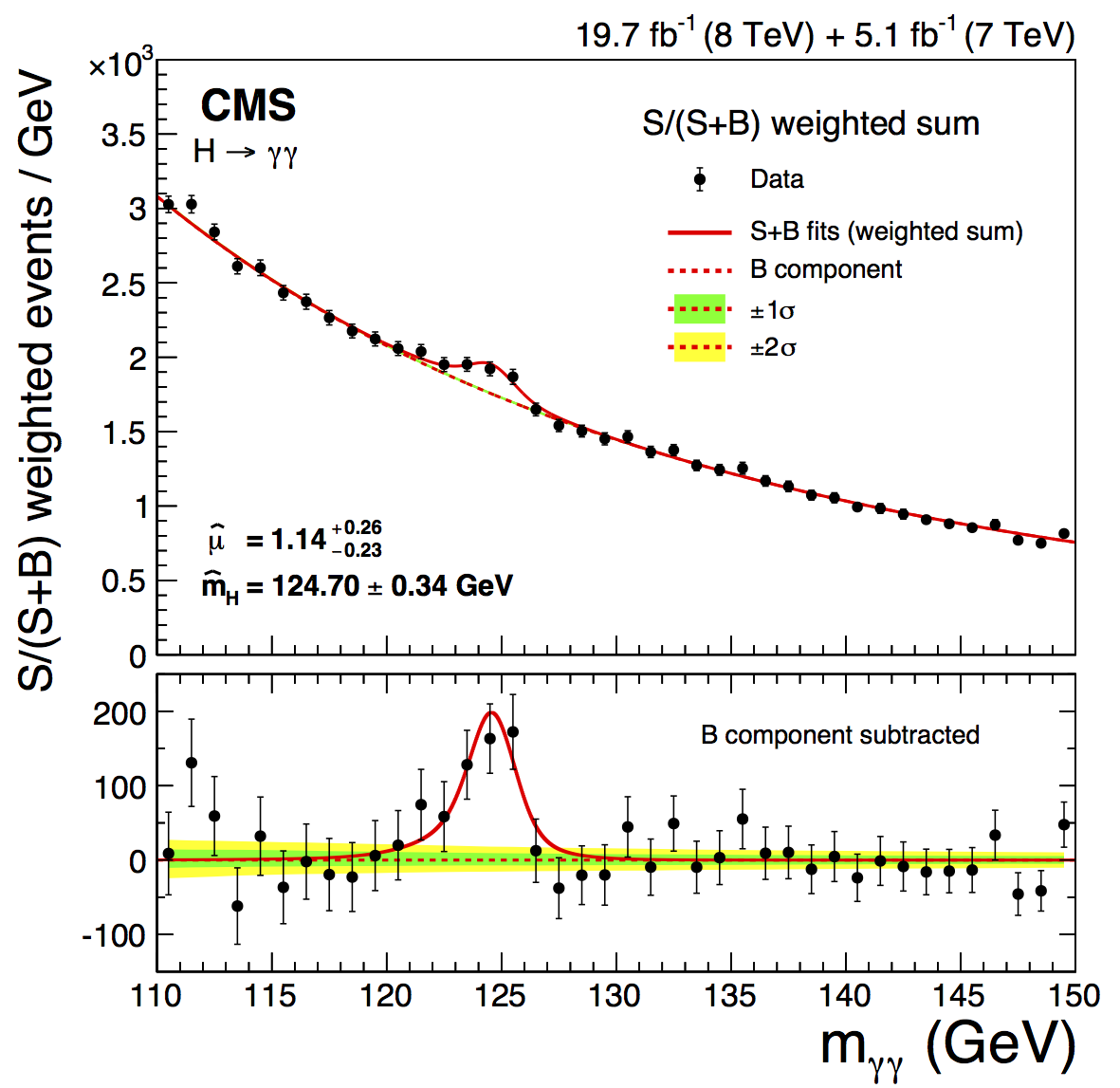}
	\caption{Experimental results for the $H \to \gamma\gamma$ decay~\cite{CMS:2014afl}.}
\end{figure}

In has been observed, that in the $\gamma\gamma$ decay channel, there is a subtle interference effect between the Higgs signal and the QCD background~\cite{Martin:2012xc,Dixon:2013haa}.
Although it changes the corresponding total cross section only by $\sim 1 \%$, this interference distorts the diphoton invariant mass distribution by effectively moving its peak to the left.
It turns out that this so called \textit{mass shift} $\Delta m_{\gamma\gamma}$ depends on $\Gamma_H$.
Under some assumptions on the structure of Higgs interactions, we approximately have $\Delta m_{\gamma\gamma} \propto \sqrt{\Gamma_H}$.
Therefore, by predicting the mass shift to a high precision, we gain information about the Higgs width.

In the NLO computation of this interference effect, large K-factors of order $0.66$ have been found~\cite{Martin:2012xc,Dixon:2013haa}.
Therefore, the NNLO study is required in order to better control the precision.
Since the Higgs production is dominated at the LHC by the gluon fusion channel, we need to consider the QCD background corrections to the $gg \to \gamma\gamma$ process.
The main difficulty stems from the fact that this process is loop-induced.
Thus, at NNLO, we need the three-loop $gg \to \gamma\gamma$ correction, two-loop $gg \to \gamma\gamma g$, as well as one-loop $gg \to \gamma\gamma gg$ and $gg \to \gamma\gamma q\bar{q}$.
Moreover, we also require an NNLO subtraction scheme to properly combine these contributions together.
By now, the NNLO subtraction for a colour singlet, such as the $\gamma\gamma$ state, is well-understood.
Due to the fact that the three-loop $gg \to \gamma\gamma$ and the two-loop $gg \to \gamma\gamma g$ amplitudes have recently become available, the NNLO study can now be performed.
We described our three-loop $gg \to \gamma\gamma$ amplitude computation earlier in this chapter, basing on our work in Ref.~\cite{Bargiela:2021wuy}, while the details of the two-loop $gg \to \gamma\gamma g$ process are given in Ref.~\cite{Agarwal:2021vdh,Badger:2021imn}.

Even though the NNLO computations are now well-understood, in practice, they may still pose a problem when numerically integrating over the degenerate kinematic regions.
A simple way to avoid this issue is to neglect the hard emissions in the so called \textit{soft-virtual approximation}~\cite{deFlorian:2012za}.
It is reliable for our process since, at lower orders, it has been shown that the corresponding cross section is dominated by the low $p_T$ region.
In this approximation we preserve all the information stemming from the virtual corrections, while considering only leading soft and collinear real contributions.
We adopted it for our process at NNLO in our work presented in Ref.~\cite{Bargiela:2022dla}. 

\begin{table}[h]
	\centering
	\renewcommand{\arraystretch}{1.5}
	\begin{tabular}{c||c|c|c|c}
		order & $\Delta m_{\gamma\gamma}^{(\text{moment})}\left[\text{MeV}\right]$ & K-factor$^{(\text{moment})}$ & $\Delta m_{\gamma\gamma}^{(\text{Gauss})}\left[\text{MeV}\right]$ & K-factor$^{(\text{Gauss})}$ \\
		\hline
		\hline
		LO & $-122.1^{+0.1\%}_{-0.3\%}$ & - & $-83.1^{+0.0\%}_{-0.3\%}$ & - \\
		\hline
		NLO & $-81.2^{+12\%}_{-12\%}$ & 0.665 & $-55.2^{+12\%}_{-12\%}$ & 0.664 \\
		\hline
		NNLOsv & $-58.0^{+16\%}_{-17\%}$ & 0.475 & $-39.4^{+16\%}_{-17\%}$ & 0.474 
	\end{tabular}
	\caption{Predictions for the mass shift at collider energy 13.6~TeV.}
	\label{tab:Kfact}
\end{table}

Even with the NNLO soft-virtual approximation, it is still nontrivial to predict the mass shift because it is highly-dependent on the detector effects.
In order to overcome this issue, we design two alternative procedures that allow us to systematically study the mass shift without simulating the detector.
Even though they both might lead to different predictions for the mass shift, as long as their corresponding K-factors match order by order, we can expect that the same K-factor would also hold the for the observable accounting for all the detector effects.
The first approach bases on the first moment property of the invariant mass distribution.
The second method relies on a Gaussian fit with standard deviation of about $1.7\GeV$~\cite{Martin:2012xc,Dixon:2013haa}.
We present the results of these two methods in Tab.~\ref{tab:Kfact}.
Consistently with the above discussion, even though the corresponding predictions for the mass shift are different, the K-factors remain similar.

Let us briefly comment on some properties of the results for the mass shift in Tab.~\ref{tab:Kfact}.
Firstly, note that the mass shift decreases in its absolute value with increasing perturbative order.
It is due to the fact that the signal K-factor is larger then the background K-factor, and the interference is considered as relative to the signal.
Secondly, the scale uncertainties of the mass shift at fixed order do not enclose the central value at higher order.
It is due to the fact that the scale uncertainties are only an approximate estimate of the missing higher order corrections.
For example, they do not account for possible new channels opening at higher orders.
However, it has been shown that starting from the NNLO, the scale uncertainties on the the Higgs production cross section encapture the corresponding central value at N3LO~\cite{Anastasiou:2015vya}.
For this reason, we may expect that our NNLO correction to the mass shift is the first convergent one in the perturbative series.

\begin{figure}[h]
	\centering
	\includegraphics[width=0.7\textwidth]{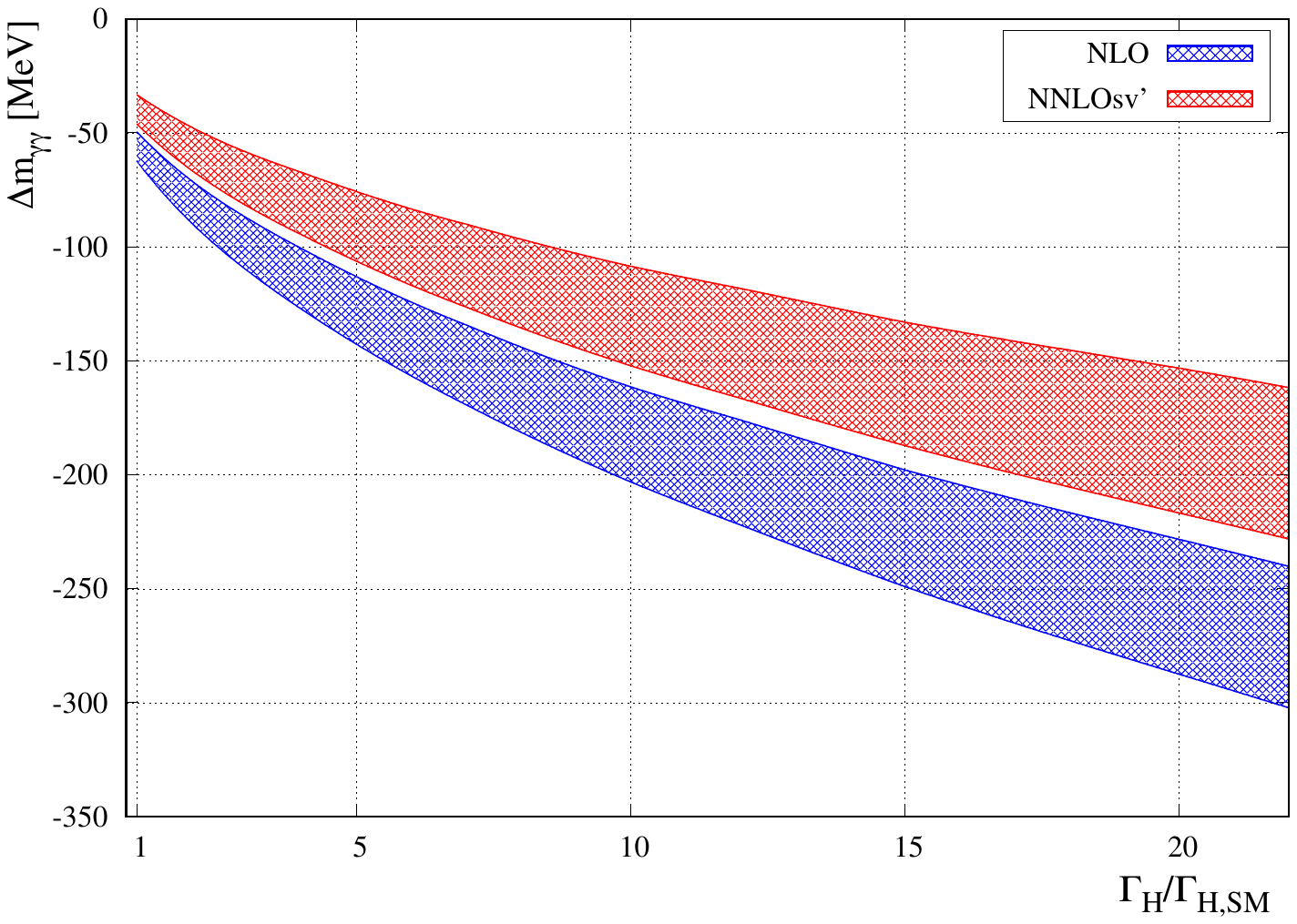}
	\caption{Mass shift as a function of the Higgs boson width.}
	\label{fig:shiftWidth}
\end{figure}

To conclude, in Fig.~\ref{fig:shiftWidth}, we plot the mass shift as a function of the Higgs width.
Note that the NNLO bounds on the Higgs widths with respect to its SM prediction are weaker then the ones obtained at NLO.
Therefore, if we assume that the current LHC resolution can reach $\sim 240 \MeV$, we cannot put stronger bounds then tens times the SM value.
However, recent projections for the resolution of the High Luminosity LHC upgrade suggest that the detector resolution may reach $\sim 70 \MeV$~\cite{Dawson:2022zbb}.
According to Fig.~\ref{fig:shiftWidth}, the corresponding bound on the Higgs width would yield $\Gamma_H < 5 \, \Gamma_{H,\text{SM}}$.
This prediction of our signal-background interference method is still not as precise as the one of the off-shell method, however it becomes more competitive.
Since these two methods rely on different assumptions, it is important to compare the results of both of them.

\section{Results for $pp \rightarrow \gamma$+jet}
\label{sec:ggaa.ppja}

In Sec.~\ref{sec:ggaa.complexity}, we focused on the details of the calculation of the three-loop amplitude for the $gg \to \gamma\gamma$ process.
As mentioned in Sec.~\ref{sec:ggaa.frontier}, we also computed in Ref.~\cite{Bargiela:2022lxz} the three-loop correction to all channels contributing to the $pp \to \gamma$+jet process.
The calculation proceeds as we described for the $gg \to \gamma\gamma$.
For further details, see Ref.~\cite{Bargiela:2022lxz}.
We omit them here in order to summarise only the results.
As mentioned in Sec.~\ref{sec:ggaa.frontier}, we also note that the all the three-loop four-point corrections to $pp\to$~2jets have already been computed in Refs~\cite{Caola:2021rqz,Caola:2021izf}.

Similarly to $gg \to \gamma\gamma$, the $gg \to g\gamma$ channel is also loop-induced.
Contrarily, the $u \bar{u} \to g\gamma$ channel starts at tree-level.
Therefore, the three-loop amplitude for $gg \to g\gamma$ starts contributing at N4LO in perturbation theory
\begin{align}
	\mathcal{V}^{(N3LO)}_{gg \to g \gamma} &=  \frac{  2 \:\text{Re}  \langle \mathcal A^{(1)} | \mathcal A^{(2)}\rangle }{\langle \mathcal A^{(1)} | \mathcal A^{(1)}\rangle } , \notag\\
	\mathcal{V}^{(N4LO)}_{gg\to g \gamma} &=  \frac{   \langle \mathcal A^{(2)} | \mathcal A^{(2)}\rangle }{\langle \mathcal A^{(1)} | \mathcal A^{(1)}\rangle } + \frac{  2 \:\text{Re}  \langle \mathcal A^{(1)} | \mathcal A^{(3)}\rangle }{\langle \mathcal A^{(1)} | \mathcal A^{(1)}\rangle } \,,
	\label{eq:ggga.pert}
\end{align}
while the three-loop amplitude for $u \bar{u} \to g\gamma$ contributes also to N3LO
\begin{align}
	\mathcal{V}^{(NLO)}_{q \bar{q} \to g \gamma} &=  \frac{  2 \:\text{Re}  \langle \mathcal A^{(0)} | \mathcal A^{(1)}\rangle }{\langle \mathcal A^{(0)} | \mathcal A^{(0)}\rangle } , \notag\\
	\mathcal{V}^{(NNLO)}_{q \bar{q} \to g \gamma}  &=  \frac{   \langle \mathcal A^{(1)} | \mathcal A^{(1)}\rangle }{\langle \mathcal A^{(0)} | \mathcal A^{(0)}\rangle } + \frac{  2 \:\text{Re}  \langle \mathcal A^{(0)} | \mathcal A^{(2)}\rangle }{\langle \mathcal A^{(0)} | \mathcal A^{(0)}\rangle }, \notag\\  
	\mathcal{V}^{(N3LO)}_{q \bar{q} \to g \gamma}  &=  \frac{   2 \:\text{Re}  \langle \mathcal A^{(1)} | \mathcal A^{(2)}\rangle }{\langle \mathcal A^{(0)} | \mathcal A^{(0)}\rangle } + \frac{  2 \:\text{Re}  \langle \mathcal A^{(0)} | \mathcal A^{(3)}\rangle }{\langle \mathcal A^{(0)} | \mathcal A^{(0)}\rangle } \,.   
\label{eq:virtualdef} 
\end{align}
We have defined the sum over colour and polarization via
\begin{align}
	\langle \mathcal A^{(\ell)} | 
	\mathcal A^{(\ell')}\rangle 
	= \sum_{\text{col},\lambda}
	\mathcal A^{(\ell, \rm fin)^*}_{\vec{\lambda}}
	\mathcal A^{(\ell', \rm fin)}_{\vec{\lambda}} \,.
	\label{eq:ggga.ampSq}
\end{align}
\begin{figure}[h]
	\centering
	\includegraphics[width=0.49\textwidth]{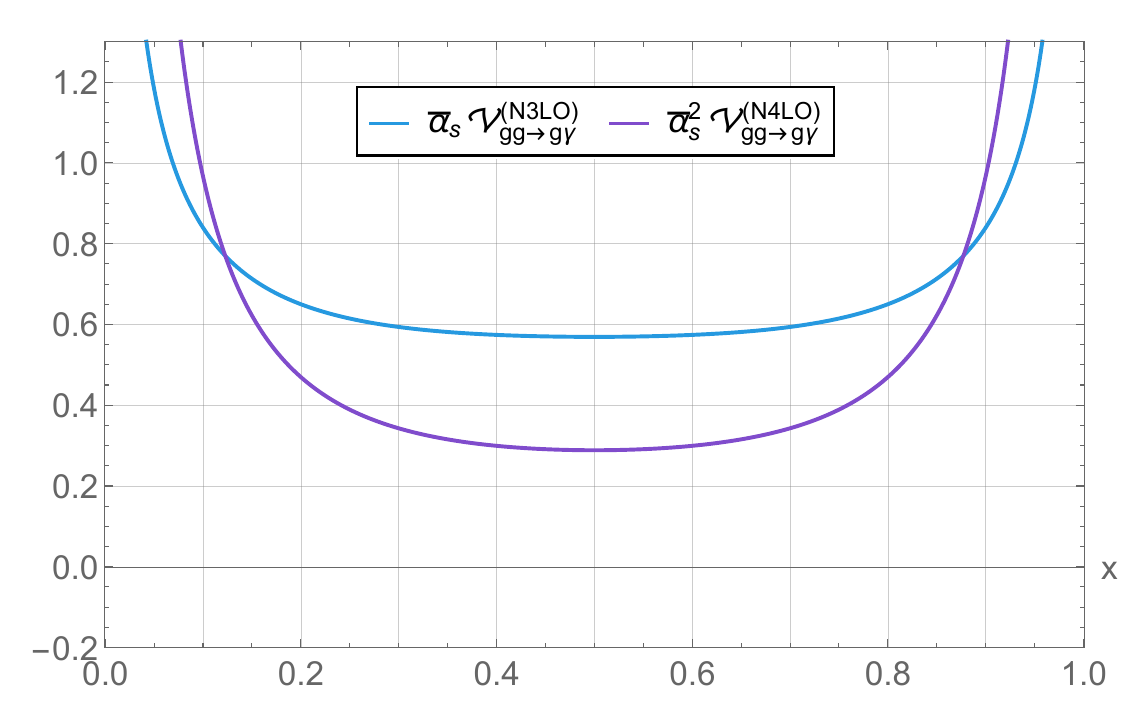}
	\hfill
	\includegraphics[width=0.49\textwidth]{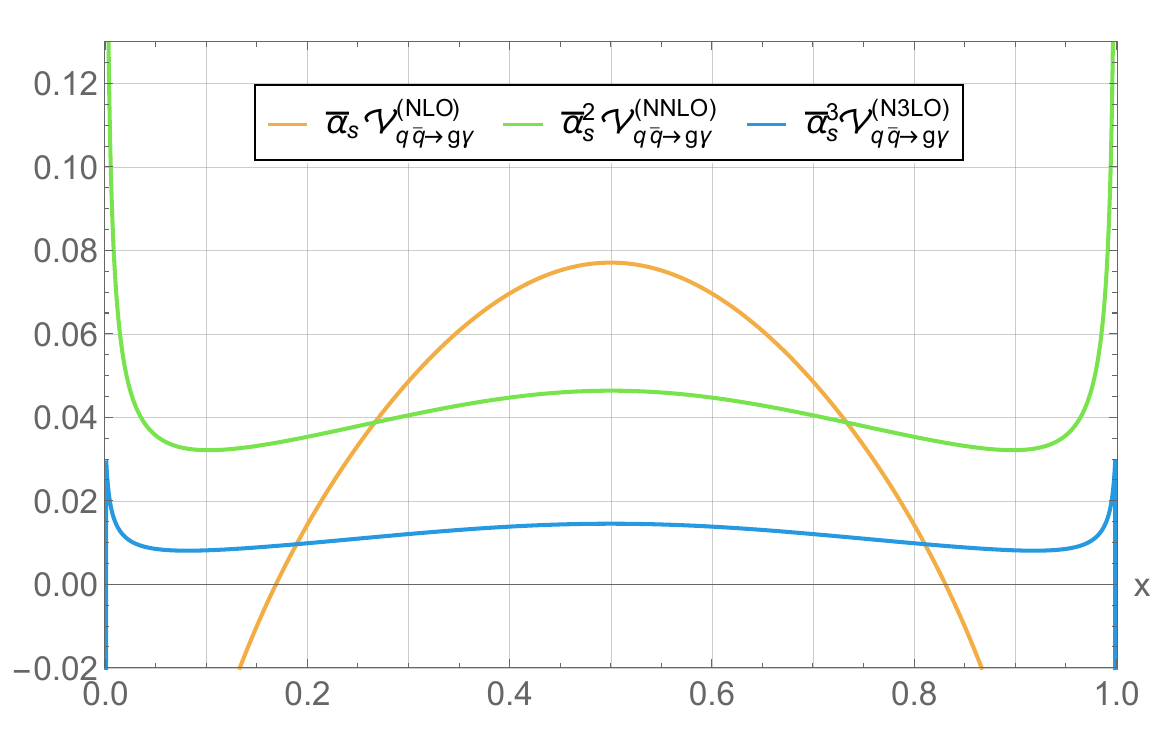}
	\caption{Perturbative expansion of the colour, helicity, and flavour summed finite part of the amplitude squared for the processes $gg \to g \gamma$ and $q\bar{q} \to g \gamma$, respectively, as a function of $x=-t/s$, normalised to the leading order.}
	\label{fig:ggga}
\end{figure}
In Fig.~\ref{fig:ggga}, we plot the ratio $\mathcal{V}$ of amplitude squared to leading order in both partonic channels.
In Ref.~\cite{Bargiela:2022lxz}, we provide analytic expressions for all helicity amplitudes for the $pp \to \gamma$+jet process.
Contrarily to the $gg \to \gamma\gamma$ process, the IR subtraction scheme follows Ref.~\cite{Becher:2009qa}.

We performed various checks on the correctness of our results.
We compared the one- and two-loop helicity amplitudes against available literature in Refs~\cite{Bern:2001df,Anastasiou:2002zn}, and we found agreement.
To validate our numerical evaluation procedure, we also checked the helicity-summed one-loop squared amplitude against \texttt{OpenLoops}~\cite{Cascioli:2011va, Buccioni:2019sur}.
We employed two independent derivations of the three-loop $\mathcal F_i$ form factors at the integrand level and verified that they agree.
Finally, we have verified that the UV and IR poles up to three loops agree with the predicted UV and IR structures.
This step provides a strong check of the correctness of the three-loop amplitudes.

Phenomenologically, this amplitude is relevant for the $pp \to \gamma$+jet process, i.e. a photon production with one associated hadronic jet.
It is one of the standard candles of the SM at the LHC.
Although we made the amplitudes available, providing beyond-NNLO cross sections for the $pp \to \gamma$+jet process would require a better understanding of the required subtraction schemes.
The three-loop amplitudes in quark-pair and quark-gluon initiated channels contribute to the N3LO cross section.
Even though the three-loop amplitude in the gluon fusion channel starts contributing only at N4LO, it is enhanced by gluon PDFs.
Therefore, our amplitudes will be necessary to reach below-percent precision predictions for the LHC in the future.

\chapter{\label{ch:2L}Two-loop mixed QCD-EWK amplitudes for $Z$+jet production}

\minitoc

\section{Motivation}
\label{sec:ppjZ.motivation}

In Ch.~\ref{ch:3L}, we have examined in detail quantum corrections to specific processes, arising from the massless QCD sector of the SM.
In this chapter, we consider also the contribution of the remaining EWK sector, introduced in Sec.~\ref{sec:intro.EWK}.
In general, it introduces in the amplitude a dependence on masses of the $W$ and $Z$ weak bosons, as well as of the Higgs boson $H$ and heavy quarks, bottom $b$ and top $t$.

In addition to the increase in amplitude complexity stemming from the number of loops and legs, as explained in Sec.~\ref{sec:ggaa.frontier}, it also arises from introducing masses.
Indeed, the Feynman integrals associated with the massive amplitudes may have a highly intricate multivariate functional dependence, as mentioned in Sec.~\ref{sec:amp.int.Symanzik}.
In practice, seeking an analytic formula is justified if either it exposes some hidden cancellations, or it allows for efficient numerical evaluation.
As we have seen in Ch.~\ref{ch:3L}, for massless QCD corrections both of these conditions are satisfied at the same time.
On the contrary, it is not clear if the analytic expressions depending on distinctly different mass parameters would have a simple structure.
Moreover, as briefly discussed in Sec.~\ref{sec:amp.int.Symanzik}, the Feynman integrals with internal masses may evaluate to unknown special functions.
For these two reasons, in the EWK sector, it is convenient to employ a numerical approach as in Refs~\cite{Bronnum-Hansen:2020mzk,Bronnum-Hansen:2021olh}.

The state-of-the-art amplitudes involving EWK corrections are the two-loop four-point mixed QCD-EWK amplitudes, see e.g. Refs~\cite{Heller:2020owb,Behring:2020cqi,Bonetti:2020hqh,Bonciani:2021zzf,Bonetti:2022lrk,Armadillo:2022bgm}.
We will discuss the comparison between the numerical and analytic approach later in Sec.~\ref{sec:qqgZ.comp.IBP}.
In this chapter, we are particularly interested in $Z$ boson production in association with one hadronic jet, which is an important process at proton colliders. 
It serves not only as a standard candle of the Standard Model, but also as a background process for Dark Matter searches, as we briefly discuss below.

The $Z$ boson is produced at the LHC at a large event rate.
In addition, a substantial decay mode of the $Z$ boson is into a leptonic pair, thus providing a clean signature.
This process is important for precision measurements of a vast variety of quantities e.g. EWK parameters, strong coupling constant, PDFs, as well as for detector calibration.
In order to inspect the dynamic properties of the $Z$ boson, we will study its transverse momentum distribution.
The simplest corresponding physical realisation is a back-to-back recoil from a single hadronic jet.
The comparison of theoretical predictions for this $Z$+jet process against LHC measurements~\cite{ATLAS:2022nrp,CMS:2022ilp} tests our understanding of the Standard Model of Particle Physics.

\begin{figure}[h]
	\centering
	\includegraphics[width=0.45\textwidth]{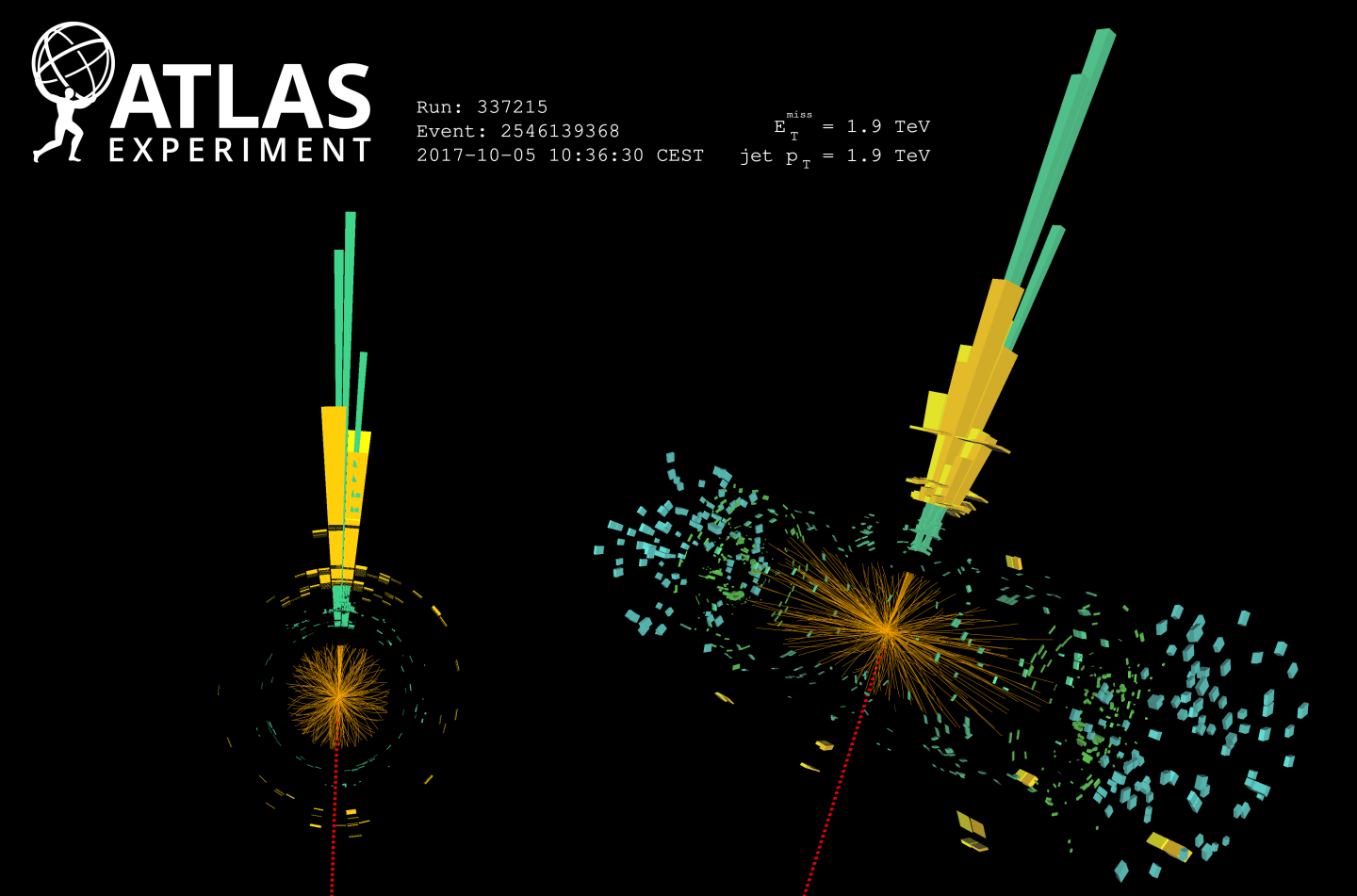}
	\quad
	\includegraphics[width=0.45\textwidth]{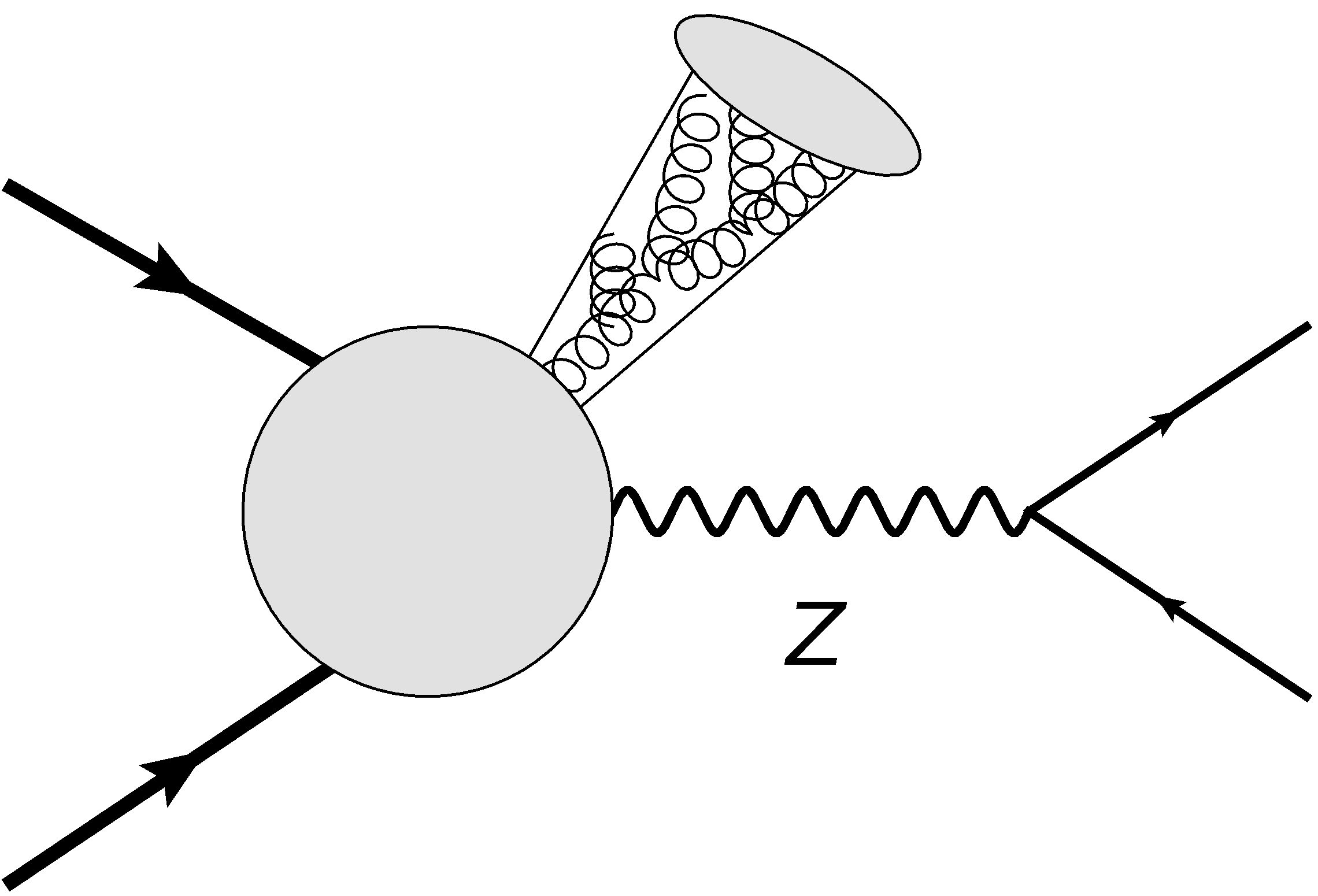}
	\caption{Detector display and a theoretical model for monojet production in association with an invisible fermion-antifermion pair.}
	\label{fig:ppjZ.monojet}
\end{figure}

The observable relevant for Dark Matter searches at the LHC~\cite{Lindert:2017olm} is a monojet with backward missing transverse energy.
This energy can be generated by Dark Matter fermions, in the BSM signal, and by invisible leptons, i.e. neutrinos, in the SM background.
The simplest production mechanism for a pair of fermions is a decay from the Z boson, see Fig.~\ref{fig:ppjZ.monojet}.
The corresponding statistical uncertainty is less then 1\% below 1 TeV of the transverse momentum of the $Z$ boson $p_{T,Z}$, and several percent until $p_{T,Z}=$ 2 TeV.
\begin{figure}[h]
	\centering
	\includegraphics[width=0.5\textwidth]{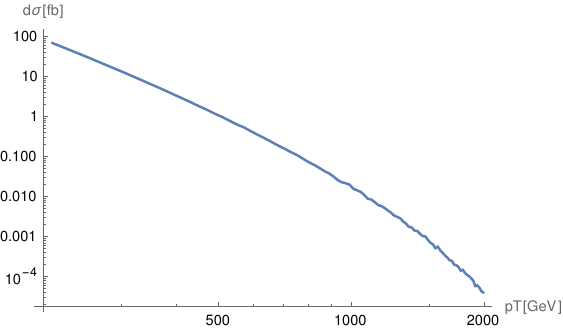}
	\caption{LO cross section differential in $p_T$ for $Z$+jet production.}
	\label{fig:ppjZ.LO}
\end{figure}
Our kinematic region of interest becomes $p_{T,Z} \in (200,2000)$~GeV, see Fig.~\ref{fig:ppjZ.LO}.
This region corresponds to $\sim 10^7-10^1$ events at luminosity $\mathcal{L}=300 \text{fb}^{-1}$ and collider energy $\sqrt{s_H}=13$~TeV.
For simplicity, we will treat $Z$ in an on-shell approximation
\begin{equation}
	\mathcal{M}(p \, p \to Z (\to\nu\bar{\nu}) + \text{jet}) \approx \mathcal{A}_\mu(p \, p \to Z \, j) \frac{1}{s - m_Z^2 + i \Gamma_Z m_Z} \mathcal{L}^\mu(Z \to\nu\bar{\nu}) \,,
	\label{eq:ppjZ.narrowWidth}
\end{equation}
which dominates up to corrections of order $\Gamma_Z/m_Z \approx 2/91 \approx 2\%$.
Since we are after a NNLO correction, which is itself of order 1\%, the off-shell $Z$ corrections are negligible.

The leading mechanism for the $Z$+jet production at the LHC is through the $qg$ channel.
It is enhanced by the gluon PDF in comparison to the $q\bar{q}$ quark annihilation channel.
Finally, the $gg$ gluon fusion channel is loop-induced and enhanced by two gluon PDFs.
The high precision of theoretical predictions for this process is achieved by considering higher order corrections in the double perturbative series, in strong and electroweak coupling
\begin{equation}
	\sigma = \sigma^{0}\left(1 + \alpha_s \, \delta^{(1,0)} + \alpha_s^2 \, \delta^{(2,0)} + \alpha \, \delta^{(0,1)} + \alpha_s \, \alpha \, \delta^{(1,1)} + \mathcal O(\alpha_s^2,\alpha^2) \right) \,.
	\label{eq:ppjZ.2series}
\end{equation}
In order to reach a percent level accuracy, NNLO QCD corrections $\delta^{(2,0)}$ were necessary, as computed in Ref.~\cite{Gehrmann-DeRidder:2015wbt,Gehrmann-DeRidder:2016cdi,Gehrmann-DeRidder:2016jns,Boughezal:2015ded,Boughezal:2016isb}.
However, the EWK corrections are required as well, because of the logarithmic Sudakov enhancement~\cite{Kuhn:1999de} in the high-energy region, which will be of interest here
\begin{equation}
	\alpha \sim 1\% \quad \longrightarrow \quad \frac{\alpha}{4 \pi s_w^2} \log^2\left(\frac{s}{m_Z^2}\right) \sim 10\% \sim \alpha_s \,.
	\label{eq:ppjZ.Sudakov}
\end{equation}
Therefore, mixed QCD-EWK correction $\delta^{(1,1)}$ is expected to be of the order of the NNLO QCD.
In order to compute it, the scattering amplitudes for the virtual corrections have to be combined with real radiation effects and PDFs.
In this work, we provide the relevant two-loop mixed QCD-EWK amplitudes.
Constructing an appropriate subtraction scheme still remains a challenge.
The resulting mixed QCD-EWK correction would complement the ones already available at NLO QCD~\cite{Giele:1993dj}, NLO EWK~\cite{Denner:2011vu}, and NNLO QCD~\cite{Gehrmann-DeRidder:2015wbt}.

Considering mixed QCD-EWK corrections can be used to test their high energy behaviour.
As mentioned above, it has been shown that NLO EWK corrections are dominated by Sudakov logarithms at high $p_{T}$~\cite{Sudakov:1954sw,Kuhn:1999de}.
This implies that the mixed QCD-EWK corrections tend to factorize at high $p_{T}$ into a product of NLO QCD and NLO EWK.
At high $p_{T}$, the nonfactorizable term is suppressed.
The calculation of mixed QCD-EWK correction would allow for a more quantitative description of the validity of this factorization.
It is also crucial for the intermediate $p_{T}$ range, where the nonfactorizable term is non-negligible.

\section{Overview of $u \bar{u} \to gZ$}
\label{sec:ppjZ.complexity}

As explained in the previous section, we consider virtual corrections to the production of the $Z$ boson in association with one hadronic jet.
As a first step towards the full result, we focus here on contributions containing neither closed fermion loops, nor massive quarks.
Our further approximations are motivated solely by phenomenology.
Since we are primarily interested in the two-loop amplitude, we treat the CKM matrix $V_{fg}$ to be unit because its off-diagonal flavour-changing terms are suppressed by at least a factor of 4.
Note that the gluon fusion channel contributes to the mixed QCD-EWK at higher perturbative order then the quark annihilation channel.
Moreover, we do not consider the bottom-quark-initiated channel because it is suppressed by the bottom PDF.
Therefore, we focus on the following process
\begin{equation}
	u(p_1) + \bar{u}(p_2)  \to g(-p_3) + Z(-p_4)
	\label{eq:qqgZ}
\end{equation}
and its crossings, as well as corresponding down-type channels.

The kinematics of this process is described as in Sec.~\ref{sec:amp.kin}.
We treat external momenta to be on-shell
\begin{equation}
	p_1^2=p_2^2=p_3^2=0 \,, \quad p_4^2=m_Z^2 \,,
	\label{eq:qqgZ.kin}
\end{equation}
with momentum conservation
\begin{equation}
	s+t+u = m_Z^2 \,.
	\label{eq:qqgZ.momCons}
\end{equation}
The physical Riemann sheet is given by $s>0$, $t<0$, and $u<0$, which are defined in Sec.~\ref{sec:amp.kin}.
The amplitude can be written as a double perturbative series in the two couplings

\scalebox{0.85}{\parbox{0.9\linewidth}{ 
\begin{align}
	\mathcal{A} =
	\sqrt{\frac{\alpha_{s,b}}{2\pi}} \sqrt{\frac{\alpha_{b}}{2\pi}}
	\left(\mathcal{A}^{(0,0)}_{b} 
	+ \frac{\alpha_{s,b}\,\mu_0^{2\ep}}{2\pi} \, \mathcal{A}^{(1,0)}_{b} 
	+ \frac{\alpha_b\,\mu_0^{2\ep}}{2\pi} \, \mathcal{A}^{(0,1)}_{b} 
	+ \frac{\alpha_{s,b}\,\mu_0^{2\ep}}{2\pi} \frac{\alpha_b\,\mu_0^{2\ep}}{2\pi} \, \mathcal{A}^{(1,1)}_{b} 
	+ \mathcal O(\alpha_{s,b}^2,\alpha_b^2) \right) \,,
\end{align}
}}

\noindent
where we explicitly indicate bare quantities with an index $b$ as well as the dimensionful reference scale $\mu_0$.

\begin{figure}
	\scalebox{0.9}{\parbox{1.0\linewidth}{
			\centering
			\includegraphics[width=0.4\textwidth]{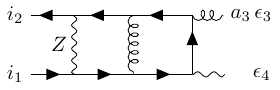}
			\begin{equation*}
				\begin{split}
					& \qquad\qquad\qquad = 
					\, -\frac{i}{2\sqrt{2}} \, g_s^3 \, e^3 \, g_{L/R}^3 \, {\color{red}\frac{1}{N_c} \, T^{a_3}_{i_1 i_2}}
					{\color{blue} \int \frac{d^dk_1}{(2\pi)^d} \, \frac{d^dk_2}{(2\pi)^d}} \\
					\times& \frac{\color{gray} \bar{u}_2
						\gamma_\mu(\slashed{k}_1+\slashed{p}_2)
						\gamma_\nu(\slashed{k}_2+\slashed{p}_3)
						\slashed{\ep}_3(\slashed{k}_2)
						\slashed{\ep}_4(\slashed{k}_2+\slashed{p}_{123})
						\gamma^\nu(\slashed{k}_1-\slashed{p}_1)
						\gamma^\mu
						u_1}
					{\color{blue} ((k_1)^2-m_Z^2) (k_2)^2 (k_1-p_1)^2 (k_1+p_2)^2 (k_2+p_3)^2 (k_2+p_{123})^2 (k_{12}+p_{23})^2}
				\end{split}
			\end{equation*}
			\\
			\raggedright
			\qquad\quad {\color{red} colour} \qquad\qquad\qquad\qquad\quad {\color{gray} tensors} \qquad\qquad\qquad\qquad\qquad {\color{blue} integrals} \\
			\centering
			\includegraphics[width=0.3\textwidth]{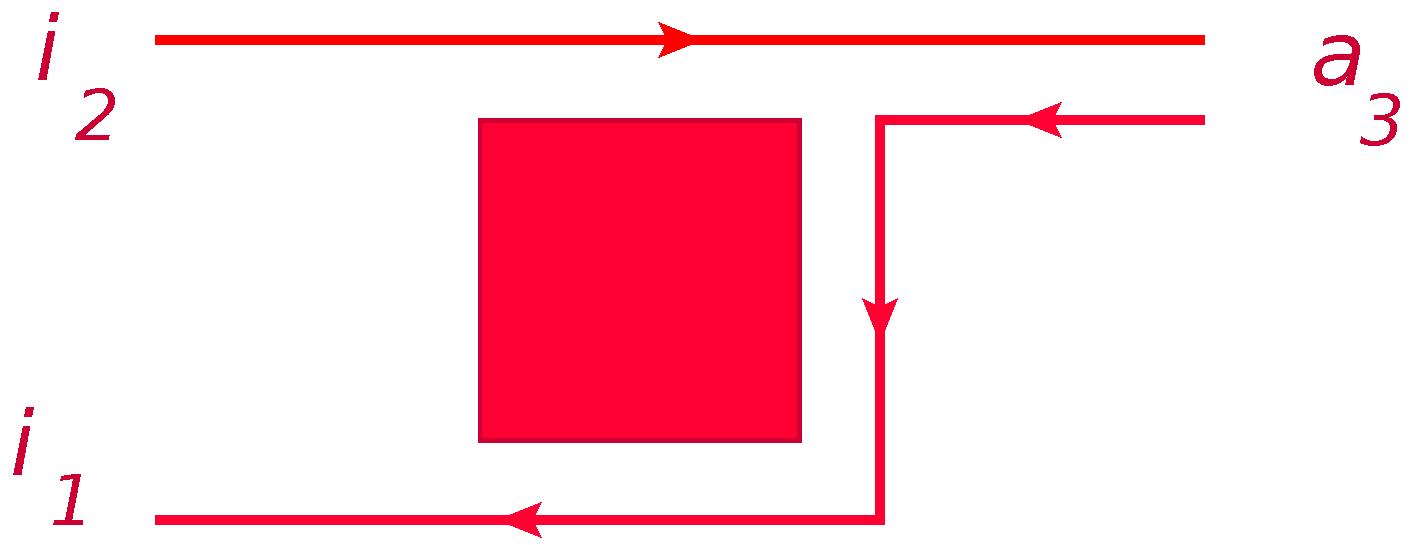}
			\hfill
			\includegraphics[width=0.2\textwidth]{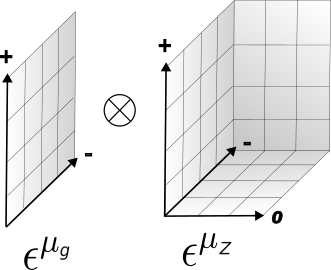}
			\hfill
			\includegraphics[width=0.3\textwidth]{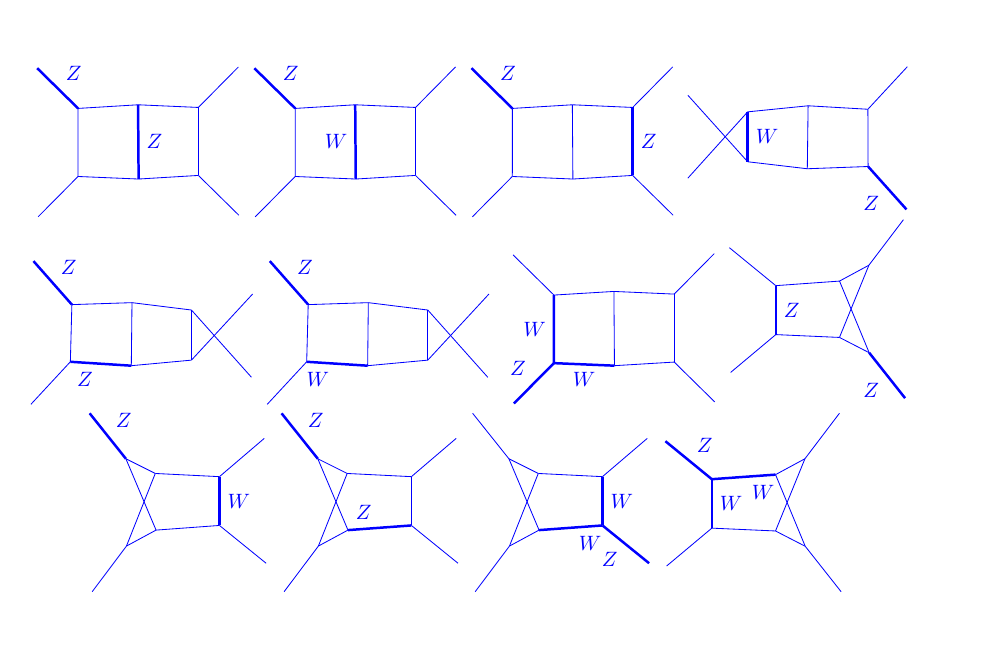}
			\begin{equation*}
				\mathcal{A}^{\vec{a},\lambdavec} = \sum_{c,t,i} {\color{red} \mathcal{C}_c^{\vec{a}}} \, {\color{gray} T_t^{\lambdavec}} \, {\color{blue} \mathcal{I}_{i}} \, r_{c,t,i}
			\end{equation*}
	}}
	\caption{Example Feynman diagram for $u \bar{u} \to gZ$ together with the three basis underlying structures.}
	\label{fig:qqgZ.diag}
\end{figure}
Consider an example Feynman diagram for the $u \bar{u} \to gZ$ process in Fig.~\ref{fig:qqgZ.diag}.
As introduced in Ch.~\ref{ch:amp}, the three underlying mathematical structures colour, tensor, and integral appear.
We schematically depict basis elements corresponding to each of them.
\begin{figure}[h]
	\centering
	\includegraphics[width=0.9\textwidth]{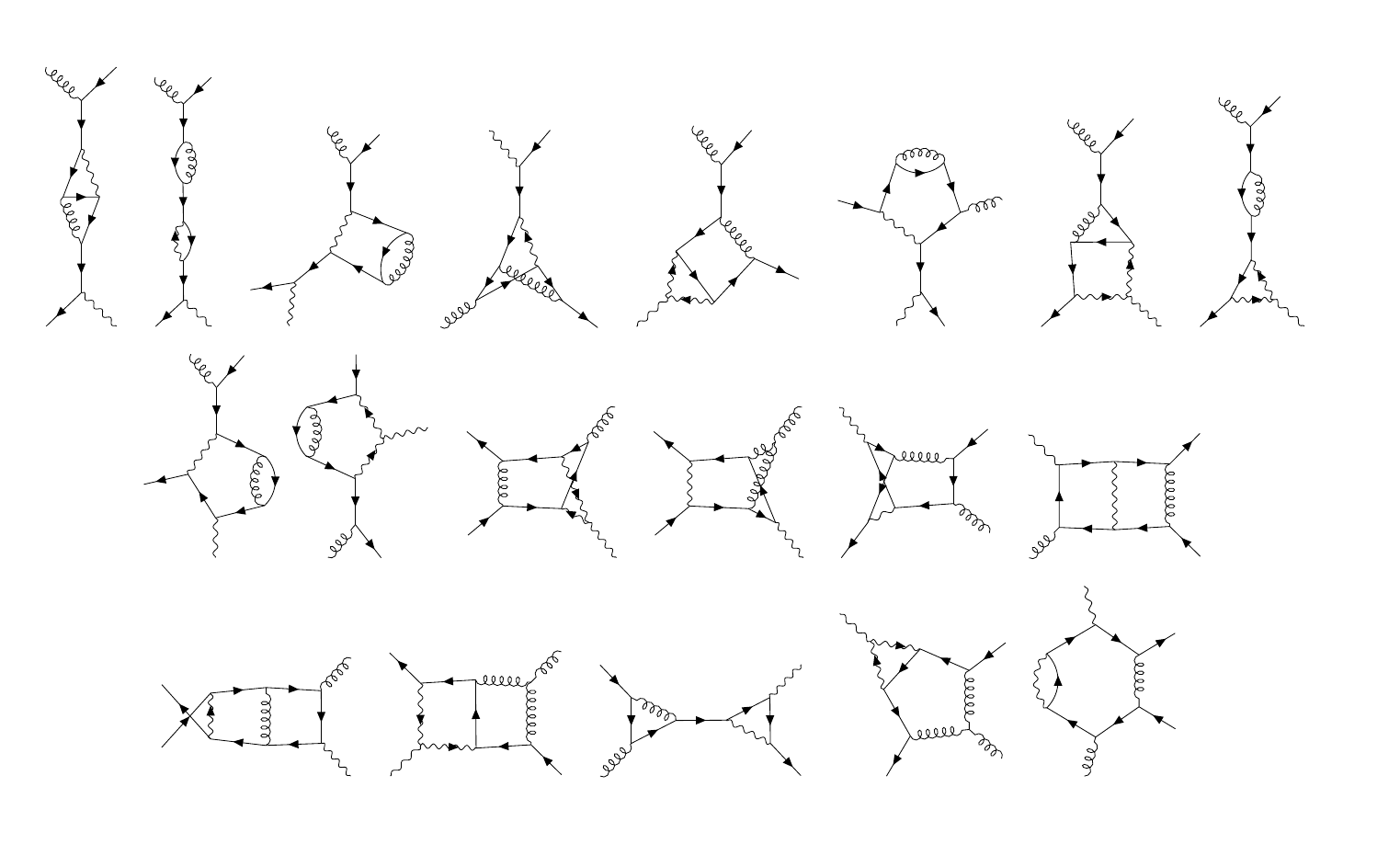}
	\caption{Example Feynman diagrams for the two-loop mixed QCD-EWK $u \bar{u} \to gZ$ process.}
	\label{fig:qqgZ.diags}
\end{figure}
Similarly to this example diagram, one has to analyse all 900 Feynman diagrams at the mixed QCD-EWK two-loop order, see for example Fig.~\ref{fig:qqgZ.diags}.
\begin{figure}[h]
	\centering
	\includegraphics[width=0.9\textwidth]{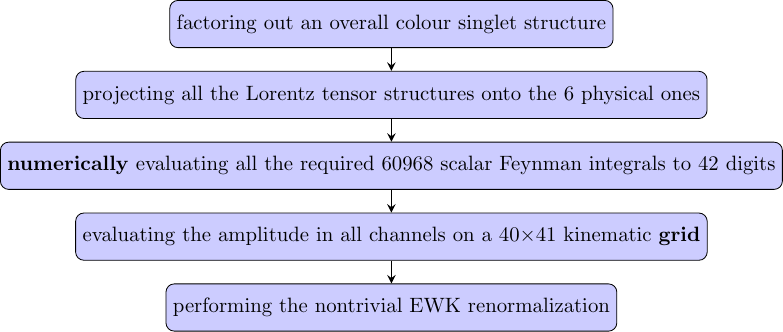}
	\caption{Computational flow chart for the amplitude.}
	\label{fig:qqgZ.flowAmp}
\end{figure}
In the following sections, we describe the computational flow outlined in Fig.~\ref{fig:qqgZ.flowAmp}, which is based on the methods introduced in Ch.~\ref{ch:amp}.
\begin{table}[t]
	\renewcommand{\arraystretch}{1.2}
	\centering
	%\begin{tabular}{ p{2.5cm}||p{0.3cm}|p{1.2cm}|p{1.2cm}|p{1.0cm} }
	\begin{tabular}{ l||r|r|r|r }
		(QCD,EWK) order & (0,0) & (1,0) & (0,1) & (1,1) \\
		\hline
		\hline
		Number of diagrams & 2 & 13 & 35 & 900 \\
		\hline
		Number of integral topologies & 0 & 1 & 4 & 18 \\
		\hline
		Number of scalar integrals & 0 & 105 & 275 & 60968 \\
		\hline
		Number of master integrals & 0 & 7 & 26 & 1202 \\
		\hline
		\hline
		Size of the Feynman diagrams list [kB] & 1 & 6 & 17 & 595 \\
		\hline
		Size before IBP reduction [kB] & 1 & 288 & 1180 & 422636 \\
		\hline
		Size of the numerical result on the grid [kB] & 908 & 3813 & 3842 & 3784 \\
	\end{tabular}
	\caption{Some indicators of the complexity of the result at various stages of the calculation at different loop orders.
	The size of the files with formulae in a consistent text format should be analysed relative to the corresponding leading order.}
	\label{tab:stats.uugZ}
\end{table}
In Tab.~\ref{tab:stats.uugZ}, we summarise the corresponding complexity of the result at various stages of the amplitude computation.
Note the growth in absolute size of the results after their numerical evaluation at lower loop orders.
Indeed, the measure of complexity is different at analytic and numerical level.
In addition, we elaborate in Sec.~\ref{sec:qqgZ.comp} on further computational resources required for the numerical evaluation.
At analytic stages of the computation, the complexity increases with corresponding orders even faster than in the $gg \to \gamma\gamma$ process summarized in Tab.~\ref{tab:stats}, due to the dependence on more mass scales.

\section{Amplitude structure}
\label{sec:qqgZ.amp}

Before discussing the details of the calculation, we describe the three mathematical structures appearing in the $u \bar{u} \to gZ$ amplitude, colour, tensor, and integral, as introduced in Ch.~\ref{ch:amp}.
Our analysis is analogous to the one in Sec.~\ref{sec:ggaa.amp} for the $gg \rightarrow \gamma\gamma$ process.
We repeat the main points here in order to help the Reader reproduce our results.

\subsection{Colour}
\label{sec:qqgZ.amp.col}

\begin{figure}[h]
	\centering
	\includegraphics[width=0.3\textwidth]{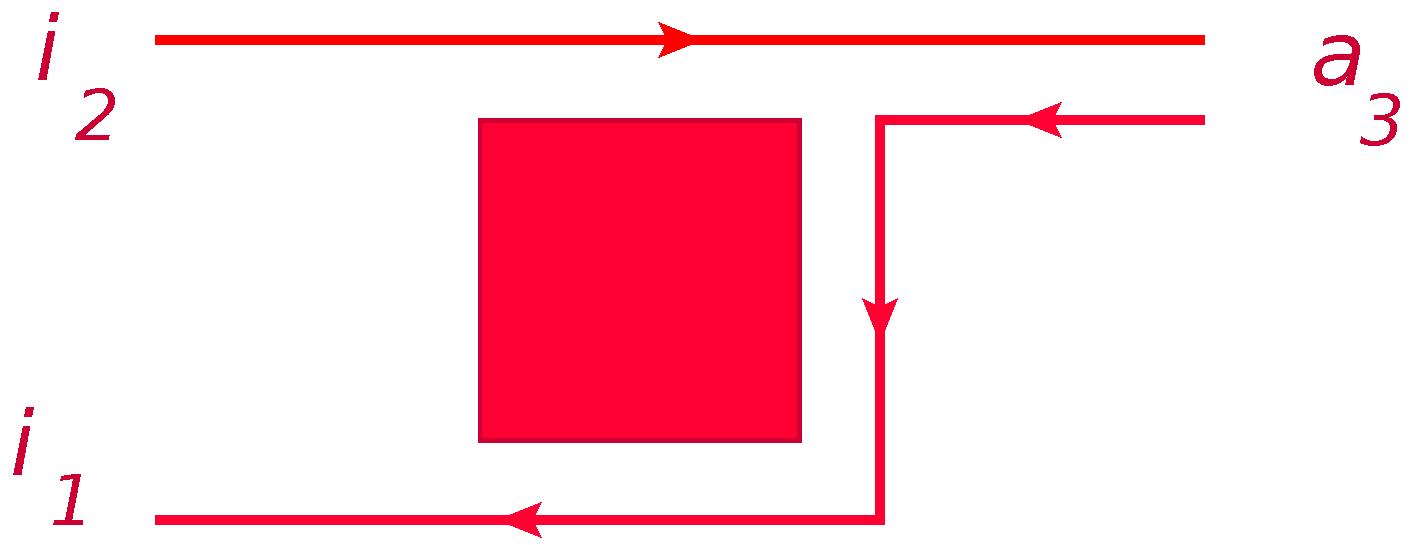}
	\caption{Schematic colour structure of the $u \bar{u} \to gZ$ amplitude.}
	\label{fig:uugZ.col}
\end{figure}

\noindent
Similarly to the $gg \rightarrow \gamma\gamma$ process discussed in Sec.~\ref{sec:ggaa.amp.col}, the $u \bar{u} \to gZ$ scattering also has a simple colour structure.
Following Sec.~\ref{sec:amp.col}, the dependence on the colour indices $i_{1,2}$ of quarks $u_1$ and $\bar{u}_2$, as well as on the index $a_3$ of the gluon $g_3$ factorizes from the amplitude
\begin{equation}
	\begin{split}
		\mathcal{A}_{i_1 i_2}^{a_3} &= T^{a_3}_{i_1 i_2} A \\
		&= T^{a_3}_{i_1 i_2} \sum_c C^{(1)}_c \, A_c \,,
	\end{split}
	\label{eq:uugZ.colDecomp}
\end{equation}
see Fig.~\ref{fig:uugZ.col}.
Note that, even though we are considering a two-loop correction, the colour corrections have a one-loop structure at the mixed QCD-EWK order.
Thus, the remaining colour-scalar structures $C^{(1)}_c$ are monomials in $\{C_A,C_F\}$ of degree 1.

\subsection{Lorentz tensors}
\label{sec:qqgZ.amp.ten}

\begin{figure}[h]
	\centering
	\includegraphics[width=0.65\textwidth]{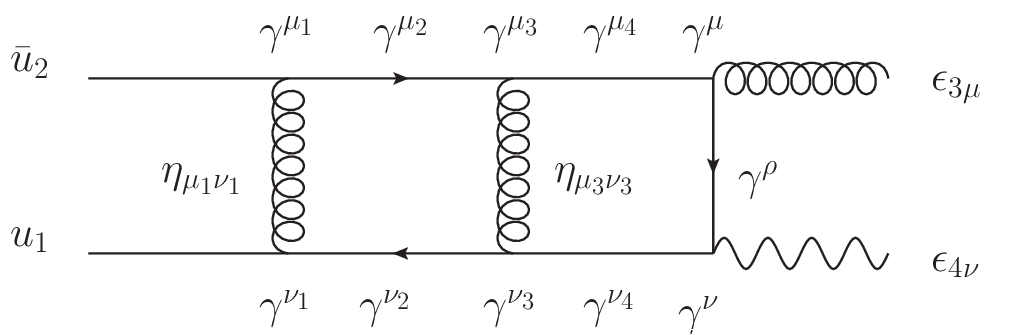}
	\caption{Lorentz tensor structures in example Feynman diagram for two-loop mixed QCD-EWK $u \bar{u} \to gZ$ amplitude. The open indices of $\gamma$ matrices are understood to be contracted with corresponding propagator momenta.}
	\label{fig:qqgZ:ten}
\end{figure}

\noindent
Similarly to Sec.~\ref{sec:ggaa.amp.ten} for the $gg \to \gamma\gamma$ process, we follow the method introduced in Sec.~\ref{sec:amp.Lorentz} in order to describe the Lorentz tensor structure of the $u \bar{u} \to gZ$ amplitude.
The colour-stripped amplitude $A_c$ for any fixed colour factor $C^{(1)}_c$ can be written as
\begin{align}
	A = \epsilon_{3,\mu}(p_3) \epsilon_{4,\nu}(p_4) \,\, \bar{u}(p_2) A^{\mu \nu} u(p_1) \,.
	\label{eq:ggaa.tenDecomp}
\end{align}
The amplitude tensor $A^{\mu \nu}$, which contains strings of Dirac $\gamma$ matrices, can be expanded in a basis of Lorentz tensors $\Gamma_i^{\mu\nu}$ as
\begin{equation}
	A^{\mu\nu} = \sum_{i=1}^{n_t} \mathcal F_i \, \Gamma_i^{\mu\nu} \,.
\end{equation}
The $39$ tensors $\Gamma_i^{\mu\nu}$ are constructed from the independent external momenta $p_{1,2,3}^\mu$, $\gamma$ matrices, as well as from the metric tensor $g^{\mu\nu}$.
Since $A^{\mu\nu}$ acts between the external on-shell quark states $u_1$ and $\bar{u}_2$, we can impose massless Dirac equations $\slashed{p}_1 u_1 = 0$ and $0 = \bar{u}_2 \, \slashed{p}_2$, which eliminate 22 tensors proportional to $\slashed{p}_1$ and $\slashed{p}_2$.
Similarly, since $A^{\mu\nu}$ is contracted with the gluon polarization vectors $\epsilon_3^\mu$, we impose the physical transversality condition
\begin{align}
	p_3 \cdot \epsilon_3 = 0 \,,\quad p_4 \cdot \ep_4=0 \,.
\end{align}
We also have a freedom to impose one constraint on the massless external gauge boson, which we choose to be
\begin{align}
	\ep_3 \cdot p_2=0 \,.
	\label{eq:uugZ.refMom}
\end{align}
This corresponds to a choice of the reference momentum $q_3 = p_2$ for the external gluon.
As a result, we are left with $n_t=7$ independent Lorentz tensors in $d$ dimensions~\cite{Gehrmann:2022vuk}
\begin{align}
	\Gamma_1^{\mu \nu} &= p_1^{\nu} \gamma^{\mu}\,, \;\;\;\;\;
	\Gamma_2^{\mu \nu} = p_1^{\mu} p_1^{\nu} \slashed{p}_3\,, \nonumber \\
	\Gamma_3^{\mu \nu} &= p_2^{\nu} \gamma^{\mu}\,, \;\;\;\;\;
	\Gamma_4^{\mu \nu} = p_1^{\mu} \gamma^{\nu}\,, \nonumber \\
	\Gamma_5^{\mu \nu} &= p_1^{\mu} p_2^{\nu} \slashed{p}_3\,, \;\;
	\Gamma_6^{\mu \nu} = g^{\mu\nu}\slashed{p}_3\,, \nonumber \\
	\Gamma_7^{\mu \nu} &= \gamma^\mu \slashed{p}_3 \gamma^\nu \,.
	\label{eq:qqgZ.Gammas}
\end{align}
After contracting with external polarization states, we arrive at
\begin{equation}
	T_i = \epsilon_{3,\mu}(p_3) \epsilon_{4,\nu}(p_4) \,\, \bar{u}(p_2) \Gamma_i^{\mu \nu} u(p_1)  \,,
	\label{eq:qqgZ.tensors}
\end{equation}
such that the colour-stripped amplitude can be decomposed as
\begin{equation}
	A = \sum_{i=1}^{7} \mathcal F_i(s,t) \, T_i \,.
	\label{eq:qqgZ.ampT}
\end{equation}

\begin{figure}[h]
	\centering
	\includegraphics[width=0.3\textwidth]{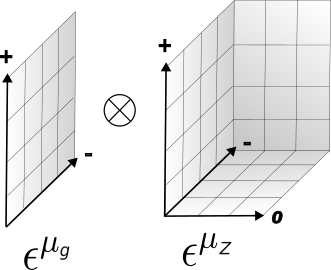}
	\caption{Spanning Lorentz tensor basis in $d=4$ for the $u \bar{u} \to gZ$ amplitude without closed fermion loops.}
	\label{fig:qqgZ.ten}
\end{figure}

In $d=4$, there are $2^2 \cdot 3/2 = 6$ independent helicity states, since gluon has 2 polarizations, $Z$ boson has 3, the massless quark pair preserves the helicity along the fermion line, and there is an overall parity symmetry for the vector current.
According to Ref.~\cite{Peraro:2020sfm}, the corresponding $\bar{n}_t=6$ tensors which span the physical four-dimensional subspace can be chosen as
\begin{equation}
	\oT_i = T_i\,,~  i=1,\dots,6 \,.
	\label{eq:uugZ.tens4}
\end{equation}
The remaining tensor can be projected out into the purely $-2\ep$ subspace with
\begin{equation}
	\oT_7 = T_7 - \sum_{j=1}^{6}(\mathcal{P}_j T_i) \oT_7 \,, \qquad
	\sum_{\text{pol}} \mathcal P_i \oT_j = \delta_{ij} \,,
	\label{eq:qqgZ.tensEps}
\end{equation}
which expands to
\begin{equation}
	\oT_7 \,\, = \bar{u}(p_2) \slashed{\epsilon}_3 \slashed{p}_3 \slashed{\epsilon}_4 u(p_1)
	- \frac{1}{s}\left( s(\oT_3+\oT_6-\oT_1) + t\oT_t + u(\oT_1+\oT_4) - 2\oT_5 \right) \,.
	\label{eq:ggaa.tensEpsExp}
\end{equation}
The explicit form of the projectors $\mathcal{P}_j$ is given in Ref.~\cite{Gehrmann:2022vuk}.

We now describe the relation of the form factors $\oF_i$ to the helicity amplitudes.
As mentioned in Sec.~\ref{sec:ppjZ.motivation}, we are interested in a further decay of the on-shell $Z$ boson into a massless fermion pair $Z(p_4) \to l(p_5) \, \bar{l}(p_6)$.
In the on-shell approximation, we construct this five-point amplitude by attaching a fermion current via the $Z$ boson propagator
\begin{equation}
	\mathcal{M} = \mathcal{A}_{\mu_4} \frac{i\left( -g^{\mu_4 \nu_4} + \frac{p_4^{\mu_4} p_4^{\nu_4}}{m_Z^2} \right)}{s - m_Z^2 + i \Gamma_Z m_Z}
	\,\, e \bar{u}(l_5) \gamma_{\nu_4} (c_{l,V} + c_{l,A} \gamma_5) u(l_6) \,.
\end{equation}
Note that the second term vanishes due to the momentum conservation $p_4=p_5+p_6$ and the Dirac equation for the massless fermionic pair $\slashed{p}_6 u(l_6) = 0$ and $0 = \bar{u}(l_5) \, \slashed{p}_5$. 
We find helicity amplitudes $\mathcal{M}_\lambdavec$ by evaluating all tensors $\oT_i$ at fixed helicity states $\lambdavec=\{\lambda_1,\lambda_2,\lambda_3,\lambda_5,\lambda_6\}$
\begin{equation}
	\mathcal{M}_\lambdavec = \sum_{i=1}^6 \oF_i \, \oT_{i,\lambdavec}\,.
\end{equation}
The independent helicity amplitudes read
\begin{equation}
	\begin{split}
		\mathcal{M}_{+-+-+} &= \,\,\,\,\, \frac{1}{\sqrt{2}} \left( \langle 12 \rangle [13]^2 \left( \alpha_1 \langle 536 ] + \alpha_2 \langle 526 ] \right) + \alpha_3 \langle 25 \rangle [13] [36] \right) \,, \\
		\mathcal{M}_{+---+} &= \,\,\,\,\, \frac{1}{\sqrt{2}} \left( \langle 23 \rangle [12]^2 \left( \gamma_1 \langle 536 ] + \gamma_2 \langle 516 ] \right) + \gamma_3 \langle 23 \rangle \langle 35 \rangle [16] \right) \,, \\
		\mathcal{M}_{-++-+} &= -\frac{1}{\sqrt{2}} \left( [23]^2 \langle 12 \rangle \left( \gamma_1 \langle 536 ] + \gamma_2 \langle 516 ] \right) + \gamma_3 [23] [36] \langle 15 \rangle \right) \,, \\
		\mathcal{M}_{-+--+} &= -\frac{1}{\sqrt{2}} \left( [12] \langle 13 \rangle^2 \left( \alpha_1 \langle 536 ] + \alpha_2 \langle 526 ] \right) + \alpha_3 [26] \langle 13 \rangle \langle 35 \rangle \right) \,,
	\end{split}
	\label{eq:hels}
\end{equation}
with little group scalar coefficients
\begin{equation}
	\begin{split}
		\alpha_1 &= -\oF_1 \,, \qquad
		\alpha_2 = \oF_2 - \oF_1 + \frac{2}{s_{23}} \oF_6 \,, \qquad
		\alpha_3 = 2 \oF_3 - \frac{2 s_{12}}{s_{23}} \oF_6 \,, \\
		\gamma_1 &= \frac{1}{s_{23}} \left( s_{13} \oF_2 + 2 (\oF_5 - \oF_3) \right) \,, \\
		\gamma_2 &= \frac{1}{s_{23}} \left( s_{13} (\oF_2 - \oF_1) - 2 (\oF_4 - \oF_5 + \oF_6) \right) \,, \\
		\gamma_3 &= -\frac{2}{s_{23}} \left( s_{23} \oF_3 + s_{12} \oF_6 \right) \,.
	\end{split}
\end{equation}
The remaining 4 helicity amplitudes can be extracted from the above by a charge conjugation transformation.

\subsection{Feynman integrals}

Finally, we describe the last remaining amplitude structure, the Feynman integrals.
For the two-loop mixed QCD-EWK $u \bar{u} \to gZ$ amplitude without top quark contributions, there are 18 independent integral topologies

\scalebox{0.60}{\parbox{0.9\linewidth}{
\begin{equation}
	\begin{split}
		\text{PL2A}&=\{k_1^2,k_2^2,(k_1-k_2)^2,(k_1-p_1)^2,(k_2-p_1)^2,(k_1-p_{12})^2,(k_2-p_{12})^2,(k_1-p_{123})^2,(k_2-p_{123})^2\}\,, \\ \text{PL2Z1}&=\{k_1^2-m_Z^2,k_2^2,(k_1-k_2)^2,(k_1-p_1)^2,(k_2-p_1)^2,(k_1-p_{12})^2,(k_2-p_{12})^2,(k_1-p_{123})^2,(k_2-p_{123})^2\}\,, \\ \text{PL2Z3}&=\{k_1^2,k_2^2,(k_1-k_2)^2-m_Z^2,(k_1-p_1)^2,(k_2-p_1)^2,(k_1-p_{12})^2,(k_2-p_{12})^2,(k_1-p_{123})^2,(k_2-p_{123})^2\}\,, \\ \text{PL2Z4}&=\{k_1^2,k_2^2,(k_1-k_2)^2,-m_Z^2+(k_1-p_1)^2,(k_2-p_1)^2,(k_1-p_{12})^2,(k_2-p_{12})^2,(k_1-p_{123})^2,(k_2-p_{123})^2\}\,, \\ \text{PL2Z6}&=\{k_1^2,k_2^2,(k_1-k_2)^2,(k_1-p_1)^2,(k_2-p_1)^2,-m_Z^2+(k_1-p_{12})^2,(k_2-p_{12})^2,(k_1-p_{123})^2,(k_2-p_{123})^2\}\,, \\ \text{PL2Z7}&=\{k_1^2,k_2^2,(k_1-k_2)^2,(k_1-p_1)^2,(k_2-p_1)^2,(k_1-p_{12})^2,-m_Z^2+(k_2-p_{12})^2,(k_1-p_{123})^2,(k_2-p_{123})^2\}\,, \\ \text{PL2W1}&=\{k_1^2-m_W^2,k_2^2,(k_1-k_2)^2,(k_1-p_1)^2,(k_2-p_1)^2,(k_1-p_{12})^2,(k_2-p_{12})^2,(k_1-p_{123})^2,(k_2-p_{123})^2\}\,, \\ \text{PL2W3}&=\{k_1^2,k_2^2,(k_1-k_2)^2-m_W^2,(k_1-p_1)^2,(k_2-p_1)^2,(k_1-p_{12})^2,(k_2-p_{12})^2,(k_1-p_{123})^2,(k_2-p_{123})^2\}\,, \\ \text{PL2W4}&=\{k_1^2,k_2^2,(k_1-k_2)^2,-m_W^2+(k_1-p_1)^2,(k_2-p_1)^2,(k_1-p_{12})^2,(k_2-p_{12})^2,(k_1-p_{123})^2,(k_2-p_{123})^2\}\,, \\ \text{PL2W6}&=\{k_1^2,k_2^2,(k_1-k_2)^2,(k_1-p_1)^2,(k_2-p_1)^2,-m_W^2+(k_1-p_{12})^2,(k_2-p_{12})^2,(k_1-p_{123})^2,(k_2-p_{123})^2\}\,, \\ \text{PL2W7}&=\{k_1^2,k_2^2,(k_1-k_2)^2,(k_1-p_1)^2,(k_2-p_1)^2,(k_1-p_{12})^2,-m_W^2+(k_2-p_{12})^2,(k_1-p_{123})^2,(k_2-p_{123})^2\}\,, \\ \text{PL2W29}&=\{k_1^2,k_2^2-m_W^2,(k_1-k_2)^2,(k_1-p_1)^2,(k_2-p_1)^2,(k_1-p_{12})^2,(k_2-p_{12})^2,(k_1-p_{123})^2,-m_W^2+(k_2-p_{123})^2\}\,, \\ \text{NPL2A}&=\{k_1^2,k_2^2,(k_1-k_2)^2,(k_1-p_1)^2,(k_2-p_1)^2,(k_1-p_{12})^2,(k_1-k_2+p_3)^2,(k_2-p_{123})^2,(k_1-k_2-p_{12})^2\}\,, \\ \text{NPL2Z1}&=\{k_1^2-m_Z^2,k_2^2,(k_1-k_2)^2,(k_1-p_1)^2,(k_2-p_1)^2,(k_1-p_{12})^2,(k_1-k_2+p_3)^2,(k_2-p_{123})^2,(k_1-k_2-p_{12})^2\}\,, \\ \text{NPL2Z4}&=\{k_1^2,k_2^2,(k_1-k_2)^2,-m_Z^2+(k_1-p_1)^2,(k_2-p_1)^2,(k_1-p_{12})^2,(k_1-k_2+p_3)^2,(k_2-p_{123})^2,(k_1-k_2-p_{12})^2\}\,, \\ \text{NPL2Z7}&=\{k_1^2,k_2^2,(k_1-k_2)^2,(k_1-p_1)^2,(k_2-p_1)^2,(k_1-p_{12})^2,-m_Z^2+(k_1-k_2+p_3)^2,(k_2-p_{123})^2,(k_1-k_2-p_{12})^2\}\,, \\ \text{NPL2Z1c13c24}&=\{k_1^2-m_Z^2,k_2^2,(k_1-k_2)^2,(k_1-p_3)^2,(k_2-p_3)^2,(k_1+p_1+p_2)^2,(k_1-k_2+p_1)^2,(k_2+p_2)^2,(k_1-k_2+p_1+p_2)^2\}\,, \\ \text{NPL2W1}&=\{k_1^2-m_W^2,k_2^2,(k_1-k_2)^2,(k_1-p_1)^2,(k_2-p_1)^2,(k_1-p_{12})^2,(k_1-k_2+p_3)^2,(k_2-p_{123})^2,(k_1-k_2-p_{12})^2\}\,, \\ \text{NPL2W4}&=\{k_1^2,k_2^2,(k_1-k_2)^2,-m_W^2+(k_1-p_1)^2,(k_2-p_1)^2,(k_1-p_{12})^2,(k_1-k_2+p_3)^2,(k_2-p_{123})^2,(k_1-k_2-p_{12})^2\}\,, \\ \text{NPL2W7}&=\{k_1^2,k_2^2,(k_1-k_2)^2,(k_1-p_1)^2,(k_2-p_1)^2,(k_1-p_{12})^2,-m_W^2+(k_1-k_2+p_3)^2,(k_2-p_{123})^2,(k_1-k_2-p_{12})^2\}\,, \\ \text{NPL2W1c13c24}&=\{k_1^2-m_W^2,k_2^2,(k_1-k_2)^2,(k_1-p_3)^2,(k_2-p_3)^2,(k_1+p_1+p_2)^2,(k_1-k_2+p_1)^2,(k_2+p_2)^2,(k_1-k_2+p_1+p_2)^2\}\,, \\ \text{NPL2W28}&=\{k_1^2,k_2^2-m_W^2,(k_1-k_2)^2,(k_1-p_1)^2,(k_2-p_1)^2,(k_1-p_{12})^2,(k_1-k_2+p_3)^2,-m_W^2+(k_2-p_{123})^2,(k_1-k_2-p_{12})^2\}
	\end{split}
\end{equation}
\label{eq:qqgZ.topos}
}}

\noindent
each with 9 generalised propagators.
They arise from the basic PL and NPL two-loop four-point massless topologies introduced in Sec.~\ref{sec:amp.int.topo}, with appropriate modifications due to external $m_Z$, as well as internal $m_Z$ and $m_W$ masses.
\begin{figure}[h]
	\centering
	\includegraphics[width=0.99\textwidth]{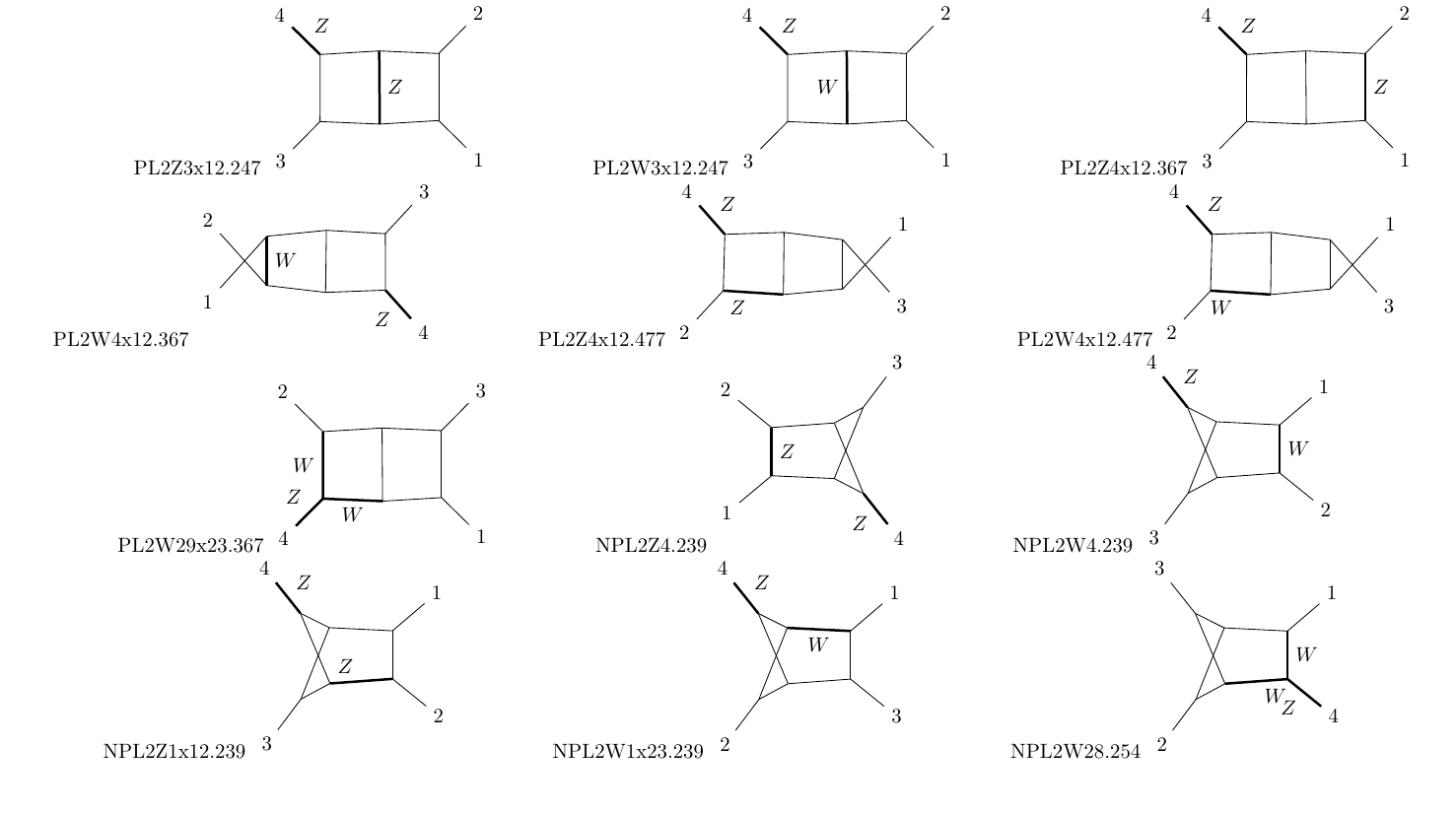}
	\caption{Example top sectors of integral topologies defined in Eq.~\ref{eq:qqgZ.topos}, in the format TOPOLOGYxCROSSING.SECTOR.
	}
	\label{fig:uugZ.topsec}
\end{figure}
See the corresponding top sectors in Fig.~\ref{fig:uugZ.topsec}.
We write a Feynman integral $\mathcal{I}_{\vec{n}}$ in the Laporta notation as
\begin{equation}
	\mathcal{I}_{\vec{n}} 
	= \int \left(\prod_{i=1}^2 \mathcal{D}^d k_i \right) \,
	\frac{\mathcal D_{8}^{-n_{8}} \mathcal D_{9}^{-n_{9}}}{\mathcal D_{1}^{n_1} \dots \mathcal D_{7}^{n_{7}}} \,,
\end{equation}
where the integration measure is
\begin{equation}
	\mathcal{D}^d k_i = \frac{d^d k_i}{(2\pi)^d} \,.
\end{equation}
Note that this Feynman measure is different then our choice in the three-loop four-point calculation in Sec.~\ref{sec:ggaa.amp.int}.

\begin{figure}[h]
	\centering
	\includegraphics[width=0.9\textwidth]{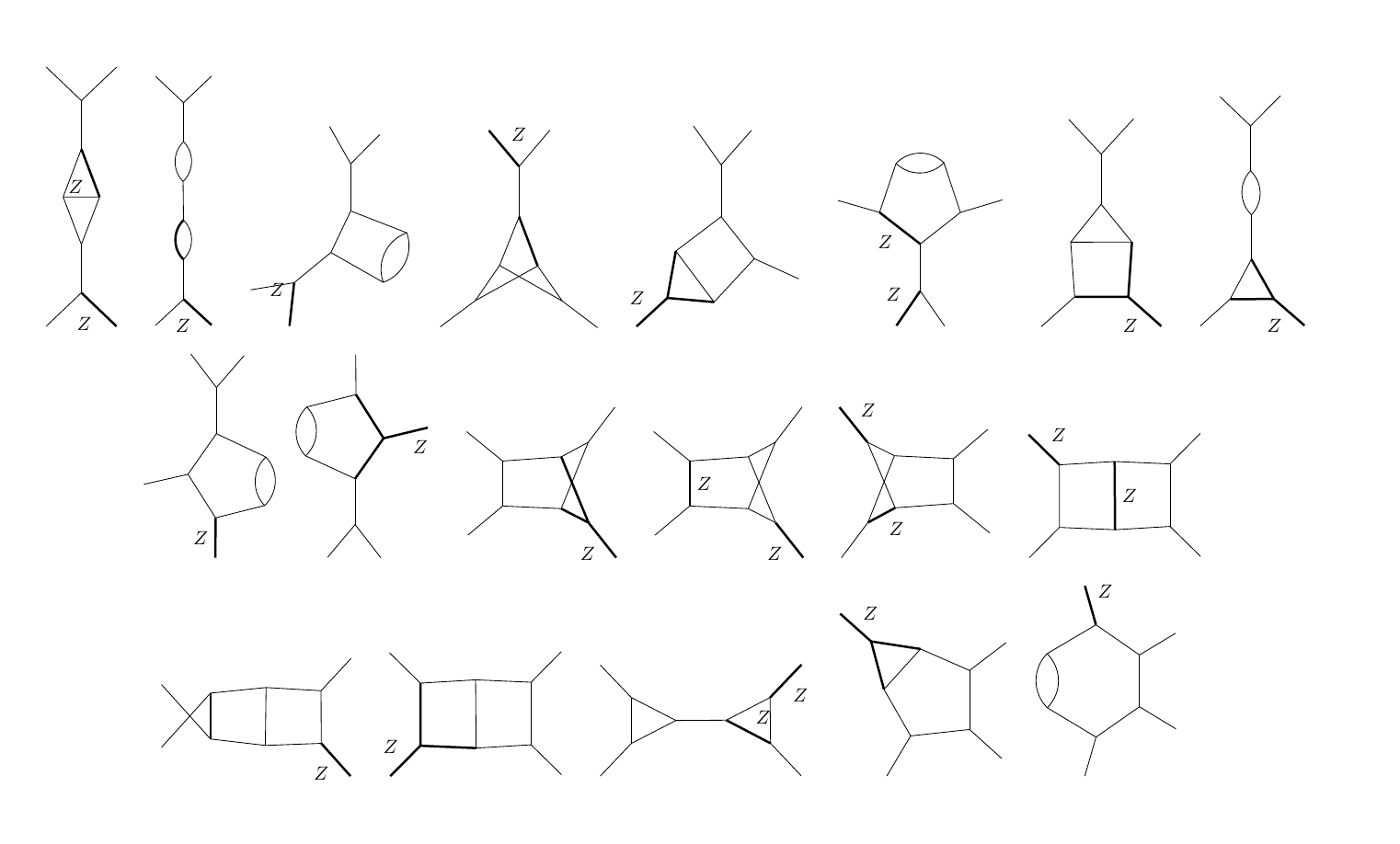}
	\caption{Example Feynman integrals for the two-loop mixed QCD-EWK $u \bar{u} \to gZ$ amplitude.}
	\label{fig:uugZ.ints}	
\end{figure}

As summarised in Tab.~\ref{tab:stats.uugZ}, after the Lorentz tensor decomposition, there are 60968 scalar Feynman integrals to be evaluated, see example graphs in Fig.~\ref{fig:uugZ.ints}.
According to the notation introduced in Sec.~\ref{sec:amp.int.topo}, their complexity reaches $r=7$, $s=4$, and $d=1$.

As described in details in Sec.~\ref{sec:ggaa.calc}, the usual computational flow at this stage would consist of IBP reducing all the required Feynman integrals to a MI basis, designing DEQs for each top sector, and solving them perturbatively in $\ep$ with appropriate BCs, like in Sec.~\ref{sec:ggaa.calc} for the QCD $gg \to \gamma\gamma$ process.
For EWK corrections, each of these three steps becomes much more intricate than in the massless QCD sector of the SM.
Since the analytic approach may not guarantee a significant decrease in the complexity of the problem, we decided to rely on a numerical approach, as mentioned in Sec.~\ref{sec:ppjZ.motivation}.
We also investigated the analytic approach, as we discuss in the Sec.~\ref{sec:qqgZ.comp.IBP}.
We found it to be suboptimal in comparison to our numerical methodology detailed in Sec.~\ref{sec:qqgZ.comp.num}.

\section{The calculation}
\label{sec:qqgZ.comp}

In this section, we provide some more extensive details about the calculation of the amplitude.
We follow an overall workflow summarised in Fig.~\ref{fig:qqgZ.flowAmp}, while introducing basic information about the exploited computer programs.
Moreover, we elaborate on the comparison of our numerical approach to a potential alternative analytic method.

\subsection{Integrand-level manipulations}
\label{sec:qqgZ.comp.intd}

\begin{figure}[h]
	\includegraphics[width=0.9\textwidth]{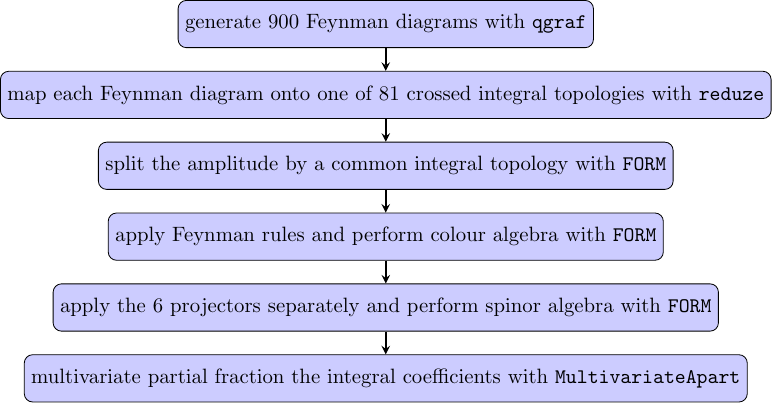}
	\caption{Computational flow for the integrand.}
	\label{fig:qqgZ.flowIntd}
\end{figure}

\noindent
Similarly to the discussion of the calculation for the $gg \to \gamma\gamma$ process in Sec.~\ref{sec:ggaa.calc.intd}, we describe the manipulations of the amplitude at the integrand level, as outlined in Fig.~\ref{fig:qqgZ.flowIntd}.
Since the programming practicalities are simple modifications of those described in Sec.~\ref{sec:ggaa.calc.intd}, we will further provide snippets of code only for conceptually new steps.

We start by generating with \texttt{qgraf} all the required 900 Feynman diagrams for the two-loop mixed QCD-EWK $u\bar{u} \to gZ$ process without top quark contributions.
We will exclude the diagrams containing closed fermion loops later, when analysing the spinor structures.
We fix this mixed perturbative order with an additional line in the run card
\begin{mycode}
true = vsum[ gs, 3, 3];
\end{mycode}
which constrains all the two-loop diagrams to contain exactly 3 strong vertices specified in the model file.
In addition to the particle content of the QCD sector of the SM with the $u$ and $d$ quark, we also need to include in the model file the necessary EWK vertices involving the gauge bosons $Z$, $W^{\pm}$, and $\gamma$.
The output of this \texttt{qgraf} job serves as an input of \texttt{FORM}.
We note that at this perturbative order, there are no Feynman diagrams involving associated ghosts $c_Z, c_{W^{\pm}}$, or scalars $\chi, \varphi^{\pm}$, and $H$.

In parallel, we generate a similar \texttt{qgraf} output appropriate for the \texttt{reduze} format.
We use it to map each Feynman diagram onto one of the 18 integral topologies, or their crossings, with a \texttt{reduze} job \texttt{find\_diagram\_shifts}.
As a result, all Feynman diagrams for this process can be mapped onto 81 crossed topologies

\medskip
\scalebox{0.95}{\parbox{0.90\linewidth}{
\{NPL2A, NPL2Ax12, NPL2Ax123, NPL2Ax132, NPL2Ax23, NPL2W1c13c24x123, NPL2W1c13c24x23, NPL2W1x123, NPL2W1x23, NPL2W28, NPL2W28x12, NPL2W4, NPL2W4x12, NPL2W4x13, NPL2W4x132, NPL2W7x123, NPL2W7x13, NPL2W7x132, NPL2W7x23, NPL2Z1, NPL2Z1c13c24x123, NPL2Z1c13c24x23, NPL2Z1x12, NPL2Z1x123, NPL2Z1x23, NPL2Z4, NPL2Z4x12, NPL2Z4x13, NPL2Z4x132, NPL2Z7x123, NPL2Z7x13, NPL2Z7x132, NPL2Z7x23, PL2A, PL2Ax12, PL2Ax123, PL2Ax13, PL2Ax132, PL2Ax23, PL2W1, PL2W1x12, PL2W1x123, PL2W1x13, PL2W1x132, PL2W1x23, PL2W29, PL2W29x12, PL2W29x123, PL2W29x13, PL2W29x132, PL2W29x23, PL2W3, PL2W3x12, PL2W3x123, PL2W3x13, PL2W3x132, PL2W3x23, PL2W4, PL2W4x12, PL2W4x123, PL2W4x13, PL2W4x132, PL2W4x23, PL2Z1, PL2Z1x12, PL2Z1x123, PL2Z1x13, PL2Z1x132, PL2Z1x23, PL2Z3, PL2Z3x12, PL2Z3x123, PL2Z3x13, PL2Z3x132, PL2Z3x23, PL2Z4, PL2Z4x12, PL2Z4x123, PL2Z4x13, PL2Z4x132, PL2Z4x23\}.
}}
\medskip

Next, we apply with \texttt{FORM} the loop momentum shift identities provided by \texttt{reduze} on the list of Feynman diagrams generated by \texttt{qgraf}.
At this stage, the resulting list of Feynman diagrams has a size of about 0.5~MB, see Tab.~\ref{tab:stats.uugZ}.
Since the further integrand-level manipulations will only expand the expressions, we parallelize our computations.
To this end, we split the amplitude $\mathcal{A}$ into 81 amplitude pieces $\mathcal{A}_{t}$ belonging to a fixed crossed integral topology.

Then, using \texttt{FORM}, we apply Feynman rules for both QCD and EWK sectors of the SM, as well as perform colour algebra at each of the 81 amplitude pieces $\mathcal{A}_{t}$ in parallel.
We note that the number of remaining Feynman diagrams which do not vanish by colour is 648.
We also combine all Dirac $\gamma$ matrices appearing along fermion lines.
For the two-loop $u\bar{u} \to gZ$ process, the amplitude contains terms with exactly one open and at most one closed fermion line.
The latter are proportional to the $n_f$ factor, and we discard these terms at this stage, as argued in Sec.~\ref{sec:ppjZ.complexity}.
Moreover, we linearly expand the spinor strings \texttt{spst(...)} with respect to the momenta.
The number of remaining Feynman diagrams without any closed fermion loops is 462.

Finally, we apply one of the 6 projectors $\mathcal{P}_i$ on each amplitude piece $\mathcal{A}_{t}$ in parallel, which leads to 486 form factors $\mathcal{F}_{t,i}$.
This steps involves summing over polarizations in $d$ dimensions.
While performing the Dirac algebra, it is useful to frequently apply the kinematic identities as in Eqs~\ref{eq:qqgZ.kin} and~\ref{eq:uugZ.refMom}.
After these manipulations, all the remaining Feynman integrals are Lorentz scalar.
Therefore, it is a natural place to write each of them in the Laporta notation corresponding to their underlying integral topology.
As a result, we arrive at a list of around $6 \times 10^4$ scalar Feynman integrals in an expression for the amplitude of a total size of about 423~MB, see Tab.~\ref{tab:stats.uugZ}.

One additional step is to simplify the coefficients of scalar integrals in each form factor $\mathcal{F}_{t,i}$ by partial fractioning them.
In the considered four-point one-mass EWK kinematics, the coefficients depend on four independent variables e.g. $t/m_Z^2$, $u/m_Z^2$, $m_W^2/m_Z^2$, and $d$.
Contrarily to the case of four-point zero-mass kinematics of the $gg \to \gamma\gamma$ process described in \ref{sec:ggaa.calc.int}, some of the denominator factors depend on more then one kinematic variable.
In order to perform in practice a multivariate partial fractioning of rational functions at hand, we exploit the \texttt{Mathematica} package \texttt{MultivariateApart}~\cite{Heller:2021qkz}.
Since all the denominators are of the simple propagator type, i.e. $\{s_{ij}^{-1}\}$, the in-built \texttt{MultivariateApart[]} function provides enough computational performance.
On the technical side, we note that it is important to apply this function on factors depending exclusively on the variables in which we are performing the partial fractioning, and not on e.g. couplings and colour factors.
After this step, the total size of the amplitude is about 368~MB.

\subsection{Preliminary integral reduction}
\label{sec:qqgZ.comp.IBP}

A natural further step in the computational flow would be to perform the analytic IBP reduction of all scalar Feynman integrals in the problem.
As mentioned earlier, in this section we provide the details of such reduction and we assess its efficiency.
In result of the following discussion, we omit this step in the final computational flow, see Fig.~\ref{fig:qqgZ.flowAmp}, and we employ a numerical approach instead.

\begin{figure}[h]
	\includegraphics[width=0.9\textwidth]{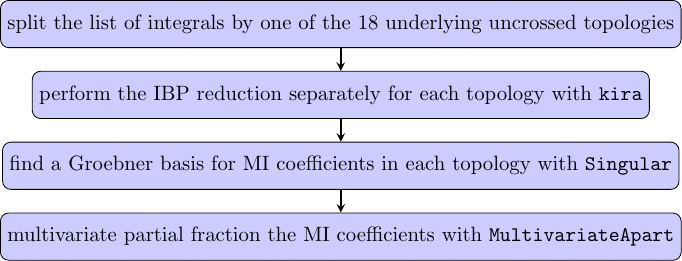}
	\caption{Computational flow for the integral reduction.}
	\label{fig:uugZ.flowInt}
\end{figure}

In order to perform the IBP reduction, we chose the most efficient publicly available reducer at the time, \texttt{kira-2.2}~\cite{Maierhofer:2017gsa,Maierhofer:2018gpa}.
This program works in a similar way to \texttt{reduze}, as described in Sec.~\ref{sec:ggaa.calc.int}, but it internally implements the finite field reconstruction of the IBP coefficients.
This procedure gave a strong computational advantage comparing to other available codes.
Since \texttt{kira} is not optimised for finding relations between different topologies, as well as their crossings, we split the list of all the relevant 60968 scalar Feynman integrals by one of the 18 underlying uncrossed topology.
As a result, we obtain 21742 uncrossed integrals collected in 18 lists of length between 695 and 2168.
Note that one has to retrieve the crossed IBP relations at the end of the reduction.

We perform the IBP reduction of each integral topology in parallel.
To this end, we execute a \texttt{kira} job of the form
\begin{mycode}
jobs:
 - reduce_sectors:
    reduce:
     - {topologies: [NPL2A], sectors: [254], r: 7, s: 4, d: 0}
     - {topologies: [NPL2A], sectors: [239], r: 7, s: 4, d: 0}
    select_integrals:
     select_mandatory_list:
      - [NPL2A,listNPL2A]
    run_symmetries: true
    run_initiate: true
    run_triangular: true
    run_back_substitution: true
    run_firefly: true
\end{mycode}
Note that \texttt{kira} requires specifying all top sectors to be reduced in a fixed topology, together with their corresponding parameters $r$, $s$, and $d$.
We explicitly show the options that we have used to improve the performance.
Importantly, we enable finite field reconstruction of the MI coefficients with \texttt{run\_firefly: true}.
Moreover, this option allows for saving intermediate results of the reduction to a local database.
It is a crucial feature for long running jobs, since we can losslessly restart the IBP reduction from its last saved status in case the job terminates unforeseeably.
We do not specify any preferred MI basis, thus, we use the default Laporta basis suggested by \texttt{kira}.
The number of MIs ranges from 27 to 98 per topology, reaching 1202 in total.
Note that, in principle, this total number of MIs does not account for any relations between the subsector MIs in different topologies.
The resulting IBP table has a size of about 1 to 640~MB per topology, which sums to around 1.6~GB of identities.

In order to simplify the IBP identities, we need to partial fraction the MI coefficients, similarly as for the integral coefficients before the IBP reduction in Sec.~\ref{sec:qqgZ.comp.intd}.
Since the denominator factors of the MI coefficients are complicated multivariate polynomials, the built-in \texttt{MultivariateApart[]} function of the \texttt{MultivariateApart} package is no longer computationally efficient enough.
In order to improve the performance of multivariate partial fractioning, we need to find an appropriate basis of independent denominator factors called the \textit{Groebner basis}.
It is an algebraic geometry concept which fixed a specific denominator basis when the multivariate partial fractioning is not unique, e.g.
\begin{equation}
	\frac{2x-y}{x(x+y)(x-y)}
	= \frac{1}{x (x+y)}+\frac{1}{(x-y) (x+y)}
	= \frac{3}{2 x (x+y)}+\frac{1}{2 x (x-y)}
\end{equation}
The problem of finding the Groebner basis can be solved in the framework of computational algebraic geometry.
Due to major recent advances in this field, there are multiple publicly available tools to efficiently solve explicit problems in computational algebraic geometry.
Here, we exploit the \texttt{Singular}~\cite{DGPS} program.

The number of denominator factors in a fixed integral topology ranges from 41 to 131, and reaches 492 distinct multivariate polynomials in all topologies.
The most complicated polynomial has 4 variables $d$, $u/m_Z^2$, $t/m_Z^2$, and $m_W^2/m_Z^2$, a degree 37, 20282 terms, 9 digit rational coefficients, and it stems from the NPL2W28 topology.
An example piece of the \texttt{Singular} job file for finding the Groebner basis in the PL2A topology reads
\begin{mycode}
ring myring=0,
(q43,q42,q41,q39,q38,q40,q37,q36,q35,q34,q32,q33,q26,q25,q24,q23,q22,q21,q20,q19,q18,q17,q16,q15,
 q14,q13,q12,q11,q10,q9,q8,q7,q6,q5,q4,q3,q2,q1,q27,q28,q29,q30,q31,d,s13,s23),
(dp(5),dp(3),dp(4),dp(26),dp(2),dp(3),dp(3));
poly mygen1=1-(-8 + d)*q1;
...
poly mygen26=1-(-1208 + 566*d - 117*d^2 + 9*d^3)*q26;
poly mygen27=1-q27*(-1 + s13);
poly mygen28=1-q28*s13;
poly mygen30=1-q30*(-1 + s23);
poly mygen31=1-q31*s23;
poly mygen32=1-q32*(-1 + s13 + s23);
poly mygen33=1-q33*(s13 + s23);
...
poly mygen38=1-q38*(10 - 3*d - 10*s13 + 3*d*s13 - 12*s23 + 4*d*s23);
...
poly mygen43=1-q43*(10 - 3*d - 30*s13 + 9*d*s13 + 30*s13^2 - 9*d*s13^2 - 10*s13^3 + 3*d*s13^3
 - 62*s23 + 19*d*s23 + 94*s13*s23 - 29*d*s13*s23 - 42*s13^2*s23 + 13*d*s13^2*s23 + 98*s23^2
  - 31*d*s23^2 - 66*s13*s23^2 + 21*d*s13*s23^2 - 46*s23^3 + 15*d*s23^3);
ideal myideal=mygen1,mygen2,mygen3,mygen4,mygen5,mygen6,mygen7,mygen8,mygen9,mygen10,mygen11,
 mygen12,mygen13,mygen14,mygen15,mygen16,mygen17,mygen18,mygen19,mygen20,mygen21,mygen22,
 mygen23,mygen24,mygen25,mygen26,mygen27,mygen28,mygen29,mygen30,mygen31,mygen32,mygen33,
 mygen34,mygen35,mygen36,mygen37,mygen38,mygen39,mygen40,mygen41,mygen42,mygen43;
ideal mygb = slimgb(myideal);
link outfile = "ASCII:w outputPL2A.m";
write(outfile, "{", mygb, "}");
close(outfile);
quit;
\end{mycode}
where ellipsis stands for lines analogous to the ones shown.
It can be generated with the function \texttt{WriteSingularBasisInput[]} of the\break\texttt{MultivariateApart} package.
In order to obtain a unique output, one has to specify an ordering of polynomials and their variables.
We choose the polynomial ordering according to the guidelines in Ref.~\cite{Heller:2021qkz}, i.e. we sort them in decreasing complexity.
We quantify high complexity based on the number of variables, number of terms, and unphysicality, respectively, in a weakening order.
In this way, together with the \texttt{slimgb()} function of \texttt{Singular}, we aim for the spurious denominators to be expressed in terms of the physical ones, e.g. of the propagator type $s_{ij}^{-1}$.
As a result, we obtain a Groebner basis of length 6320 for the PL2A topology.
Overall, we have obtained increasingly more complicated Groebner bases for 5 topologies PL2A, NPL2Z1c13c24, NPL2A, NPL2Z4, and NPL2Z1 respectively, reaching up to 82862 polynomials for the last topology.
Due to the high complexity of the remaining 13 topologies, their corresponding \texttt{Singular} jobs did not terminate in a reasonable CPU time.
We discuss potential ways to resolve this issue later in this subsection.

We finally turn to the description of the multivariate partial fractioning of the MI coefficients in the IBP table.
We perform it using the \texttt{ApartReduce[expression, GroebnerBasis, ordering, UseFormProgram->True]} function of the\break\texttt{MultivariateApart} package.
We note again that it is important to apply this method on expressions depending only on the variables in which the partial fractioning is performed.
For better performance, it is useful to enable \texttt{FORM} to be a back end for the computation of polynomial reductions.
This option works only if the \texttt{FORM} path is appropriately linked to the \texttt{Mathematica} .wl script with
\begin{mycode}
$HistoryLength = 1;
OrigPATH=Environment["PATH"];
SetEnvironment["PATH"->OrigPATH<>":/FORMexecPath"];
$ApartTemporaryDirectory=(path<>"temp");
\end{mycode}
The first line above assures that \texttt{Mathematica} does not store more then 1 past evaluated expression in its internal memory.
In general, the default value \texttt{Infinity} may quickly lead to out-of-memory issues.
Since we have not obtained the Groebner basis for all topologies, we performed the partial fractioning only on a subset of the corresponding IBPs.
For example, after partial fractioning, the 1.2~MB IBP table in the PL2A topology becomes 1.4~MB in size.
Even though the IBP identities do not shrink, they are now in a form which allows for easy application at the amplitude level.
Moreover, there is only one spurious denominator factor $(10 - 3d - 10t + 3dt - 12u + 4du)$ remaining after the multivariate partial fractioning.
Unfortunately, it still appears after substituting the IBPs into the amplitude piece for the PL2A topology $\mathcal{A}_{\text{PL2A}}$.
Since it does not correspond to any physical singularity, it should eventually cancel when combining with contributions from all the topologies.
In order to see it explicitly, one would have to either find all relations between MIs in different topologies, or expand all the MIs in terms of special functions up to fixed order in $\ep$.
We now discuss the efficiency of these potential approaches.

After elaborating on the IBP reduction intricacies of the multiscale multiloop process at hand, let us summarise its efficiency.
Firstly, the sheer size of the resulting IBP relations casts doubt on their applicability.
Indeed, the 640~MB IBP table for the NPL2W28 topology seems to increase the complexity of the original 5~MB integrand expression.
Secondly, any potential simplifications in the MI coefficients require finding an appropriate Groebner basis.
This highly nonlinear algebraic problem is very difficult to solve for multiscale topologies.
Thirdly, even if it is possible to perform the multivariate partial fractioning of the MI coefficients, some spurious denominator factors still persist at the amplitude level in fixed topology.
In order to resolve these issues, there are two alternative ways to proceed, analytic and numerical.

Analytically, one may choose a different MI basis such that the most complex denominators no longer appear in MIs coefficients.
This approach has been recently applied e.g. to the two-loop mixed QCD-EWK $pp \to H$+jet amplitudes in Refs~\cite{Bonetti:2020hqh,Bonetti:2022lrk}, and it is based on algorithms in Refs~\cite{Smirnov:2020quc,Usovitsch:2020jrk}.
Even though applying this method for our process might be possible, extending it to account for massive top loop contributions seems very difficult at the moment.
Indeed, in addition to the 4 variables $d$, $u/m_Z^2$, $t/m_Z^2$, and $m_W^2/m_Z^2$ considered here, a dependence on the top quark mass $m_t$ and the Higgs boson mass $m_H$ would be introduced.
Similar difficulty may arise when applying the direct integration method of Refs~\cite{Bonetti:2020hqh,Bonetti:2022lrk} using \texttt{HyperInt}~\cite{Panzer:2014caa} to all the MIs in the problem, as long as they are linearly reducible.
This criterion may no longer be valid for the integrals involving multiple internal masses.
Moreover, the iterated integrals appearing in the result may not be easy to evaluate numerically, which is required for phenomenological applications.
Indeed, even without top loops, the final $\ep$ expanded result of Refs~\cite{Bonetti:2020hqh,Bonetti:2022lrk} is expressed in terms of GPLs with letters involving square roots of multivariate polynomials.
As a consequence, we propose to employ a numerical approach, as described in Sec.~\ref{sec:qqgZ.comp.num}, which is more easily generalisable to include the effects of massive quark loops.

\subsection{Numerical evaluation}
\label{sec:qqgZ.comp.num}

\begin{figure}[h]
	\includegraphics[width=0.9\textwidth]{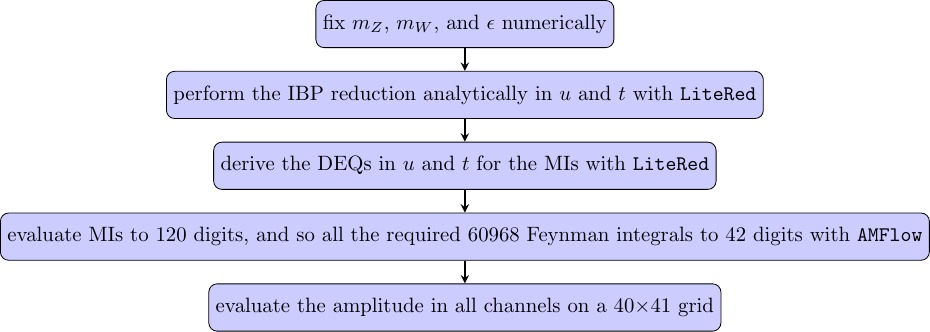}
	\caption{Computational flow for the numerical evaluation.}
	\label{fig:uugZ.flowNum}
\end{figure}

\noindent
After elaborating on the motivation for our numerical approach, we proceed to its description.
We follow the procedure outlined in Fig.~\ref{fig:uugZ.flowNum}.
We start by fixing the vector boson masses $m_Z$ and $m_W$ to numerical values.
For precision motivated reasons, we choose the ratio of squares of the two masses to be a simple rational number, while the reference mass $m_Z$ can be chosen in agreement with its Particle Data Group (PDG) value~\cite{ParticleDataGroup:2022pth}, i.e.
\begin{equation}
	\begin{split}
		m_Z   &= m_{Z,\text{PDG}} = 91.1876 \GeV \,, \\
		m_W^2 &= \frac{7}{9} m_Z^2 = (80.4199)^2 \GeV^2 \,.
	\end{split}
\label{eq:ppjZ.mZmW}
\end{equation}
Our approximation for the $m_W$ mass value agrees with the PDG value $m_{W,\text{PDG}} = 80.377\GeV$ to 3 digits, which accurately reproduces the leading order total cross section.

Moreover, we can also fix the value of $\ep$ to a small positive number.
Indeed, since we seek a result expanded in powers of $\ep$, it is possible to reconstruct its coefficients from several numerical samples~\cite{Liu:2022chg}.
Conservatively, we choose 10 values
\begin{align}
	\epsilon_n = 10^{-7}+n\times 10^{-9} \,, \quad n = 1,\ldots,10 \,,
\end{align}
in order to reconstruct the series coefficients from $\epsilon^{-4}$ to $\epsilon^0$, while accounting for any potential $\ep$ divergent prefactors of the integrals.

With $m_Z$, $m_W$, and $\epsilon$ fixed numerically, we have vastly reduced the parametric complexity of the integral structure.
For this reason, we are able to perform the IBP reduction analytically in $u$ and $t$ in a much more efficient manner than in Sec.~\ref{sec:qqgZ.comp.IBP}.
It is convenient to keep the exact dependence in $u$ and $t$ at this stage in order not to repeat the reduction when sampling over the phase space of the process.
We perform the IBP reduction separately for each of the 57 top sectors per crossing.
We generate the identities with \texttt{LiteRed}~\cite{Lee:2012cn, Lee:2013mka} and solve them using finite field arithmetic~\cite{vonManteuffel:2014ixa, Peraro:2016wsq} implemented in \texttt{FiniteFlow}~\cite{Peraro:2019svx}.
The size of the typical linear system of IBP relations to be solved is $\mathcal{O}(10^4)$.
The powers of Mandelstam variables $u$ and $t$ appearing in resulting rational function coefficients of MIs reach 28.
Since these powers are so large, there will be a vast numerical hierarchy between different MIs and their coefficients at phase space points of unequal order.
In order to control the potential resulting large cancellations at the amplitude level, high-precision MI evaluation is required.

Next, we construct two power-expansion-based DEQs~\cite{Kotikov:1990kg, Czakon:2008zk} in $u$ and $t$ for MIs in each top sector.
They allow us to evolve the solution from some boundary condition to an arbitrary phase space point in $u$ and $t$.
We perform this evolution within the in-built DEQ solver of \texttt{AMFlow}, introduced in Sec.~\ref{sec:amp.num.AMFlow}.
For massive integral topologies, the auxiliary mass $\eta$ is naturally defined as one of the mass scales.
For massless topologies, it is enough to promote 1 propagator to have an $\eta$ dependence, which results in introducing at most 4 more MIs per topology, in comparison to the $\eta$-independent basis in Sec.~\ref{sec:qqgZ.comp.IBP}.
We compute the BCs with 120 digit precision at a rational point
\begin{align}
	\frac{t}{m_Z^2}=-\frac{31}{11} \,, \quad \frac{u}{m_Z^2} = -\frac{13}{7} \,,
\end{align}
using the \texttt{AMFlow}~\cite{Liu:2022chg} package of \texttt{Mathematica}.
It implements the auxiliary mass flow method~\cite{Liu:2017jxz,Bronnum-Hansen:2020mzk,Liu:2021wks}, introduced in Sec.~\ref{sec:amp.num.AMFlow}.
In order to achieve such a high precision, it is enough to expand by regions the BC at $\eta=-i\infty$ to leading order, and analytically continue it to $\eta=-i0^-$ along an $\eta_i$ path of length not exceeding 20.

The resulting MIs inherit the 120 digit precision on the whole numerical grid.
Together with the IBP table, we can evaluate all the required 60968 Feynman integrals to 10 orders in $\ep$ with a precision of around $70-7n$ digits at the $n^{\text{th}}$ sub-leading $\ep$ pole.
It means, that there are $\sim$~50 digit cancellations when combining the MIs with their IBP coefficients.
Therefore, by evaluating MIs at very high precision, we control the large cancellation effects appearing in the result for integral evaluation.
With the analytic IBP table, all the integrals can be evaluated in $\sim$~1 CPUh per a phase space point.
At a fixed kinematic point, the result has a size of about 60~MB.
Therefore, it is useful not to store evaluated lists of integrals at different phase space points, but rather the whole numerically evaluated amplitude.
As an example of a high precision evaluation, we present here the result for the top sector integral in the most complicated topology in the process, i.e. the last diagram in Fig.~\ref{fig:uugZ.topsec}.
At mass values as in Eq.~\ref{eq:ppjZ.mZmW} and at an example kinematic point
\begin{equation}
	\frac{u}{m_Z^2} = -\frac{4212389009}{875622495} \,, \qquad
	\frac{t}{m_Z^2} = -\frac{185568373013477}{1751244990} \,,
\end{equation}
the result reads

\scalebox{0.55}{\parbox{0.9\linewidth}{
\begin{equation}
\begin{split}
	&\mathcal{I}_{\text{NPL2W28x12, 254}} = \\
	&\frac{3.0906596244443939233735868243720876301611376323059 \cdot 10^{-33}}{\epsilon^4} \\ &-\frac{8.6982149550765304704269203265996575847651084073412 \cdot 10^{-32} + i \, 1.9544714863999450257859339734280327013888767054745 \cdot 10^{-32}}{\epsilon^3} \\
	&+\frac{1.2687966449353006102942224933760780663861340457210 \cdot 10^{-30} + i \, 5.6436126481241928121313821894114179219856434323968 \cdot 10^{-31}}{\epsilon^2} \\
	&-\frac{1.2789521457906368621593037090148843633179169961641 \cdot 10^{-29} + i \, 8.819609492466885812112902878890860383979377590370 \cdot 10^{-30}}{\epsilon} \\
	&+9.983567007812284762268669560040490673920606641026 \cdot 10^{-29} + i \, 9.758036266276083395230075358873187122450506262448 \cdot 10^{-29}
\end{split}
\end{equation}
}}

\noindent
with $\sim$~50 digit precision.

Finally, having chosen a numerical approach, we need to numerically parametrize the whole physical kinematic phase space in an extensive way.
To this end, we construct a two-dimensional grid in Mandelstam invariants $u$ and $t$ such that the amplitude is appropriate for promoting to virtual hadronic differential distributions.
We elaborate on the validity of our procedure further in Sec.~\ref{sec:ppjZ.res} when discussing checks.
The physical variables to consider are the transverse momentum of the $Z$ boson
\begin{align}
	p_{T,Z}^2 = (p_Z^1)^2 + (p_Z^2)^2 \,,
\label{eq:ppjZ.pTZ}
\end{align}
and the rapidity of the $Z$ boson
\begin{align}
	y_Z = \frac{1}{2}\log\left(\frac{p_Z^0+p_Z^3}{p_Z^0-p_Z^3}\right) \,.
\label{eq:ppjZ.yZ}
\end{align}
They are related to our Mandelstam parametrization through
\begin{equation}
	\begin{split}
		p_{T,Z}^2 &= \frac{ut}{s} \,, \\
		y_Z &= \frac{1}{2} \log\left(\frac{s+u}{s+t}\right) \,.
	\end{split}
\label{eq:ppjZ.pTyTOsij}
\end{equation}
Note that the corresponding kinematic variables for the gluon are $p^2_{T,g}=p^2_{T,Z}$ and $y_g=\frac{1}{2}\log\left(\frac{u}{t}\right)$.
These variables correspond to the partonic center of mass frame. See Sec.~\ref{sec:ppjZ.res} for the discussion on the transformation to the hadronic lab frame.
Since there are no thresholds in the relevant high $p_{T,Z}$ region, $p_{T,Z} \in (200,2000)$~GeV, we simply choose our grid to be logarithmically uniform in $p_{T,Z}$ and linearly uniform in $y_Z$, with dimensions 40$\times$41, i.e.
\begin{equation}
	\begin{split}
		p_{T,Z,n} &= 200 \cdot 10^{n/39} \,, \qquad n \in [0,39] \cap \mathbb{Z} \,, \\
		y_{Z,m} &= \frac{m}{4}-5 \,, \qquad\qquad m \in [0,40] \cap \mathbb{Z} \,.
	\end{split}
\label{eq:ppjZ.grid}
\end{equation}
Note that both the grid parametrization in Eq.~\ref{eq:ppjZ.grid} as well as the change of variables in Eq.~\ref{eq:ppjZ.pTyTOsij} are not rational transformations.
In order to control the precision of the AMFlow method, we evaluate all the required Feynman integrals at the resulting numerical $u$ and $t$ grid rationalized within 8 digit agreement.
We will discuss the effect of this rationalization on phenomenological results in Sec.~\ref{sec:ppjZ.res}.

The above computation flow applies to the $u\bar{u} \to gZ$ channel, as well as all its crossings, and down-type counterparts, in order to complete the study of the $pp \to Z$+jet process, introduced in Sec.~\ref{sec:ppjZ.motivation}.
Contrarily to the case of a fully analytic result, in general, a numerical evaluation on a grid cannot be kinematically crossed.
Therefore we provide 6 separate results for all independent partonic channels relevant for this process, i.e.
$u \, \bar{u} \to g \, Z$,
$u \, g \to u \, Z$,
$g \, u \to u \, Z$,
$d \, \bar{d} \to g \, Z$,
$d \, g \to d \, Z$,
$g \, d \to d \, Z$.
All the 6 numerical grids are defined in the same way as in Eq.~\ref{eq:ppjZ.grid}.
We also note that it is much more convenient to evaluate the original analytic integrand for the uncrossed $u\bar{u} \to gZ$ scattering on a crossed numerical grid, rather then crossing the integrand and evaluating it on the original grid in Eq.~\ref{eq:ppjZ.grid}.
We extract the down-type channels from the up-type ones via a combination of a simple up-down exchange in all the associated couplings, together with an effective sign change of the $W^+W^-Z$ coupling, since the corresponding vertex is antisymmetric under the electric charge sign change of the attached fermion line.

\section{UV renormalization and IR regularization}
\label{sec:ppjZ.uvir}

The bare two-loop mixed QCD-EWK amplitude $\mathcal{A}^{(1,1)}_{b}$ for the $u\bar{u} \to gZ$ process has pole divergences in $\ep$ starting at $\ep^{-4}$, as argued in Sec.~\ref{sec:amp.int.div}.
The IR divergences stem exclusively from the mixed QCD-QED sector, since massive EWK bosons do not contribute to the IR real radiation.
Contrarily, the UV divergences arise from contributions of the whole QCD-EWK sector, including the Higgs sector, which does not enter the considered amplitude at the bare level.
In order to extract the physical information encaptured at $\ep^0$ in the finite part of the amplitude, we need to subtract these divergences with the UV renormalization and IR regularization procedure.

We start with the QCD renormalization of the bare strong coupling $\alpha_{s,b}$.
Since we consider only massless QCD corrections, we choose the $\overline{ \rm MS}$ renormalization scheme, as defined in Sec.~\ref{sec:amp.div.UV}.
Note that the mixed two-loop QCD-EWK correction has a UV QCD structure of a one-loop correction.
Since we do not consider here the contributions from closed fermion loops, the leading coefficient of the QCD beta function reads
\begin{equation}
	\beta_0 = \frac{11}{6}C_A \,. 
\end{equation}

The EWK renormalization is much more involved because of the contributions from all the EWK particles.
We follow here the detailed discussion in Refs~\cite{Denner:1991kt,Denner:2019vbn}.
Due to multiple relations between EWK parameters, any renormalization scheme requires choosing a set of independent parameters.
For our process, we choose $\{G_{\mu},m_Z,m_W\}$ to be independent, where $G_\mu = 1.16639 \times 10^{-5}$~\cite{ParticleDataGroup:2022pth}. We also work in the $G_\mu$ \textit{on-shell scheme}.
In comparison to other renormalization schemes, it absorbs some universal corrections into the LO contribution~\cite{Denner:2019vbn}.
Analogously to Eq.~\ref{eq:ampRenDef} in Sec.~\ref{sec:amp.div.UV}, we write
\begin{equation}
	\mathcal{A}_b(g_{s,0},g_{u,L/R,0}) \sqrt{Z_u Z_{\bar{u}} Z_g Z_Z} = 
	\mathcal{A}(g_{s},g_{u,L/R})\, Z_{\mathcal{A}} \,,
\end{equation}
where the combined renormalization factor expands to
\begin{equation}
\begin{split}
	Z_{\mathcal{A}}
	&= 1 + \frac{1}{2}\left(\delta_{Z_g}+\delta_{Z_u}+\delta_{Z_{\bar{u}}}+\delta_{Z_{ZZ}}-\frac{Q_{up}}{g_{u,L/R}}\delta_{Z_{\gamma Z}}\right)
	+\delta_{g_{u,L/R}}+\delta_{g_s} \\
	&= 1 + \delta_{\mathcal{A}}
	= 1 + \frac{\alpha}{2\pi} \delta_{\mathcal{A}}^{(0,1)} + \frac{\alpha_s}{2\pi} \frac{\alpha}{2\pi} \delta_{\mathcal{A}}^{(1,1)}
	\,.
\end{split}
\label{eq:EWKren}
\end{equation}
Notice the $\gamma Z$ kinetic mixing term $\delta_{Z_{\gamma Z}}$ in Eq.~\ref{eq:EWKren}, which for an off-shell photon can be accounted for by a simple refactoring of the overall coupling.
As a result of this mixed QCD-EWK renormalization procedure, we arrive at an expression for the amplitude $\mathcal{A}$ in a renormalized perturbative series
\begin{align}
	\mathcal{A} = \sqrt{\frac{\alpha_s}{2\pi}} \sqrt{\frac{\alpha}{2\pi}} T^{a_3}_{i_1 i_2} \left(
	A^{(0,0)} + \frac{\alpha_s}{2\pi} \, A^{(1,0)} + \frac{\alpha}{2\pi} \, A^{(0,1)} + \frac{\alpha_s}{2\pi} \frac{\alpha}{2\pi} \, A^{(1,1)} + \mathcal O\left(\alpha_s^2,\alpha^2\right)
	\right) \,,
\end{align}
where the renormalized fixed order amplitudes are related to their bare counterparts via
\begin{equation}
\begin{split}
	A^{(0,0)} &= S_\ep^{-1/2} A^{(0,0)}_b \,, \\
	A^{(1,0)} &= S_\ep^{-1/2} \left( S_\ep^{-1} \mu^{2\ep} A^{(1,0)}_b
	- \frac{\beta_0}{2\ep} A^{(0,0)}_b \right) \,, \\
	A^{(0,1)} &= S_\ep^{-1/2} \left( \mu^{2\ep} A^{(0,1)}_b
	+ \delta_{\mathcal{A}}^{(0,1)} A^{(0,0)}_b \right) \,, \\
	A^{(1,1)} &= S_\ep^{-1/2} \left( S_\ep^{-1} \mu^{4\ep} A^{(1,1)}_b
	- \frac{\beta_0}{2\ep} \mu^{2\ep} A^{(0,1)}_b
	+ \delta_{\mathcal{A}}^{(0,1)} S_\ep^{-1} \mu^{2\ep} A^{(1,0)}_b \right) \\
	&+ S_\ep^{-1/2} \left( S_\ep^{-1} \delta_{\mathcal{A}}^{(1,1)}
	- \frac{\beta_0}{2\ep} \delta_{\mathcal{A}}^{(0,1)} \right) A^{(0,0)}_b \,,
\end{split}
\end{equation}
at the renormalization scale $\mu = \mu_0 = \mu_R$.
Note the absence of the $\overline{ \rm MS}$ $S_\ep$ factor in the $G_{\mu}$ on-shell EWK renormalization scheme.

\begin{figure}[h]
	\centering
	\includegraphics[width=0.9\textwidth]{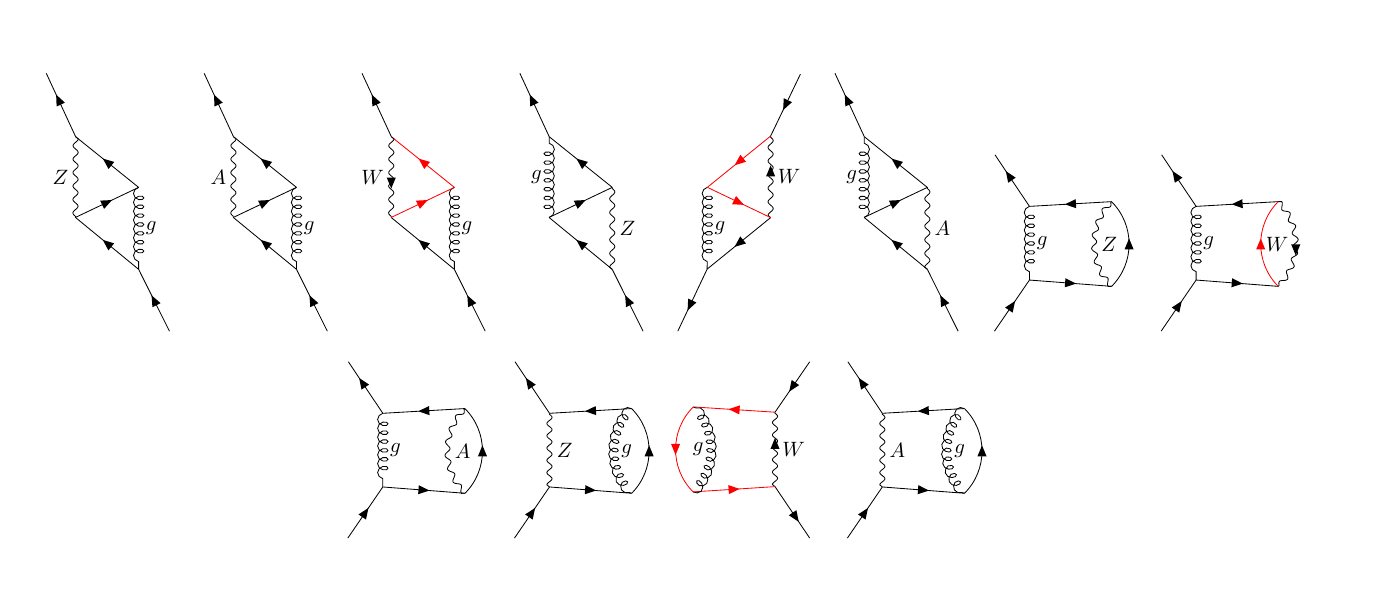}
	\caption{Feynman integrals for the two-loop mixed QCD-EWK UV renormalization of the $u \bar{u}$ two-point function. Red lines denote down-type quarks.}
	\label{fig:uugZ.ren}	
\end{figure}

The renormalization factors $\delta_{Z}$ in Eq.~\ref{eq:EWKren} are related to the two-point functions $\Sigma$ of appropriate propagators, as reviewed in Ref.~\cite{Denner:1991kt}.
For example, the wavefunction renormalization factors read
\begin{equation}
\begin{split}
	\delta_{Z_u} &= \delta_{Z_{\bar{u}}} = -\Re \, \Sigma_{u\bar{u}}(0) \,, \\
	\delta_{Z_{ZZ}} &= -\Re \, \frac{\partial}{\partial p^2} \Sigma_{ZZ,T} |_{p^2=m_Z^2} \,, \\
	\delta_{Z_{\gamma Z}} &= -\Re \, \frac{2}{m_Z^2} \Sigma_{\gamma Z,T}(m_Z^2) \,, 
\end{split}
\end{equation}
while $\delta_{Z_g}=0$ since we do not take into account the contributions from closed fermion loops.
We define the transverse $\Sigma_{T}$ and longitudinal $\Sigma_{L}$ part of a two-point function $\Sigma$ through
\begin{equation}
	\Sigma_{ZZ}^{\mu\nu}(p) = -ig^{\mu\nu}(p^2-m_Z^2)
	-i\left( g^{\mu\nu} - \frac{p^\mu p^\nu}{p^2} \right) \Sigma_{ZZ,T}(p^2)
	-i\frac{p^\mu p^\nu}{p^2} \Sigma_{ZZ,L}(p^2)
\end{equation}
The two-point functions $\Sigma$ can be computed perturbatively, for example see the Feynman diagrams for $\Sigma_{ZZ}$ at mixed QCD-EWK order in Fig.~\ref{fig:uugZ.ren}.
Since the two-loop mixed QCD-EWK correction has a UV EWK structure of a one-loop correction, the corresponding divergences in $\ep$ start at $\ep^{-1}$ order.
For example, the divergent EWK renormalization factors for the whole right-handed amplitude read
\begin{equation}
\begin{split}
	\delta_{\mathcal{A},R}^{(0,1)} &= -\frac{Q_{uo}^2 s_w^2}{8 \pi c_w^2} \frac{1}{\ep} + \mathcal{O}(\ep^0) \,, \\
	\delta_{\mathcal{A},R}^{(1,1)} &= C_F \frac{3 Q_{up}^2 s_w^2}{128 \pi^2 c_w^2} \frac{1}{\ep} + \mathcal{O}(\ep^0) \,,
\end{split}
\end{equation}
in agreement with the one-loop result in Ref.~\cite{Kuhn:2005az}.
Note that, contrarily to the $\overline{ \rm MS}$ QCD renormalization scheme, in the EWK $G_{\mu}$ on-shell scheme, there is a nontrivial contribution to the finite part at $\ep^0$ in $\delta_{\mathcal{A}}$.
We take the value of the Higgs mass, which contributes to this finite part, to be
\begin{equation}
	m_H^2 = \frac{47}{25} m_Z^2 = (125.03)^2 \GeV^2 \,,
\label{eq:ppjZ.mH}
\end{equation}
which agrees with the PDG value $m_{H,\text{PDG}} = 125.25\GeV$ to 3 digits.

Our final result is the IR and UV finite amplitude $A^{(i,j,{\rm fin})}$ defined via
\begin{equation}
	\begin{split}
		A^{(0,0)} &=
		A^{(0,0,{\rm fin})} \,, \\
		A^{(1,0)} &=
		\mathcal I_{1,0}\, A^{(0,0)} +
		A^{(1,0,{\rm fin})} \,, \\
		A^{(0,1)} &=
		\mathcal I_{0,1}\, A^{(0,0)} +
		A^{(0,1,{\rm fin})} \,, \\
		A^{(1,1)} &= \mathcal I_{1,1}\, A^{(0,0)} + 
		\mathcal I_{1,0}\, A^{(0,1,{\rm fin})} +
		\mathcal I_{0,1}\, A^{(1,0,{\rm fin})} +
		A^{(1,1,{\rm fin})} \,.
		\label{eq:catani}
	\end{split}
\end{equation}
In order to extract the finite part of renormalized perturbative corrections $A^{(i,j)}$, we regularize the remaining IR singularities with the Catani operator $\mathcal{I}_{i,j}$ introduced in Sec.~\ref{sec:amp.div.IR}.
Since the process has a simple colour structure, the corresponding Catani operator yields
\begin{equation}
	\begin{split}
		\frac{2\Gamma(1-\epsilon)}{e^{\gamma_E\epsilon}}
		\mathcal I_{1,0}(\epsilon) &= 
		\left(\frac{\mu^2}{-s-i\varepsilon}\right)^\epsilon
		(C_A-2C_F)\left(\frac{1}{\epsilon^2}+\frac{3}{2\epsilon}\right) \\
		&- \left(\left(\frac{\mu^2}{-u-i\varepsilon}\right)^\epsilon + \left(\frac{\mu^2}{-t-i\varepsilon}\right)^\epsilon\right)
		\left(C_A\left(\frac{1}{\epsilon^2}+\frac{3}{4\epsilon}\right) + \frac{\beta_0}{2\epsilon}\right)
		\,, \\
		-\frac{\Gamma(1-\epsilon)}{e^{\gamma_E\epsilon}}
		\mathcal I_{0,1}(\epsilon) &= 	\left(\frac{\mu^2}{-s-i\varepsilon}\right)^\epsilon
		S_\ep Q_{up}^2 \left(\frac{1}{\epsilon^2}+\frac{3}{2\epsilon}\right)
		\,, \\
		\mathcal I_{1,1}(\epsilon) - \mathcal I_{1,0}(\epsilon) \, \mathcal I_{0,1}(\epsilon) &=
		\frac{e^{\gamma_E\epsilon}}{\Gamma(1-\epsilon)}
		\left(\frac{\mu^2}{-s-i\varepsilon}\right)^{2\epsilon}
		\frac{1}{\ep}S_\ep
		C_F Q_{up}^2
		\left(\frac{\pi^2}{2} - 6 \zeta_3 - \frac{3}{8}\right) \,,
		\label{eq:poles}	
	\end{split}
\end{equation}
in analogy to Ref.~\cite{Buccioni:2022kgy}.
Note the appearance of the $S_\ep$ factor in the EWK IR regularization due to its lack in the on-shell EWK renormalization scheme.

\section{Checks and results}
\label{sec:ppjZ.res}

Before presenting our final results, we describe various checks that we have performed on our results.
We start with the check of the validity of our choice of the kinematic grid in Eq.~\ref{eq:ppjZ.grid}.
We argue below that it is appropriate for reliably reproducing differential hadronic cross section distributions.

Following Eq.~\ref{eq:intro.fact}, we seek the virtual hadronic cross section
\begin{equation}
\begin{split}
	\sigma^{(1,1,\text{fin})}(E_h^2) &= \sum_{i,j=-4}^4 \int_0^1 dx_1 \int_0^1 dx_2 \, f_i(x_1,\mu_F) f_j(x_2,\mu_F) \, \hat{\sigma}^{(1,1,\text{fin})}(s,\mu_R) \,, \\
	\hat{\sigma}^{(1,1,\text{fin})} &=	\frac{1}{2s} \int d\text{LIPS}
	\, 2\Re \, \left(\mathcal{A}^{(0,0,\text{fin})*}_{ij} \mathcal{A}^{(1,1,\text{fin})}_{ij}\right) \,,
\label{eq:ppjZ.hadr}
\end{split}
\end{equation}
arising from our new two-loop mixed QCD-EWK amplitude $\mathcal{A}^{(1,1,\text{fin})}_{ij}$, numerically evaluated on a grid in all $ij$ channels, as well as the associated differential distributions.
$f_i(x)$ denotes a PDF of flavour $i$ at momentum fraction $x$, and $s$ is related to the hadronic energy $E_h$ via $s=x_1x_2E_h^2$.
The lack of $\mu_F$ dependence in $\hat{\sigma}^{(1,1,\text{fin})}$ in comparison to Eqs~\ref{eq:intro.fact} and \ref{eq:intro.xsec} is reflecting the purely virtual nature of the correction and the absence of PDF counterterms appearing in Eq.~\ref{eq:IRrealPert}.
In order to probe the amplitude $\mathcal{A}$ at phase space points intermediate to the grid evaluation, we have to rely on interpolation.
In principle, one can design an interpolating function based on the grid rationalized to 8 digits on which we evaluated the amplitude.
Since our primary purpose here are checks, we can rely on a simpler, but possibly less precise method of interpolating the amplitude on the regular original grid in Eq.~\ref{eq:ppjZ.grid} e.g. with the function \texttt{Interpolation[]} of \texttt{Mathematica}.
We find this approach to be sufficient for our purpose.

In order to proceed with our numerical approach, we need to evaluate the PDFs at enough numerical points to reproduce the integral in Eq.~\ref{eq:ppjZ.hadr} from a set of samples.
To this end, similarly as in partonic kinematic variables, we create a grid in the momentum fraction e.g.
$x_{n} = 10^{-n/100} \,, n \in [1,500] \cap \mathbb{Z}$.
We evaluate PDFs in \texttt{Mathematica} using the package \texttt{ManeParse}~\cite{Clark:2016jgm}, which links to the usual PDF interface \texttt{LHAPDF}~\cite{Buckley:2014ana}.
Its in-built function \texttt{pdfFunction[PDFset, flavour, x, Q]} evaluates a fixed flavour PDF from a given PDF set at momentum fraction $x$ and energy $Q$.
Similarly, \texttt{pdfAlphaS[PDFset, Q]} evaluates the strong coupling constant $\alpha_s$ at energy $Q$ according to the PDF fit in a given set.
Following Ref.~\cite{Lindert:2017olm}, we evaluate the PDFs and $\alpha_s$ at dynamic scale appropriate for this process i.e. $\mu=\mu_{F/R}=H_T'/2$, where
\begin{equation}
	H_T' = \sqrt{p_{T,Z}^2+m_{Z}^2} + |p_{T,Z}| \,.
\end{equation}

\begin{figure}[h]
	\centering
	\includegraphics[width=0.45\textwidth]{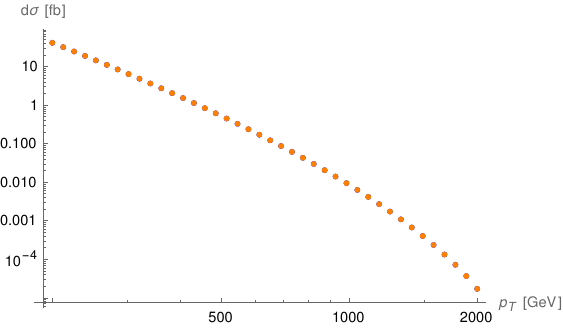}
	\includegraphics[width=0.45\textwidth]{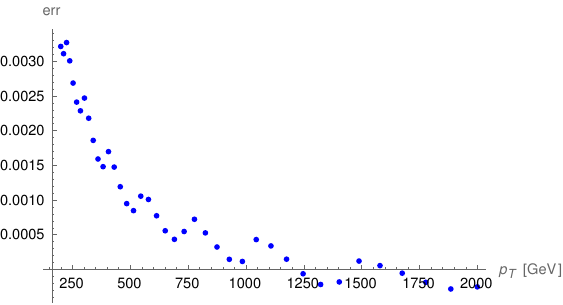}
	\includegraphics[width=0.45\textwidth]{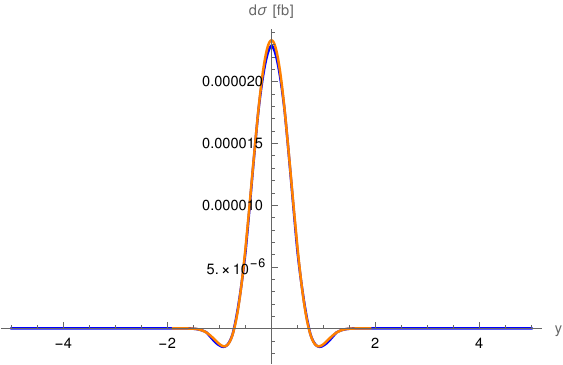}
	\includegraphics[width=0.45\textwidth]{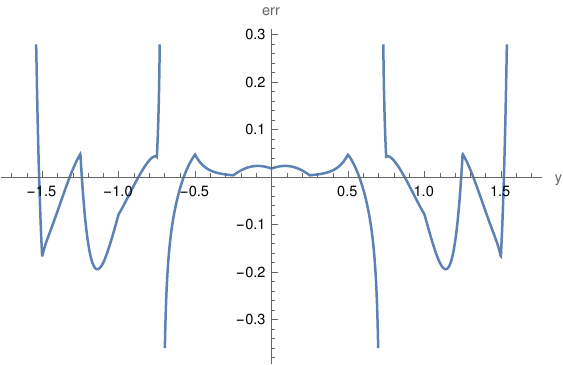}
	\caption{Comparison of our results and the ones generated with \texttt{MCFM} for the LO $u\bar{u} \to gZ$ differential hadronic distributions $\frac{d\sigma}{dp_{T,\text{j}}}$ and $\frac{d\sigma}{dp_{T,\text{j}} dy_{\text{j}}}|_{p_{T,\text{j}}=2000\GeV}$, together with the corresponding relative difference.}
	\label{fig:ppjZ.hadr}
\end{figure}

We are interested in the differential distributions $\frac{d\sigma}{dy_{\text{j}} dp_{T,\text{j}}}$, $\frac{d\sigma}{dp_{T,\text{j}}}$, and $\frac{d\sigma}{dy_{\text{j}}}$.
We perform the integration over a function interpolating over the grid with \texttt{Integrate[]}.
For better performance, it is useful to perform the integration over one variable at a time, i.e. first compute $\frac{d\sigma}{dy_{\text{j}} dp_{T,\text{j}}}$, and then integrate over either $y_{\text{j}}$ or $p_{T,\text{j}}$.
We compared our results against \texttt{MCFM 6.8}~\cite{Campbell:2015qma} and found e.g. at least a 3 digit agreement at LO in the $p_{T,\text{j}}$ distribution of the $u\bar{u} \to gZ$ cross section, see Fig.~\ref{fig:ppjZ.hadr}.
The 2 digit agreement at central rapidity in the corresponding $y_{\text{j}}$ distribution does not persist beyond the unphysical zero-crossing effect starting at $|y_{\text{j}}| \approx 0.75$, see Fig.~\ref{fig:ppjZ.hadr}.
This effect should vanish when combining all the contributions from relevant partonic channels.
It stems from negative values of PDFs, which no longer have a probabilistic interpretation beyond LO~\cite{Collins:2011zzd}.
The crossing of zero at $|y_{\text{j}}| \approx 0.75$ and $|y_{\text{j}}| \approx 1.5$ is reflected in infinite 0/0 error in Fig.~\ref{fig:ppjZ.hadr}.
In the intermediate negative region, the relative difference does not exceed 20\%.
The decrease in precision in comparison to the $p_{T,\text{j}}$ distribution stems from the change of variables comparing to the original grid parametrization.
Indeed, the transverse momentum of the $Z$ boson is the same as of the jet, which is not the case for rapidity.
When comparing the rapidity distribution against \texttt{MCFM}, it is important to notice that since only a difference of two rapidities is boost invariant, one needs to account for it by a further transformation $y \to y + \frac{1}{2}\log\left(\frac{x_1}{x_2}\right)$.
We perform our check at high-luminosity energy of the LHC collision $E_h = 13600$~GeV, and with PDF set \texttt{NNPDF31\_nnlo\_as\_0118\_luxqed\_0000}~\cite{NNPDF:2017mvq,Bertone:2017bme}.
To complete the analytis, these checks should be performed at virtual NLO QCD and NLO EWK order in all partonic channels for all distributions.

\begin{figure}[h]
	\centering
	\includegraphics[width=0.5\textwidth]{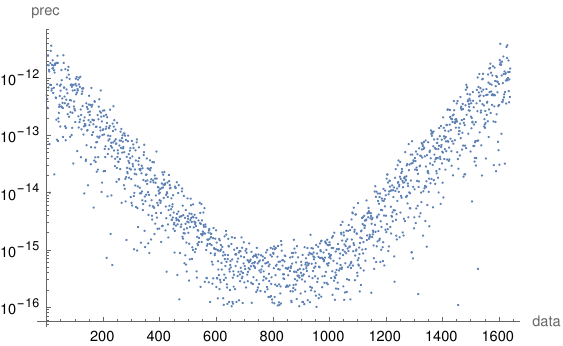}
	\caption{Relative absolute difference between our results and the ones generated with \texttt{OpenLoops} for the tree level $u\bar{u} \to gZ$ amplitude squared for all grid points. The grid points are collected along the $x$ axis in correspondence to the rapidity $y_Z \in [-5,5]$.}
	\label{fig:OpenLoops}
\end{figure}

As a byproduct of our computation of the mixed QCD-EWK amplitude $A^{(1,1,{\rm fin})}$, we have computed all relevant lower-loop results, i.e. tree level $A^{(0,0,{\rm fin})}$, one-loop QCD $A^{(1,0,{\rm fin})}$, and one-loop EWK $A^{(0,1,{\rm fin})}$.
In order to check our lower-loop results against available literature, we have benchmarked the amplitude squared against \texttt{OpenLoops 2}~\cite{Buccioni:2019sur}.
We found at least 12 digit agreement on the whole grid in all partonic channels.
We note here, that even when comparing a fully analytic tree level result against \texttt{OpenLoops}, there is a precision loss of the program at high absolute value of the rapidity, see example in Fig.~\ref{fig:OpenLoops}.
Nonetheless, it provides a conclusive enough check of our one-loop results.

\begin{figure}[h]
	\centering
	\includegraphics[width=0.5\textwidth]{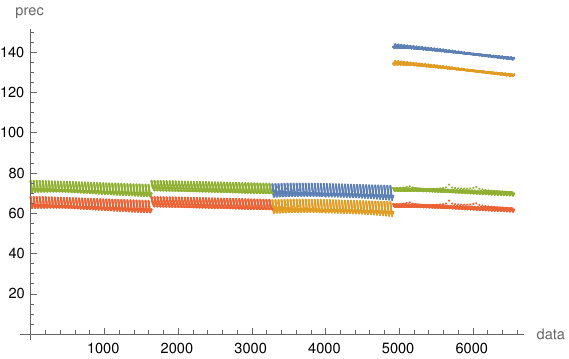}
	\caption{Number of digits to which the $\ep$ poles cancel in the real part of the finite part of the $u\bar{u} \to gZ$ amplitude squared on the whole grid for the 4 contributions, $\mathcal{A}^{(0,0,\text{fin})*} \mathcal{A}^{(1,0,\text{fin})}$, $\mathcal{A}^{(0,0,\text{fin})*} \mathcal{A}^{(0,1,\text{fin})}$, $\mathcal{A}^{(0,0,\text{fin})*} \mathcal{A}^{(1,1,\text{fin})}$, and $\mathcal{A}^{(1,0,\text{fin})*} \mathcal{A}^{(0,1,\text{fin})}$, respectively. The poles $\ep^{-4}$, $\ep^{-3}$, $\ep^{-2}$, and $\ep^{-1}$ correspond to the blue, orange, green, and red colour, respectively.}
	\label{fig:ppjZ.poles}
\end{figure}

At the mixed QCD-EWK order, we have performed some strong self-consistency checks.
One of them is to check the cancellation of poles in $\ep$ against universal UV and IR structures, as described in Sec.~\ref{sec:ppjZ.uvir}.
With our high precision numerical evaluation of $A^{(1,1,\text{fin})}$, we found $\ep$ pole cancellation to $\sim$~50 digits on the whole grid in all 6 partonic channels, see example in Fig.~\ref{fig:ppjZ.poles}.

As a check of our computational framework beyond the universal singular structure, we have computed the two-loop massless QCD correction $A^{(2,0,\text{fin})}$ and compared it against available literature~\cite{Gehrmann:2022vuk,Garland:2002ak}.
We found 16 digit agreement at the $\ep^0$ finite part level at one representative kinematic point in the $u\bar{u}\to gZ$ channel.

Finally, we present our results for the finite part of the mixed QCD-EWK amplitude $A^{(1,1,\text{fin})}$.
We start by showing a high precision evaluation
\\
\scalebox{0.9}{\parbox{1.0\linewidth}{
\begin{equation}
	\begin{split}
		A^{(1,1,\text{fin})}_L &=
		\oT_1 \, (-1.67762319126822917\times 10^{-4}-\left(8.9124020261485948\times 10^{-5}\right) i) \GeV^{-2} \\
		&+ \oT_2 \, (-6.14083995697083227\times 10^{-10}-\left(1.15078349742596179\times 10^{-10}\right) i) \GeV^{-4} \\
		&+ \oT_3 \, (4.9873142274505437+2.6245263772911926 i) \GeV^{-2} \\
		&+ \oT_4 \, (-4.3253096831118915-2.4559371163486791 i) \GeV^{-2} \\
		&+ \oT_5 \, (2.361437106860579067\times 10^{-7}+\left(5.30188993175405567\times 10^{-8}\right) i) \GeV^{-4} \\
		&+ \oT_6 \, (4.6415462269520851+2.5128183816883992 i) \GeV^{-2} \,, \\
		A^{(1,1,\text{fin})}_R &=
		\oT_1 \, (-1.9707761016083767\times 10^{-5}+\left(1.0965436511377151\times 10^{-5}\right) i) \GeV^{-2} \\
		&+ \oT_2 \, (1.65144181401856366\times 10^{-11}+\left(2.08626789202156372\times 10^{-11}\right) i) \GeV^{-4} \\
		&+ \oT_3 \, (6.1391040692819743\times 10^{-1}-\left(4.1040250855245174\times 10^{-1}\right) i) \GeV^{-2} \\
		&+ \oT_4 \, (-5.9029051802906398\times 10^{-1}+\left(4.0662570124472741\times 10^{-1}\right) i) \GeV^{-2} \\
		&+ \oT_5 \, (5.4350465607404134\times 10^{-9}-\left(6.527583146744599\times 10^{-10}\right) i) \GeV^{-4} \\
		&+ \oT_6 \, (5.8134142630821290\times 10^{-1}-\left(4.1777025512005486\times 10^{-1}\right) i) \GeV^{-2} \,,
	\end{split}
\label{eq:ppjZ.ampnum}
\end{equation}
}}
\\
\noindent
at one of the rationalized grid points defined in Eq.~\ref{eq:ppjZ.grid}
\\
\scalebox{0.9}{\parbox{1.0\linewidth}{
\begin{equation}
	\begin{split}
		s_{12} &= \,\,\,\,\, \frac{200343109174296505501}{188240625000} \GeV^2 \,, \qquad
		s_{23} = -\frac{2206428746331193800294949}{2073235183593750} \GeV^2 \,, \\
		s_{13} &= -\frac{663464134484282958883}{16585881468750000} \GeV^2 \,, \qquad \mu^2 = s_{12} \,,
	\end{split}
\end{equation}
}}
\\
\noindent
in the $u\bar{u}\to gZ$ channel.
Note that the uniform mass dimension of the amplitude is recovered when combining the units of numerical form factors in Eq.~\ref{eq:ppjZ.ampnum} and the tensors defined in Eqs~\ref{eq:qqgZ.Gammas}, \ref{eq:qqgZ.tensors}, and \ref{eq:uugZ.tens4}.
The final result of this work is to provide the finite part of form factors $\oF_{i}^{(1,1,\text{fin})}$, as well as their lower loop counterparts evaluated to higher $\ep$ on the whole grid in all partonic channels.
\begin{figure}[h]
	\centering
	\includegraphics[width=0.45\textwidth]{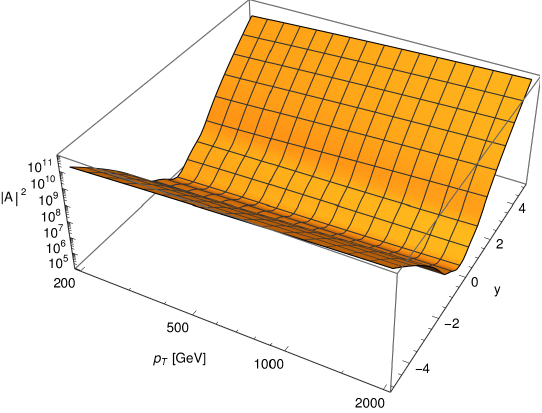}
	\includegraphics[width=0.45\textwidth]{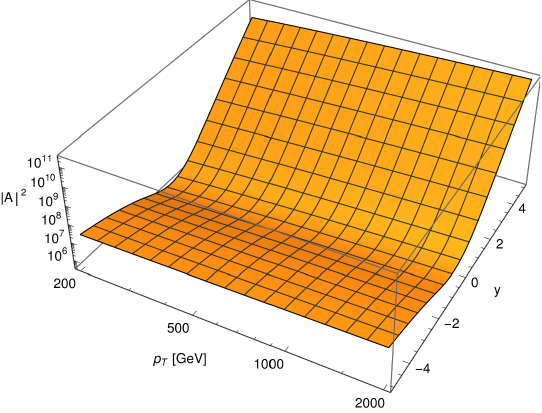}
	\includegraphics[width=0.45\textwidth]{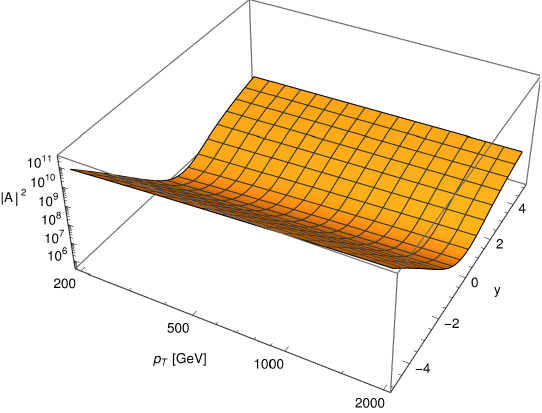}
	\includegraphics[width=0.45\textwidth]{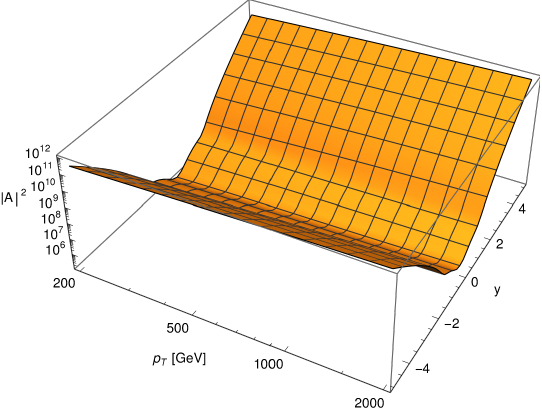}
	\includegraphics[width=0.45\textwidth]{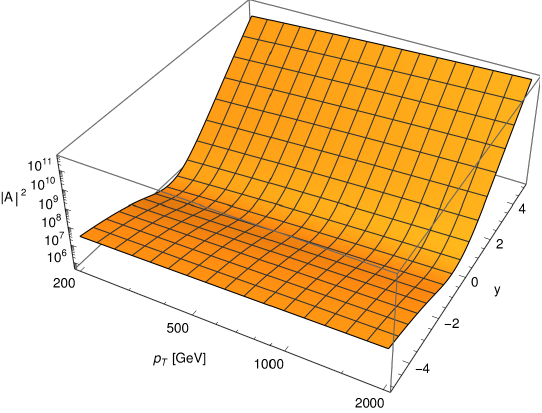}
	\includegraphics[width=0.45\textwidth]{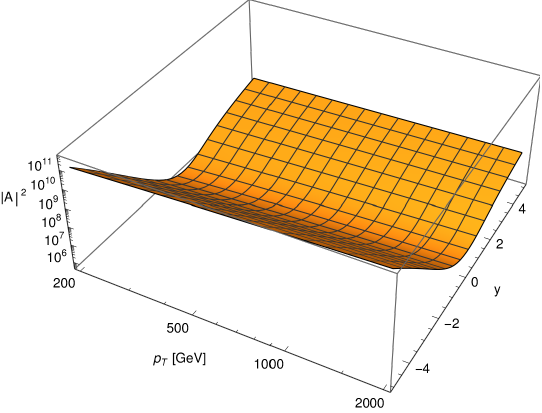}
	\caption{Absolute value of the virtual NNLO finite remainders
	$2\Re\left(\mathcal{A}^{(0,0,\text{fin})*}_L \mathcal{A}^{(1,1,\text{fin})}_L + \mathcal{A}^{(0,0,\text{fin})*}_R \mathcal{A}^{(1,1,\text{fin})}_R\right)$
	in all partonic channels
	$u \, \bar{u} \to g \, Z$,
	$u \, g \to u \, Z$,
	$g \, u \to u \, Z$,
	$d \, \bar{d} \to g \, Z$,
	$d \, g \to d \, Z$,
	and $g \, d \to d \, Z$,
	as functions of transverse momentum $p_T$ and rapidity $y$ of the $Z$ boson.}
\label{fig:ppjZ.num}
\end{figure}
In Fig.~\ref{fig:ppjZ.num}, we plot our results on a two-dimensional grid in the transverse momentum $p_{T,Z}$ and rapidity $y_Z$.
For the sake of brevity, we show the absolute value of the virtual NNLO finite remainders
$2\Re\left(\mathcal{A}^{(0,0,\text{fin})*}_L \mathcal{A}^{(1,1,\text{fin})}_L + \mathcal{A}^{(0,0,\text{fin})*}_R \mathcal{A}^{(1,1,\text{fin})}_R\right)$
in all partonic channels.

\chapter{\label{ch:concl}Conclusions}

In this thesis, we have considered scattering amplitudes relevant for high-precision LHC phenomenology.
In Ch.~\ref{ch:intro}, we have shown how they enter the hadronic cross section calculations.
In Ch.~\ref{ch:amp}, we analysed the general structure of amplitudes, and we reviewed state-of-the-art methods for computing them.
We discussed advantages and shortcomings of these methods, as well as we pointed out the bottlenecks in modern amplitude computations.
The main results of our work are the frontier applications to the analytic three-loop four-point massless QCD amplitudes in Ch.~\ref{ch:3L}, as well as to the numerical two-loop four-point one-mass mixed QCD-EWK corrections in Ch.~\ref{ch:2L}.

As a result of this work, we have computed the helicity amplitudes for the processes $gg\to\gamma\gamma$ and $pp\to\gamma$+jet at the state-of-the-art order of three-loop massless QCD.
Phenomenologically, these processes are among the standard candles of the SM at the LHC.
For our analytical three-loop calculation, we have adopted a new projector-based prescription to compute helicity amplitudes in the 't Hooft-Veltman scheme.
We also rederived all the MIs for this problem using the differential equations method, and we confirmed that it requires only one boundary condition integral.
The expressions at the intermediate stages of our amplitude calculation were quite sizable, thus we employed recent ideas for the demanding IBP reduction.
Our final results are remarkably compact.
They can be expressed in terms of HPLs of weight up to six.
This makes the numerical evaluation of our result both fast and numerically stable.
Our analytic results are provided in Refs~\cite{Bargiela:2021wuy,Bargiela:2022lxz}.
We checked the pole structure of our results, as well as we have correctly reproduced all the relevant lower-order results.

Theoretically, the last missing three-loop four-point massless QCD correction remaining to be computed is the $\gamma\gamma\to\gamma\gamma$ amplitude.
On the more formal side, it would be also interesting to understand why only one boundary condition integral it enough to derive all the MIs for this process.
Phenomenologically, we have already applied our result for the $gg\to\gamma\gamma$ amplitude to compute the NNLO soft-virtual corrections to signal-background interference for gluon-fusion Higgs production in the diphoton channel at the LHC.
More specifically, we focused on the peak shift in the diphoton invariant mass distribution.
We used our results to obtain an updated prediction for the bounds that can be put on the Higgs width.
In the future, it would be interesting to perform an exact NNLO study to improve the precision of this theoretical prediction even more.

Beyond massless QCD, we have computed the two-loop mixed QCD-EWK amplitudes for $pp \to Z$+jet process in light-quark-initiated channels, without closed fermion loops.
This process provides important insight into the high-precision studies of the SM, as well as into Dark Matter searches at the LHC.
We have employed a numerical approach based on the high-precision evaluation of Feynman integrals in the problem with the modern AMFlow method~\cite{Liu:2022chg}.
We obtained results evaluated on a two-dimensional grid in all relevant partonic channels.
We chose this kinematic grid such that it is appropriate for phenomenological applications.
We performed multiple checks of our computational framework, lower-order results, as well as of properties of the two-loop result.

A natural extension of our work is to include the effects of both massless and massive closed fermion loops.
In fact, our preliminary study has shown that our numerical framework can be easily generalised to include these effects.
Moreover, it would be important to combine our QCD-EWK virtual correction with real radiation effects to obtain hadron-level differential cross section distributions.
It still requires designing an appropriate IR subtraction scheme.
Phenomenologically, this calculation would provide a better understanding of $Z\to$invisible recoiling against a jet, which is a key background for Dark Matter searches at the LHC.
Theoretically, it would allow us to test the factorization at high energies due to the presence of Sudakov logarithms.

Finally, it seems that the current methods for amplitude computation are still not optimal in their efficiency.
Indeed, even relying on the state-of-the-art approaches led to sizable expressions at the intermediate level, which collapsed to compact results in their final form.
This calls for a deeper understanding of the amplitude structure, in order to unveil its hidden simplicity.
It would be interesting to investigate it in the future.

\startappendices

\chapter{\label{app:a} Summary of packages}

\begin{table}[h]
	\renewcommand{\arraystretch}{1.5}
	\centering
	%\begin{tabular}{ p{2.5cm}||p{0.3cm}|p{1.2cm}|p{1.2cm}|p{1.0cm} }
	\begin{tabular}{ l||l }
		purpose & packages \\
		\hline
		\hline
		Feynman diagrams generation & \texttt{qgraf}~\cite{Nogueira:1991ex} \\
		Feynman diagrams drawing & \texttt{qgraf-xml-drawer}~\cite{qraf:drawer}, \texttt{TikZ-Feynman}~\cite{Ellis:2016jkw}, \texttt{JaxoDraw}~\cite{Vermaseren:1994je,Binosi:2003yf} \\
		Dirac $\gamma$ algebra & \texttt{FORM}~\cite{Vermaseren:2000nd}, \texttt{Tracer}~\cite{Jamin:1991dp} \\
		spinor-helicity & \texttt{Spinors}~\cite{Maitre:2007jq} \\
		IBP reduction & \texttt{reduze}~\cite{Studerus:2009ye,vonManteuffel:2012np}, \texttt{kira}~\cite{Maierhofer:2017gsa,Maierhofer:2018gpa}, \texttt{FIRE}~\cite{Smirnov:2019qkx}, \texttt{LiteRed}~\cite{Lee:2013mka} \\
		Groebner basis & \texttt{Singular}~\cite{DGPS} \\
		multivariate partial fractioning & \texttt{MultivariateApart}~\cite{Heller:2021qkz} \\
		one-loop Feynman integrals & \texttt{QCDLoop}~\cite{Ellis:2007qk}, \texttt{CollierLink}~\cite{Denner:2014gla}, \texttt{Package-X}~\cite{Shtabovenko:2016whf} \\
		numerical Feynman integrals & \texttt{pySecDec}~\cite{Borowka:2017idc,Borowka:2018goh}, \texttt{AMFlow}~\cite{Liu:2022chg}  \\
		expanding Hypergeometrics & \texttt{HypExp}~\cite{Huber:2005yg} \\
		direct Feynman integration & \texttt{MB}~\cite{Czakon:2005rk}, \texttt{HyperInt}~\cite{Panzer:2014caa}, \texttt{FeynGKZ}~\cite{Ananthanarayan:2022ntm} \\
		Polylogarithmic algebra & \texttt{HPL}~\cite{Maitre:2005uu}, \texttt{PolyLogTools}~\cite{Duhr:2019tlz} \\
		numerical NLO amplitudes & \texttt{MadGraph}~\cite{Alwall:2014hca}, \texttt{OpenLoops}~\cite{Cascioli:2011va, Buccioni:2019sur} \\
		numerical NLO cross sections & \texttt{MCFM}~\cite{Campbell:2015qma} \\
		PDFs & \texttt{LHAPDF}~\cite{Buckley:2014ana}, \texttt{ManeParse}~\cite{Clark:2016jgm} \\
	\end{tabular}
\end{table}

\chapter{\label{app:b} Spinor-helicity identities}

We provide here a brief summary of useful spinor-helicity identities, following a review in Ref.~\cite{Dixon:1996wi} in the notation introduced in Sec.~\ref{sec:spinhel}
\begin{itemize}
\item antisymmetry
\begin{equation}
	\langle i j \rangle = - \langle j i \rangle \,, \qquad
	[ i j ]= - [ j i ] \,, \qquad
	\langle i j ] = [ i j \rangle = 0 \,,
\end{equation}
\item polarization sum
\begin{equation}
	|i\rangle [i| = \frac{1+\gamma_5}{2} \slashed{p}_i \,, \qquad |i] \langle i| = \frac{1-\gamma_5}{2} \slashed{p}_i \,,
\end{equation}
\item Gordon identity
\begin{equation}
	\langle i \gamma^\mu i ] = 2 p_i^\mu \,,
\end{equation}
\item Mandelstam invariant
\begin{equation}
	\langle i j \rangle [ j i ] = \langle i j i ] = \Tr\left(\frac{1-\gamma_5}{2}\slashed{p}_i\slashed{p}_j\right) = 2 p_i \cdot p_j = s_{ij} \,,
\end{equation}
\item trace over all internal $\gamma$ matrices, e.g.
\begin{equation}
	\begin{split}
		\langle i j l m i ] &= \langle i j \rangle [ j l ] \langle l m \rangle [ m i ] =  \Tr\left(\frac{1-\gamma_5}{2}\slashed{p}_i\slashed{p}_j\slashed{p}_l\slashed{p}_m\right) \\
		&= \frac{1}{2}\left( s_{ij}s_{lm} - s_{il}s_{jm} + s_{im}s_{jl} - 4i\varepsilon_{\mu\nu\rho\sigma} p_i^\mu p_j^\nu p_l^\rho p_m^\sigma \right) \,,
	\end{split}	
\end{equation}
where $\varepsilon_{\mu\nu\rho\sigma}$ is the four-dimensional Levi-Civita tensor,
\item charge conjugation
\begin{equation}
	\langle i \gamma^\mu j ] = [ j \gamma^\mu i \rangle \,,
\end{equation}
\item complex conjugation
\begin{equation}
	\left([ij]\right)^*=\langle ij \rangle \,,
\end{equation}
\item Fierz rearrangement
\begin{equation}
	[ i \gamma^\mu j \rangle [ k \gamma_\mu l \rangle = 2 [ ik ] \langle lj \rangle
\end{equation}
relying on a basis
\begin{equation} \{\mathbb{I},\gamma^\mu,[\gamma^\mu,\gamma^\nu],\gamma^5,\gamma^5\gamma^\mu \}
\end{equation}
for expressions involving $\gamma$ matrices in $d=4$ spacetime dimensions,
\item Schouten identity
\begin{equation}
	|j \rangle \langle kl \rangle + |k \rangle \langle lj \rangle + |l \rangle \langle jk \rangle = 0 \,,
\end{equation}
\item and some properties of polarization vectors
\begin{equation}
	\begin{split}
		\epsilon_\pm(p,q) \cdot p &= 0 \,, \\
		\epsilon_\pm(p,q) \cdot q &= 0 \,, \\
		\epsilon_\pm(p_i,q) \cdot \epsilon_\pm(p_j,q) &= 0 \,, \\
		\epsilon_\pm(p_i,p_j) \cdot \epsilon_\mp(p_j,q) &= 0 \,, \\
		\epsilon_\pm(p_i,p_j) | j^\pm \rangle &= 0 \,, \\
		\langle j^\mp | \epsilon_\pm(p_i,p_j) &= 0 \,.
	\end{split}	
\end{equation}
\end{itemize}

{\doublespacing
\renewcommand*\MakeUppercase[1]{#1}%
\printbibliography[heading=bibintoc,title={\bibtitle}]}

\end{document}